\documentclass[11pt,letterpaper]{article}
\usepackage{iftex}
\ifPDFTeX
  \usepackage[utf8]{inputenc}
\fi
\usepackage[T1]{fontenc}
\usepackage[margin=1in]{geometry}
\usepackage{setspace,amsmath,amssymb,amsthm,mathtools,array}
\usepackage{lmodern,graphicx}
\usepackage{caption}
\ifXeTeX
  \usepackage[protrusion=true]{microtype}
\else
  \usepackage[tracking=true,protrusion=true]{microtype}
\fi
\usepackage{xcolor}
\usepackage[normalem]{ulem}
\definecolor{Navy}{RGB}{30,70,235}
\definecolor{Green}{RGB}{0,128,0}
\definecolor{CompanionAddition}{RGB}{155,30,45}
\newif\ifshowcompanionadditions
\showcompanionadditionstrue

\definecolor{EnergyChange}{RGB}{0,92,170}
\newif\ifshowenergychanges
\showenergychangesfalse
\DeclareRobustCommand{\energychange}[1]{\ifshowenergychanges\textcolor{EnergyChange}{#1}\else#1\fi}
\usepackage{tikz}
\usetikzlibrary{arrows.meta}
\usepackage[breakable]{tcolorbox}
\usepackage{float}
\usepackage{todonotes}
\usepackage[boxed, commentsnumbered, linesnumbered, ruled, vlined]{algorithm2e}
\usepackage{titlesec}
\usepackage[plainpages=false,colorlinks=true,linkcolor=Navy,
  citecolor=Green,urlcolor=Green]{hyperref}
\usepackage{etoolbox,thmtools}
\makeatletter
\ifXeTeX\else
\ifdefstring{\MT@version}{2025/02/11 v3.2a}{\def\ManuscriptPatchShowHyphens#1#2#3#4#5{\MT@exp@cs\patchcmd{showhyphens }
      {\color@begingroup}{\color@begingroup\pdfadjustspacing\z@}
      {}{\PackageError{manuscript}{Unable to patch showhyphens}{Update microtype to version 3.2b or later.}}}
  \patchcmd{\MT@setup@expansion}
    {\MT@ifdefined@n@TF{showhyphens }}
    {\ManuscriptPatchShowHyphens}
    {}{\PackageError{manuscript}{Unable to patch microtype expansion setup}{Update microtype to version 3.2b or later.}}}{}\fi
\makeatother
\makeatletter
\@ifundefined{newcounteralias}{}{\renewcommand\thmt@autorefsetup{\@xa\def\csname\thmt@envname autorefname\@xa\endcsname
      \@xa{\thmt@thmname}}}
\makeatother
\declaretheorem[name=Theorem,numberwithin=section]{theorem}
\declaretheorem[name=Lemma,sibling=theorem,
  refname={Lemma,Lemmas},Refname={Lemma,Lemmas}]{lemma}
\declaretheorem[name=Observation,sibling=theorem,
  refname={Observation,Observations},Refname={Observation,Observations}]{observation}

\makeatletter
\newcommand*{\FixTheoremHyperAnchor}[1]{\@namedef{theH#1}{\theHsection.\arabic{theorem}}}
\forcsvlist{\FixTheoremHyperAnchor}{theorem,lemma,observation,proposition,corollary,claim}
\makeatother
\makeatletter
\patchcmd{\algocf@caption@algo}
  {\hyper@refstepcounter{algocf}}
  {\hyper@makecurrent{algocf}}
  {}{\PackageError{manuscript}{Unable to defer algorithm anchors}{}}
\pretocmd{\algocf@captiontext}
  {\leavevmode\Hy@raisedlink{\hyper@anchorstart{\@currentHref}\hyper@anchorend}}
  {}{\PackageError{manuscript}{Unable to anchor algorithm captions}{}}
\makeatother
\makeatletter
\renewenvironment{proof}[1][\proofname]{\par
  \pushQED{\qed}\normalfont \topsep6\p@\@plus6\p@\relax
  \trivlist
  \item[\hskip\labelsep
              \phantomsection
              \itshape
              #1\@addpunct{.}]\ignorespaces
}{\popQED\endtrivlist\@endpefalse
}
\makeatother

\renewcommand{\epsilon}{\varepsilon}
\newcommand{\E}{\mathbb E}
\usepackage{enumitem}
\usepackage{needspace}
\usepackage{booktabs,tabularx}
\titlespacing*{\section}{0pt}{2.2ex plus .6ex minus .3ex}{1.1ex plus .2ex}
\titlespacing*{\subsection}{0pt}{1.8ex plus .5ex minus .3ex}{.8ex plus .2ex}
\titlespacing*{\subsubsection}{0pt}{1.5ex plus .4ex minus .2ex}{.6ex plus .2ex}
\setlist[itemize]{itemsep=0pt, parsep=0pt, topsep=2pt, partopsep=0pt}
\setlist[enumerate]{itemsep=0pt, parsep=0pt, topsep=2pt, partopsep=0pt}
\AtBeginDocument{\setlength{\abovedisplayskip}{5pt plus 2pt minus 2pt}\setlength{\belowdisplayskip}{5pt plus 2pt minus 2pt}\setlength{\abovedisplayshortskip}{2pt plus 2pt}\setlength{\belowdisplayshortskip}{4pt plus 2pt minus 2pt}}
\allowdisplaybreaks[1]
\AtBeginEnvironment{proof}{\Needspace{4\baselineskip}}
\providecommand{\ee}{\mathrm{e}}
\title{Single-Variable Connectivity in Random $k$-SAT}
\author{Chunyang Wang\thanks{National Institute of Informatics, Tokyo, Japan.
    Email: \href{mailto:c_wang@nii.ac.jp}{\texttt{c\_wang@nii.ac.jp}}.}
  \qquad
  Yitong Yin\thanks{State Key Laboratory for Novel Software Technology,
    New Cornerstone Science Laboratory,
    Nanjing University, Nanjing, China.
    Emails: 
    \href{mailto:yinyt@nju.edu.cn}{\texttt{yinyt@nju.edu.cn}} (Yitong Yin),
    \href{mailto:yixiaoyu@smail.nju.edu.cn}{\texttt{yixiaoyu@smail.nju.edu.cn}} (Yixiao Yu),
    \href{mailto:zhangyiyao@smail.nju.edu.cn}{\texttt{zhangyiyao@smail.nju.edu.cn}} (Yiyao Zhang).}
  \qquad
  Yixiao Yu\footnotemark[2]
  \qquad
  Yiyao Zhang\footnotemark[2]}

\date{}
\numberwithin{equation}{section}

\hypersetup{pdftitle={Single-Variable Connectivity in Random k-SAT},pdfauthor={Chunyang Wang, Yitong Yin, Yixiao Yu, Yiyao Zhang}}
\newcommand{\Ham}{d_{\mathrm H}}
\newcommand{\sech}{\operatorname{sech}}

\patchcmd{\thebibliography}
  {\section*{\refname}}
  {\section*{\phantomsection\refname}\addcontentsline{toc}{section}{\refname}}
  {}{\PackageError{manuscript}{Unable to add bibliography to contents}{}}

\usepackage{prettyref,dsfont}
\let\brainberrymathbb\mathbb

\newcommand{\dd}{\,\mathrm{d}}
\newcommand{\defeq}{:=}
\renewcommand{\mathbb}[1]{\ifstrequal{#1}{1}{\mathds{1}}{\brainberrymathbb{#1}}}

\declaretheorem[name=Definition,style=definition,sibling=theorem]{definition}

\makeatletter
\forcsvlist{\FixTheoremHyperAnchor}{definition,remark}
\makeatother
\makeatletter
\newcommand*{\brainberry@newnamedref}[2]{\newrefformat{#1}{\hyperref[##1]{#2~\ref*{##1}}}}
\brainberry@newnamedref{thm}{Theorem}
\brainberry@newnamedref{lem}{Lemma}
\brainberry@newnamedref{ppt}{Property}
\brainberry@newnamedref{pps}{Proposition}
\brainberry@newnamedref{def}{Definition}
\brainberry@newnamedref{problem}{Problem}
\brainberry@newnamedref{cor}{Corollary}
\brainberry@newnamedref{fig}{Figure}
\brainberry@newnamedref{tab}{Table}
\brainberry@newnamedref{fact}{Fact}
\brainberry@newnamedref{claim}{Claim}
\brainberry@newnamedref{alg}{Algorithm}
\brainberry@newnamedref{rem}{Remark}
\brainberry@newnamedref{obs}{Observation}
\brainberry@newnamedref{item}{Item}
\brainberry@newnamedref{step}{Step}
\brainberry@newnamedref{cons}{Constraint}
\brainberry@newnamedref{app}{Appendix}
\brainberry@newnamedref{line}{Line}

\definecolor{wcy}{RGB}{255, 165, 0}
\definecolor{ConnectivityOverview}{RGB}{0, 96, 64}
\definecolor{zyy}{RGB}{112, 48, 160}
\makeatother
\newrefformat{eq}{\hyperref[#1]{(\ref*{#1})}}

\ExplSyntaxOn
\regex_const:Nn \c__brainberry_appendix_anchor_regex
  { \A ( appendix \. | (sub)* section \. [A-Z]+ ( \. | \Z ) ) }
\cs_new_protected:Npn \brainberrysectionrefname #1
  {
    \exp_args:NNe \regex_match:NnTF
      \c__brainberry_appendix_anchor_regex
      { \getrefbykeydefault{#1}{anchor}{} }
      { Appendix }
      { Section }
  }
\ExplSyntaxOff
\newrefformat{sec}{\hyperref[#1]{\brainberrysectionrefname{#1}~\ref*{#1}}}

\newcommand{\pref}{\prettyref}
\newcommand{\Cref}{\pref}
\newcommand{\cref}{\pref}

\begin{document}
\pagenumbering{roman}
\maketitle
\begin{abstract}
We resolve longstanding questions about connectivity, reconstruction,
and looseness in random $k$-SAT, for sufficiently large fixed $k$ as the
number $n$ of variables tends to infinity.

\textbf{Single-variable connectivity up to reconstruction.}
At clause densities up to
\[
\frac{2^k}{k}
\bigl(\ln k+\ln\ln k+\gamma_*-o_k(1)\bigr),
\]
a polynomial-time algorithm joins a typical pair of satisfying
assignments with high probability, using $O_k(n)$ single-variable
flips that preserve satisfiability.
Here $\gamma_*\approx0.871$ arises from a refined
tree-reconstruction analysis.

\textbf{Settling the reconstruction threshold.}
Building on the tools developed for connectivity,
we determine the clustering threshold defined by the
\emph{reconstruction} criterion to be
\[
\frac{2^k}{k}
\bigl(\ln k+\ln\ln k+\gamma_*+o_k(1)\bigr),
\]
below the classical cavity prediction with constant term $1+O(\ln\ln k/\ln k)$.

\textbf{Shattering before rigidity.}
We locate the looseness--rigidity transition at a higher density,
matching the classical cavity prediction for clustering:
\[
\frac{2^k}{k}
\left(\ln k+\ln\ln k+1+\Theta\left(\frac{\ln\ln k}{\ln k}\right)\right).
\]
Below the transition, we establish a stronger form of
\emph{looseness}: any variable of a typical solution can be reversed
by logarithmically many single-variable flips that preserve
satisfiability.
Above it, reversing most variables of a typical solution requires
changing linearly many variables (\emph{rigidity}).
Yet throughout an interval between reconstruction and rigidity,
the solution space already splits into groups of exponentially small
relative size separated by linear Hamming distance
and linear energy barriers
(\emph{shattering}), while logarithmic looseness persists.

These results
resolve the giant-component and logarithmic-looseness conjectures
of Chen--Mani--Moitra (SODA 2023), the sublinear-looseness
conjecture of Achlioptas--Coja-Oghlan (FOCS 2008), and
the reconstruction formulation of the clustering problem
of Krzakala--Montanari--Ricci-Tersenghi--Semerjian--Zdeborov\'a
(PNAS 2007).
\end{abstract}
\clearpage
\begingroup\small\setstretch{1}
\makeatletter
\patchcmd{\l@section}{1.0em \@plus\p@}{.5em \@plus\p@}{}{}
\makeatother
\tableofcontents\endgroup
\clearpage
\pagenumbering{arabic}
\section{Introduction}\label{sec:introduction}
Understanding how simple local constraints give rise to global structure
and computational difficulty is a central challenge at the interface of
theoretical computer science, probability, and statistical physics.
Random constraint satisfaction problems bring these questions together in
a common mathematical setting: as constraints accumulate, the space of
solutions can change its organization before solutions disappear
altogether. Random $k$-SAT is a canonical model for studying these
phenomena, connecting typical computational complexity with the phase
transitions of disordered systems~\cite{MSL92,KS94,MZ96}.

Two questions are central to this picture: when do satisfying assignments
exist, and how accessible are they from one another through local updates?
The first concerns the satisfiability threshold; the second matters for
algorithms that explore the solution space.

The satisfiability threshold $\alpha_{\mathrm{sat}}$ illustrates the success
of this interaction between physics and mathematics. For $m$ clauses on $n$ variables, write
$\alpha=m/n$ for the clause density. Early experiments linked the cost of
search to the transition~\cite{MSL92,KS94}. Friedgut~\cite{Fri99} proved
sharpness, and rigorous bounds~\cite{AP04} established the leading density
$2^k\ln2$. 
The cavity method predicted the precise limiting
threshold~\cite{MPZ02,MMZ06}, established up to an additive $o_k(1)$
by Coja-Oghlan--Panagiotou~\cite{COP16} and exactly for all sufficiently
large $k$ by Ding--Sly--Sun~\cite{DSS22}.

\begin{figure}[htbp]
\centering
\includegraphics[width=.70\linewidth]{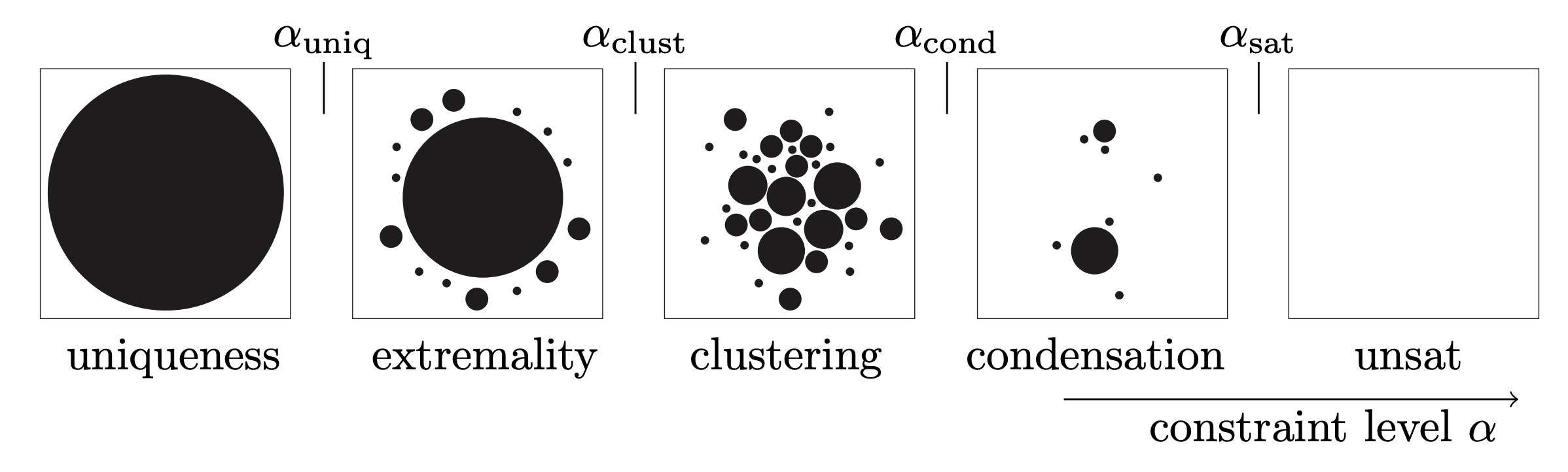}
\captionsetup{font=small,width=.9\linewidth}
\caption{Predicted phase diagram for random $k$-SAT at large $k$,
from~\cite[Figure~1]{DSS22}, adapted there from~\cite{KMRTSZ07}.
We locate $\alpha_{\mathrm{clust}}$ under the reconstruction criterion in
\pref{thm:main-reconstruction}.}
\label{fig:phase-picture}
\end{figure}

Moving between solutions may become difficult long before they
disappear. Statistical physics predicts a sequence of transitions
within the satisfiable phase~\cite{KMRTSZ07,MRTS08}.
We focus on the \emph{clustering}, or \emph{dynamic}, transition
($\alpha_{\mathrm{clust}}$ in \pref{fig:phase-picture}), at which
solutions are predicted to organize into exponentially many groups.

A central characterization of the predicted clustering transition
is the onset of \emph{reconstruction}: distant variable values in a
uniformly sampled solution retain nonvanishing information about a
typical variable beyond what the formula alone reveals~\cite{MM06,KMRTSZ07}.
Under \emph{non-reconstruction}, this extra information vanishes
with distance.
Locating this information transition rigorously is a key step
toward understanding the predicted clustered phase.

This picture has also shaped algorithm design. The cavity method led to
survey propagation, whose messages describe the competing
groups of solutions~\cite{MPZ02,BMZ05}. On the rigorous side,
Coja-Oghlan's \textsc{Fix} algorithm finds a satisfying assignment in
polynomial time up to \mbox{$(1-o_k(1))2^k\ln k/k$}, the leading predicted
clustering scale~\cite{CO10}.
Low-degree polynomial algorithms face a provable obstruction at a
constant multiple of this density~\cite{BH22}.

Krzakala et al.~\cite{KMRTSZ07} asked to determine this reconstruction
threshold rigorously.
The predicted reconstruction threshold is expressed on the
large-$k$ density scale
\begin{equation}\label{eq:offset-density}
 \alpha=\frac{2^k}{k}\bigl(\ln k+\ln\ln k+c\bigr).
\end{equation}
Classical cavity calculations predicted
$c=1+O(\ln\ln k/\ln k)$ for random
SAT~\cite{KMRTSZ07,MRTS08}.
A later heuristic tree-reconstruction analysis by
Budzynski--Semerjian~\cite{BS20} predicted the smaller constant
$\gamma_*\approx0.871$ for graph coloring,
leaving the corresponding SAT calculation open.
Chen--Lonkar--Wang--Yang--Yin~\cite{CLWYY25} proved non-reconstruction
for finite random SAT formulas up to $2^k/\operatorname{poly}(k)$,
far below this scale.

The geometric picture concerns both global connectivity and local freedom.
\emph{Connectivity} asks whether almost all solutions can be joined
by small satisfying moves;
\emph{shattering} partitions the solution space into groups of
exponentially small relative size separated by linear Hamming
distance and linear energy barriers.
For a \emph{typical solution}, sampled uniformly from the satisfying
assignments of the given formula, \emph{looseness} asks whether any
variable can be reversed in a nearby satisfying assignment,
without requiring a satisfying path
(logarithmic looseness means Hamming distance $O_k(\ln n)$);
a variable is \emph{rigid} if every satisfying assignment reversing
its value lies at linear Hamming distance from the original
solution.

The prediction of a dominant connected phase dates back to early cavity studies~\cite{MZ02}. 
Later accounts describe almost all solutions as connected by single-variable flips below the clustering transition~\cite{Ach09,Zhou10}.
Rigorous results established separation and rigidity~\cite{MMZ05,AR06,AC08}, but positive connectivity guarantees remained incomplete.
Below $(1-o_k(1))2^k\ln k/k$, Achlioptas--Coja-Oghlan~\cite[Conjecture~1]{AC08} conjectured that every variable of a typical solution can be reversed by changing $o(n)$ variables.
Chen--Mani--Moitra~\cite[Conjectures~1.8 and~1.9]{CMM23} conjectured logarithmic looseness and connectivity among almost all solutions using moves changing $O_k(\ln n)$ variables.
Near this scale, logarithmic looseness was proved for $99\%$ of variables~\cite{CPP12}; 
for every variable, only at lower densities~\cite{CLWYY25}, while connectivity was proved at exponentially smaller densities~\cite{CMM23,CGGGHMM24}.

These conjectures leave open the finer density bounds
and whether single-variable flips suffice.
More broadly, the central question is:
\begin{center}
\emph{How does reconstruction relate to global connectivity
and local freedom in the solution space?}
\end{center}

\subsection{Main results}\label{sec:subsec-main-results}
We resolve the connectivity and looseness conjectures and determine the
reconstruction threshold.

\textbf{Single-variable connectivity up to reconstruction.}
Single-variable flips, the smallest possible moves, connect almost all
solutions up to the asymptotic reconstruction threshold while preserving
satisfiability throughout. A polynomial-time algorithm connects a typical
pair in $O_k(n)$ flips.

\textbf{Settling the reconstruction threshold.}
Reconstruction begins at $c=\gamma_*+o_k(1)$, below the classical
prediction $c=1+O(\ln\ln k/\ln k)$ and matching the refined constant
predicted for colorings.

\textbf{Shattering before rigidity.}
We locate the looseness--rigidity transition at
$c=1+\Theta(\frac{\ln\ln k}{\ln k})$.
Below it, any variable of a typical solution is reversible in $O_k(\ln n)$ satisfying single-variable flips;
above it, most variables are rigid.
Yet for $0.9615\le c<1$, shattering with linear energy barriers
coexists with looseness, realizing the predicted clustered but
unfrozen regime~\cite{Sem08,MRTS08}.

\paragraph{Model and conventions.}\label{sec:subsec-model}
The formula $\Phi=\Phi(k,n,m)$ has $m$ independent clauses on $[n]$.
Each chooses a uniform $k$-element variable set and independent fair
literal signs; complete clauses may repeat. Write $\alpha=m/n$.
The mass of a set of solutions is its fraction of all solutions.

A satisfying path is a sequence of satisfying assignments in which
consecutive assignments differ at one variable; its length is the
number of flips. All logarithms are natural.
All results fix sufficiently large $k$ and then sufficiently large $n$,
uniformly over the stated density range.

\paragraph{Algorithmic single-variable connectivity.}
The constant $\gamma_*\approx0.871$, numerically estimated
in~\cite{BS20}, is defined in \pref{eq:critical-definition} as the
critical value of a refined tree-reconstruction recursion.

\begin{theorem}[Algorithmic single-variable connectivity]
\label{thm:main-theorem}\label{thm:algorithmic}\label{thm:main}
There is a randomized polynomial-time algorithm such that, if
\begin{equation}\label{eq:main-connectivity-range}
 0\le\alpha\le\frac{2^k}{k}
 \left(\ln k+\ln\ln k+\gamma_*
       -\frac{1}{(\ln\ln k)^{1/16}}\right),
\end{equation}
then with probability at least $1-1/n$ over $\Phi$, the formula
is satisfiable and the algorithm joins two supplied independent uniform solutions by a
satisfying path of $O_k(n)$ single-variable flips, with conditional
success probability at least $1-1/n$ over the endpoints and its random bits.

Consequently, on the same event, a \emph{giant component}
contains a set of mass at least $1-1/n$ whose members can be joined
within the set by $O_k(n)$ satisfying single-variable flips.
\end{theorem}

The algorithm has an absolute polynomial exponent and always
returns a verified path or failure.
Averaging over the first endpoint gives a solution
$x$ such that the solutions within $L=O_k(n)$ satisfying flips
of $x$ have mass at least $1-1/n$.
Joining shortest paths through $x$ connects any two of them
within this set in at most $2L$ flips.

\Needspace{8\baselineskip}
\paragraph{The clustering threshold under reconstruction.}
We determine the clustering threshold $\alpha_{\mathrm{clust}}(k)$
under the \emph{reconstruction} criterion as defined in~\cite{KMRTSZ07,MRT11}.

For satisfiable~$\Phi$, let $\mu_\Phi$ be uniform on its satisfying
assignments. Conditional on $\Phi$, choose $v$ uniformly from $[n]$
and independently sample $\sigma\sim\mu_\Phi$.
Write $\sigma_B$ for its restriction to $B\subseteq[n]$.
Let $\overline B_r(v)$ be the variables at distance at least $r$
from $v$, where two variables are adjacent if they appear in a
common clause.
Write $p_v=\mu_\Phi(\sigma_v=1)$ and
$q_{v,r}=\mu_\Phi(\sigma_v=1\mid\sigma_{\overline B_r(v)})$.
On unsatisfiable formulas, we set both quantities and prediction errors to zero.

Non-reconstruction and reconstruction mean $R_k(\alpha)=0$
and $R_k(\alpha)>0$, respectively, where
\[
 R_k(\alpha):=\lim_{r\to\infty}\limsup_{n\to\infty}
 \mathbb E[|q_{v,r}-p_v|].
\]
Here and in the theorem below, expectations are over the random
formula $\Phi = \Phi(k, n, \lfloor \alpha n \rfloor)$, the uniform variable $v$, and the uniform solution
$\sigma\sim\mu_\Phi$.
The limits are taken with $k$ and $\alpha$ fixed,
first $n\to\infty$, then $r\to\infty$.
The formula is known in both marginals; only the sampled exterior is
additional information.

Among densities at which $\Phi$ is satisfiable with high probability,
define the clustering threshold as the first reconstruction onset:
\[
 \alpha_{\mathrm{clust}}(k):=\inf\{\alpha:R_k(\alpha)>0\}.
\]
This first-onset definition does not assume that reconstruction
is monotone in the clause density.

\Needspace{9\baselineskip}
\begin{theorem}[Reconstruction onset]\label{thm:main-reconstruction}
Non-reconstruction holds throughout \pref{eq:main-connectivity-range}.

For every fixed $\varepsilon>0$ and all sufficiently large $k$,
reconstruction holds throughout
\begin{equation}\label{eq:main-reconstruction-bounds}
 \frac{2^k}{k}
 \bigl(\ln k+\ln\ln k+\gamma_*+\varepsilon\bigr)
 \le\alpha\le{2^k\left(\ln2-\frac1{\ln k}\right)}.
\end{equation}
At each such density, for every fixed $r\ge1$, the formula and
$\sigma_{\overline B_r(v)}$ predict $\sigma_v$ with expected error
$O(1/(k\ln k))+o_n(1)$, where $O(\cdot)$ hides an absolute constant.
Consequently,
\begin{equation}\label{eq:main-reconstruction-headline}
 \alpha_{\mathrm{clust}}(k)=\frac{2^k}{k}
 \bigl(\ln k+\ln\ln k+\gamma_*+o_k(1)\bigr).
\end{equation}
\end{theorem}

\Needspace{8\baselineskip}
\paragraph{Shattering before rigidity.}
We locate the looseness--rigidity transition at
$c=1+\Theta(\ln\ln k/\ln k)$, matching the classical
clustering prediction~\cite{KMRTSZ07}.
Local single-variable connectivity holds throughout the looseness
range. Shattering nevertheless occurs below this transition.

\begin{theorem}[Looseness--rigidity transition]
\label{thm:main-looseness}
For every sufficiently large $k$, there exists $n_0(k)$ such that,
for every $n\ge n_0(k)$, the following hold on opposite sides of
the two density bounds
\begin{equation}\label{eq:main-looseness-headline}
 \frac{2^k}{k}\left(\ln k+\ln\ln k+1+
 \frac{\ln\ln k+1/2}{\ln k}
 \ \pm \frac{8(\ln\ln k)^2}{(\ln k)^2}\right).
\end{equation}
In each density range below, with probability at least $1-1/n$
over $\Phi$, the formula is satisfiable and the corresponding
conclusions hold.
\begin{enumerate}[label=(\alph*),leftmargin=*]
\item\label{item:main-looseness-part}
\textup{(Looseness)}
Up to the lower bound, a deterministic $O_k(n)$-time algorithm
can reverse any requested variable of any of at least a $1-1/n$ fraction
of solutions along a satisfying path of at most $32(\ln k)^5\ln n$ flips.

\item\label{item:main-local-connectivity-part}
\textup{(Local connectivity)}
Under the same density bound and on the same event,
each such solution can be connected to every solution
in a Hamming ball of radius $\Theta_k(n)$ around it
by a satisfying path of length $O(n)$,
in deterministic $O_k(n^2)$ time.

\item\label{item:main-rigidity-part}
\textup{(Rigidity)}
From the upper bound up to $2^k(\ln2-1/\ln k)$,
all but an $\ee^{-\Omega_k(n)}$ fraction
of solutions have $(1-o_k(1))n$ variables such that every satisfying
assignment reversing any one of them differs from the original
solution on $\Omega_k(n)$ variables.
\end{enumerate}
\end{theorem}

\begin{theorem}[Shattering]
\label{thm:main-shattering}
If
\begin{equation}\label{eq:main-shattering-range}
 \frac{2^k}{k}
 \bigl(\ln k+\ln\ln k+\ln(34/13)\bigr)
 \le\alpha\le
 2^k\left(\ln2-\frac1{\ln k}\right),
\end{equation}
then with probability at least $1-1/n$, $\Phi$ is satisfiable
and its solutions partition into regions of mass at most
$\ee^{-\Omega_k(n)}$ at pairwise Hamming distance $\Omega_k(n)$.
Every single-variable-flip path in $\{0,1\}^n$
between distinct regions visits an assignment violating
$\Omega_k(n)$ clauses.
\end{theorem}

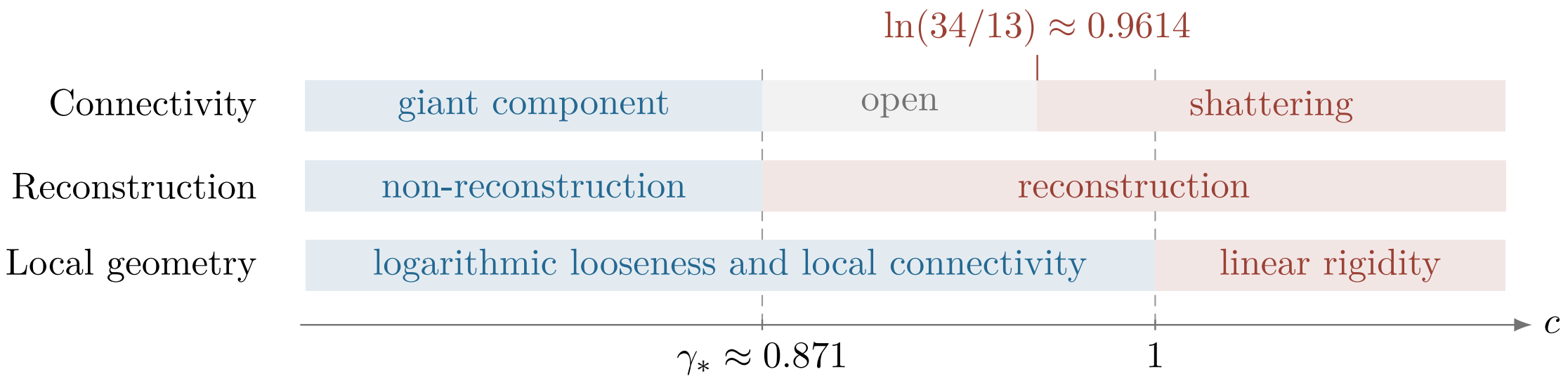
\begin{figure}[H]
\centering
\begin{tikzpicture}[x=1cm,y=1cm,font=\small,
  every node/.style={inner sep=2pt}]
  \definecolor{proved}{RGB}{36,104,145}
  \definecolor{obstructed}{RGB}{155,65,53}
  \def\gammax{4.3}
  \def\shatterx{6.89}
  \def\onex{8.0}
  \def\rightx{11.3}

  \fill[proved!13]
    (0,1.26) rectangle (\gammax,1.74);
  \fill[black!5]
    (\gammax,1.26) rectangle (\shatterx,1.74);
  \fill[obstructed!13]
    (\shatterx,1.26) rectangle (\rightx,1.74);
  \node[anchor=east] at (-.38,1.5) {Connectivity};
  \node[text=proved] at (2.15,1.5) {giant component};
  \node[text=black!55] at (5.595,1.5) {open};
  \node[text=obstructed] at (9.095,1.5) {shattering};

  \draw[obstructed,line width=.45pt]
    (\shatterx,1.74)--(\shatterx,1.98);
  \node[text=obstructed,anchor=south]
    at (\shatterx,1.98)
    {$\ln(34/13)\approx0.9614$};

  \fill[proved!13]
    (0,.51) rectangle (\gammax,.99);
  \fill[obstructed!13]
    (\gammax,.51) rectangle (\rightx,.99);
  \node[anchor=east] at (-.38,.75) {Reconstruction};
  \node[text=proved] at (2.15,.75) {non-reconstruction};
  \node[text=obstructed] at (7.8,.75) {reconstruction};

  \fill[proved!13]
    (0,-.24) rectangle (\onex,.24);
  \fill[obstructed!13]
    (\onex,-.24) rectangle (\rightx,.24);
  \node[anchor=east] at (-.38,0) {Local geometry};
  \node[text=proved] at (4,0) {logarithmic looseness and local connectivity};
  \node[text=obstructed] at (9.65,0) {linear rigidity};

  \foreach \x in {\gammax,\onex}{
    \foreach \lo/\hi in {
      -.56/-.24,.24/.51,.99/1.26,1.74/1.90}
      \draw[black!40,densely dashed,line width=.4pt]
        (\x,\lo)--(\x,\hi);
  }

  \draw[-{Latex[length=1.7mm]},black!55,line width=.45pt]
    (-.05,-.56)--(11.55,-.56);
  \foreach \x in {\gammax,\onex}
    \draw[black!55,line width=.45pt]
      (\x,-.51)--(\x,-.61);
  \node[anchor=north] at (\gammax,-.66)
    {$\gamma_*\approx0.871$};
  \node[anchor=north] at (\onex,-.66) {$1$};
  \node[anchor=west] at (11.58,-.56) {$c$};
\end{tikzpicture}
\captionsetup{font=small,width=.9\linewidth}
\caption{Proved regimes for
$\alpha=\tfrac{2^k}{k}(\ln k+\ln\ln k+c)$
as $k\to\infty$.
Shattering with linear energy barriers coexists with logarithmic looseness and local
connectivity for $\ln(34/13)\le c<1$.}
\label{fig:thresholds}
\end{figure}

\paragraph{Summary of the results.}
\pref{fig:thresholds} summarizes the proved regimes on the
density scale \pref{eq:offset-density}.

Together, our results establish a striking coexistence of
shattering and local accessibility, giving a rigorous realization
of the clustered but unfrozen regime predicted in~\cite{Sem08,MRTS08}.
For $\ln(34/13)\le c<1$, reconstruction holds and solutions
form regions of exponentially small mass separated by linear
Hamming distance and linear energy barriers.
Yet, from almost every solution, polynomial-time algorithms find
logarithmic satisfying paths reversing any variable and
$O(n)$-length paths to every sufficiently nearby solution,
all within its single-variable component.
Below this regime, however, the shattering onset and loss of a
giant component remain open between $\gamma_*$ and $\ln(34/13)$.

\begin{table}[p]
\centering
\setlength{\tabcolsep}{4pt}
\renewcommand{\arraystretch}{1.05}
\begin{tabularx}{\linewidth}{@{}
  >{\raggedright\arraybackslash}p{.17\linewidth}
  >{\raggedright\arraybackslash}p{.42\linewidth}
  >{\raggedright\arraybackslash}X@{}}
\toprule
Source & Clause density & Guarantee\\
\midrule

\multicolumn{3}{@{}l}{\textit{Connectivity}}\\
\addlinespace[2pt]
Pure-literal rule\newline\cite{Mol05,Ach09}
& $\alpha<\alpha_{\rm PL}(k)\sim\frac{2\ln k}{k}$
& Single-variable moves\\
\addlinespace[2pt]
\cite[Theorem~1.4]{CMM23}
& $\alpha\le2^{k/6}$
& $O(\alpha k^6\ln n)$-variable moves\\
\addlinespace[2pt]
\cite[Theorem~1.3]{CGGGHMM24}
& $\alpha\le2^{0.227k}$
& $O(k^5\ln n)$-variable moves\\
\addlinespace[2pt]
\textbf{This work}\newline\pref{thm:main}
& $\alpha\le\frac{2^k}{k}
   (\ln k+\ln\ln k+\gamma_*-o_k(1))$
& Single-variable paths of length $O_k(n)$\\
\addlinespace[2pt]
\textbf{This work} (local)\newline
\pref{thm:main-looseness}\ref{item:main-local-connectivity-part}
& $\alpha\le\frac{2^k}{k}
   \Bigl(\ln k+\ln\ln k+1+
   \Theta\bigl(\frac{\ln\ln k}{\ln k}\bigr)\Bigr)$
& Joins every nearby solution by $O(n)$
  satisfying single-variable flips\\

\cmidrule(lr){1-3}
\multicolumn{3}{@{}l}{\textit{Shattering}}\\
\addlinespace[2pt]
\cite[Theorem~3]{CPP12}
& $\alpha\ge\frac{2^k}{k}
   (\ln k+\ln\ln k+2+o_k(1))$
& Groups of exponentially small mass
  at linear Hamming distance\\
\addlinespace[2pt]
\textbf{This work}\newline
\pref{thm:main-shattering}
& $\alpha\ge\frac{2^k}{k}
   (\ln k+\ln\ln k+\ln(34/13))$
& Exponentially small mass; linear Hamming
  separation and energy barriers\\

\midrule
\multicolumn{3}{@{}l}{\textit{Looseness}}\\
\addlinespace[2pt]
\cite[Theorem~2]{CPP12}
& $\frac{2^k}{k}\lesssim\alpha
   \le\frac{2^k}{k}(\ln k-\ln\ln\ln k-10)$
& At least $99\%$ of variables are $(\ln n)$-loose\\
\addlinespace[2pt]
\cite[Theorem~1.6]{CMM23}
& $\alpha\le2^{k/6}$
& Every variable is $O(\alpha k^6\ln n)$-loose\\
\addlinespace[2pt]
\cite[Theorem~1.5]{CGGGHMM24}
& $\alpha\le2^{0.227k}$
& Every variable is $\operatorname{poly}(k)\ln n$-loose\\
\addlinespace[2pt]
\cite[Theorem~4.13]{CLWYY25}
& $\alpha\le\frac{2^k}{\operatorname{poly}(k)}$
& Every variable is $\operatorname{poly}(k)\ln n$-loose\\
\addlinespace[2pt]
\textbf{This work}\newline
\pref{thm:main-looseness}\ref{item:main-looseness-part}
& $\alpha\le\frac{2^k}{k}
   \Bigl(\ln k+\ln\ln k+1+
   \Theta\bigl(\frac{\ln\ln k}{\ln k}\bigr)\Bigr)$
& Every variable is $O(\ln^5 k\ln n)$-loose\newline
  via satisfying single-variable paths\\

\cmidrule(lr){1-3}
\multicolumn{3}{@{}l}{\textit{Rigidity}}\\
\addlinespace[2pt]
\cite[Theorem~4]{AC08}
& $\alpha\ge(1+o_k(1))\frac{2^k\ln k}{k}$
& A $1-o_k(1)$ fraction of variables are rigid\\
\addlinespace[2pt]
\cite[Theorem~3]{CPP12}
& $\alpha\ge\frac{2^k}{k}
   (\ln k+\ln\ln k+2+o_k(1))$
& At least $99\%$ of variables are rigid\\
\addlinespace[2pt]
\textbf{This work}\newline
\pref{thm:main-looseness}\ref{item:main-rigidity-part}
& $\alpha\ge\frac{2^k}{k}
   \Bigl(\ln k+\ln\ln k+1+
   \Theta\bigl(\frac{\ln\ln k}{\ln k}\bigr)\Bigr)$
& A $1-o_k(1)$ fraction of variables are rigid\\
\bottomrule
\end{tabularx}

\captionsetup{font=small,skip=6pt}
\caption{Density bounds for connectivity, shattering, looseness,
and rigidity in random $k$-SAT.
All guarantees hold with high probability.
Connectivity, looseness, and rigidity concern almost all solutions;
pure-literal connectivity holds for all.
For local connectivity, ``nearby'' means differing on at most
a sufficiently small fraction of variables, depending only on $k$.
All shattering and rigidity bounds assume
$\alpha\le2^k(\ln2-1/\ln k)$.}
\label{tab:comparison}
\end{table}

\subsection{Comparison with previous results}
\pref{tab:comparison} compares density ranges for connectivity, shattering,
looseness, and rigidity. Our positive results improve density bounds
and construct satisfying single-variable paths in polynomial time.
For looseness, this strengthens the guarantee of a nearby solution
reversing the requested variable.

\pref{tab:reconstruction-comparison} compares non-reconstruction and
reconstruction bounds. The bounds in~\cite[Theorem~3(1),(3)]{CPP12}
imply reconstruction: a positive fraction of variables have marginals
bounded away from $0$ and $1$ yet, by linear rigidity, are determined
by the sampled exterior of a fixed-radius ball.

\begin{table}[htbp]
\centering
\setlength{\tabcolsep}{6pt}
\renewcommand{\arraystretch}{1.05}
\begin{tabularx}{\linewidth}{@{}
  >{\raggedright\arraybackslash}p{.31\linewidth}
  >{\raggedright\arraybackslash}X@{}}
\toprule
Source & Clause density\\
\midrule

\multicolumn{2}{@{}l}{\textit{Non-reconstruction}}\\
\addlinespace[2pt]
\cite[Theorem~1.11]{CLWYY25}
& $\alpha\le\frac{2^k}{\operatorname{poly}(k)}$\\
\addlinespace[2pt]
\textbf{This work}\newline\pref{thm:main-reconstruction}
& $\alpha\le\frac{2^k}{k}
   (\ln k+\ln\ln k+\gamma_*-o_k(1))$\\

\midrule
\multicolumn{2}{@{}l}{\textit{Reconstruction}}\\
\addlinespace[2pt]
Consequence of\newline
\cite[Theorem~3(1),(3)]{CPP12}
& $\alpha\ge\frac{2^k}{k}
   (\ln k+\ln\ln k+2+o_k(1))$\\
\addlinespace[2pt]
\textbf{This work}\newline\pref{thm:main-reconstruction}
& $\alpha\ge\frac{2^k}{k}
   (\ln k+\ln\ln k+\gamma_*+o_k(1))$\\
\bottomrule
\end{tabularx}

\captionsetup{font=small,skip=6pt}
\caption{Density ranges for non-reconstruction and reconstruction
in random $k$-SAT.
The reconstruction ranges additionally assume
$\alpha\le2^k(\ln2-1/\ln k)$.}
\label{tab:reconstruction-comparison}
\end{table}

\subsection{Technical overview}\label{sec:overview}

\begin{figure}[ht]
\centering
\begin{tikzpicture}[
  box/.style={
    rounded corners=2pt,align=center,text width=3.85cm,
    inner sep=4pt,font=\small,line width=.45pt},
  method/.style={box,minimum height=1.45cm,
    draw=Navy!55,fill=Navy!4},
  result/.style={box,minimum height=1.45cm,
    draw=black!40,fill=black!3},
  link/.style={-{Stealth[length=1.7mm]},
    draw=black!65,line width=.55pt}
]
\node[method] (trees) at (0,0)
  {Complete forcing trees\\\pref{sec:trees}};
\node[method] (flow) at (4.45,0)
  {Soft-field recursion\\and reversible updates\\
   \pref{sec:local}};
\node[method] (law) at (8.9,0)
  {Local weak convergence\\of the Gibbs distribution\\
   \pref{sec:reconstruction-overview}};

\node[result] (loose) at (0,-2.1)
  {Looseness and rigidity\\
   \pref{sec:looseness-rigidity}};
\node[result] (conn) at (4.45,-2.1)
  {Single-variable\\connectivity\\
   \pref{sec:transfer-section}};
\node[result] (rec) at (8.9,-2.1)
  {Reconstruction\\transition\\\pref{sec:reconstruction}};

\node[result,text width=3cm,minimum height=2.45cm]
(shatter) at (12.75,-1.05)
  {Shattering\\[4pt]
{\footnotesize\mbox{Auxiliary cores}}\\
{\footnotesize\mbox{and witness counts}}\\[2pt]
   \pref{sec:shattering-below-one}};

\draw[link] (trees.south)--(loose.north);
\draw[link] (trees.south east)--(conn.north west);
\draw[link] (flow.south)--(conn.north);
\draw[link] (flow.south east)--(rec.north west);
\draw[link] (law.south)--(rec.north);

\end{tikzpicture}
\captionsetup{font=small,skip=6pt}
\caption{Main proof dependencies.}
\label{fig:proof-routes}
\end{figure}
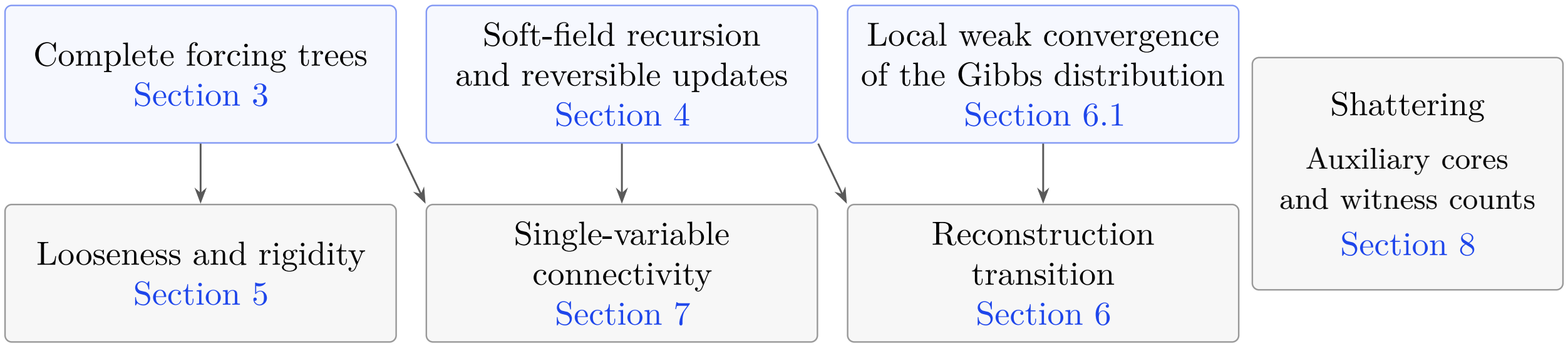

Our proofs pin down two phase transitions governed by two different tree
recursions:
\begin{itemize}
\item \textbf{Looseness--rigidity transition at $c=1+o_k(1)$ (hard-field recursion).}
The hard-field recursion tracks whether the boundary forces the root to
a single value. Below this threshold, complete forcing trees yield
logarithmic looseness and, via common repairs, local connectivity; above
it, the $*$-core yields linear rigidity and geometric shattering
(\pref{sec:trees}, \pref{sec:looseness-rigidity},
\pref{sec:companion-local-connectivity}).
\item \textbf{Clustering transition under reconstruction at $c=\gamma_*+o_k(1)$ (soft-field recursion).}
The soft-field recursion tracks soft boundary-conditioned marginals on
trees. Its transition at $\gamma_*$ governs both the reconstruction
threshold (via local weak convergence; \pref{sec:local},
\pref{sec:reconstruction}) and, through reversible local updates
combined with common repairs, single-variable connectivity below
$\gamma_*$ (\pref{sec:local}, \pref{sec:transfer-section}).
\end{itemize}
Finally, auxiliary cores and a prefix formula establish shattering
with linear energy barriers throughout
\eqref{eq:main-shattering-range}
(\pref{sec:shattering-below-one}).
The proof dependencies are summarized in~\pref{fig:proof-routes}.

The algorithms and proofs were developed with substantive AI
assistance under the authors' direction and review; details
appear in the \hyperref[sec:ai-methodology]{AI methodology statement}.

\paragraph{Forcing trees and the $*$-core threshold.}
Reversing a variable in a solution may require other variables to change as well.
A clause \emph{forces} a variable if its literal is the only true literal, so reversing this variable requires another literal in the clause to become true.
In \pref{sec:trees}, we represent these recursive repairs by \emph{complete forcing trees}~\cite{Sem08}.
Each variable has all its forcing clauses as children, and each clause has one child variable whose flip provides a replacement true literal.

Such repairs can be obstructed by a set in which every variable is forced by a clause entirely within that set.
The $*$-core consists of the largest such set together with all clauses involving only its variables~\cite[Definition~4.1]{MR12}.
This formulation corresponds to the core assignments and coarsening processes studied in~\cite{MMW07,AR06}.
Reversing any of its variables requires another $*$-core variable to change.
On the corresponding infinite Poisson tree, the number of clauses forcing the root is Poisson with mean $\lambda_+=k\alpha/(2^k-1)$, where $\alpha$ is the clause density.
The \emph{hard-field} recursion tracks the probability that the boundary forces the root to a single value.
The $*$-core threshold is the smallest forcing mean for which this probability has a positive limit as the boundary depth tends to infinity, namely
\[
 \ln k+\ln\ln k+1+\frac{\ln\ln k+1/2}{\ln k}
 +O\left(\frac{(\ln\ln k)^2}{(\ln k)^2}\right).
\]
Below this threshold, the root admits a finite repair almost surely.
Above it, the root lies in the $*$-core with positive probability and then admits no finite repair.
Returning to the finite model, we show that below the lower bound in~\eqref{eq:main-looseness-headline}, with high probability, for each typical solution and small forbidden set, at least half the variables root complete forcing trees of size $O_k(1)$ avoiding that set (\pref{lem:set-avoiding-trees}).
These small trees are used in the looseness and connectivity repairs.

\paragraph{Soft-field recursion and local updates.}
In \pref{sec:local}, we analyze tree reconstruction for planted $k$-SAT
formulas through a \emph{soft-field recursion} $\mathcal R_\beta$
(\pref{def:soft-recursion}) motivated by~\cite{BS20}, which arises as
the large-$k$ limit of suitably rescaled weighted marginals of the
one-step planted-tree update (\pref{def:tree-updates}).

We prove that $\mathcal R_\beta$ undergoes a sharp phase transition at
the critical threshold $\gamma_*$ (\pref{thm:critical-threshold}):
starting from $\nu_0=0$ with $\beta=\ee^{-\gamma}$, its masses diverge
as the depth tends to infinity for $\gamma<\gamma_*$, and remain bounded
for $\gamma\ge\gamma_*$. At a gap $\delta>0$ below $\gamma_*$, for
sufficiently small $\delta$, the potential reaches a fixed large level
in $O(\delta^{-1})$ steps (\pref{lem:quant-bands}).

In the supercritical regime $\gamma<\gamma_*$, we use this quantitative
growth to design a reversible local-update algorithm on trees
(\pref{alg:local-update}) that reveals selected subtrees and resamples
the remaining variables.

In the stated finite-$k$ density range below $\gamma_*$, for
appropriately chosen parameters, the \emph{local-update flow}---the
root-flip indicator weighted by the reciprocal number of changed
variables, with denominator at least one---has expectation at least
$(\ln k)^{10}/k$ over the planted tree and update randomness
(\pref{thm:local}). Its input and output are connected by a satisfying
path of single-variable flips. Extending this flow to locally
tree-structured neighborhoods yields an exponentially likely lower bound
on the sum of flows in planted random formulas
(\pref{lem:root-flow}, \pref{sec:local-formulas}).

\paragraph{Looseness and rigidity.}
In \pref{sec:looseness}, we grow a forcing tree from any prescribed variable to prove logarithmic looseness up to the lower bound in~\eqref{eq:main-looseness-headline}.
Earlier local repairs established logarithmic looseness for $99\%$ of variables~\cite{CPP12}.
At lower densities, proofs for every variable used marking~\cite{CMM23,CGGGHMM24} or recursive coupling with witness-tree counts~\cite{CLWYY25}.
The marking approach chooses variables so that fixing their values according to a uniformly random solution leaves the remaining dependency graph with components of size $O_k(\ln n)$, with high probability.
Repairs are confined to these components, leaving the requested variable unpinned if it is marked.
In our construction, we process clauses that force selected variables but have not yet been repaired.
Whenever possible, we attach a preselected small complete forcing tree rooted at another variable of the clause.
Otherwise, we add the clause with one of its other variables as a child.
Additions use new variables and prevent selected variables from sharing a clause unless it joins them in the tree, keeping other clauses satisfied during flips.
For typical solutions, few candidates are excluded while the repair is small, so an admissible extension always exists.
Only single-variable additions leave new forcing clauses to process.
The root and these variables form a subtree whose clauses have few small-tree candidates.
Such clauses are rare in our range below the $*$-core threshold, where small trees are abundant (\pref{lem:set-avoiding-trees}).
Using this rarity, we prove that for typical solutions, only $O_k(\ln n)$ clauses in total force the variables of any such subtree (\pref{lem:forcing-tree-counts}).
Each step processes a new such clause and adds $O_k(1)$ variables, so the whole repair has $O_k(\ln n)$ variables.
Flipping the completed tree from leaves to root produces a satisfying single-variable path.

For rigidity, earlier proofs used expansion within large internally forced subformulas~\cite{AC08,CPP12}.
Molloy and Restrepo~\cite{MR12} studied these forcing dependencies within the $*$-core and determined the freezing threshold for a class of symmetric CSPs.
In \pref{sec:shattering}, we adapt their argument to show that a $1-o_k(1)$ fraction of variables lie in the $*$-core and to rule out solutions changing an intermediate number of its variables.
Both properties hold with high probability in the planted model and transfer to typical solutions in the rigidity range of~\pref{thm:main-looseness}.
All smaller changes within the $*$-core are confined to a common small exceptional set, so reversing any other core variable requires linearly many changes.
Geometric shattering in the same range follows as a direct consequence of the above argument and the solution-count lower bound.

\paragraph{Single-variable connectivity and local connectivity.}
With high probability, the marking approach also connects two independent uniform solutions by changing marked variables one at a time, repairing within the affected components, and then adjusting the unmarked variables component by component.
This gives $O_k(\ln n)$-variable connectivity for $\alpha\le2^{k/6}$~\cite{CMM23}, with the density bound later improved to $\alpha\le2^{0.227k}$~\cite{CGGGHMM24}.

Unlike marking-based methods, our algorithm for \pref{thm:main-theorem} connects two uniform solutions $\sigma,\tau$ using coupled Glauber walks followed by common repairs. 
The algorithm first samples an independent auxiliary formula $\Phi_{\mathrm{aux}}$ and adds its clauses falsified by either $\sigma$ or $\tau$ to $\Phi$, forming $\Phi_+$.
For fixed $\sigma,\tau$, this augmentation enables a comparison between the law of $\Phi_+$ conditional on $\sigma,\tau\in\Omega_\Phi$ and the unconditioned Poisson formula law (\pref{lem:transfer-extension}), and $\Phi_+$ has local tree structure and sparsity with high probability, simultaneously for all $\sigma,\tau$ (\pref{lem:unsigned-envelope}).
We derive failure bounds for fixed $\sigma,\tau$ and then use solution-count bounds to transfer them to independent uniform solutions of $\Phi$ (\pref{lem:norm-transfer}).

We interpolate between $\sigma$ and $\tau$ by changing small blocks of variables, obtaining $z_0=\sigma,\ldots,z_{b_\star}=\tau$ with $b_\star=O_k(1)$. The intermediate assignments may violate $\Phi$. For each $z_i$, let $\Phi_i$ consist of the clauses of $\Phi_+$ satisfied by $z_i$. Starting from $z_i$, we run $\Theta_k(n)$ steps of Glauber dynamics on $\Phi_i$ to obtain $x_i$, using the same random updates across all walks (\pref{alg:connect}). The shared updates keep consecutive states close.
The local-update flow from \pref{sec:local} shows that each walk moves far from $z_i$ in the $\gamma_*$ range~\eqref{eq:main-connectivity-range} (\pref{lem:escape-signed}).
In the Poisson model, clauses in $\Phi_+\setminus\Phi_i$ are false under $z_i$ and have uniform scopes independent of the walk given $z_i$, so this distance ensures that few have at most one true literal under $x_i$, with high probability (\pref{lem:coupled-terminal}).

We next use a procedure called $\mathsf{Repair}$ to turn these states into nearby solutions.
Given $\Phi_0\subseteq\Phi_+$, $x\in\Omega_{\Phi_0}$, and a small set $W$, $\mathsf{Repair}$ seeks a solution $y\in\Omega_{\Phi_+}$ agreeing with $x$ outside a small set $U\supseteq W$.
On success, every solution of $\Phi$ agreeing with $x$ outside $U$ can reach $y$ by satisfying single-variable flips, changing each disagreeing variable once.

The procedure handles clauses whose true literals under $x$ involve only variables in $U$, using a small complete forcing tree in $\Phi_0$ or a pair matching when no suitable tree is available (\pref{alg:common-repair}).
We bound the number of tree additions and pair-matching steps to keep $U$ small; the matching can be extended because no small variable set contains too many clauses (\pref{lem:repair-deterministic}).
These estimates hold up to the lower bound in~\eqref{eq:main-looseness-headline}, with constant term $1$ (\pref{lem:set-avoiding-trees}, \pref{lem:repair-small-sets}).
Together with the preceding bound on clauses in $\Phi_+\setminus\Phi_i$, they guarantee successful repair at the walk states with high probability (\pref{lem:transfer-terminal}).

We first repair the interior states $x_i$ into nearby solutions $y_i$, and set $y_0=x_0$ and $y_{b_\star}=x_{b_\star}$ since the walks from $\sigma$ and $\tau$ stay in $\Omega_\Phi$.
For each consecutive pair, a second repair starts from the small set containing the variables changed by the first repairs and those on which $x_i,x_{i+1}$ disagree.
It supplies a common target through which $\mathsf{FindPath}$ connects $y_i$ and $y_{i+1}$ within $\Omega_\Phi$ (\pref{lem:repair-regularity}).
Concatenating these paths with the walks from $\sigma$ and $\tau$ gives a satisfying path of $O_k(n)$ single-variable flips.

For local connectivity, we apply $\mathsf{Repair}$ directly to a typical solution $x$ with $\Phi_0=\Phi_+=\Phi$, initializing $W$ with the variables on which $x$ and a nearby solution $y$ disagree.
Here $\Phi_+\setminus\Phi_0=\varnothing$, so the clause bound holds automatically and no Glauber walk is needed.
The common target connects both solutions by satisfying single-variable flips throughout the looseness range with constant term $1$ (\pref{thm:main-looseness}\ref{item:main-local-connectivity-part}).

\paragraph{Reconstruction in finite formulas.}
Because $\mu_\Phi$ is supported on globally satisfying assignments,
tree-like neighborhoods are not sufficient to determine its local law.
Related prior analyses of local marginals and reconstruction used
overlap-concentration or replica-symmetry hypotheses~\cite{GM07,MRT11,COP18},
positive-temperature replica symmetry~\cite{CMR22}, or correlation
decay at lower clause densities~\cite{CLWYY25,CCGPRZZ26}.

In \pref{sec:ordinary-local-law}, we establish local weak convergence of
$\mu_\Phi$ throughout $0\le\alpha\le2^k(\ln2-1/\ln k)$
(\pref{lem:ordinary-local-law}), identifying the local limit as an
ordinary signed Poisson tree with i.i.d.\ boundary weights drawn from
a deterministic stationary marginal law $P$. We prove this by combining
the one-variable cavity deletion~\cite{MZ02,MMZ06,KMRTSZ07} with random
pinning~\cite{Mon08} and a one-step Hellinger comparison
(\pref{lem:local-star}) against free-boundary tree marginals, using the
solution-count lower bound (\pref{lem:norm-lower}) and entropy
subadditivity to supply the required average conditional variance.

This local limit reduces reconstruction on $\Phi$ to the large-depth
behavior of the ordinary tree updates. Under the stated marginal
regularity condition, the laws of their one-step boundary-conditioned
output marginals are within $\ee^{-\Omega(k)}$ in total variation of the
corresponding planted outputs (\pref{lem:tree-update-properties}):
\begin{itemize}
\item In the non-reconstruction range
\eqref{eq:main-connectivity-range} (\pref{sec:boundary-transition}) with $\alpha>0$,
the supercritical local-update flow (\pref{thm:local}) transfers to an
expected conditional root variance lower bound at a depth $L_*=O((\ln\ln k)^{1/16})$
independent of $n$ (\pref{lem:depth-L-seed}), after which the Hellinger comparison
(\pref{lem:local-star}) contracts boundary influence geometrically with
depth. The case $\alpha=0$ is immediate.
\item In the reconstruction range
\eqref{eq:main-reconstruction-bounds}
(\pref{sec:reconstruction-beyond}), a comparison in the square-root
order dominates the planted update by the subcritical recursion
$\mathcal R_\beta$ and absorbs the $\ee^{-\Omega(k)}$
ordinary-to-planted error into a shift of $\beta$, bounding the tree
boundary-prediction error by $O(1/(k\ln k))$ at every depth
(\pref{lem:tree-prediction}).
\end{itemize}

\paragraph{Shattering below the $*$-core threshold.}
Earlier shattering proofs established
gaps in Hamming distances and bounded the sizes of nearby solution
sets~\cite{AC08,CPP12}, while in the rigidity range
\pref{sec:shattering} (\pref{thm:rigidity}) uses a large $*$-core, most of whose variables are rigid~\cite{MR12}. Below the $*$-core threshold,
however, the $*$-core of a typical solution is empty in the looseness
range (\pref{sec:looseness}), and near the threshold unforced variables
can drive acyclic forcing cascades among forced variables.

In \pref{sec:shattering-below-one}, we prove shattering with
\energychange{linear energy barriers} throughout
\eqref{eq:main-shattering-range} (\pref{thm:main-shattering},
\pref{thm:shattering-below-one}) by confining low-energy components of
typical solutions using the energy-sublevel graph~\cite{AC08,BDH16} of
the prefix formula at the lower endpoint density. Within this prefix, we
assign an \emph{auxiliary core} (\pref{def:sbo-core}) to each forced
disagreement set; it retains cyclic and multi-variable dependencies
while leaving an acyclic complement.
If the auxiliary core grows large along a low-energy path of small steps,
DAG trimming extracts an intermediate-size witness without increasing
the energy (\pref{lem:sbo-confinement}), which a planted first-moment
bound rules out with high probability (\pref{lem:sbo-census}).

Combining this core bound with planted forcing expansion
(\pref{lem:sbo-planted-expansion}) and planted-to-uniform transfer
(\pref{lem:norm-transfer}) confines the low-energy component of every
typical solution to a small Hamming ball, outside a set of solutions of
exponentially small mass. The full formula's solution-count lower bound
(\pref{lem:norm-lower}) then bounds the solution mass of each such ball,
while the atypical solution set controls the remaining components
(\pref{sec:sbo-energy-barriers}).

\section{Preliminaries}\label{sec:preliminaries}
\subsection{Notation and geometry}\label{sec:preliminaries-geometry}
Write $\Phi=(V,\+C)$, where $V=[n]$ and $\+C$ is the indexed
clause collection. Let $\Omega_\Phi$ be its solution set,
$Z_\Phi=|\Omega_\Phi|$, and $\mu_\Phi$ the uniform measure on
$\Omega_\Phi$, set to zero when $Z_\Phi=0$.
Throughout, $\alpha=m/n$ denotes the clause density of a full formula.
Poisson means such as $\lambda$ and $\lambda_+$ are specified with their auxiliary models.
We identify a formula with its indexed clause collection when writing
$C\in\Phi$ or $\Psi\subseteq\Phi$, and write $V(H)$ for the variable
vertices of a formula, factor graph, or tree $H$.

\paragraph{Geometry of the solution space.}
Write $\Delta(x,y)=\{v\in[n]:x(v)\ne y(v)\}$ for the disagreement set
and $\Ham(x,y)=|\Delta(x,y)|$ for Hamming distance.
Let $G_D(\Phi)$ join distinct solutions at Hamming distance at most $D$;
$G_1(\Phi)$ is the usual solution graph~\cite{GKMP09}.
Connectivity of almost all solutions means that $G_1(\Phi)$ has a
component of mass $1-o_n(1)$.
A variable is \emph{$r$-loose} under $\sigma$ if a solution within
Hamming distance $r$ reverses it, and \emph{$r$-rigid}
otherwise~\cite{AC08,Sem08}.
Logarithmic looseness uses $r=O_k(\ln n)$, and linear rigidity uses
$r=cn$ for some $c=c(k)>0$.
\emph{Shattering} means a partition of $\Omega_\Phi$ into nonempty
regions, each of mass at most $\ee^{-cn}$ and at pairwise Hamming
distance at least $cn$, for some fixed $c=c(k)>0$.
Moreover, every single-variable path in $\{0,1\}^n$ between
distinct regions must visit an assignment violating at least
$cn$ clause occurrences.
This last requirement gives \emph{linear energy barriers}
as in~\cite[Definition~2]{AC08}.
Without the energy-barrier requirement, we call this
\emph{geometric shattering}.

\begin{definition}[Factor graph]
\label{def:factor-graph}
The \emph{factor graph} $\+G(\Phi)$ of $\Phi=(V,\+C)$ is the bipartite graph on
$V\cup\+C$ with edges $\{v,C\}$ between each clause $C$ and its variable set
(\emph{scope}) $\operatorname{vbl}(C)$. Write $B_r(\Phi,v)$ for the radius-$2r$
ball around $v\in V$ in $\+G(\Phi)$; in a tree rooted at $v$, the \emph{depth}
of a variable is half its distance from $v$.
\end{definition}

For $R\subseteq V$ and $x_R\in\{0,1\}^R$ extending to $\Omega_\Phi$, write
$\mu_{\Phi,x_R}:=\mu_\Phi(\cdot\mid\sigma_R=x_R)$ for the conditional
distribution.

\subsection{Formula laws and local limits}

\paragraph{Planted and Poissonized formula laws.}
Under the \emph{planted} law, $\Phi\sim\Phi(k,n,m)$ is conditioned
on a fixed $x$ satisfying it: clauses have independent uniform scopes
and signs uniform among the $2^k-1$ patterns satisfied by $x$.
\pref{lem:norm-transfer} relates this law to a uniform solution.
Poissonizing the clause count with mean $m'$ makes signed-clause
counts independent by Poisson marking~\cite[Chapter~5]{LP18}.
Conditioning back costs at most $3\sqrt{m'}$
when $m'\ge1$ is an integer.
The one-true-literal clauses form the following model with parameter
$m'k/[n(2^k-1)]$, independently of the other clauses.
Thus $m'=m$ gives forcing mean $\alpha k/(2^k-1)$;
retaining the satisfied clauses of $\operatorname{Pois}(m)$ fair-sign
clauses gives $m'=m(1-2^{-k})$ and forcing mean $\alpha k2^{-k}$.

\begin{definition}[Poisson clauses with one true literal]
\label{def:poisson-support-model}
For an assignment $x\in\{0,1\}^n$ and $\lambda\ge0$, let $\mathsf{Pois}_1(x,\lambda)$ be the random formula $\Phi$ obtained by generating, independently for each $v\in[n]$, $\operatorname{Pois}(\lambda)$ clauses whose scope consists of $v$ and a uniform $(k-1)$-subset of $[n]\setminus\{v\}$, with literal signs chosen so that $v$ has the unique true literal under $x$.
\end{definition}

\begin{definition}[Poisson $k$-SAT trees]
\label{def:poisson-tree}\label{def:planted-root-rate}
For $k\ge2$ and $D\ge0$, a \emph{Poisson factor tree} $T$ rooted at $o$ gives each variable independently $\operatorname{Pois}(D)$ child clauses, each with $k-1$ new child variables; write $T_r=B_r(T,o)$ for its depth-$r$ truncation, with boundary at depth $r$. Equipping $T$ (and $T_r$) with literal signs defines three tree models:
\begin{itemize}
\item \emph{Ordinary tree $\mathsf{Tree}_{\mathrm{ord}}(D)$:} all literal signs are i.i.d.\ uniform in $\{\pm1\}$.
\item \emph{Planted tree $\mathsf{Tree}_{\mathrm{pl}}(\lambda_+)$:} set $D=(2^k-1)\lambda_+$, draw i.i.d.\ uniform spins $x_v\in\{0,1\}$, and given $x$, draw each clause's signs independently and uniformly among the $2^k-1$ patterns satisfied by $x$.
\item \emph{One-true-literal tree $\mathsf{Tree}_{\mathrm{one}}(\lambda)$:} set $D=k\lambda$ and, given an assignment $x$, choose in each clause a uniform variable to have a true literal under $x$ and set its other $k-1$ literals false under $x$.
\end{itemize}
In the last two models, retaining only child clauses whose parent has the unique true literal
gives the \emph{forward Poisson tree}, with independent
$\operatorname{Pois}(\lambda_+)$ or $\operatorname{Pois}(\lambda)$
clause lists, respectively.
\end{definition}

\begin{lemma}[Poisson local limit]
\label{lem:poisson-local-limit}
Fix integers $k\ge2$, $r\ge0$, and distinct roots $v,w\in[n]$.
As $n\to\infty$, their depth-$r$ balls, up to rooted isomorphism,
converge jointly in total variation to independent depth-$r$
truncations of the following trees:
\begin{enumerate}[label=\textup{(\roman*)}]
\item For $\Phi\sim\Phi(k,n,m)$
(or its $\operatorname{Pois}(m)$ clause-count version)
with $m/n\to\alpha<\infty$:
$T\sim\mathsf{Tree}_{\mathrm{ord}}(k\alpha)$.

\item For $\Phi\sim\Phi(k,n,m)$ conditioned on $x\models\Phi$,
or its Poissonized planted law with mean clause count $m$,
where $m/n\to\alpha<\infty$
(uniformly in $x\in\{0,1\}^n$ when clauses are marked by literal
truth values under $x$, or jointly with
$x\sim\operatorname{Unif}(\{0,1\}^n)$):
$T\sim\mathsf{Tree}_{\mathrm{pl}}(\lambda_+)$ with
$\lambda_+=k\alpha/(2^k-1)$.

\item For $\Phi\sim\mathsf{Pois}_1(x,\lambda)$ from
\pref{def:poisson-support-model}
(marked by literal truth values under $x$, uniformly in $x$):
$T\sim\mathsf{Tree}_{\mathrm{one}}(\lambda)$.
\end{enumerate}
Consequently, for every bounded function $f$ of a rooted (marked) ball,
\[
 \frac1n\sum_{v\in[n]}f(B_r(\Phi,v))
 \xrightarrow{\mathbb P}\E[f(T_r)].
\]
\end{lemma}
\begin{proof}
Explore the two balls breadth-first, stopping at $M$ vertices.
Collisions have probability $O_{k,M}(1/n)$, and fresh clause counts
converge to independent Poisson counts with the displayed means and
marks. In case~\textup{(iii)}, the true- and false-literal counts
have means $\lambda$ and $(k-1)\lambda$.
Take $n\to\infty$ and then $M\to\infty$; the empirical assertion
follows by the second-moment method.
\end{proof}

\begin{lemma}[Sampling without replacement]\label{lem:sampling-products}
For a uniform $d$-subset $I\subseteq[N]$ and $a_1,\ldots,a_N\ge0$,
\[
 \E\left[\prod_{i\in I}a_i\right]\le\left(\frac1N\sum_{i=1}^Na_i\right)^d.
\]
\end{lemma}
\begin{proof}
Averaging the largest and smallest entries preserves contributions
containing exactly one and increases those containing both.
Iteration converges to the constant list, giving the bound.
\end{proof}

The fixed-count bound below is McDiarmid's inequality~\cite{McD89}.
\begin{lemma}[Bounded differences for clause counts]\label{lem:prob-poisson-lipschitz}
Suppose $F(\mathcal C)\in[0,n]$ changes by at most $L\ge1$ under
one clause insertion or deletion. For $j\ge1$ independent marked
clauses from a fixed law and $u>0$,
$\Pr(F\le\E[F]-u)\le\exp(-u^2/(2jL^2))$.
If instead $|\mathcal C|\sim\operatorname{Pois}(\nu n)$ with the same
independent marks, then, for $0<a\le1$ and all sufficiently
large $n$ depending only on $\nu,L,a$,
\[
 \Pr(F\le\E[F]-an)\le
 3\exp\left(-\frac{a^2n}{128(\nu+1)L^2}\right).
\]
\end{lemma}
\begin{proof}[Poisson bound]
Write $N=|\mathcal C|$.
Let $g_j$ be the expectation of $F$ for $j$ independent clauses.
Coupling by appending clauses gives
\[
 |g_j-\E[F]|\le L\E[|j-N|]
 \le L\bigl(|j-\nu n|+\sqrt{\nu n}\bigr).
\]
On $|N-\nu n|\le an/(8L)$, for large $n$ we have
$|g_N-\E[F]|\le an/2$ and $N\le(\nu+1)n$.
The fixed-count bound and the Poisson tail bound give
\[
 \Pr(F\le\E[F]-an)
 \le\ee^{-a^2n/(8(\nu+1)L^2)}
     +2\ee^{-a^2n/(128(\nu+1)L^2)},
\]
which implies the claim. Zero clause counts are deterministic,
also covering $\nu=0$.
\end{proof}

\begin{lemma}[Bounded-degree ball sensitivity]\label{lem:ball-sensitivity}
For integers $r,D\ge1$, call $v$ \emph{$(r,D)$-locally tree-structured}
if $B_r(\Phi,v)$ is a tree and every variable in it has degree at
most $D$ in $\Phi$. Let $\widetilde B_{r,D}(\Phi,v)$ be this ball at
such roots and $\varnothing$ otherwise, and put
$M_{r,D}=\sum_{h=0}^{r+1}(kD)^h$. Then
\begin{equation}\label{eq:ball-sensitivity}
 |V(\widetilde B_{r,D}(\Phi,v))|\le M_{r,D},\qquad
 |\{v:u\in V(\widetilde B_{r,D}(\Phi,v))\}|\le M_{r,D}.
\end{equation}
One clause insertion or deletion changes at most $2kM_{r,D}$ of these
balls, including fixed variable and clause marks. The same bound
holds for the change in a sum of $[0,1]$-valued functions of these balls.
\end{lemma}
\begin{proof}
Count paths from a root or back from $u$, using at most $D$ incident
clauses and $k$ variables at each step. A changed nonempty ball,
before or after the operation, contains an endpoint of the changed
clause. Count from its at most $k$ endpoints in the two formulas.
\end{proof}

\subsection{Solution counts}\label{sec:normalization}
Write
\[
 \epsilon_0=2^{-k},\qquad M=\mathbb E[Z_\Phi]=2^n(1-\epsilon_0)^m,
 \qquad \Gamma=\alpha k^3\epsilon_0^2,
\]
and let $L_Z=40n\ee^{\Gamma n}$ and
$\mathcal Z=\{Z_\Phi\ge M/L_Z\}$.
The following bound is proved in \pref{app:normalization}.

\begin{lemma}[Lower bound on the number of solutions]\label{lem:norm-lower}
If $\alpha\le2^k(\ln2-1/\ln k)+o_n(1)$, then
\[
 \Pr_\Phi(\mathcal Z)\ge1-\frac1{2n},\qquad
 M/L_Z\ge\exp\bigl(n/(2\ln k)\bigr)
\]
for sufficiently large $k$ and $n$.
\end{lemma}

The next lemma adapts the planted-to-uniform argument
of~\cite{AC08}, including a two-assignment extension.

\begin{lemma}[Fixed-assignment to uniform-solution transfer]
\label{lem:norm-transfer}
Let $r\in\{1,2\}$. Suppose an event
$\mathcal B(\Phi,x_1,\ldots,x_r)$ has conditional probability at most
$b$ for every fixed ordered $r$-tuple, given that all its assignments
satisfy $\Phi$, whenever this conditioning event has positive probability.
Then
\[
 \mathbb E[\mathbb{1}\{\mathcal Z\}\cdot\mu_\Phi^{\otimes r}(\mathcal B)]
 \le\begin{cases}
 bL_Z,&r=1,\\
 2bL_Z^2\exp(2\alpha\epsilon_0(2/3)^kn),&r=2,
 \end{cases}
\]
where the pair bound assumes $\alpha\le2^k/512$.
If instead one draws $\operatorname{Pois}(m)$ clauses from each fixed
tuple's conditional clause law, with integer mean $m\ge1$, replace
$b$ by $3\sqrt m\,b$.
Here $\mu_\Phi^{\otimes r}(\mathcal B)$ also averages any auxiliary
randomness under the same conditional law as in the hypothesis.
\end{lemma}
\begin{proof}
On $\mathcal Z$, $Z_\Phi^{-r}\le(L_Z/M)^r$.
Summing the conditional bound over fixed tuples gives
\[
 \mathbb E[\mathbb{1}\{\mathcal Z\}\cdot\mu_\Phi^{\otimes r}(\mathcal B)]
 \le b\left(\frac{L_Z}{M}\right)^r
     \sum_{\boldsymbol x}\Pr(\boldsymbol x\in\Omega_\Phi^r)
 =bL_Z^r\frac{\mathbb E[Z_\Phi^r]}{M^r}.
\]
For $r=1$ the ratio is one; the pair bound follows from the
overlap calculation in \pref{app:normalization}.
The Poissonized version follows by conditioning on its clause count,
since $\Pr(\operatorname{Pois}(m)=m)\ge1/(3\sqrt m)$.
\end{proof}

\section{Complete forcing trees}
\label{sec:trees}

To flip a variable $v$ in a satisfying assignment $\sigma$ without violating any clause, one must also flip a variable in every clause that currently has $v$ as its sole true literal, and continue recursively. 
Such a cascade can form a \emph{complete forcing tree} rooted at $v$.
If the full restricted factor graph on its variables is acyclic, the tree can be flipped bottom-up along a satisfying path (\pref{obs:flip-complete-tree}).
Together with its dual obstruction---the $*$-core of persistent forcing dependencies~\cite{AR06,MMW07,AC08,MR12}---this combinatorial structure governs the looseness--rigidity transition~\cite{Sem08,CPP12}.

\subsection{Small complete forcing trees}
\label{sec:forcing-poisson}

\begin{definition}[Forcing clauses and trees]
\label{def:support}\label{def:forcing-tree}
Fix a CNF formula $\Phi=(V,\+C)$ and an assignment $\sigma$.
Let $\operatorname{sat}_\sigma(C)$ be the variables whose literals in $C$ are true under $\sigma$.
A clause $C$ \emph{forces} $v$ if $\operatorname{sat}_\sigma(C)=\{v\}$.
We write $\+C_\sigma^{(1)}=\{C\in \+C:|\operatorname{sat}_\sigma(C)|=1\}$ and $\Phi_\sigma^{(1)}=(V,\+C_\sigma^{(1)})$.

A \emph{forcing tree} is a variable-rooted subtree of the factor graph in which each clause forces its parent and has a unique child.
It is \emph{complete} if each of its variables has all clauses forcing it as children.
Its \emph{size} and \emph{depth} are the count and maximum depth of its variable vertices.
\end{definition}

For a subset of variables $U\subseteq V$, the \emph{restricted factor graph} on $U$ is the subgraph induced by $U$ and all clauses containing variables in $U$.

\begin{observation}[Flipping a complete forcing tree]
\label{obs:flip-complete-tree}
Let $\sigma\in\Omega_\Phi$, and let $T$ be a complete forcing tree under $\sigma$ with variable set $U=V(T)$.
If the restricted factor graph on $U$ is acyclic, then flipping the variables of $U$ in non-increasing order of depth in $T$ defines a satisfying path of length $|U|$ from $\sigma$ to the solution $\tau\in\Omega_\Phi$ with $\Delta(\sigma,\tau)=U$.
\end{observation}
\begin{proof}
By acyclicity of the restricted factor graph, each clause $C\in T$ meets $U$ only at its parent and child in $T$; since the child has greater depth than the parent, it is flipped first and satisfies $C$ before the parent is falsified.
Every clause $C\notin T$ meets $U$ in at most one variable, which $C$ cannot force under $\sigma$ by completeness of $T$; thus $C$ retains a true literal outside $U$ throughout.
\end{proof}

We next show that in the 1-forced Poisson model (\pref{def:poisson-support-model}) below the looseness threshold of \pref{thm:main-looseness}, a macroscopic fraction of variables root complete forcing trees of size at most $(\ln k)^6$ avoiding any small forbidden set.

\begin{lemma}[Small complete forcing trees]\label{lem:set-avoiding-trees}
Fix an assignment $\sigma$, let $k$ and $n\ge n_0(k)$ be
sufficiently large, and put $s=\lceil(\ln k)^6\rceil$.
If $\Phi\sim\mathsf{Pois}_1(\sigma,\lambda)$ has forcing mean $\lambda$ and
\[
 0\le\lambda\le\ln k+\ln\ln k+1+
 \frac{\ln\ln k+1/2}{\ln k}
 -\frac{6(\ln\ln k)^2}{(\ln k)^2},
\]
then, with probability at least $1-\exp(-n/(32k^{8s}))$,
for every $\Psi\subseteq\Phi$ and
$B\subseteq[n]$ with $|B|\le n/(16k^{2s})$, at least $n/2$
variables root complete forcing trees in $\Psi$ of size at most
$(\ln k)^6$ avoiding $B$.
\end{lemma}
\begin{proof}
For $\Psi\subseteq\Phi$, pruning from a complete forcing tree every clause absent from $\Psi$ and its descendants preserves the root, completeness in $\Psi$, avoidance of $B$, and the size bound. Thus it suffices to prove the assertion for $\Psi=\Phi$. Since $\Phi'\sim\mathsf{Pois}_1(\sigma,\lambda')$ with $\lambda'\le\lambda$ couples as a subformula $\Phi'\subseteq\Phi$, the same pruning also lets us take $\lambda$ at the displayed upper bound.

\paragraph{Probability of a finite complete forcing tree.}
Put $\varepsilon=(\ln\ln k)^2/(\ln k)^2$ and $r=1/[k(\ln k+\ln\ln k)]$, and write $x=-\ln r$.
For sufficiently large $k$, Taylor expansion implies
\begin{equation}\label{eq:forcing-explicit-expansion}
 \left|\frac{x}{(1-r)^{k-1}}-\left(\ln k+\ln\ln k+1+
 \frac{\ln\ln k+1/2}{\ln k}\right)\right|\le\varepsilon,
 \qquad \frac{(k-1)xr}{1-r}=1+o(\varepsilon).
\end{equation}
In the forward Poisson tree, let $p_i$ be the probability that the root admits a complete forcing tree of depth at most $i$.
Independence of the child subtrees implies $p_0=\ee^{-\lambda}$ and $p_{i+1}=F(p_i)$, where $F(p)=\exp(-\lambda(1-p)^{k-1})$.
Set $z_i=p_i/r$ and $H(z)=\ln(F(rz)/r)$.
By \eqref{eq:forcing-explicit-expansion}, we have $4\varepsilon\le H(1)\le7\varepsilon$ and $H'(1)=1+o(\varepsilon)$.
Moreover, $-2/\ln k\le H''(z)\le0$ on $[1/3,3\ln\ln k]$.
Since $z_{i+1}=\exp(H(z_i))$, Taylor's theorem implies
\[
 z_{i+1}-z_i\ge\frac{\varepsilon+(z_i-1)^2}{8},\qquad
 z_{i+1}\le4\qquad(1/3\le z_i\le2).
\]
The same expansion implies $H(z)=z-1+o_k(1)$ uniformly for $2\le z\le3\ln\ln k$.
Let $J$ be the first index with $z_J\ge2$ and $K$ the first with $z_K\ge3\ln\ln k$.
Since $z_0\in[1/3,1/2]$ for large $k$, the drift bound allows $O(\varepsilon^{-1/2})$ steps before $J$ with $|z_i-1|\le\sqrt\varepsilon$.
For each integer $j\ge0$, the range $2^j\sqrt\varepsilon<|z_i-1|\le2^{j+1}\sqrt\varepsilon$ contributes $O(2^{-j}\varepsilon^{-1/2})$ steps.
Summing these bounds yields $J=O(\varepsilon^{-1/2})=O(\ln k)$.
For $J\le i<K$, we have $z_{i+1}\ge4z_i/3$, so $K-J=O(\ln\ln\ln k)$.
The expansion at $z=3\ln\ln k$ and monotonicity of $F$ imply $p_{K+1}\ge(\ln k)^2/(6k)$.
For $t=K+2=O(\ln k)$, the next update satisfies $1-p_t\le2\ln k\exp(-(k-1)(\ln k)^2/(6k))=o_k(1)$.

\paragraph{Expected size of the selected tree.}
At each clause, we choose the first child that admits a complete forcing tree within the remaining depth.
Let $W_i$ be the expected size of the selected tree, conditional on its existence at depth at most $i$.
Poisson splitting makes the clause families with and without
a successful child independent; conditioning on root success
forces the latter to be empty and leaves the former unchanged.
The selected successful child has conditional expected tree
size $W_i$, so
\[
 W_0=1,\qquad W_{i+1}=1+a_iW_i,\qquad
 a_i=\lambda[1-(1-p_i)^{k-1}]\le\min\{2z_i,2\ln k\}.
\]
Let $\Delta_i=p_{i+1}-p_i$.
Convexity of $F$ on $[0,2r]$ implies
\[
 a_i\le\lambda(k-1)p_i
 =F'(p_{i-1})(1-p_{i-1})^{-(k-2)}
 \le\frac{\Delta_i}{\Delta_{i-1}}\ee^{2(k-1)p_{i-1}}
 \qquad(1\le i<J).
\]
Here $\Delta_i\ge\varepsilon r/8$ for $0\le i<J$, $p_J\le4r$, and $(k-1)rJ=O(1)$.
Multiplying the bounds above therefore shows that $\prod_{i=j}^{J-1}a_i=O(\varepsilon^{-1})$ for every $1\le j<J$.
Since $a_0<1$, the recursion for $W_i$ implies $W_J=O(J/\varepsilon)=O((\ln k)^3)$.
The next $K-J$ factors are at most $6\ln\ln k$, and the last two are at most $2\ln k$.
Consequently, $W_t\le(\ln k)^{5+o_k(1)}=o((\ln k)^6)$.
By Markov's inequality, the root admits a complete tree of depth at most $t$ and size at most $(\ln k)^6$ with probability at least $p_t(1-W_t/(\ln k)^6)\ge15/16$ for sufficiently large $k$.

\paragraph{Concentration and set avoidance in the finite formula.}
In the finite formula, we say a variable is good if it is $(t+1,k^4)$-locally tree-structured and admits such a complete tree.
Let $G$ count these variables.
In the limiting Poisson tree, the probability of a degree above $k^4$ within depth $t+1$ is at most $k^{O(t)}\ee^{-\Omega(k^4\ln k)}\le1/32$ for sufficiently large $k$.
Revealing one further generation determines the boundary degrees.
For each fixed $k$, \pref{lem:poisson-local-limit} bounds the total variation error by $1/32$ for sufficiently large $n$.
Hence $\E[G]\ge n(15/16-1/32-1/32)=7n/8$.
By \pref{lem:ball-sensitivity}, one clause changes $G$ by at most $A_k=2kM_{t+1,k^4}=k^{O(t)}$.
The total clause count has law $\operatorname{Pois}(\lambda n)$, so \pref{lem:prob-poisson-lipschitz} implies
\[
 \Pr(G<3n/4)
 \le3\exp\!\left(-\frac{n}{8192(\lambda+1)A_k^2}\right)
 \le\exp(-n/(32k^{8s})).
\]
The last inequality holds for sufficiently large $k,n$, since $t=O(\ln k)$ and $s=\lceil(\ln k)^6\rceil$.

We fix one such tree for each good variable.
By \pref{lem:ball-sensitivity}, any variable belongs to at most $M_{t+1,k^4}\le k^{2s}$ of these trees.
Every allowed $B$ therefore meets at most $|B|k^{2s}\le n/16$ selected trees.
On $G\ge3n/4$, at least $3n/4-n/16=11n/16>n/2$ selected trees avoid $B$, simultaneously for every allowed $B$.
The failure probability is bounded by $\Pr(G<3n/4)$, as required.
\end{proof}

\subsection{Obstructed explorations}\label{sec:repair-witness-counts}

To grow complete forcing trees of size $O_k(\ln n)$ from arbitrary roots by attaching the small trees of \pref{lem:set-avoiding-trees}, we must rule out large forcing trees composed of clauses with too few small-tree candidates. We bound the Gibbs weight of such trees by resampling literal signs on their forcing clauses.

Fix a formula $\Phi_0$, let $C_1,\ldots,C_r$ be $k$-clauses with fixed variable sets and independent uniform random signs, and fix a designated variable $v_i\in\operatorname{vbl}(C_i)$ for each $i$.
Write $\Phi=\Phi_0\land\bigwedge_{i=1}^r C_i$ and let $S=\{\sigma:\text{each }C_i\text{ forces }v_i\}$.

\begin{lemma}[Averaging literal signs]\label{lem:support-signs}
For any $A\subseteq\Omega_{\Phi_0}$, $B\supseteq\{v_1,\ldots,v_r\}$, and assignment $\tau$ on $B$ with $\mu_{\Phi_0}(\sigma_B=\tau)>0$,
\[
 \E_{\rm signs}[\mu_\Phi(S\cap A\cap\{\sigma_B=\tau\})] \le2^{-kr}\mu_{\Phi_0}(A\mid\sigma_B=\tau),
 \qquad
 \E_{\rm signs}[\mu_\Phi(S\cap A)]\le2^{|B|-kr}.
\]
\end{lemma}
\begin{proof}
Let $Z_0(\tau)$ count solutions of $\Phi_0$ that agree with $\tau$ on $B$.
If $S\cap A\cap\{\sigma_B=\tau\}$ is nonempty, the literal of each $v_i$ in $C_i$ is true under $\tau$.
Hence all $Z_0(\tau)$ such solutions also satisfy $\Phi$, so $Z_\Phi\ge Z_0(\tau)$.
Each fixed assignment in $A\cap\{\sigma_B=\tau\}$ belongs to $S$ for exactly one of the $2^{kr}$ sign choices.
Therefore,
\[
 \E_{\rm signs}[\mu_\Phi(S\cap A\cap\{\sigma_B=\tau\})]
 \le\frac{2^{-kr}|A\cap\{\sigma_B=\tau\}|}{Z_0(\tau)}
 =2^{-kr}\mu_{\Phi_0}(A\mid\sigma_B=\tau).
\]
Summing over the at most $2^{|B|}$ feasible assignments $\tau$ yields the second bound.
\end{proof}

In the following, let $\Phi\sim\Phi(k,n,m)$ with $m=\lfloor\alpha n\rfloor$, and set $s=\lceil(\ln k)^6\rceil$.
Under a solution $\sigma$, a clause is \emph{obstructed} if, among its variables whose literals are false under $\sigma$, fewer than $1024(\ln k)^6$ root complete forcing trees in $\Phi$ of size at most $(\ln k)^6$.
For a forcing tree $T$, let $D_\sigma(T)$ be the number of clauses of $\Phi$ that force variables of $T$, and let $\mathcal B_t(\Phi,v)$ be the set of solutions $\sigma\in\Omega_\Phi$ under which some forcing tree $T$ rooted at $v$ has all its clauses obstructed and satisfies $D_\sigma(T)\ge t$.

\begin{lemma}[Obstructed forcing trees]\label{lem:forcing-tree-counts}
If $k$ is sufficiently large and
\[
 \alpha\le\frac{2^k-1}{k}\left(\ln k+\ln\ln k+1+
 \frac{\ln\ln k+1/2}{\ln k}
 -\frac{6(\ln\ln k)^2}{(\ln k)^2}\right),
\]
then for every root $v$ and integer $t$ with
$2\ee\sqrt k\ln k\le t\le\ln n$,
\[
 \E_\Phi[\mathbb{1}\{\mathcal Z\}\cdot\mu_\Phi(\mathcal B_t(\Phi,v))] \le2k^{-t/2}+\ee^{-n/k^{9s}}.
\]
\end{lemma}
\begin{proof}
The case $m=0$ is immediate, so assume $m\ge1$.
Let $\mathcal H_\Phi\subseteq\Omega_\Phi$ be the set of solutions $\sigma$ for which $\Phi_\sigma^{(1)}$ satisfies the conclusion of \pref{lem:set-avoiding-trees}.
For fixed $\sigma$, if we sample $\operatorname{Pois}(m)$ clauses independently and uniformly from those satisfied by $\sigma$, then $\Phi_\sigma^{(1)}\sim\mathsf{Pois}_1(\sigma,mk/[n(2^k-1)])$ as defined in \pref{def:poisson-support-model}.
By \pref{lem:set-avoiding-trees} and the Poissonized transfer bound in \pref{lem:norm-transfer} (with $3\sqrt m\le n$),
\[
 \E_\Phi[\mathbb{1}\{\mathcal Z\}\cdot\mu_\Phi(\mathcal H_\Phi^c)] \le40n^2\exp(\alpha k^34^{-k}n-n/(32k^{8s})) \le \ee^{-n/k^{9s}}.
\]

Write $a=\ee\sqrt k$.
We claim that, for $1\le j\le\ln n+1$,
\begin{equation}\label{eq:weighted-tree-count}
 \E_\Phi\left[\E_{\mu_\Phi}\left[\mathbb{1}\{\sigma\in\mathcal H_\Phi\}\cdot
       \sum_{|T|=j}a^{D_\sigma(T)}\right]\right]
 \le k^{2a}\rho_k^{j-1},\quad \text{where }\rho_k=a k^{2a}\ee^{-k/4}<1/2,
\end{equation}
and the sum is over forcing trees $T$ of size $j$ rooted at $v$ whose clauses are all obstructed under $\sigma$.
Fix the tree shape and labels of $T$, and write $U$ for its variable set.
Each of the $j-1$ clauses of $T$ forces a variable in $U$, so
\[
 a^{D_\sigma(T)}
 =a^{j-1}\prod_{C\notin T}
       (1+(a-1)\mathbb{1}\{C\text{ forces a variable of }U\}).
\]
Expanding the product over subsets of $q$ additional clauses outside $T$ that force variables in $U$, let $\Phi_0$ be obtained by deleting those $q$ clauses together with the $j-1$ clauses of $T$.
Designate a forced variable in $U$ for each additional clause, and let $S$ be the event that each deleted clause forces its designated variable, with the parent designated for each clause of $T$.
Let $A\subseteq\Omega_{\Phi_0}$ consist of solutions with at least $n/2$ roots of complete forcing trees in $\Phi_0$ of size at most $(\ln k)^6$ avoiding $U$.
Since $|U|=j\le\ln n+1\le n/(16k^{2s})$, we have $\mathcal H_\Phi\subseteq A$.
On $S$, every deleted clause forces a variable in $U$ and therefore none outside $U$, since a clause forces at most one variable. Thus $(\Phi_0)_\sigma^{(1)}$ and $\Phi_\sigma^{(1)}$ agree on all clauses forcing variables outside $U$, so any complete forcing tree avoiding $U$ in $\Phi_0$ remains complete in $\Phi$.

On $S$, each tree clause forces its parent, so all its other literals are false under $\sigma$. Hence, for fixed $\Phi_0$ and $\sigma\in A$, obstruction requires each tree clause's remaining $k-2$ variables to contain fewer than $1024(\ln k)^6$ of these roots.
Conditional on the specified incidences, these $(k-2)$-tuples are sampled without replacement independently across clauses and contain at least $k/3$ such roots in expectation.
By \pref{lem:sampling-products} and exponential Markov's inequality, this event has probability at most $\ee^{-k/4}$ per clause.
For each realization of these variable sets, $A$ intersected with the required root-count events depends only on $\Phi_0$ and those sets, not on the deleted clauses' signs, so \pref{lem:support-signs} applies with $B=U$.
Since $\mu_{\Phi_0}(\cdot\mid\sigma_U=\tau)$ does not depend on the deleted clauses' variable sets, averaging over those variable sets and summing over the at most $2^j$ feasible assignments $\tau$ on $U$ bounds the averaged contribution by $2^{j-k(j-1+q)}\ee^{-k(j-1)/4}$.

There are at most $4^{2(j-1)}n^{j-1}m^{j-1}$ choices for $T$ and $m^qj^q/q!$ choices for the $q$ additional clauses and their forced variables in $U$.
The probabilities that the clauses contain their specified tree and forced variables are at most $(k/n)^{2(j-1)}$ and $(k/n)^q$, respectively.
Combining these factors and using $mk/(n2^k)\le2\ln k$ yields \eqref{eq:weighted-tree-count}, since the total is at most
\[
 2(64k\ln k)^{j-1}a^{j-1}\ee^{-k(j-1)/4}
 \sum_{q\ge0}\frac{1}{q!}\left(\frac{(a-1)mjk}{n2^k}\right)^q \le k^{2j}a^{j-1}\ee^{-k(j-1)/4}\ee^{2(a-1)j\ln k} =k^{2a}\rho_k^{j-1}.
\]

Truncating a tree witnessing $\sigma\in\mathcal B_t(\Phi,v)$ if necessary, we obtain a subtree $T$ rooted at $v$ with $|T|\le t+1$ and $D_\sigma(T)\ge t$.
Indeed, if the witnessing tree is larger, a rooted subtree with exactly $t+1$ variables contains $t$ clauses, each forcing its parent in the subtree.
Thus Markov's inequality and \eqref{eq:weighted-tree-count} imply
\[
 \E_\Phi[\mu_\Phi(\mathcal H_\Phi\cap\mathcal B_t(\Phi,v))] \le a^{-t}k^{2a}\sum_{j=1}^{t+1}\rho_k^{j-1} \le2k^{-t/2}\ee^{2\ee\sqrt k\ln k-t}\le2k^{-t/2}.
\]
Adding the earlier bound on $\E_\Phi[\mathbb{1}\{\mathcal Z\}\cdot\mu_\Phi(\mathcal H_\Phi^c)]$ proves the lemma.
\end{proof}

\begingroup
\section{Soft-field recursion and local updates}\label{sec:local}

The forcing-tree estimates in \pref{sec:trees} are governed by the $*$-core threshold
$\lambda_+=\ln k+\ln\ln k+1+o_k(1)$~\cite{KMRTSZ07,Sem08,MRTS08}.
In cavity terminology, \emph{hard fields} force a variable to one value, whereas \emph{soft fields} leave both values possible.

Reconstruction and reversible updates also depend on the strength of these soft biases: they can accumulate to make a variable nearly determined without forcing it.

For proper $q$-colourings of regular trees with branching factor $q(\ln q+\ln\ln q+c)$, non-reconstruction holds up to $c=1-\ln 2-o_q(1)$ and reconstruction holds at a constant $c<1$~\cite{sly2009reconstruction,MRT11,sly2016reconstruction}. Budzynski and Semerjian~\cite{BS20} derived a limiting \emph{soft-field recursion} that heuristically locates the transition at $\gamma_*\approx 0.871<1$, leaving its threshold analysis and the asymmetric case of random $k$-SAT open.

In \pref{sec:critical}, we formulate the one-step planted tree update (\pref{def:tree-updates}), derive its limiting soft-field recursion $\mathcal R_\beta$ (\pref{def:soft-recursion}), and establish its phase transition at $\gamma_*\in[2(\sqrt2-1),1)$ (\pref{thm:critical-threshold}) together with a near-critical escape bound (\pref{lem:quant-bands}). In \pref{sec:local-updates}, we use this supercritical escape to construct a reversible local update on trees (\pref{alg:local-update}) with planted root flow at least $\epsilon_k=(\ln k)^{10}/k$ below $\gamma_*$ (\pref{thm:local}). Finally, in \pref{sec:local-formulas}, we extend this update to locally tree-structured neighborhoods in general formulas and prove exponential concentration of the empirical root flow in planted random formulas (\pref{lem:root-flow}).

\subsection{The one-step tree update and the soft-field recursion}\label{sec:critical}

\paragraph{The one-step planted tree update.}

Set $d=k-1$ and consider a root variable joined to $Nd$ child variables via $N$ clauses $C\in[N]$, where $\xi_C\in\{0,1\}$ denotes the root assignment that falsifies its literal in $C$.
Suppose the child subtrees are independent, and let $s_C\in[0,1]$ be the probability that all child literals of clause $C$ are false.
Define the root weights
\[
 W_b(s)\defeq\prod_{C:\xi_C=b}(1-s_C)\quad(b=0,1),\qquad
 F(s)\defeq W_0(s)+W_1(s).
\]
$W_b(s)$ is the probability that all $N$ clauses are satisfied when the root is assigned $b$.
Under a uniform root prior conditioned on satisfying all clauses, the root takes value $b$ with probability $W_b(s)/F(s)$ when $F(s)>0$.
Let $\bar q_{C,i}$ be the probability that the literal of the $i$th child of clause $C$ is false.
Substituting $s_C=\prod_{i=1}^d\bar q_{C,i}$, we write $W_b(\bar q)=\prod_{C:\xi_C=b}(1-\prod_{i=1}^d\bar q_{C,i})$ and $F(\bar q)=W_0(\bar q)+W_1(\bar q)$, suppressing their dependence on $N$ and $\xi$.

We now formalize the resulting one-step recursion for the boundary-conditioned marginal $Q$ on the planted tree (\pref{def:planted-root-rate}) by starting from auxiliary fair literal signs and conditioning on clause satisfaction.

\begin{definition}[One-step planted tree update]\label{def:tree-updates}
Let $Q\in[0,1]$ be a random variable, interpreted as a child's conditional probability of taking value $1$ given its subtree formula and boundary values.
Set $D=2^k\lambda\ge0$.
The update from $Q$ to the root marginal $Q^+$ consists of two stages:
\begin{enumerate}
\setlength{\itemsep}{2pt}\setlength{\parskip}{0pt}
    \item \textbf{Local tree and child-marginal generation:}
    Sample $N\sim\operatorname{Pois}(D)$ and, conditional on $N$, independent fair bits $\xi_C\in\{0,1\}$ ($C\in[N]$).
    Independently of these bits, sample mutually independent child marginals $Q_{C,i}$ ($C\in[N],i\in[d]$), each with the same distribution as $Q$.
    Incorporate the child literal signs by setting, independently for each $(C,i)$,
    \[
        \bar Q_{C,i} =
        \begin{cases}
        Q_{C,i} & \text{with probability } 1/2, \\
        1-Q_{C,i} & \text{with probability } 1/2.
        \end{cases}
    \]
    \item \textbf{Planted reweighting and marginal recurrence:}
    The planted law is obtained by accepting a Stage~1 sample $(\bar Q,N,\xi)$ with probability $F(\bar Q)/2$ and redrawing the entire sample after rejection.
    Here $F(\bar Q)/2$ is the conditional probability that all $N$ clauses are satisfied under a uniform root prior.
    For an accepted sample, set
    \[
        (1-Q^+,Q^+)\propto\bigl(W_0(\bar Q),W_1(\bar Q)\bigr),
    \]
    with $Q^+$ arbitrary on the null event $F(\bar Q)=0$ under the reweighted law.
\end{enumerate}
\end{definition}

Since $\E[\bar Q_{C,i}]=1/2$, multilinearity gives $\E[W_b(\bar Q)\mid N,\xi]=(1-2^{-d})^{|\{C\in[N]:\xi_C=b\}|}$, and averaging over $\xi$ and $N\sim\operatorname{Pois}(2^k\lambda)$ yields $\E[F(\bar Q)]=2\ee^{-\lambda}$.
The mean acceptance probability is therefore $\ee^{-\lambda}$.
Conditional on acceptance, the joint distribution of $(\bar Q,N,\xi)$ has density $(F(\bar Q)/2)/\ee^{-\lambda}=\ee^\lambda F(\bar Q)/2$ relative to its Stage~1 distribution.
Under the Stage~1 law, $\E[F(\bar Q)/2\mid N]=(1-2^{-k})^N$, so the accepted degree has law $\operatorname{Pois}((2^k-1)\lambda)$.
Starting from $Q_0\sim\operatorname{Bernoulli}(1/2)$ on the pinned boundary, iterating \pref{def:tree-updates} at $\lambda=\lambda_+$ produces the depth-$r$ boundary-conditioned root marginal of $(T_r,x)\sim\mathsf{Tree}_{\mathrm{pl}}(\lambda_+)$ (\pref{def:planted-root-rate}).

\paragraph{The $f$-weighted measure and soft-field recursion.}
We now introduce a weighted measure of the unfrozen marginals to study the large-$k$ limit of this update.
For $q\in(0,1)$ and $z\in\mathbb R$, define the half-log-odds (cavity field) and its inverse by
\[
 z(q)\defeq\frac12\ln\frac q{1-q},\qquad
 q(z)\defeq\frac{\ee^{2z}}{1+\ee^{2z}}=\frac{1+\tanh z}2,
\]
with $z(0)=-\infty$ and $z(1)=+\infty$.
Then $z(Q^+)=\frac12\ln\bigl(W_1(\bar Q)/W_0(\bar Q)\bigr)$ is additive across clauses.

To factorize the planted weight across clauses, we use the weight~\cite[Sections~4.3--4.6]{BS20}
\[
 f(q)\defeq 2\sqrt{q(1-q)}=\sech z(q).
\]
Let $q\in[0,1]$ be a symmetric random marginal ($q\stackrel{d}{=}1-q$), and let $S>0$ be a scale parameter.
Each marginal value $q$ receives weight $f(q)$, with zero weight at $q=0,1$.
The measure $\nu$ records this weighted mass at $z(q)$ and scales it by $S$, without normalization to total mass one.
For Borel $A\subseteq\mathbb R$ and bounded Borel $\psi:[0,1]\to\mathbb R$ with $\psi(0)=\psi(1)=0$, define the $f$-weighted measure $\nu$ on $\mathbb R$ and weighted pairing $\langle\psi,\nu\rangle$ by
\begin{equation}\label{eq:posterior-functionals}
\begin{gathered}
 \nu(A)\defeq S\cdot\E\bigl[\mathbb{1}\{0<q<1\text{ and }z(q)\in A\}\cdot f(q)\bigr],\\
 \langle\psi,\nu\rangle\defeq\int_{\mathbb R}\cosh (z)\,\psi(q(z))\nu(\mathrm{d}z)=S\cdot\E[\psi(q)].
\end{gathered}
\end{equation}
Define the \emph{$f$-weighted mass} $m(\nu)$ and \emph{scaled unfrozen probability} $u(\nu)$ by
\[
 m(\nu) \defeq\langle f,\nu\rangle=\nu(\mathbb R), \qquad u(\nu) \defeq\langle\mathbb{1}\{0<\cdot<1\},\nu\rangle=\int_{\mathbb R}\cosh z\,\nu(\mathrm{d}z)=S \cdot \Pr(0<q<1).
\]
Multiplication by $\cosh z=1/f(q(z))$ in the pairing reverses the $f$-weighting.
For fixed $S$, taking $\psi=\mathbb{1}\{\cdot\in B\}$ for Borel $B\subseteq(0,1)$ in \eqref{eq:posterior-functionals} recovers $\Pr(q\in B)$ from $\nu$.
Symmetry determines the remaining masses as $\Pr(q=0)=\Pr(q=1)=(1-u(\nu)/S)/2$.
For fixed $S$, $\nu$ determines the distribution of $q$, including its masses at $0$ and $1$.
The one-step planted update can therefore be expressed as an update of $\nu$.

A \emph{single-soft-child} clause has $0<s_C<1$, one unfrozen child, and $d-1$ child literals fixed to false.
If the unfrozen child's literal-true marginal is $q(z)$, this clause shifts $z(Q^+)$ by $\pm a(z)$, where
\[
 a(z)\defeq-\frac12\ln q(z)=\frac12\ln(1+\ee^{-2z}).
\]
Clauses with $0<s_C<1$ and at least two unfrozen children are called \emph{multi-soft-child} clauses.

To describe the total mass in the limiting recursion, define
\[
 g(q)\defeq 2\left(\sqrt q+\sqrt{1-q}-1\right),\qquad h(\nu)\defeq\langle g,\nu\rangle.
\]
For $q\in[0,1]$, the functions $g$ and $f$ satisfy
\[
 g(q)=\frac{2f(q)}{1+\sqrt q+\sqrt{1-q}},\qquad 1\le\sqrt q+\sqrt{1-q}\le\sqrt2.
\]
Thus $c_0f(q)\le g(q)\le f(q)$, where $c_0\defeq 2(\sqrt2-1)$.
Integration against the nonnegative measure $\cosh z\,\nu(\mathrm{d}z)$ preserves these inequalities, so $c_0m(\nu)\le h(\nu)\le m(\nu)$.

\begin{definition}[Soft-field recursion]\label{def:soft-recursion}
For a symmetric measure $\nu$ on $\mathbb R$ with $u(\nu)<\infty$ and $\beta>0$, the \emph{soft-field recursion} maps $\nu$ to the finite measure
\begin{equation}\label{eq:limiting-map}
 \mathcal R_\beta\nu\defeq\beta\ee^{h(\nu)}K_\nu,
\end{equation}
where $K_\nu$ is the law of $X-X'$ for independent copies $X,X'$ of $\sum_{z\in\Pi} a(z)$.
The Poisson point process $\Pi$ has intensity $w(z)\nu(\mathrm{d}z)$ on $\mathbb R$, with
\[
 w(z)\defeq 2\sqrt{q(z)} \cdot \cosh z=\sqrt{1+\ee^{2z}}.
\]
\end{definition}

By Poisson tilting (since $\int(w-2\cosh z)\dd\nu=\frac12h(\nu)-u(\nu)$), for a Poisson process $s=(s_1,\ldots,s_N)$ on $(0,1)$ with intensity $\rho_\nu(A)\defeq 4\int_{1-q(z)\in A}\cosh z\,\nu(\mathrm{d}z)$ and independent fair bits $\xi_1,\ldots,\xi_N\in\{0,1\}$, every bounded Borel $\psi:[0,1]\to\mathbb R$ with $\psi(0)=\psi(1)=0$ satisfies
\begin{equation}\label{eq:reference-clause-average}
 \langle\psi,\mathcal R_\beta\nu\rangle
 =\beta\ee^{2u(\nu)}\E_{\rho_\nu}\!\left[\frac{F(s)}2\psi\!\left(\frac{W_1(s)}{F(s)}\right)\right],
\end{equation}
with $m(\mathcal R_\beta\nu)=\beta \ee^{h(\nu)}$ and $u(\mathcal R_\beta\nu)=\beta \ee^{u(\nu)}$.

For the one-step update $Q\mapsto Q^+$ at parameter $\lambda>0$, set $S\defeq d\lambda$ and $\beta\defeq S\ee^{-\lambda}$.
Then $\beta\to\ee^{-\gamma}$ as $k\to\infty$ when $\lambda=\ln k+\ln\ln k+\gamma$ with fixed $\gamma$.
Let $\nu,\nu^+$ denote the $f$-weighted measures~\eqref{eq:posterior-functionals} at scale $S$ of independent fair symmetrizations of $Q,Q^+$.
A child is unfrozen with probability $u(\nu)/S$, which controls how often a clause has one or several unfrozen children.
The next lemma bounds the error in approximating $\nu^+$ by $\mathcal R_\beta\nu$ in terms of $u(\nu)$.

\begin{lemma}[Soft-field limit of the one-step planted update]\label{lem:planted-soft-limit}
For every input $Q\in[0,1]$ and $\lambda>0$,
\begin{equation}\label{eq:planted-scalar-envelopes}
 m(\nu^+)\le\beta\ee^{h(\nu)},\qquad u(\nu^+)\le\beta\ee^{u(\nu)}.
\end{equation}
If $\lambda\ge1$, then for every $U>0$, uniformly over inputs with $u(\nu)\le U$ and Borel $\psi:[0,1]\to[-1,1]$ vanishing at $0,1$,
\begin{equation}\label{eq:planted-one-step-error}
 \bigl|\langle\psi,\nu^+\rangle-\langle\psi,\mathcal R_\beta\nu\rangle\bigr|
 =O_{\beta,U}\!\left(\frac{u(\nu)^2}{\lambda}\right).
\end{equation}
The map $\cosh z\,\nu(\mathrm{d}z)\mapsto\cosh z\,\mathcal R_\beta\nu(\mathrm{d}z)$ is locally Lipschitz in total variation and in $\beta$ when $u(\nu)$ and $\beta$ are bounded.
\end{lemma}
\begin{proof}
In \pref{def:tree-updates}, a clause $C$ has i.i.d.\ child false-literal probabilities $\bar Q_{C,1},\ldots,\bar Q_{C,d}$ with product $s_C=\prod_{i=1}^d\bar Q_{C,i}$, mean $\E[s_C]=2^{-d}$, and $\Pr(s_C=1)=2^{-d}(1-u(\nu)/S)^d$.
Since $1-\sqrt{1-x}=\sum_{\ell\ge1}c_\ell x^\ell$ has $c_\ell>0$ and $1-a^d\le d(1-a)$ for $a=2\E[\bar Q_{C,1}^\ell]\in[0,1]$, we have
\[
\E\left[\sqrt{1-s_C}-1+s_C\right]\le d2^{1-d}\E\left[\sqrt{1-\bar Q_{C,1}}-1+\bar Q_{C,1}\right]=d2^{-d-1}\E\left[g(Q)\right]=h(\nu)/D.
\]
By the Poisson product formula and $\frac F2f(Q^+)=\prod_{C=1}^N\sqrt{1-s_C}$,
\begin{align*}
 m(\nu^+)&=S\exp(\lambda+D(\E[\sqrt{1-s_C}]-1))
 =\beta\exp(D\E[\sqrt{1-s_C}-1+s_C])\le\beta\ee^{h(\nu)}.
\end{align*}
Write $u=u(\nu)$ and let $\rho_h=2\lambda(1-u/S)^d$ be the hard-clause rate under the auxiliary Poisson law.
Only configurations with no hard clause contribute to $u(\nu^+)$; their auxiliary probability is $\ee^{-\rho_h}$, and the conditional mean of $F/2$ over the remaining clauses is $\ee^{-\lambda+\rho_h/2}$.
Thus
\[
 u(\nu^+)=S\ee^\lambda\ee^{-\rho_h}\ee^{-\lambda+\rho_h/2}
 =S\ee^{-\lambda(1-u/S)^d}\le S\ee^{-\lambda+u}=\beta\ee^u,
\]
where the inequality follows from $(1-u/S)^d\ge1-du/S=1-u/\lambda$. This proves \eqref{eq:planted-scalar-envelopes} for all $\lambda>0$.

For \eqref{eq:planted-one-step-error}, assume $\lambda\ge1$ and $u\le U$, and set $\theta\defeq(1-u/S)^{d-1}=1-O(u/\lambda)$.

Before planted reweighting, split the clauses with $s_C>0$ into hard clauses ($s_C=1$), single-soft-child clauses ($0<s_C<1$ with one child in $(0,1)$ and $d-1$ children fixed to false), and multi-soft-child clauses ($0<s_C<1$ with at least two children in $(0,1)$).

A single-soft-child clause has $d$ choices for the child in $(0,1)$, probability $\bigl(\frac12(1-u/S)\bigr)^{d-1}=2^{-(d-1)}\theta$ that the remaining $d-1$ children are false, and arrival rate $D=2^{d+1}\lambda$, so $D\cdot d\cdot 2^{-(d-1)}\theta=4S\theta$ identifies its intensity on $(0,1)$ as $\rho_s=\rho_{\theta\nu}$.

Writing $\rho_m$ for the intensity of multi-soft-child clauses on $(0,1)$, the three intensities satisfy
\begin{equation}\label{eq:clause-intensities}
 \rho_m((0,1))\le\frac{4u^2\ee^{u/\lambda}}{\lambda},\qquad
 \int s\dd\rho_s=2\theta u,\qquad
 \rho_h+\int s\dd\rho_s+\int s\dd\rho_m=2\lambda.
\end{equation}
Clauses with $s_C=0$ have no effect, and hard clauses contribute zero whenever $\psi$ vanishes at $0,1$.
Let $M_m$ count the clauses in $\rho_m$ and put $a_m=\int(1-s)\rho_m(\mathrm{d}s)\le\rho_m((0,1))$.
On $\{M_m=0\}$ with no hard clauses (auxiliary probability $\ee^{-\rho_h-\rho_m((0,1))}$), since $\rho_h+\rho_m((0,1))=2\lambda-2\theta u+a_m$, multiplying the planted prefactor $S\ee^\lambda$ by $\ee^{-\rho_h-\rho_m((0,1))}$ gives $\ee^{-a_m}\beta\ee^{2u(\theta\nu)}$.

For any Borel $0\le\psi\le1$ vanishing at $0,1$, independence of the Poisson processes and \eqref{eq:reference-clause-average} therefore give the exact finite-$k$ identity
\begin{equation}\label{eq:single-child-exact}
 S\E_{\rm pl}\bigl[\psi(Q^+)\mathbb{1}\{M_m=0\}\bigr]
 =\ee^{-a_m}\langle\psi,\mathcal R_\beta(\theta\nu)\rangle
\end{equation}
(in particular, when $\rho_m=0$, the finite-$k$ update is governed \emph{exactly} by $\mathcal R_\beta(\theta\nu)$),
while on $\{M_m\ge1\}$, using $F/2\le1$, absence of hard clauses on $\{0<Q^+<1\}$, and $2\lambda-\rho_h\le2u$,
\[
 S\E_{\rm pl}\bigl[\psi(Q^+)\mathbb{1}\{M_m\ge1\}\bigr]
 \le S\ee^{\lambda-\rho_h}\Pr_{\rm Pois}(M_m\ge1)
 \le\beta\ee^{2u}\rho_m((0,1)).
\]
Since $u(\mathcal R_\beta(\theta\nu))=\beta\ee^{\theta u}$ and $\ee^{u/\lambda}\le\ee^U$, these bounds imply $|\langle\psi,\nu^+\rangle-\langle\psi,\mathcal R_\beta(\theta\nu)\rangle|\le 2\beta\ee^{2u}\rho_m((0,1))=O_{\beta,U}(u^2/\lambda)$.

The same order holds for signed tests by separating their positive and negative parts.
Finally, replacing $\rho_s=\theta\rho_\nu$ by $\rho_\nu$ in \eqref{eq:reference-clause-average} changes the Poisson intensity by total mass $4u(1-\theta)=O(u^2/\lambda)$; coupling the two Poisson processes and comparing their prefactors gives both \eqref{eq:planted-one-step-error} and the local Lipschitz property of $\cosh z\,\mathcal R_\beta\nu(\mathrm{d}z)$.
\end{proof}

\paragraph{Phase transition of the soft-field recursion.}
Starting from the frozen boundary $\nu_0^\gamma=0$, define $\nu_{r+1}^\gamma=\mathcal R_{\ee^{-\gamma}}\nu_r^\gamma$, write $m_r^\gamma=m(\nu_r^\gamma)$ and $h_r^\gamma=h(\nu_r^\gamma)$, and define the critical threshold
\begin{equation}\label{eq:critical-definition}
 \gamma_*\defeq\inf\bigl\{\gamma\in\mathbb R:\sup_{r\ge0} m_r^\gamma<\infty\bigr\}.
\end{equation}

The following theorem bounds $\gamma_*$ and characterizes the transition in $m_r^\gamma$ at this threshold.

\begin{theorem}[Critical threshold of the soft-field recursion]\label{thm:critical-threshold}
The critical threshold $\gamma_*$ of the soft-field recursion satisfies $c_0\le\gamma_*<1$. Specifically,
\begin{itemize}
    \item Subcritical regime and criticality: for $\gamma\ge\gamma_*$, $m_r^\gamma\le1$ for every $r\ge0$, with $m_r^{\gamma_*}\uparrow1$ and $h_r^{\gamma_*}\uparrow\gamma_*$ as $r\to\infty$;
    \item Supercritical regime: for $\gamma<\gamma_*$, both $m_r^\gamma\uparrow\infty$ and $h_r^\gamma\uparrow\infty$ as $r\to\infty$.
\end{itemize}
\end{theorem}

To prove \pref{thm:critical-threshold}, we first establish stochastic monotonicity of $\mathcal R_\beta$ against concave test functions, which yields monotonicity of the iterates $\nu_r^\gamma$ in $r$ and $-\gamma$ together with Lipschitz control of $h(\mathcal R_\beta\nu)$.
For symmetric measures $\nu,\nu'$ with finite $u$, write $\nu'\preceq\nu$ if $\langle\psi,\nu'\rangle\le\langle\psi,\nu\rangle$ for every continuous concave $\psi:[0,1]\to\mathbb R$ with $\psi(0)=\psi(1)=0$.
\begin{lemma}[Poisson comparison]\label{lem:information}
The map $\mathcal R_\beta$ preserves $\preceq$, and for $\nu'\preceq\nu$,
\begin{equation}\label{eq:potential-contraction}
 0\le h(\mathcal R_\beta\nu)-h(\mathcal R_\beta\nu')
 \le\beta \ee^{h(\nu)}[h(\nu)-h(\nu')].
\end{equation}
The iterates satisfy $\nu_r^\gamma\preceq\nu_{r+1}^\gamma$ and $\nu_r^{\gamma'}\preceq\nu_r^\gamma$ whenever $\gamma'\ge\gamma$.
Moreover, $m_r^\gamma-h_r^\gamma$ is nondecreasing in $r$, and for fixed $r$, both $m_r^\gamma$ and $h_r^\gamma$ are continuous in $\gamma$.
\end{lemma}
\begin{proof}
For a continuous concave $\psi:[0,1]\to\mathbb R$ with $\psi(0)=\psi(1)=0$, set $\Psi(v_0,v_1)=(v_0+v_1)\psi(v_1/(v_0+v_1))$ with $\Psi(0,0)=0$. Adding a clause of parameter $s\in(0,1)$ with fair root sign $\xi\in\{0,1\}$ changes $\exp(\int s\dd\rho)\Psi(v_0,v_1)$ in expectation by
\[
 B_s\Psi(v)\defeq\frac12 \left[\Psi((1-s)v_0,v_1)+\Psi(v_0,(1-s)v_1)\right]-(1-s)\Psi(v).
\]
Concavity and $1$-homogeneity of $\Psi$ imply that $s\mapsto B_s\Psi(v)$ is concave on $[0,1]$ and vanishes at $0,1$.
Along $\rho_t=(1-t)\rho_{\nu'}+t\rho_\nu$, differentiating the Poisson expectation in \eqref{eq:reference-clause-average} gives
\[
 \frac{\dd}{\dd t}\left(\ee^{\int s\dd\rho_t}\E_{\rho_t}[\Psi(v_0W_0,v_1W_1)]\right)
 =\ee^{\int s\dd\rho_t}\E_{\rho_t}\!\left[\int B_s\Psi(v_0W_0,v_1W_1)(\rho_\nu-\rho_{\nu'})(\mathrm{d}s)\right]\ge0,
\]
so $\mathcal R_\beta$ preserves $\preceq$.
Testing $\mathcal R_\beta\nu'\preceq\mathcal R_\beta\nu$ with the concave functions $g$ and $f-g$ bounds $h(\mathcal R_\beta\nu)-h(\mathcal R_\beta\nu')$ between $0$ and $m(\mathcal R_\beta\nu)-m(\mathcal R_\beta\nu')=\beta(\ee^{h(\nu)}-\ee^{h(\nu')})\le\beta\ee^{h(\nu)}[h(\nu)-h(\nu')]$, proving \eqref{eq:potential-contraction}.
Monotonicity of $\nu_r^\gamma$ and $m_r^\gamma-h_r^\gamma$ follows by induction from $\nu_0^\gamma=0$ and concavity of $f-g$, while continuity in $\gamma$ follows from \eqref{eq:potential-contraction} and $m_{r+1}^\gamma=\ee^{h_r^\gamma-\gamma}$.
\end{proof}

We next use a Fourier test function $F_t(z)=1-\cos(tz)$ controlled by $g$ to show that any supercritical mass $m_{r+1}^\gamma\ge 1+\varepsilon$ forces a quadratic potential increment $h_{r+1}^\gamma-h_r^\gamma\ge\varepsilon^2/7$.

\begin{lemma}[Supercritical drift]\label{lem:quant-drift}
There is an absolute constant $\varepsilon_0\in(0,1/10)$ such that for any $\gamma\le1$, $0<\varepsilon\le\varepsilon_0$, and $r\ge1$, if $m_r^\gamma\le16$ and $m_{r+1}^\gamma\ge1+\varepsilon$, then $h_{r+1}^\gamma-h_r^\gamma\ge\varepsilon^2/7$.
\end{lemma}
\begin{proof}
Fix $\gamma\le1$ and write $\nu_r=\nu_r^\gamma$, $m_r=m_r^\gamma$, $h_r=h_r^\gamma$, and $\beta=\ee^{-\gamma}\ge\ee^{-1}>1/3$.

For $F_t(z)\defeq 1-\cos(tz)$ ($0<t\le 1/10$), the symmetric compound Poisson law $K_{\nu_r}$ in \eqref{eq:limiting-map} has characteristic function $\int\cos(tz)\,K_{\nu_r}(\mathrm{d}z)=\ee^{-A(t)}$ with $A(t)\defeq 2\int w(z)F_t(a(z))\nu_r(\mathrm{d}z)$.
Since $\nu_{r+1}=m_{r+1}K_{\nu_r}$, integrating $F_t$ against $\nu_{r+1}$ gives the identity
\begin{equation}\label{eq:fourier-mass-identity}
 \int F_t\dd\nu_{r+1}=m_{r+1}\bigl(1-\ee^{-A(t)}\bigr).
\end{equation}

We bound the increment $\int F_t\dd\nu_{r+1}-\int F_t\dd\nu_r$ from above by $3(h_{r+1}-h_r)$ and from below by showing that $A(t)>\int F_t\dd\nu_r$ for small $t>0$.

For the upper bound, set $\psi_t(q)\defeq f(q)F_t(z(q))$, so that $\psi_t(q(z))=\sech z\,F_t(z)$ and $\langle\psi_t,\nu\rangle=\int F_t\dd\nu$ by \eqref{eq:posterior-functionals}.
Using $q'(z)=2q(1-q)=\frac12 f(q)^2$ and the chain rule $4q^2(1-q)^2\phi''(q)=(\partial_z^2+2\tanh z\,\partial_z)\phi(q(z))$, where $(\partial_z^2+2\tanh z\,\partial_z)\sech z=-\sech z$ and the first-derivative cross terms $2(\sech z)'F_t'(z)+2\tanh z\sech z\,F_t'(z)=0$ cancel, direct differentiation gives
\[
 \begin{aligned}
 4q^2(1-q)^2g''(q)&=-f(q)\bigl(q^{3/2}+(1-q)^{3/2}\bigr)\le-\frac1{\sqrt 2}f(q),\\
 4q^2(1-q)^2\psi_t''(q)&=\sech z\bigl(F_t''(z)-F_t(z)\bigr)=f(q)\bigl[(1+t^2)\cos(tz)-1\bigr].
 \end{aligned}
\]
For $0<t\le1/10$, we have $|\psi_t''(q)|\le(2+t^2)\sqrt 2\,|g''(q)|\le 3|g''(q)|$, so $3g\pm\psi_t$ are concave on $(0,1)$.
Testing $\nu_r\preceq\nu_{r+1}$ (\pref{lem:information}) against $3g-\psi_t$ yields
\begin{equation}\label{eq:fourier-potential-upper}
 \int F_t\dd\nu_{r+1}-\int F_t\dd\nu_r\le 3(h_{r+1}-h_r).
\end{equation}

For the lower bound, symmetry of $\nu_r$ averages the integrand of $A(t)-\int F_t\dd\nu_r$ over $\pm z$:
\[
 A(t)-\int F_t\dd\nu_r
 =\int_{\mathbb R}\bigl(w(z)F_t(a(z))+w(-z)F_t(a(-z))-F_t(z)\bigr)\nu_r(\mathrm{d}z).
\]
Fix $z\ge0$, and note that $a(-z)=z+a(z)$, $\ee^{-2z}/4\le a(z)\le\ee^{-2z}/2\le 1/2$, and $w(\pm z)>1$:
\begin{itemize}
\item If $z\le 1/t$, then $tz\le ta(-z)\le 3/2$. Since $F_t$ is increasing on $[0,3/(2t)]$ with $F_t(x)\ge t^2x^2/4$, the $-z$ term satisfies $w(-z)F_t(a(-z))\ge F_t(z)$, while the $+z$ term gives $w(z)F_t(a(z))\ge t^2a(z)^2/4\ge t^2\ee^{-4z}/64$.
\item If $z>1/t$, then $|F_t'|\le t$ gives $w(-z)F_t(a(-z))-F_t(z)\ge F_t(z+a(z))-F_t(z)\ge -ta(z)\ge-\frac12 t\ee^{-2/t}$, and $t^2\ee^{-4z}/64\le\frac12 t\ee^{-2/t}$.
\end{itemize}
Combining both cases shows that for all $z\ge0$ (and hence all $z\in\mathbb R$ by symmetry),
\[
 w(z)F_t(a(z))+w(-z)F_t(a(-z))-F_t(z)
 \ge \frac{t^2}{64}\ee^{-4|z|}-t\ee^{-2/t}.
\]
For the concave test function $\psi_0(q)\defeq 2\min\{q,1-q\}$, we have $\cosh z\,\psi_0(q(z))=\ee^{-|z|}$ and hence $\langle\psi_0,\nu\rangle=\int\ee^{-|z|}\dd\nu$ by \eqref{eq:posterior-functionals}.
Testing $\nu_1=\beta\delta_0\preceq\nu_r$ against $\psi_0$ gives $\int \ee^{-|z|}\dd\nu_r\ge\beta\ge1/3$, and Hölder's inequality $(\int\ee^{-|z|}\dd\nu_r)^4\le m_r^3\int\ee^{-4|z|}\dd\nu_r$ yields $\int \ee^{-4|z|}\dd\nu_r\ge\beta^4/m_r^3\ge(1/3)^4/16^3$.
Integrating the pointwise bound against $\nu_r$ (with $m_r\le 16$) therefore gives $A(t)-\int F_t\dd\nu_r\ge c t^2$ for an absolute constant $c>0$ and all sufficiently small $t>0$, uniformly in $r,\gamma$.

Choosing $\varepsilon_0$ small enough, continuity of $A$ gives $A(t)=\varepsilon/(1+\varepsilon)>\int F_t\dd\nu_r$ for some $0<t\le1/10$. Combining \eqref{eq:fourier-mass-identity} and \eqref{eq:fourier-potential-upper} with $\int F_t\dd\nu_r<A(t)$, $m_{r+1}\ge 1+\varepsilon$, and $1-\ee^{-x}\ge x-x^2/2$ at $x=A(t)=\varepsilon/(1+\varepsilon)$ then gives
\begin{align*}
 3(h_{r+1}-h_r)&\ge\int F_t\dd\nu_{r+1}-\int F_t\dd\nu_r \ge(1+\varepsilon)(1-\ee^{-A(t)})-A(t)\\
 &\ge(1+\varepsilon)\Bigl(A(t)-\frac{A(t)^2}{2}\Bigr)-A(t) =\frac{\varepsilon^2}{2(1+\varepsilon)}\ge\frac{3\varepsilon^2}{7}.\qedhere
\end{align*}
\end{proof}

\begin{proof}[Proof of \pref{thm:critical-threshold}]
For $\gamma\ge1$, $m_{r+1}^\gamma\le\ee^{m_r^\gamma-\gamma}$ keeps $m_r^\gamma\le1$ by induction from $m_0^\gamma=0$.

Whenever $m_r^\gamma$ is bounded, the monotone limits $M=\lim_r m_r^\gamma$ and $H=\lim_r h_r^\gamma$ satisfy $M=\ee^{H-\gamma}$ and $c_0M-\ln M\le H-\ln M=\gamma$.

If $m_r^\gamma$ is bounded and $m_{r_1}^\gamma>1$ for some $r_1$, then $\gamma<1$ and $M<4$: the function $x\mapsto c_0x-\ln x$ is increasing for $x\ge4$ and $4c_0-\ln4>1$.
Monotonicity of $m_r^\gamma$ gives $m_{r+1}^\gamma\ge m_{r_1}^\gamma>1$ for all $r\ge r_1$, so \pref{lem:quant-drift} forces $h_r^\gamma$ to grow by at least a positive constant per step, contradicting $H<\infty$.

Thus boundedness holds on a closed interval $[\gamma_*,\infty)$ (\pref{lem:information}), is equivalent to $m_r^\gamma\le1$ for all $r$, and requires $\gamma\ge c_0M-\ln M\ge c_0$, while for $\gamma<\gamma_*$ both $m_r^\gamma$ and $h_r^\gamma\ge c_0m_r^\gamma$ diverge.

At $\gamma=\gamma_*$, suppose $M<1$. For $\gamma=\gamma_*-\eta$ with $\eta>0$, set $e_r\defeq h_r^\gamma-h_r^{\gamma_*}$.
By \eqref{eq:potential-contraction}, for sufficiently small $\epsilon,\eta>0$, whenever $e_r\le\epsilon$ we have
\[
 e_{r+1}
 \le M\ee^{\eta+e_r}e_r+(\ee^\eta-1)H
 \le\frac{1+M}{2}e_r+2\eta
 \le\epsilon
\]
provided $\eta\le(1-M)\epsilon/4$.
By induction from $e_0=0$, we obtain $e_r\le\epsilon$ for all $r\ge0$, contradicting the divergence of $h_r^\gamma$ at $\gamma=\gamma_*-\eta<\gamma_*$.
Hence $M=1$ and $H=\gamma_*$, and \pref{lem:information} gives
\begin{equation}\label{eq:critical-gap}
 1-\gamma_*
 =\lim_r(m_r^{\gamma_*}-h_r^{\gamma_*})
 \ge m_1^{\gamma_*}-h_1^{\gamma_*}
 =(1-c_0)\ee^{-\gamma_*}>0.\qedhere
\end{equation}
\end{proof}

Finally, iterating the quadratic drift of \pref{lem:quant-drift} yields a quantitative near-critical escape bound for $\gamma\le\gamma_*-\delta$, which will govern the depth $L=O(\delta^{-1})$ of our local update in \pref{sec:local-updates}.

\begin{lemma}[Near-critical escape]\label{lem:quant-bands}
For sufficiently small $\delta>0$ and $\gamma\le\gamma_*-\delta$, the first depth $r_0=\min\{r\ge1:h_r^\gamma>13\}$ satisfies
\[
 r_0=O(\delta^{-1}),\qquad
 \prod_{r=1}^{r_0-1}\max\{1,m_{r+1}^\gamma\}=O(\delta^{-8}).
\]
\end{lemma}
\begin{proof}
Write $m_r=m_r^\gamma$ and $h_r=h_r^\gamma$.
We may assume $\ee^{-\gamma}<16$, since otherwise $h_1>13$ and $r_0=1$.
For $1\le r<r_0$, put $\ell_r\defeq\ln m_{r+1}=h_r-\gamma<16$.
Since $m_r\le13/c_0<16$, by \pref{lem:quant-drift}, for $\ell_r>0$,
\begin{equation}\label{eq:quant-variable-drift}
 \ell_{r+1}-\ell_r\ge \min\{\ell_r,\varepsilon_0\}^2 / 7.
\end{equation}

At criticality, set $e_r\defeq\gamma_*-h_r^{\gamma_*}$.
By the monotonicity in \pref{lem:information} and the critical limits in \pref{thm:critical-threshold},
$m_{r+1}^{\gamma_*}-h_{r+1}^{\gamma_*}\le1-\gamma_*$ and $0\le e_r\le e_0=\gamma_*<1$.
Hence
\[
 e_{r+1}=\gamma_*-h_{r+1}^{\gamma_*}
 \le1-m_{r+1}^{\gamma_*}
 =1-\ee^{-e_r}
 \le e_r-e_r^2/3,
\]
where the last inequality holds for $0\le e_r\le1$.
Thus $e_r\le3/(r+3)$ and $\ell_r\ge\delta-3/(r+3)$.
Outside $\delta/2\le\ell_r<\varepsilon_0$, there are $O(\delta^{-1})$ steps, contributing $O(1)$ to $\sum_{r<r_0}(\ell_r)_+$.

For $\delta/2\le\ell_r<\varepsilon_0$, \eqref{eq:quant-variable-drift} implies $1/\ell_r-1/\ell_{r+1}\ge1/8$ and $\ln(\ell_{r+1}/\ell_r)\ge\ell_r/8$.
Telescoping with $\ell_{r+1}\le\ee^{\varepsilon_0}+\ln16$ yields $O(\delta^{-1})$ further steps and $\sum_{r<r_0}(\ell_r)_+\le8\ln(1/\delta)+O(1)$, as required.
\end{proof}

\subsection{Local update on trees}\label{sec:local-updates}

We now use the supercritical growth of the soft-field recursion (\pref{def:soft-recursion}) to construct a reversible local update on trees that flips the root along a short path of single-variable flips.

Fix an integer $L\ge1$, and let $\Phi=(V,\+C)$ be a tree formula with root variable $o\in V$ and depth at most $L$.
Its depth-$L$ boundary consists of the variables at depth $L$ and may be empty.
For each $v\in V$, let $V_v$ consist of $v$ and its descendant variables, and let $\Phi_v$ be the subformula on $V_v$.

Given a retention parameter $\vartheta\in[0,1]$ and an input solution $x\in\Omega_\Phi$, our local update $\mathsf{LocalUpdate}(x)$ (\pref{alg:local-update}) initializes a revealed (frozen) set $R\subseteq V$ with the depth-$L$ boundary, traverses the tree bottom-up enlarging $R$ by three rules, and resamples $y\sim\mu_{\Phi,x_R}$:
\begin{itemize}
\setlength{\itemsep}{2pt}\setlength{\parskip}{0pt}
\item \emph{Multi-soft-child clause pruning (\pref{line:reveal-children}):} freezes all child subtrees of any child clause with at least two children whose literals have positive probability of being true ($q_{C,u}>0$), eliminating multi-soft-child clauses so that the finite-$k$ marginal update is governed directly by $\mathcal R_\beta$ (\eqref{eq:single-child-exact});
\item \emph{Marginal window test (\pref{line:reveal-marginal}--\pref{line:reveal-retain}):} freezes $V_v$ when $q_v\notin[a_k,1-a_k]$, preventing extreme half-log-odds from inflating unrevealed subtrees, with negligible truncation loss;
\item \emph{Depth-$1$ subtree retention (\pref{line:initial-reveal}):} independently freezes each depth-$1$ subtree with probability $1-\vartheta$, which tunes the potential entering the root to a controlled level so that the update achieves the required normalized expected root flow.
\end{itemize}

\begin{algorithm}[ht]
\caption{$\mathsf{LocalUpdate}(x)$}\label{alg:local-update}
\SetKwInOut{Data}{Data}
\SetKwInOut{Input}{Input}
\SetKwInOut{Output}{Output}
\Data{An integer $L\ge1$, a tree formula $\Phi=(V,\+C)$ with root variable $o\in V$ and depth at most $L$, and a retention parameter $\vartheta\in[0,1]$.}
\Input{A solution $x\in\Omega_\Phi$.}
\Output{A resampled solution $y\in\Omega_\Phi$.}
$a_k\gets\bigl(1+\exp(2(\max\{1,\ln\ln k\})^{3/4})\bigr)^{-1}$\;
Draw independent $\eta_v\gets\operatorname{Bernoulli}(\vartheta)$ for all $v$ at depth $1$\;
$R\gets\{v\in V: \operatorname{depth}(v)=L\}\cup\displaystyle\bigcup_{\operatorname{depth}(v)=1, \eta_v=0}V_v$\label{line:initial-reveal}\;
\ForEach{variable $v\in V$ in decreasing order of depth}{
    \ForEach{child clause $C$ of $v$}{
        \ForEach{$u\in\operatorname{vbl}(C)\setminus\{v\}$}{
            $q_{C,u}\gets\mu_{\Phi_u,x_{R\cap V_u}}(\text{$u$'s literal in $C$ is true})$\;
        }
        \lIf{$\bigl|\{u\in\operatorname{vbl}(C)\setminus\{v\}:q_{C,u}>0\}\bigr|\ge2$\label{line:reveal-children}}{$R\gets R\cup\bigcup_{u\in\operatorname{vbl}(C)\setminus\{v\}}V_u$}
    }
    $q_v\gets\mu_{\Phi_v,x_{R\cap V_v}}(\sigma_v=1)$\label{line:reveal-marginal}\;
    \lIf{$q_v\notin[a_k,1-a_k]$\label{line:reveal-retain}}{$R\gets R\cup V_v$}
}
Sample $y\sim\mu_{\Phi,x_R}$ and \Return $y$\;
\end{algorithm}

For a formula $\Phi$ and $x\in\Omega_\Phi$, let $\mathcal L_x\subseteq\Phi$ consist of the clauses with one or two $x$-true literals.
The next theorem establishes the main properties of \pref{alg:local-update}: reversibility and single-flip reachability, dependence only on $\mathcal L_x$, and a \emph{planted root flow} lower bound showing that below $\gamma_*$, the update flips the root with relatively high probability per changed variable.

\begin{theorem}[Local update on a tree]\label{thm:local}
For any integer $L\ge1$, tree formula $\Phi$ rooted at $o$ with depth at most $L$, retention probability $\vartheta\in[0,1]$, and $x\in\Omega_\Phi$, let $y\gets\mathsf{LocalUpdate}(x)$ (\pref{alg:local-update}).
\begin{enumerate}[label=\textup{(\roman*)}]
\item \emph{(Reachability and reversibility.)} It always holds that $\Delta(x,y)\cap\{v\in V:\operatorname{depth}(v)=L\}=\varnothing$, $x$ and $y$ are joined by a path of length $\Ham(x,y)$ in $G_1(\Phi)$, and
\[
 \Pr(\mathsf{LocalUpdate}(x')=y')=\Pr(\mathsf{LocalUpdate}(y')=x'), \quad \forall x',y'\in\Omega_\Phi.
\]
\item \emph{(Clause pruning.)} The law of $\Delta(x,y)$ is unchanged when $\Phi$ is replaced by $B_L(\mathcal L_x,o)$, keeping $L,\vartheta$ fixed.
\item \emph{(Planted root flow.)} For sufficiently large $k$ and every
\[
 0\le\lambda_+\le\ln k+\ln\ln k+\gamma_*-\frac{1}{(\ln\ln k)^{1/16}}+\frac{1}{\ln k},
\]
there exist $L=O((\ln\ln k)^{1/16})$ and $\vartheta\in[0,1]$ such that for $(\Phi,x)$ sampled to depth $L$ from $\mathsf{Tree}_{\mathrm{pl}}(\lambda_+)$ (\pref{def:planted-root-rate}),
\[
 \E\left[\frac{\mathbb{1}\{y_o\ne x_o\}}{\max\{1,\Ham(x,y)\}}\right]
 \ge\epsilon_k:=\frac{(\ln k)^{10}}k.
\]
\end{enumerate}
\end{theorem}

For the analysis of \pref{alg:local-update} on $(\Phi,x)\sim\mathsf{Tree}_{\mathrm{pl}}(\lambda_+)$ with $\lambda_+>0$, set $T\defeq(\max\{1,\ln\ln k\})^{3/4}$ (so $a_k=(1+\ee^{2T})^{-1}$ and the marginal test in \pref{line:reveal-retain} is equivalent to $|z(q_v)|> T$), $S\defeq d\lambda_+$, and $\beta\defeq S\ee^{-\lambda_+}$.

For a variable $v$, put $r=L-\operatorname{depth}(v)\in\{0,1,\dots,L\}$ and 
let $R_v\subseteq V_v$ and $q_v=\mu_{\Phi_v,x_{R_v}}(\sigma_v=1)$ denote its revealed set and marginal in $\Phi_v$ after running the bottom-up reveal (\pref{line:reveal-children}--\pref{line:reveal-retain}) through $v$ and, at depth $1$ ($r=L-1$), freezing $V_v$ if $\eta_v=0$ (so $q_v\in\{0,1\}$ when $v\in R_v$).

By sign symmetry of $\mathsf{Tree}_{\mathrm{pl}}(\lambda_+)$, $q_v\stackrel{d}{=}1-q_v$; let $\widehat\nu_r$ be its $f$-weighted measure at scale $S$ from \eqref{eq:posterior-functionals} and define the $f$-weighted unrevealed size
\[
 \mathcal V_r\defeq S\,\E\bigl[f(q_v)\,|V_v\setminus R_v|\bigr].
\]
For $0\le r<L$, set $\theta_r\defeq(1-u(\widehat\nu_r)/S)^{d-1}$ and $\bar\nu_{r+1}\defeq\mathcal R_\beta(\theta_r\widehat\nu_r)$, and let $\vartheta_r=\vartheta$ for $r=L-2$ (the transition to depth $1$) and $\vartheta_r=1$ otherwise.

The next lemma connects $\mathsf{LocalUpdate}$ with the soft-field recursion (\pref{def:soft-recursion}) through the operator $\mathcal R_\beta(\theta_r\widehat\nu_r)$.

\begin{lemma}[Recursion for $\mathsf{LocalUpdate}$]\label{lem:actual-recursion}
Starting from $\widehat\nu_0=0$ and $\mathcal V_0=0$, for each $0\le r<L$, the parent's $f$-weighted measure before its marginal test and possible depth-$1$ freeze is $\bar\nu_{r+1}$, and after both is
\begin{align}
 \widehat\nu_{r+1}&=\vartheta_r\mathbb{1}_{[-T,T]}\bar\nu_{r+1},\label{eq:actual-weighted-recursion}\\
 \mathcal V_{r+1}&\le m(\widehat\nu_{r+1})+4\ee^T\vartheta_r\theta_r\,m(\bar\nu_{r+1})\,\mathcal V_r.\label{eq:size-recursion}
\end{align}
\end{lemma}
\begin{proof}
Postpone each initial depth-$1$ freeze until after that subtree root's marginal test. For fixed coins, retained subtrees evolve identically, and failed subtrees are still fully revealed before $o$ is processed, so the final $R$ is unchanged.
Work under the auxiliary law of \pref{def:tree-updates}, before the parent's planted reweighting. A child transcript records its formula, literal sign in $C$, coins, revealed set and revealed values. Every reveal decision uses only previously recorded values; hence, conditional on these transcripts, child assignments have independent uniform conditional laws. Clause pruning either keeps the child transcripts or reveals all their values. These operations act separately on independent clause marks, preserving their marked Poisson law before the parent's marginal test.

Let $\rho$ be the resulting intensity of $s=\prod_{u\in\operatorname{vbl}(C)\setminus\{v\}}(1-q_{C,u})$, with $q_{C,u}$ recomputed after clause pruning.
Only single-soft-child clauses remain soft, with intensity $\rho_{\theta_r\widehat\nu_r}$ as in \eqref{eq:clause-intensities}.
Before and after pruning, $s$ is the posterior probability of the same all-false event. The tower property therefore gives $\int_{[0,1]}s\dd\rho=2\lambda_+$, so the new forcing rate is $\rho(\{1\})=2\lambda_+-2u(\theta_r\widehat\nu_r)$.
The planted density remains $\ee^{\lambda_+}F/2$, since $F/2$ is the satisfaction probability conditional on all enlarged transcripts. Thus \eqref{eq:single-child-exact}, with $\rho_m=0$, gives $S\E_{\rm pl}[\psi(Q^+)]=\langle\psi,\bar\nu_{r+1}\rangle$ for every bounded Borel $\psi$ vanishing at $0,1$; the marginal window and independent depth-$1$ retention give \eqref{eq:actual-weighted-recursion}.

When $v\notin R_v$, $|V_v\setminus R_v|=1+\sum_{u\notin R_u}|V_u\setminus R_u|$ over the active single-soft-child clauses of $v$.
Before clause pruning, attach the size mark $A_u=|V_u\setminus R_u|$ to the symmetrized child marginal $q$, giving the joint measure $S\E[f(q)\delta_{(z(q),A_u)};\,0<q<1]$ with $z$-marginal $\widehat\nu_r$.
The parent's $f$-weight satisfies $\frac F2f(Q^+)=\prod_C\sqrt{1-s_C}$, so Poisson tilting multiplies this joint measure by $\theta_rw(z)$ for each root sign, with total prefactor $m(\bar\nu_{r+1})$. Thus the conditional size law given $z$ is unchanged.
Dropping the parent's marginal test and applying Campbell's formula, while retaining its independent depth-$1$ retention factor, gives
\[
 \begin{aligned}
 \mathcal V_{r+1}-m(\widehat\nu_{r+1})
 &\le2\vartheta_r\theta_r m(\bar\nu_{r+1})\int w(z)\E[|V_u\setminus R_u|\mid z(q_{C,u})=z]\widehat\nu_r(\mathrm{d}z)\\
 &\le4\ee^T\vartheta_r\theta_r m(\bar\nu_{r+1})\mathcal V_r.
 \end{aligned}
\]
Here the factor $2$ counts the two root signs; the last inequality uses $w(z)\le2\ee^T$ on $[-T,T]$ and \eqref{eq:posterior-functionals} to identify the integral without $w$ as $\mathcal V_r$.
\end{proof}

Before proving \pref{thm:local}, we bound the first absolute moment and central mass of the compound-Poisson difference law $K_\nu$ (\pref{def:soft-recursion}) to control the marginal-window truncation.

\begin{lemma}[Half-log-odds absolute moment and central mass]\label{lem:output}
Let $\nu$ be a symmetric finite measure supported on $[-T,T]$ with $T\ge2$, $m=m(\nu)$, and $h=h(\nu)$.
If $Z\sim K_\nu$, then
\begin{align}
 \E[|Z|]&\le2\int w(z)a(z)\nu(\mathrm{d}z)\le\int(|z|+2)\nu(\mathrm{d}z),\label{eq:jump-first-moment}\\
 K_\nu([-T,T])&\ge\frac1{10(1+h)}.\label{eq:central-mass}
\end{align}
\end{lemma}
\begin{proof}
The triangle inequality for $Z=X-X'$ and the bound $w(z)a(z)+w(-z)a(-z)\le z+2$ for $z\ge0$ (from $a(-z)=z+a(z)$ and $a(z)\le\ee^{-2z}/2$) give \eqref{eq:jump-first-moment}.
For \eqref{eq:central-mass}, if $\E[X]=0$ then $K_\nu([-T,T])=1$; otherwise Markov's inequality gives $\Pr(X\in[0,2\E[X]])\ge1/2$.
Partitioning $[0,2\E[X]]$ into at most $1+2\E[X]/T$ subintervals $I$ of length at most $T$ and using $2\E[X]=2\int w(z)a(z)\nu(\mathrm{d}z)\le(T+2)m$, Cauchy--Schwarz, $T\ge2$, and $c_0m\le h$ gives
\[
 K_\nu([-T,T])\ge\sum_I\Pr(X\in I)^2
 \ge\frac1{4(1+2\E[X]/T)}
 \ge\frac1{4(1+2m)}\ge\frac1{10(1+h)}. \qedhere
\]
\end{proof}

\begin{proof}[Proof of \pref{thm:local}]
For (i), since $\{v\in V:\operatorname{depth}(v)=L\}\subseteq R$, $\Delta(x,y)\cap\{v\in V:\operatorname{depth}(v)=L\}=\varnothing$.
Because \pref{alg:local-update} inspects $x$ only on variables added to $R$, for fixed coins every solution extending $x_R$ produces the same revealed set $R$, which gives symmetry of the uniform conditional resampling $y\sim\mu_{\Phi,x_R}$ for fixed coins; averaging over the coins preserves symmetry.

If $\Delta(x,y)\ne\varnothing$ and no $v\in\Delta(x,y)$ could be flipped in $x$ within $\Omega_\Phi$, each $v\in\Delta(x,y)$ would have a distinct clause $C_v$ whose unique $x$-true literal is $v$ and which contains another variable of $\Delta(x,y)$ (since $y\models C_v$), giving $2|\Delta(x,y)|$ vertices and at least $2|\Delta(x,y)|$ edges in the induced factor subgraph, contradicting acyclicity; repeating single-variable flips connects $x$ to $y$ by a path of length $\Ham(x,y)$ in $G_1(\Phi)$.

For (ii), any clause $C$ with at least three $x$-true literals has at least two on children $u$, each with $q_{C,u}>0$, so \pref{line:reveal-children} freezes all child subtrees of $C$ at $x$; deleting $C$ and its child subtrees leaves the run on $B_L(\mathcal L_x,o)$ unchanged.

For (iii), if $\lambda_+=0$, take $L=1$ and $\vartheta=1$:
the tree consists only of its root, which flips with probability
$1/2\ge\epsilon_k$.
Now assume $\lambda_+>0$ and sample $(\Phi,x)$ to depth $L$
from $\mathsf{Tree}_{\mathrm{pl}}(\lambda_+)$
(\pref{def:planted-root-rate}).
Conditional on the unsigned tree, every satisfying pair $(\Phi,x)$
has probability $2^{-|V|}(2^k-1)^{-|\+C|}$, so $x$ has conditional
law $\mu_\Phi$ given $\Phi$ (and is conditionally uniform given
the depth-$L$ variables).
Thus, when $x\sim\mu_\Phi$ independently of the coins, $x$ and $y\sim\mu_{\Phi,x_R}$ are conditionally independent and uniform given $(\Phi,R,x_R)$ (also after conditioning on the depth-$L$ variables), so with $q_o\defeq\mu_{\Phi,x_R}(\sigma_o=1)$ we have
\begin{equation}\label{eq:root-flip-identity}
 \Pr(y_o\ne x_o\mid\Phi,R,x_R)=2q_o(1-q_o).
\end{equation}
If $\lambda_+\le\ln k-10\ln\ln k-1$, take $L=1$ and $\vartheta=1$: only the root can change, its marginal belongs to $\{0,1/2,1\}$, and the marginal window retains every unfrozen root, so it flips with probability $\ee^{-\lambda_+}/2\ge\epsilon_k$.
Otherwise set $\delta=(\ln\ln k)^{-1/16}$ and introduce the target potential $H=12\ln\ln k$, so $\frac12\ln k\le\lambda_+\le2\ln k$ and $\beta\ge\ee^{-\gamma_*+\delta}>\ee^{-1}$.

The window parameter $T=\delta^{-12}=o(\ln\ln k)$ is large enough relative to the $O(\delta^{-1})$ bottleneck steps and $O(\delta^{-8})$ amplification of \pref{lem:quant-bands} to make the cumulative truncation loss $O(\delta^{-10}/T)=O(\delta^2)$ negligible, yet small enough that the sibling-pruning error $O(\delta^{-9}\ee^T/\ln k)$ is $o(1)$ and the update-size overhead $\exp(O(\delta^{-1}T))=(\ln k)^{o(1)}$ remains manageable. The target potential $H=12\ln\ln k$ then supplies the root-mass lower bound $m_L\ge\beta\ee^{H-o(1)}/[10(1+H)]\ge(\ln k)^{12-o(1)}$, large enough to overcome this overhead and achieve normalized root flow at least $\epsilon_k$.

Set $\widetilde\nu_{r+1}\defeq\mathbb{1}_{[-T,T]}\bar\nu_{r+1}$ (so $\widehat\nu_{r+1}=\vartheta_r\widetilde\nu_{r+1}$). Starting from $\widehat\nu_0=\widetilde\nu_0=0$, run the recursion in \pref{lem:actual-recursion} with $\vartheta_r=1$ up to the first candidate $\widetilde\nu_{r+1}$ with potential at least $H$, stopping before its retention factor is applied; we first prove that this occurs at a finite height.
We will use the common estimate, with $h=h(\widehat\nu_r)$,
\begin{equation}\label{eq:sibling-potential-loss}
 0\le h-h(\theta_r\widehat\nu_r)
 \le\frac{h\,u(\widehat\nu_r)}{\lambda_+}
 \le\frac{\ee^T h^2}{c_0\lambda_+}.
\end{equation}
This follows from $1-\theta_r\le u(\widehat\nu_r)/\lambda_+$ and $u(\widehat\nu_r)\le\ee^T h/c_0$; the loss is $O(\ee^T/\ln k)$ when $h\le13$ and $o(1)$ when $h\le H$.

\paragraph{Step 1: Comparison up to potential $13$.}
Write $\nu_r=\mathcal R_\beta^r0$ and $r_0=\min\{r\ge1:h(\nu_r)>13\}$.
By \pref{lem:information}, $\widetilde\nu_r\preceq\nu_r$ up to and including the first hit of $H$. Since $H>13$, this hit cannot precede $r_0$.
Set $E_r\defeq h(\nu_r)-h(\widetilde\nu_r)+\frac{r_0-r}{T}\int|z|\widetilde\nu_r(\mathrm{d}z)$ for $1\le r\le r_0$.
The coefficient $(r_0-r)/T$ reserves one first-absolute-moment bound for each remaining truncation, whose potential loss is at most its $|z|/T$-moment; since $\widetilde\nu_1=\nu_1=\beta\delta_0$, we have $E_1=0$ (including when $r_0=1$).
For $1\le r<r_0$, split $h(\nu_{r+1})-h(\widetilde\nu_{r+1})$ at the untruncated iterate $\bar\nu_{r+1}=\mathcal R_\beta(\theta_r\widetilde\nu_r)$ into the propagated input error and the step-$(r+1)$ truncation loss:
\[
 E_{r+1}
 =\bigl(h(\nu_{r+1})-h(\bar\nu_{r+1})\bigr)
  +\bigl(h(\bar\nu_{r+1})-h(\widetilde\nu_{r+1})\bigr)
  +\frac{r_0-r-1}{T}\int_{|z|\le T}|z|\bar\nu_{r+1}(\mathrm{d}z).
\]
For the propagated input error, $h(\widetilde\nu_r)\le13$, so \eqref{eq:sibling-potential-loss} and \eqref{eq:potential-contraction} give
\[
 h(\nu_{r+1})-h(\bar\nu_{r+1})
 \le m(\nu_{r+1})\bigl(h(\nu_r)-h(\theta_r\widetilde\nu_r)\bigr)
 \le m(\nu_{r+1})\left[h(\nu_r)-h(\widetilde\nu_r)+O\!\left(\frac{\ee^T}{\ln k}\right)\right].
\]
For the truncation loss, $\cosh z\,g(q(z))\le1\le|z|/T$ on $\{|z|>T\}$ gives $h(\bar\nu_{r+1})-h(\widetilde\nu_{r+1})\le\frac1T\int_{|z|>T}|z|\bar\nu_{r+1}(\mathrm{d}z)$.
Combining this tail integral with the central integral $\frac{r_0-r-1}{T}\int_{|z|\le T}|z|\bar\nu_{r+1}(\mathrm{d}z)$ and applying \eqref{eq:jump-first-moment} to $\bar\nu_{r+1}=m(\bar\nu_{r+1})K_{\theta_r\widetilde\nu_r}$ (using $m(\bar\nu_{r+1})\le m(\nu_{r+1})$ and $m(\widetilde\nu_r)=O(1)$) yields
\[
 \begin{aligned}
 h(\bar\nu_{r+1})-h(\widetilde\nu_{r+1})+\frac{r_0-r-1}{T}\int_{|z|\le T}|z|\bar\nu_{r+1}(\mathrm{d}z)&\le\frac{r_0-r}{T}\int|z|\bar\nu_{r+1}(\mathrm{d}z)\\
 &\le m(\bar\nu_{r+1})\frac{r_0-r}{T}\int(|z|+2)\theta_r\widetilde\nu_r(\mathrm{d}z)\\
 &\le m(\nu_{r+1})\left[\frac{r_0-r}{T}\int|z|\widetilde\nu_r(\mathrm{d}z)+O\!\left(\frac{r_0-r}{T}\right)\right].
 \end{aligned}
\]
Adding the propagated input error gives the one-step error recurrence
\[
 E_{r+1}\le m(\nu_{r+1})\left[E_r+O\!\left(\frac{\ee^T}{\ln k}+\frac{r_0-r}{T}\right)\right],
\]
and unfolding from $E_1=0$ to $r_0$ via \pref{lem:quant-bands} ($r_0=O(\delta^{-1})$ and $\prod_{s=r}^{r_0-1}m(\nu_{s+1})=O(\delta^{-8})$) gives
\[
 h(\nu_{r_0})-h(\widetilde\nu_{r_0})
 =E_{r_0}
 \le\sum_{r=1}^{r_0-1}\left(\prod_{s=r}^{r_0-1}m(\nu_{s+1})\right)O\!\left(\frac{\ee^T}{\ln k}+\frac{r_0-r}{T}\right)
 =O\!\left(\frac{\delta^{-9}\ee^T}{\ln k}+\delta^2\right)=o(1),
\]
so $h(\widetilde\nu_{r_0})\ge12$.

\paragraph{Step 2: Growth from $12$ to $H$.}
For any input with $h=h(\widehat\nu_r)\le H$, \eqref{eq:sibling-potential-loss}, \eqref{eq:limiting-map}, and \eqref{eq:central-mass} give
\begin{equation}\label{eq:truncated-mass}
 m(\widetilde\nu_{r+1})=m(\bar\nu_{r+1})K_{\theta_r\widehat\nu_r}([-T,T])\ge\frac{m(\bar\nu_{r+1})}{10(1+h)}=\frac{\beta\ee^{h-o(1)}}{10(1+h)}.
\end{equation}
Before the first hit, if $12\le h<H$, this and $h(\nu)\ge c_0m(\nu)$ show that the next potential is at least $(c_0/(10\ee)-o(1))\ee^h/(1+h)\ge2h$.
Thus either $H$ has been reached by $r_0$, or it is reached in $O(\ln H)$ further steps.
We may now choose
\[
 L-1=\min\{r\ge1:h(\widetilde\nu_r)\ge H\},\qquad
 \vartheta=\frac{H}{h(\widetilde\nu_{L-1})}\in(0,1].
\]
Thus $r_0\le L-1\le r_0+O(\ln H)$, so $L=O(\delta^{-1})$, and \eqref{eq:actual-weighted-recursion} gives $h(\widehat\nu_{L-1})=H$, also when $L=2$.
Since $\vartheta_{L-1}=1$, \eqref{eq:truncated-mass} at $h=H$ gives the final root mass $m_L\defeq m(\widehat\nu_L)=S\E[f(q_o)]\ge\beta\ee^{H-o(1)}/[10(1+H)]$.

\paragraph{Step 3: Update size and root-flip bound.}
Since $h(\widehat\nu_r)\le H$ for $0\le r<L$, \eqref{eq:truncated-mass} and \eqref{eq:actual-weighted-recursion} give $\vartheta_r m(\bar\nu_{r+1})\le10(1+H)m(\widehat\nu_{r+1})$.
Before the final root ($r<L-1$), the output potential is also at most $H$, so this bound is at most $10H(1+H)/c_0$, and \eqref{eq:size-recursion} gives $\ln(1+\mathcal V_{L-1})=O(L(T+\ln(H+1)))$.
At the root, \eqref{eq:size-recursion} yields $\ln(\mathcal V_L/m_L)\le\ln(1+40\ee^T(1+H)\mathcal V_{L-1})=O(\delta^{-13})=o(\ln\ln k)$.
When $o\in R$, we have $f(q_o)=0$, whereas when $o\notin R$, we have $|V\setminus R|\ge1$, $|z(q_o)|\le T$, and $2q_o(1-q_o)=\frac12f(q_o)^2\ge\frac12\ee^{-T}f(q_o)$.
Applying Cauchy--Schwarz to $f(q_o)=\bigl(f(q_o)/|V\setminus R|\bigr)^{1/2}\bigl(f(q_o)|V\setminus R|\bigr)^{1/2}$ on $\{o\notin R\}$ gives
\[
 \Bigl(\frac{m_L}{S}\Bigr)^2
 =\bigl(\E[f(q_o)]\bigr)^2
 \le\E\left[\frac{f(q_o)}{|V\setminus R|}\mathbb{1}\{o\notin R\}\right]\E\bigl[f(q_o)|V\setminus R|\bigr]
 =\E\left[\frac{f(q_o)}{|V\setminus R|}\mathbb{1}\{o\notin R\}\right]\frac{\mathcal V_L}{S}.
\]
Combining this with $\Ham(x,y)\le|V\setminus R|$, \eqref{eq:root-flip-identity}, the root mass bound above, and $S\le2k\ln k$ yields
\begin{align*}
 \E\left[\frac{\mathbb{1}\{y_o\ne x_o\}}{\max\{1,\Ham(x,y)\}}\right]
 &\ge\E\left[\frac{2q_o(1-q_o)}{\max\{1,|V\setminus R|\}}\right]\\
 &\ge\frac{\ee^{-T}}{2}\E\left[\frac{f(q_o)}{|V\setminus R|}\mathbb{1}\{o\notin R\}\right]
 \ge\frac{\ee^{-T}m_L^2}{2S \mathcal V_L}
 \ge\frac{(\ln k)^{11-o(1)}}k\ge\epsilon_k.\qedhere
\end{align*}
\end{proof}

\subsection{Extending to locally tree-structured formulas}\label{sec:local-formulas}

We now extend the tree update of \pref{sec:local-updates} to locally tree-structured neighborhoods in general formulas and establish concentration of the empirical root flow in planted random formulas.

Fix $L\ge1$ and $\vartheta\in[0,1]$, and extend any local update output on a ball to $V(\Phi)$ by keeping $x$ unchanged outside the ball.
For a formula $\Phi$, $x\in\Omega_\Phi$, and $v\in V(\Phi)$, when $v$ is $(L,k^4)$-locally tree-structured in $\mathcal L_x$ (\pref{lem:ball-sensitivity}), let $y\gets\mathsf{LocalUpdate}(x)$ be the output of \pref{alg:local-update} on $B_L(\mathcal L_x,v)$ and define the \emph{local-update flow}
\[
 g_v(\Phi,x)\defeq\E\left[\frac{\mathbb{1}\{y_v\ne x_v\}}{\max\{1,\Ham(x,y)\}}\right],
\]
setting $g_v(\Phi,x)\defeq0$ at all other variables.

\begin{lemma}[Comparison and concentration of local-update flow]\label{lem:root-flow}
\leavevmode
\begin{enumerate}[label=\textup{(\roman*)}]
\item \emph{(Comparison.)} For any formula $\Phi$, $x\in\Omega_\Phi$, and $v\in V(\Phi)$ such that $B_L(\Phi,v)$ is a tree, let $y'\gets\mathsf{LocalUpdate}(x)$ be the output of \pref{alg:local-update} on $B_L(\Phi,v)$. Then
\[
 \E\left[\frac{\mathbb{1}\{y'_v\ne x_v\}}{\max\{1,\Ham(x,y')\}}\right]\ge g_v(\Phi,x),
\]
with equality when $v$ is $(L,k^4)$-locally tree-structured in $\mathcal L_x$.
\item \emph{(Concentration.)} Given $x\in\{0,1\}^n$ and $\alpha\ge0$, let $\Phi_x$ consist of $\operatorname{Pois}(\alpha n)$ clauses sampled independently and uniformly among $k$-clauses satisfied by $x$, with forcing mean $\lambda_+=\alpha k/(2^k-1)$.
Assume $k$ is sufficiently large and $\alpha$ is such that $\lambda_+$ lies in the range of \pref{thm:local}(iii), and choose $(L,\vartheta)$ according to \pref{thm:local}(iii).
Then there is an absolute constant $C>0$ such that for $b_k=\exp(-C(L+1)\ln k)$ and all $n\ge n_0(k)$,
\[
 \Pr\left(\sum_{v\in[n]}g_v(\Phi_x,x)<\frac{\epsilon_kn}{2}\right)
 \le3\exp(-b_kn).
\]
\end{enumerate}
\end{lemma}
\begin{proof}
If $v$ is $(L,k^4)$-locally tree-structured in $\mathcal L_x$, \pref{thm:local}(ii) gives identical laws for $\Delta(x,y')$ from $B_L(\Phi,v)$ and $\Delta(x,y)$ from $B_L(\mathcal L_x,v)$, proving equality. Otherwise $g_v=0$, so the claimed inequality is immediate.

For the second, each clause of $\Phi_x$ belongs to $\mathcal L_x$ with probability $(k+\binom{k}{2})/(2^k-1)$, so $\mathcal L_x$ has $\operatorname{Pois}(\lambda_+(k+1)n/2)$ i.i.d.\ clauses, and in the corresponding tree from $\mathsf{Tree}_{\mathrm{pl}}(\lambda_+)$ each variable has $\operatorname{Pois}(\lambda_+k(k+1)/2)$ child clauses in $\mathcal L_x$.
Since $\lambda_+\le2\ln k$ and the expected number of variables through depth $L$ is $k^{O(L+1)}$, a union bound over degrees through depth $L+1$ gives
\[
 k^{O(L+1)}\Pr(\operatorname{Pois}(\lambda_+k(k+1)/2)\ge k^4)
 \le\exp(-k^4\ln k)\le\epsilon_k/8.
\]
By \pref{thm:local}(ii)--(iii) and \pref{lem:poisson-local-limit}(ii), $\E[\sum_{v\in[n]}g_v(\Phi_x,x)]\ge(7/8-1/8)\epsilon_kn=3\epsilon_kn/4$ for $n\ge n_0(k)$ (uniformly in $x$ and $\alpha$).
Since $g_v\in[0,1]$ depends only on $\widetilde B_{L,k^4}(\mathcal L_x,v)$, \pref{lem:ball-sensitivity} bounds the effect of changing one clause on $\sum_v g_v$ by $A_k=2kM_{L,k^4}=k^{O(L+1)}$, and \pref{lem:prob-poisson-lipschitz} at deviation $\epsilon_kn/4$ yields the failure bound $3\exp(-\Omega(\epsilon_k^2n/((1+\lambda_+k)A_k^2)))\le3\exp(-b_kn)$.
\end{proof}

\endgroup
\section{Proofs of looseness and rigidity}
\label{sec:looseness-rigidity}
We prove parts~\ref{item:main-looseness-part} and~\ref{item:main-rigidity-part} of \pref{thm:main-looseness} using the forcing tree methods
developed in \pref{sec:trees}.
We defer the proof of part~\ref{item:main-local-connectivity-part} to \pref{sec:companion-local-connectivity}.
Below the $*$-core threshold, complete forcing trees give logarithmic
satisfying paths reversing any variable of a typical solution.
Above it, we use a large $*$-core to prove linear rigidity and
exponentially small solution components.

\subsection{Looseness below the \texorpdfstring{$*$}{*}-core threshold}
\label{sec:looseness}

\begin{proof}[Proof of looseness in \pref{thm:main-looseness}\ref{item:main-looseness-part}]
We prove the looseness conclusion with probability at least $1-1/n$ over $\Phi$, for at least a $1-1/(2n)$ fraction of its solutions.
Assume the stated density range and write $m=\lfloor\alpha n\rfloor$.
For sufficiently large $k$, the assumed density bound satisfies the
hypothesis of \pref{lem:forcing-tree-counts}, including the factor
$2^k/(2^k-1)$.
Put $q=\lfloor16\ln n/\ln k\rfloor$.
For nonempty $U\subseteq V$, write
\[
 \mathsf{exc}_{\Phi}(U)
 =\sum_{C\in\+C}(|\textnormal{vbl}(C)\cap U|-1)_+-|U|+1.
\]
For a connected full restricted factor graph, this is the number of
edges outside a spanning tree.
Let $\mathcal E_\Phi$ consist of solutions under which some set $U$
of at most $32(\ln k)^5\ln n$ variables has connected restriction
in $\Phi_\sigma^{(1)}$ and $\mathsf{exc}_\Phi(U)\ge256(\ln k)^6$.

\paragraph{A failed exploration gives a witness.}
The algorithm first checks whether every variable has degree at most
$\lceil\ln n\rceil$ in $\Phi$ and returns \textnormal{\textsc{Fail}}
otherwise. We bound the probability of this event below.
Assume this check passes.
Fix $\sigma\notin\mathcal E_\Phi$ and a root $v$.
Start with the forcing tree $T=\{v\}$ and its variable set $U$.
At a clause forcing a used variable but absent from $T$, attach a
false child and its first complete tree of size at most $(\ln k)^6$
whenever possible. Test these fixed candidates for all false children
before attaching a single false child without its complete tree. Use fixed orders
for every choice, and permit only additions disjoint from $U$ that
keep its full restricted factor graph acyclic. Stop when $T$ is
complete or after $q$ additions. Return \textnormal{\textsc{Fail}}
if no addition is possible or if $T$ remains incomplete at the cutoff.

An addition is always possible before this cutoff. An uncovered clause
$C$ meets $U$ only at its forced variable. For each of its $k-1$ false
variables $u$, let $W_u$ be the variable set of the first complete
forcing tree of size at most $(\ln k)^6$ rooted at $u$, if one exists,
and put $W_u=\{u\}$ otherwise. The set
$U^+=U\cup\bigcup_u W_u$ has connected forcing restriction and
\[
 |U^+|\le1+(q+k-1)(\ln k)^6\le32(\ln k)^5\ln n.
\]
Thus its full excess is less than $256(\ln k)^6$.
The full restriction on $U$, together with all edges of $C$, is a tree.
Contract this tree, choose a spanning tree of the resulting graph,
and lift it to a spanning tree $R$ of the full restriction on $U^+$.
Removing $C$ from $R$ leaves one component containing $U$ and one
component for each candidate $u$. Mark every candidate component
containing an endpoint of an edge omitted from $R$.
Each omitted edge marks at most two components, so fewer than
$512(\ln k)^6$ candidates are marked.

An unmarked component has no edge to another component except through
$C$. A forcing path from its candidate cannot cross $C$, which forces
the old variable outside that component; hence its chosen complete
tree, if any, lies entirely inside the component. It is disjoint from
$U$, and adjoining it to $U$ leaves the full restriction a subgraph
of $R$, hence acyclic. The same holds for a singleton candidate.
Since $k-1>512(\ln k)^6$ for large $k$, an addition exists.

If $C$ is unobstructed, at least $1024(\ln k)^6$ candidates admit small
complete trees, so one of these remains unmarked. Thus an addition
without a complete tree occurs only at an obstructed clause.

Let $B$ consist of $v$ and the variables attached without a complete
tree. Every uncovered
forcing clause forces a variable of $B$, because variables added
inside a complete tree already have all their forcing clauses in $T$.
The additions without complete trees and their attachment
clauses form a forcing subtree on $B$ satisfying the clause
restriction defining $\mathcal B_t$.
If $T$ is incomplete after $q$ additions, its $q$ selected clauses
and one uncovered clause all force $B$. Thus
$\sigma\in\mathcal B_{q+1}(\Phi,v)$.
The solutions with a failed root exploration are therefore contained in
\[
 \mathcal B_\Phi
 =\mathcal E_\Phi\ \cup\ \bigcup_{v\in[n]}\mathcal B_{q+1}(\Phi,v).
\]
Successful explorations give the required acyclic complete trees,
with at most $1+q(\ln k)^6\le32(\ln k)^5\ln n$ variables.

\paragraph{Probability of excess.}
Fix $2\le\ell\le32(\ln k)^5\ln n$.
From a spanning tree of the forcing restriction on $\ell$ variables,
remove the leaf clauses. The remaining tree has $1\le r\le\ell-1$
clauses and $\ell+r-1$ required incidences.
There are at most $4^{\ell+r-1}n^\ell m^r$ choices, with incidence
probability at most $(k/n)^{\ell+r-1}$.
Conditional on these incidences, write $X=\mathsf{exc}_\Phi(U)$.
Remaining slots in the selected clauses hit $U$ with probability
at most $2\ell/n$. For an outside clause with $J$ hits,
$\Pr(J\ge j)\le\E\bigl[\binom{J}{j}\bigr]\le(k\ell/n)^j/j!$, and its
contribution to $X$ is $(J-1)_+$. Consequently,
\[
 \E[n^{(J-1)_+/2}]
 \le1+\sum_{j\ge2}n^{(j-1)/2}\frac{(k\ell/n)^j}{j!}
 \le1+\frac{k^2\ell^2}{2n\sqrt n}\ee^{k\ell/\sqrt n}
 \le1+\frac{k^2\ell^2}{n\sqrt n},
\]
uniformly over the stated range of $\ell$ for sufficiently large $n$.
These bounds give
\[
 \E[(\sqrt n)^X]
 \le(1+2\ell/\sqrt n)^{kr}
       \left(1+\frac{k^2\ell^2}{n\sqrt n}\right)^m\le2,
 \qquad
 \Pr(X\ge256(\ln k)^6)\le2n^{-128(\ln k)^6}.
\]
For fixed scopes, choose each selected clause's forced variable in
at most $k^r$ ways. By \pref{lem:support-signs}, their total expected
solution mass is at most $(2k2^{-k})^r$.
Summing the shapes, incidences, and forcing choices gives
\[
 \sum_{r=1}^{\ell-1}4^{\ell+r-1}n^\ell m^r
       (k/n)^{\ell+r-1}(2k2^{-k})^r\le nk^{3\ell},
\]
using $\alpha k2^{-k}\le2\ln k$.
The excess depends only on scopes, so multiplying and summing yields
\[
 \E_\Phi[\mu_\Phi(\mathcal E_\Phi)]
 \le2n^{-128(\ln k)^6}
       \sum_{2\le\ell\le32(\ln k)^5\ln n}nk^{3\ell}
 \le n^{-3}.
\]

\paragraph{Probability of a forcing witness.}
Let $\mathcal Z$ be the event from \pref{sec:normalization}, with
$\Pr_\Phi(\mathcal Z^c)\le1/(2n)$ by \pref{lem:norm-lower}.
Apply \pref{lem:forcing-tree-counts} with $t=q+1$.
For large $k,n$, $2\ee\sqrt k\ln k\le q+1\le\ln n$
and $k^{-(q+1)/2}\le n^{-8}$.
A union bound therefore gives
\[
 \E_\Phi\!\left[\mathbb{1}\{\mathcal Z\} \cdot
 \mu_\Phi\!\left(\bigcup_v\mathcal B_{q+1}(\Phi,v)\right)\right]
 \le n\left(2n^{-8}
       +\exp(-n/k^{9s})\right)\le n^{-3},
\]
where $s=\lceil(\ln k)^6\rceil$.
Each variable's degree has law $\operatorname{Bin}(m,k/n)$ with
mean $O_k(1)$. A Chernoff bound and a union bound show that the
degree check fails with probability at most $n^{-4}$ for sufficiently
large $n$.
On $\mathcal Z$, if this check passes and $\mu_\Phi(\mathcal B_\Phi)\le1/(2n)$, the construction succeeds for every variable of each solution outside $\mathcal B_\Phi$.
By Markov's inequality, the probability that $\mathcal Z$ holds and either the degree check fails or $\mu_\Phi(\mathcal B_\Phi)>1/(2n)$ is at most
\begin{equation}\label{eq:looseness-failure-probability}
 n^{-4}+2n\E_\Phi[\mathbb{1}\{\mathcal Z\}\cdot\mu_\Phi(\mathcal B_\Phi)] \le n^{-4}+\frac4{n^2}.
\end{equation}
Adding $\Pr_\Phi(\mathcal Z^c)\le1/(2n)$ gives total failure probability at most $1/(2n)+n^{-4}+4/n^2\le1/n$.

\paragraph{Linear-time implementation.}
Build the variable--clause incidence lists and forcing lists in
$O_k(n+m)$ time. For each variable, precompute its first complete
tree with at most $b=\lfloor(\ln k)^6\rfloor$ variables, if one exists.
Such a tree has at most $b-1$ forcing clauses in total. Reject any
search branch that repeats a variable or exceeds this clause budget;
reject variables with longer forcing lists immediately.
At each discovered variable include all its forcing clauses and
enumerate their false-child choices.
There are at most $b-1$ child choices,
each with at most $k-1$ options, so each root takes $O_k(1)$ time.
Thus all candidates are computed in $O_k(n+m)$ time.

During exploration, find the next uncovered forcing clause by
scanning the forcing lists of $U$. For each candidate $W$, test
disjointness from $U$, scan the clauses meeting $U\cup W$, and use
BFS to test acyclicity of its full restriction.
Since $|U\cup W|=O_k(\ln n)$ and all degrees are at most
$\lceil\ln n\rceil$, each scan or trial takes $O_k((\ln n)^2)$ time.
Allocate marking arrays once and reset only visited entries.
There are $O(k)$ trials per addition and at most $q=O_k(\ln n)$
additions, so exploration takes $O_k((\ln n)^3)=o(n)$ time.
Together with preprocessing, this gives $O_k(n+m)=O_k(n)$ time in the stated density range.
Flipping the completed tree bottom-up by \pref{obs:flip-complete-tree}
gives the required variable flipping path.
The path is represented by its initial assignment and the ordered list of variables to flip.
\end{proof}

\subsection{Rigidity and geometric shattering above the
\texorpdfstring{$*$}{*}-core threshold}
\label{sec:shattering}
To prove rigidity, we use the $*$-core to identify variables that remain fixed along paths whose steps change few variables. This also controls the masses of the components of $G_D(\Phi)$.

\begin{theorem}[Quantitative rigidity and geometric shattering]
\label{thm:rigidity}
Let $k$ and $n\ge n_0(k)$ be sufficiently large, and let
$\Phi\sim\Phi(k,n,m)$ with
$m=\lfloor\alpha n\rfloor$ and
\begin{equation}\label{eq:main-shatter-sharp}
 \frac{2^k-1}{k}\left(\ln k+\ln\ln k+1+
       \frac{\ln\ln k+1/2}{\ln k}
       +\frac{7(\ln\ln k)^2}{(\ln k)^2}\right)
 \le\alpha\le2^k\left(\ln2-\frac1{\ln k}\right).
\end{equation}
With probability at least $1-1/n$, $\Phi$ is satisfiable and the following hold:
\begin{enumerate}
    \item At least a $1-\exp(-n/k^{40})$ fraction of its solutions $\sigma$ have at least $(1-2/(k\ln k))n$ variables that are not $(n/k^{40})$-loose under $\sigma$;
    \item For every integer $1\le D\le n/k^{40}$, every component of $G_D(\Phi)$ has mass at most $\exp(-n/k^{40})$.
\end{enumerate}
\end{theorem}
The second conclusion gives geometric shattering in the rigidity
range. The stronger conclusion including energy barriers is proved
in \pref{sec:shattering-below-one} throughout
\eqref{eq:main-shattering-range}.
The upper bound in \eqref{eq:main-looseness-headline} exceeds
the lower bound in \eqref{eq:main-shatter-sharp}.
Combining the rigidity conclusion, this proves
\pref{thm:main-looseness}\ref{item:main-rigidity-part}.

\Needspace{6\baselineskip}
We use the $*$-core formulation of Molloy and Restrepo~\cite[Section~4]{MR12}.
\begin{definition}[$*$-core]\label{def:star-core}
Fix a CNF formula $\Phi=(V,\+C)$ and an assignment $\sigma$. Let $V_*\subseteq V$ be the largest set such that every $v\in V_*$ is forced under $\sigma$ by a clause $C$ with $\operatorname{vbl}(C)\subseteq V_*$. The $*$-core under $\sigma$ is the subformula of $\Phi$ with variable set $V_*$ and clause set $\{C\in\+C:\operatorname{vbl}(C)\subseteq V_*\}$.
\end{definition}

Fix a formula $\Phi$ and a solution $\sigma$, and let $V_*$
be the variable set of its $*$-core under $\sigma$.
A set $S\subseteq V_*$ is \emph{$*$-flippable} under $\sigma$
if every core clause forcing a variable $v\in S$ contains
another variable in $S$.

\begin{observation}\label{obs:core-difference-flippable}
For every solution $\tau$ of $\Phi$, the set
$V_*\cap\Delta(\sigma,\tau)$
is $*$-flippable under $\sigma$.
\end{observation}
Indeed, a core clause whose unique true literal changes
must acquire another true literal, so another variable
of that clause also changes.

Achlioptas and Coja-Oghlan~\cite[full version,
Appendices~B.3--B.4]{AC08} give earlier proofs of rigidity and
shattering.
We follow the $*$-core approach of Molloy and
Restrepo~\cite[Section~6]{MR12}:
a large core and a gap in the sizes of $*$-flippable sets keep
most variables fixed along satisfying paths with small steps.
We prove core bounds with exponential failure under the Poisson
fixed-assignment law, then transfer them to uniformly sampled solutions.

\begin{definition}[$*$-deletion process]\label{def:core-deletion}
Fix a CNF formula $\Phi=(V,\+C)$, an assignment $\sigma$, and an order on $V$. Initially, all variables and clauses of $\Phi$ remain. At each step, the first remaining variable forced by no remaining clause under $\sigma$ is deleted, together with all clauses containing it. If no such variable exists, the process stops.
\end{definition}
No variable of $V_*$ can be deleted, and the final variable set satisfies
its defining property, so this process leaves exactly the $*$-core.

Fix $\sigma$ and a forcing mean $\lambda_+\ge0$, and sample
$\Phi\sim\mathsf{Pois}_1(\sigma,\lambda_+)$. Put
$\varepsilon=(\ln\ln k)^2/(\ln k)^2$,
$r=1/[k(\ln k+\ln\ln k)]$, and $x=-\ln r$. Assume
\begin{equation}\label{eq:core-density-estimate}
 \lambda_+\le k,\qquad \lambda_+(1-r)^{k-1}\ge x+4\varepsilon.
\end{equation}
By \eqref{eq:forcing-explicit-expansion}, the density range
\eqref{eq:main-shatter-sharp} gives this estimate for sufficiently
large $k,n$.
Let $H_1\subseteq V_*$ consist of the variables forced by exactly
one clause in the $*$-core.
We first use the $*$-deletion process in \pref{def:core-deletion}
to show that the core is large and $H_1$ is small enough that
a small backward exploration through these forcing clauses
adds fewer than one variable per step on average.
After controlling directed cycles, we combine this bound with
the two independent clause requirements at each variable in
$V_*\setminus H_1$ to exclude intermediate-size $*$-flippable sets.

For comparison, independently thin the initial labeled forcing clauses:
at step $j+1\le n-k+1$, delete each remaining clause with probability
$(k-1)/(n-j-1)$; after that all clauses are gone.
At each fixed step $j$, the survival probability is
$s_j=\binom{n-1-j}{k-1}/\binom{n-1}{k-1}$, with $s_n=0$,
so the forcing counts are independent $\operatorname{Pois}(\lambda_+ s_j)$
variables. Letting $Z_{j,0}$ and $Z_{j,1}$ count variables in $V$
with respectively zero and one remaining forcing clauses gives
\[
 Z_{j,0}\sim\operatorname{Bin}(n,\exp(-\lambda_+ s_j)),\qquad
 Z_{j,1}\sim\operatorname{Bin}(n,\lambda_+ s_j\exp(-\lambda_+ s_j)).
\]

Couple this thinning with the $*$-deletion process by initially revealing
only the label and forced variable of each clause.
After $j$ deletions, the other $k-1$ variables of each remaining clause
are still independent uniform subsets of the undeleted variables
excluding their forced variable.
Reveal only which clauses contain the next deleted variable.
The revealed history records clause labels and forced variables,
the deleted variables, and which clauses each deletion hits,
but no other positions in surviving clauses.
Since no remaining clause forces that variable, each is hit with
probability $(k-1)/(n-j-1)$, independently of the others.
Thus both processes can delete the same clauses until the $*$-deletion
process stops at $J=n-|V_*|$; only the thinning continues thereafter.
At that time exactly the deleted variables have forcing degree zero, so
$Z_{J,0}=J$ and $Z_{J,1}=|H_1|$.
More generally, $Z_{j,0}\ge j$ for $j\le J$.
Consequently, finding $j_0$ with $Z_{j_0,0}<j_0$ forces $J<j_0$,
giving $|V_*|>n-j_0$ and $|H_1|\le\max_{j\le j_0}Z_{j,1}$.

\begin{lemma}\label{lem:core-size}
With probability at least $1-\exp(-n/k^{30})$ over $\Phi$,
\begin{equation}\label{eq:core-finite-bounds}
 |V_*|>\left(1-\frac{3}{2k\ln k}\right)n,\qquad
 |H_1|\le\left(1-\frac1{8k}\right)\frac{|V_*|-1}{k-1}.
\end{equation}
\end{lemma}
\begin{proof}
Put $d=k-1$ and $j_0=\lfloor nr\rfloor$.
For $j\le j_0$, $s_j\ge(1-r)^d-k^2/n$, so
\eqref{eq:core-density-estimate} gives $\lambda_+ s_j\ge x+3\varepsilon$
for sufficiently large $n$. Hence
\[
 \E[Z_{j_0,0}]\le nr\ee^{-3\varepsilon}\le nr-n/k^5,\qquad
 \frac{d\E[Z_{j,1}]}{n(1-r)}
 \le\frac{d(x+3\varepsilon)r}{1-r}\ee^{-3\varepsilon}
 \le1-\frac1{4k}.
\]
The second bound uses the monotonicity of $u\ee^{-u}$ for $u\ge1$
and $dxr/(1-r)=1+o(\varepsilon)$ from
\eqref{eq:forcing-explicit-expansion}.
Hoeffding and a union bound give deviations above these means of
at most $n/k^6$, simultaneously, except with probability
$(n+1)\ee^{-2n/k^{12}}\le\ee^{-n/k^{30}}$.
Then $Z_{j_0,0}<j_0$, so $J<j_0$ and
$|V_*|>n(1-r)>(1-3/(2k\ln k))n$. Moreover,
\[
 |H_1|\le\frac{n(1-r)}d\left(1-\frac1{4k}\right)+\frac n{k^6}
 \le\left(1-\frac1{8k}\right)\frac{|V_*|-1}{d}. \qedhere
\]
\end{proof}

Condition on the $*$-deletion history described above with
\eqref{eq:core-finite-bounds} satisfied.
It determines $V_*$, the surviving clause labels, and $H_1$.
For each surviving clause forcing $v$, its other $k-1$ variables
form a uniform subset of $V_*\setminus\{v\}$,
independently across clauses.

A $*$-flippable set containing $v\in H_1$ must contain another variable in the unique clause in the $*$-core that forces $v$.
We record these possible dependencies in a directed graph on $V_*$, with an edge $v\to w$ whenever $v\in H_1$ and $w\ne v$ belongs to this clause.

For $A\subseteq V_*$, let $R(A)$ consist of $A$ and all variables
in $H_1$ that can reach $A$ along directed paths.
The following estimate uses only clauses forcing variables in
$H_1$; after these are fixed, the clauses forcing variables
outside $H_1$ remain independent.

\begin{lemma}\label{lem:core-reach}
With probability at least $1-\exp(-n/k^{30})$ over these clause variable choices, for every $A\subseteq V_*$ with $n/k^{24}\le|A|\le2n/k^{12}$, we have $|R(A)|<2k^4|A|$.
\end{lemma}
\begin{proof}
Fix $A$ in the stated range and set $a=|A|$ and
$t=2k^4a\le4n/k^8$. Starting from $A$, variables are processed in a fixed order as they are discovered.
Each step adds every undiscovered variable in $H_1$ whose unique forcing clause contains the current variable.
After $j$ steps, each clause still being tested is uniform subject to avoiding the $j$ processed variables. At each of the first $t$ steps, the number added is a sum of conditionally independent indicators with mean at most $(k-1)|H_1|/(|V_*|-t-1)\le1-1/(16k)$.
Let $S$ be the total number added in the first $t$ steps, counting zero additions after the exploration stops.
Applying $1+x\le\ee^x$ to the independent indicators at each step and iterating conditional expectations, we obtain $\mathbb E[\ee^{S/(32k)}] \le\exp((1-1/(16k))t(\ee^{1/(32k)}-1))$.
Reaching $t$ variables requires $S\ge t-|A|$. Thus, by Markov's inequality,
\[
 \Pr(|R(A)|\ge t)
 \le\exp\left(-\frac{t-|A|}{32k}+\left(1-\frac1{16k}\right)t(\ee^{1/(32k)}-1)\right) \le\exp(-t/k^3),
\]
where the last inequality follows from $|A|\le t/(2k^4)$ and $\ee^u-1\le u+u^2$ for $u=1/(32k)$.
Finally, $\ln(\ee n/a)\le1+24\ln k\le k$ in the stated range,
so a union bound gives failure probability at most
\[
 \sum_{a=\lceil n/k^{24}\rceil}^{\lfloor2n/k^{12}\rfloor}
 \binom na\exp(-2ka)
 \le\sum_{a\ge\lceil n/k^{24}\rceil}\exp(-ka)
 \le\exp(-n/(2k^{23}))\le\exp(-n/k^{30}).\qedhere
\]
\end{proof}

\begin{lemma}\label{lem:flippable-gap}
With probability at least $1-3\exp(-n/k^{30})$ over these clause variable choices, no $*$-flippable set has size in $[n/k^{13},2n/k^{13}]$.
\end{lemma}
\begin{proof}
We first control directed cycles in the directed graph defining $R(A)$.
Since only variables in $H_1$ have outgoing edges, every such cycle lies in $H_1$.
There are at most $\binom{|H_1|}{s}s!\le|H_1|^s$ collections
of vertex-disjoint directed cycles containing $s$ variables in total.
Each specified collection has probability
$((k-1)/(|V_*|-1))^s$, by independence of the clauses forcing
distinct variables.
By a union bound, the probability that any collection contains at least $n/k^{24}$ variables in total is at most $\sum_{s\ge n/k^{24}}(1-1/(8k))^s\le\exp(-n/k^{30})$.

We now reveal the clauses forcing variables in $H_1$.
We condition on a realization in which every collection of vertex-disjoint directed cycles contains fewer than $n/k^{24}$ variables and assume that the conclusion of \pref{lem:core-reach} holds.
Let $B$ be the set of variables in a maximal collection of vertex-disjoint directed cycles. Then $|B|<n/k^{24}$ and every directed cycle intersects $B$.

For a fixed $A\subseteq V_*\setminus H_1$, we consider $*$-flippable sets $S$ with $S\setminus H_1=A$.
Each variable in $S\cap H_1$ has an outgoing edge within $S$.
Following these edges reaches $A$ or a cycle intersecting $B$, so $S\subseteq R(A\cup B)$.
The revealed clauses already determine $R(A\cup B)$.

When $n/k^{24}\le s=|A|\le2n/k^{13}$, the set $A\cup B$ lies in the range of \pref{lem:core-reach}, which gives $|R(A\cup B)|<2k^4|A\cup B|<4k^4s$.
Each $v\in A$ is forced by at least two clauses in the $*$-core; we select the two smallest such clauses by the clause ordering.
Flippability requires each selected clause forcing $v$ to contain a variable in $R(A\cup B)\setminus\{v\}$.
Its other $k-1$ variables form a uniform subset of $V_*\setminus\{v\}$, so this has probability at most
\[
 \frac{(k-1)(|R(A\cup B)|-1)}{|V_*|-1}
 \le\frac{k\cdot4k^4s}{n/2}
 =\frac{8k^5s}{n},
\]
where $|V_*|-1\ge n/2$ follows from \eqref{eq:core-finite-bounds} for large $k$ and $n$.
The $2s$ selected clauses are distinct, so their remaining variable choices are independent.
Multiplying these bounds and summing over $A$, the probability that such an $S$ exists with $|S\setminus H_1|=s$ is at most
\[
 \binom ns\left(\frac{8k^5s}{n}\right)^{2s}
 \le\left(\frac{k^{11}s}{n}\right)^s\le k^{-s}.
\]
Summing over $n/k^{24}\le s\le2n/k^{13}$, the probability that any such $S$ exists is at most $\exp(-n/k^{30})$.
Including the reachability and cycle bounds, the total failure probability is at most $3\exp(-n/k^{30})$.
Otherwise, every $*$-flippable set $S$ with $|S|\le2n/k^{13}$ has $|S\setminus H_1|<n/k^{24}$, since larger values are excluded above.
Together with $|B|<n/k^{24}$, this allows us to choose
$A_0\subseteq V_*$ containing $(S\setminus H_1)\cup B$ with
$|A_0|=\lceil2n/k^{24}\rceil$.
The reachability bound then yields, for sufficiently large $n$,
\[
 |S|\le |R(A_0)|<2k^4\left\lceil\frac{2n}{k^{24}}\right\rceil
 <\frac{5n}{k^{20}}<\frac{n}{k^{13}}. \qedhere
\]
\end{proof}

\begin{proof}[Proof of \pref{thm:rigidity}]\label{sec:core-geometry}
Fix $\sigma$ and use the Poissonized planted law of mean clause count $m$.
Its forcing mean $\lambda_+=mk/[n(2^k-1)]$ satisfies
\eqref{eq:core-density-estimate} throughout \eqref{eq:main-shatter-sharp}
for sufficiently large $k,n$, including the rounding in $m$.
Apply \pref{lem:core-size} and \pref{lem:flippable-gap}.
Clauses with at least two true literals affect neither $V_*$ nor its
$*$-flippable sets, so the two core conclusions fail with probability
at most $4\exp(-n/k^{30})$.
Let $\mathcal B_\Phi$ consist of solutions for which the core has at most
$(1-3/(2k\ln k))n$ variables or a $*$-flippable set of size in
$[n/k^{13},2n/k^{13}]$.
The Poissonized form of \pref{lem:norm-transfer} gives
\[
 \mathbb E_\Phi[\mathbb{1}\{\mathcal Z\}\cdot\mu_\Phi(\mathcal B_\Phi)]
 \le480n\sqrt m\exp\bigl(\Gamma n-n/k^{30}\bigr)
 \le\exp(-n/(2k^{35})).
\]
By \pref{lem:norm-lower} and Markov's inequality, with probability
at least $1-1/n$, both $Z_\Phi\ge\exp(n/(2\ln k))$ and
$\mu_\Phi(\mathcal B_\Phi)\le\exp(-n/k^{40})$ hold.
We prove both conclusions on this event.

We first identify the variables that remain fixed in a component. For $\sigma\notin\mathcal B_\Phi$, let $U$ be a largest $*$-flippable set under $\sigma$ of size less than $n/k^{13}$. By the size gap, every $*$-flippable set of size less than $2n/k^{13}$ has size less than $n/k^{13}$. Unions of two such sets are again $*$-flippable and hence also have size less than $n/k^{13}$. Maximality of $U$ therefore implies that all these sets lie in $U$.
By \pref{obs:core-difference-flippable} and the size gap,
any solution disagreeing with $\sigma$ on $V_*\setminus U$
differs from $\sigma$ on more than $2n/k^{13}$ core variables,
hence on more than $n/k^{13}$ variables in $V_*\setminus U$.
It therefore has Hamming distance greater than $n/k^{13}$
from every solution agreeing with $\sigma$ on $V_*\setminus U$.
Consequently, for every integer $1\le D\le n/k^{40}$,
the component of $\sigma$ in $G_D(\Phi)$ fixes $V_*\setminus U$,
and every variable in this set is not $(n/k^{40})$-loose
under $\sigma$.
Their number is
$|V_*\setminus U|>(1-3/(2k\ln k)-1/k^{13})n
\ge(1-2/(k\ln k))n$.

For the second conclusion, any component of $G_D(\Phi)$ containing a solution outside $\mathcal B_\Phi$ has at most $2^{2n/(k\ln k)}$ solutions. Its mass is at most $2^{2n/(k\ln k)}\exp(-n/(2\ln k))\le\exp(-n/k^{40})$. Every other component has mass at most $\mu_\Phi(\mathcal B_\Phi)\le\exp(-n/k^{40})$, completing the proof.
\end{proof}

\section{Proof of the reconstruction threshold}\label{sec:reconstruction}

While prior reductions from graph to tree reconstruction required overlap-concentration or replica-symmetry hypotheses~\cite{GM07,MRT11,COP18}, positive temperature~\cite{CMR22}, or lower densities~\cite{CCGPRZZ26}, we prove \pref{thm:main-reconstruction} by establishing \emph{local weak convergence} of $\mu_\Phi$ throughout the density range $0\le\alpha\le 2^k(\ln2-\frac1{\ln k})$ (\pref{lem:ordinary-local-law}): the local law on a depth-$r$ ball converges to an ordinary Poisson tree with independent boundary weights drawn from a deterministic stationary marginal law $P$~\cite{KMRTSZ07,MRTS08}. We prove this limit in \pref{sec:ordinary-local-law} by combining random-pinning factorization~\cite{Mon08} with a one-step Hellinger contraction against free-boundary tree marginals (\pref{lem:local-star}).

In \pref{sec:reconstruction-overview}, we state \pref{lem:ordinary-local-law} and formulate the one-step ordinary tree update (\pref{def:ordinary-tree-update}) together with its planted approximation (\pref{lem:tree-update-properties}) and one-step Hellinger comparison (\pref{lem:local-star}). In \pref{sec:boundary-transition}, we transfer the planted root-flow bound of \pref{thm:local}(iii) to a depth-$L_*$ expected conditional root variance lower bound on the ordinary tree (\pref{lem:depth-L-seed}), after which \pref{lem:local-star} contracts boundary information with depth to establish non-reconstruction below $\gamma_*$. In \pref{sec:reconstruction-beyond}, a one-sided comparison bounds the ordinary recursion by subcritical soft-field iterates at every depth (\pref{lem:tree-prediction}) to establish reconstruction above $\gamma_*$.

\subsection{Local weak convergence and the tree recursion}\label{sec:reconstruction-overview}

Throughout this section, we write
\[
 d=k-1,\qquad D=k\alpha,\qquad \lambda=D2^{-k}.
\]
We assume $k$ is sufficiently large and work throughout the density range stated below:
\[
 0\le\alpha\le 2^k\left(\ln2-\frac1{\ln k}\right).
\]
For $\Phi\sim\Phi(k,n,m)$ with $m/n\to\alpha$, a variable $v\in[n]$, a radius $r\ge 0$, and $\sigma\sim\mu_\Phi$ (with any convention for $\sigma$ when $\Omega_\Phi=\varnothing$), recall from \pref{sec:subsec-main-results} the unconditioned root marginal $p_v=\mu_\Phi(\sigma_v=1)$ and the boundary-conditioned marginal $q_{v,r}=\mu_\Phi(\sigma_v=1\mid\sigma_{\overline B_r(v)})$, with the same zero convention on unsatisfiable formulas.
Define the empirical marginal distribution $P_n=n^{-1}\sum_{v\in[n]}\delta_{p_v}$.

To model the local law of $\mu_\Phi$ on $B_r(\Phi, v)$, let $T_r$ be the depth-$r$ truncation of $T\sim\mathsf{Tree}_{\mathrm{ord}}(D)$ (\pref{def:poisson-tree}), with depth-$r$ variable boundary $\partial T_r$, and represent the influence of the formula outside the ball by independent boundary weights $p_u \in (0,1)$ on $\partial T_r$ drawn from a marginal distribution $P$.

\Needspace{6\baselineskip}
\begin{definition}[Tree Gibbs distribution with boundary weights]\label{def:ordinary-tree}
Given a distribution $P$ on $(0,1)$ and $T_r$ as above, draw independent boundary weights $\boldsymbol p=(p_u)_{u\in\partial T_r}\sim P^{\otimes\partial T_r}$ and define the weighted Gibbs distribution $\mu_{T_r,\boldsymbol p}$ on $\{0,1\}^{V(T_r)}$ by
\[
 \mu_{T_r,\boldsymbol p}(\sigma)
 \propto \mathbb{1}\{\sigma\models T_r\}\cdot\prod_{u\in\partial T_r} p_u^{\sigma_u}(1-p_u)^{1-\sigma_u}.
\]
\end{definition}

For $\sigma\sim\mu_{T_r,\boldsymbol p}$, write
\[
 p_{r,P}=\mu_{T_r,\boldsymbol p}(\sigma_o=1),\qquad
 Q_{r,P}=\mu_{T_r,\boldsymbol p}(\sigma_o=1\mid\sigma_{\partial T_r})
\]
for the unconditioned and boundary-conditioned root marginals, which serve as the tree counterparts of the finite-formula marginals $p_v$ and $q_{v,r}$ (when $B_r(\Phi,v)$ is a tree, its set $\partial B_r(\Phi,v)$ of depth-$r$ variables separates $v$ from $\overline B_r(v)$, so $q_{v,r}=\mu_\Phi(\sigma_v=1\mid\sigma_{\partial B_r(\Phi,v)})$).

The next lemma identifies a deterministic limit $P$ of $P_n$ and the joint asymptotic law of $(p_v,q_{v,r})$.

\begin{lemma}[Local weak convergence of the Gibbs distribution]\label{lem:ordinary-local-law}
For each $\alpha$ in the range above, there is a deterministic distribution $P$ on $(0,1)$ depending only on $(k,\alpha)$ such that as $n\to\infty$ with $m/n\to\alpha$, we have $P_n\to P$ in probability and:
\begin{enumerate}
\setlength{\itemsep}{2pt}\setlength{\parskip}{0pt}
\item[\textup{(i)}] \emph{(Marginal regularity.)} With $G(p)\defeq\min\{p,1-p\}^{-1}$, the law $P$ satisfies
\begin{equation}\label{eq:graph-state-bias}
 \E_P[G(p)]\le4.
\end{equation}
\item[\textup{(ii)}] \emph{(Reduction to the tree.)} For a uniform $v\in[n]$ and each fixed $r\ge0$, $\operatorname{Law}(p_{r,P})=P$ and $(p_v,q_{v,r})\Rightarrow(p_{r,P},Q_{r,P})$.
\end{enumerate}
\end{lemma}

We prove \pref{lem:ordinary-local-law} in \pref{sec:ordinary-local-law}. The relation $\operatorname{Law}(p_{r,P})=P$ shows that $P$ is a stationary law of the tree marginal recursion~\cite{KMRTSZ07,MRTS08}.
By \pref{lem:ordinary-local-law}(ii), both regimes of \pref{thm:main-reconstruction} reduce to the large-$r$ behavior of $(p_{r,P},Q_{r,P})$, which evolves across depths according to the ordinary counterpart of \pref{def:tree-updates}.

\begin{definition}[One-step ordinary tree update]\label{def:ordinary-tree-update}
Let $(p,Q)\in(0,1)\times[0,1]$ be a pair of random variables satisfying $\E[Q\mid p]=p$, and let $D\ge0$. Sample $N\sim\operatorname{Pois}(D)$, and conditioned on $N$, sample independent fair bits $\xi=(\xi_C)_{C\in[N]}\in\{0,1\}^N$ and i.i.d.\ copies $(p_{C,i},Q_{C,i})$ of $(p,Q)$ ($C\in[N],i\in[d]$), incorporating child literal signs by setting, independently for each $(C,i)$,
\[
    (\bar p_{C,i},\bar Q_{C,i}) =
    \begin{cases}
    (p_{C,i},Q_{C,i}) & \text{with probability } 1/2, \\
    (1-p_{C,i},1-Q_{C,i}) & \text{with probability } 1/2.
    \end{cases}
\]
With $\bar p=(\bar p_{C,i})_{C\in[N],i\in[d]}$, $\bar Q=(\bar Q_{C,i})_{C\in[N],i\in[d]}$, and $W_b,F$ as in \pref{sec:critical}, preserve the law of $(\bar p,N,\xi)$, reweight the conditional law of $\bar Q$ given $(\bar p,N,\xi)$ by $F(\bar Q)/F(\bar p)$, and define $(p^+,Q^+)$ by
\begin{equation}\label{eq:tree-pair-update}
    (1-p^+,p^+)\propto\bigl(W_0(\bar p),W_1(\bar p)\bigr), \qquad
    (1-Q^+,Q^+)\propto\bigl(W_0(\bar Q),W_1(\bar Q)\bigr),
\end{equation}
with $Q^+=1/2$ when $F(\bar Q)=0$ (which has probability zero under the reweighted law).
\end{definition}

Since $p\in(0,1)$, we have $W_b(\bar p)>0$, $F(\bar p)>0$, and $p^+=W_1(\bar p)/F(\bar p)\in(0,1)$ pointwise; in particular, the unconditioned marginal update $p\mapsto p^+$ depends only on $p$.
By multilinearity and $\E[Q\mid p]=p$, we have $\E[W_b(\bar Q)\mid\bar p,N,\xi]=W_b(\bar p)$ for $b\in\{0,1\}$, so the conditional density $F(\bar Q)/F(\bar p)$ has conditional mean one and satisfies $\E[Q^+\mid p^+]=p^+$.
Flipping all root signs $\xi_C\mapsto 1-\xi_C$ exchanges $W_0$ and $W_1$ while preserving $F(\bar p)$ and $F(\bar Q)$, so $(p^+,Q^+)\stackrel{d}{=}(1-p^+,1-Q^+)$.

By Bayes' rule on $T_{r+1}$ (\pref{def:ordinary-tree}), the pair $(p_{r+1,P},Q_{r+1,P})$ is obtained by applying \pref{def:ordinary-tree-update} to $(p_{r,P},Q_{r,P})$ starting from $Q_{0,P}\sim\operatorname{Bernoulli}(p_{0,P})$ with $p_{0,P}\sim P$.
Since $\operatorname{Law}(p_{r,P})=P$, the law $P$ is symmetric around $1/2$, so $(p_{r,P},Q_{r,P})\stackrel{d}{=}(1-p_{r,P},1-Q_{r,P})$ for every $r\ge0$ (with $Q_{0,P}\sim\operatorname{Bernoulli}(1/2)$).
Moreover, viewing $T_r$ as the depth-$r$ truncation of $T_{r+1}$, each $u\in\partial T_r$ roots an independent depth-$1$ subtree $T^{(u)}$ of $T_{r+1}$ whose unconditioned root marginal $p'_u\defeq\mu_{T^{(u)},\boldsymbol p|_{\partial T^{(u)}}}(\sigma_u=1)$ has law $\operatorname{Law}(p_{1,P})=P$; hence the restriction of $\mu_{T_{r+1},\boldsymbol p}$ to $\{0,1\}^{V(T_r)}$ is $\mu_{T_r,\boldsymbol p'}$ with $\boldsymbol p'=(p'_u)_{u\in\partial T_r}\sim P^{\otimes\partial T_r}$. Because $\partial T_r$ separates the root $o$ from $\partial T_{r+1}$, under $\mu_{T_{r+1},\boldsymbol p}$ we have $Q_{r+1,P}=\E[\mu_{T_r,\boldsymbol p'}(\sigma_o=1\mid\sigma_{\partial T_r})\mid\sigma_{\partial T_{r+1}}]$, and Jensen's inequality gives
\begin{equation}\label{eq:tree-posterior-monotonicity}
 \E\bigl[Q_{r+1,P}(1-Q_{r+1,P})\bigr]\ge\E\bigl[Q_{r,P}(1-Q_{r,P})\bigr]
 \qquad(r\ge0).
\end{equation}

To connect \pref{def:ordinary-tree-update} with the planted tree recursion studied in \pref{sec:local}, write $Q_{\mathrm{pl}}^+$ for the planted output of \pref{def:tree-updates} applied to $Q$.
Note that for $p\in(0,1)$, since $G(p)-2=2|p-1/2|/\min\{p,1-p\}$ and $|p-1/2|\min\{p,1-p\}\le1/16$, we have $(p-1/2)^2\le(G(p)-2)/32$ pointwise, so any law with $\E[G(p)]\le4$ satisfies $\E[(p-1/2)^2]\le1/16$.
The next lemma shows that this bound concentrates the ordinary normalizer $F(\bar p)$ around $2\ee^{-\lambda}$, giving an exponentially accurate planted approximation alongside $1$-Wasserstein stability.

\begin{lemma}[Planted approximation of the tree update]\label{lem:tree-update-properties}
There is $\eta_k=\ee^{-\Omega(k)}$ depending only on $k$ such that for all $0<\lambda\le k$:
\begin{enumerate}
\setlength{\itemsep}{2pt}\setlength{\parskip}{0pt}
\item[\textup{(i)}] For any input $(p,Q)$ in \pref{def:ordinary-tree-update} with $\E[G(p)]\le4$,
\[
 \|\operatorname{Law}(Q^+)-\operatorname{Law}(Q_{\mathrm{pl}}^+)\|_{\mathrm{TV}}\le\eta_k.
\]
\item[\textup{(ii)}] The planted update $\operatorname{Law}(Q)\mapsto\operatorname{Law}(Q_{\mathrm{pl}}^+)$ is $4d\lambda$-Lipschitz in $\mathsf W_1$.
\end{enumerate}
\end{lemma}
\begin{proof}
For each clause $C$, the signed child marginal $\bar p_{C,i}$ has $\E[\bar p_{C,i}]=1/2$ and $\E[\bar p_{C,i}^2]=1/4+\E[(p-1/2)^2]$, so $s_C=\prod_{i=1}^d\bar p_{C,i}$ has $\E[s_C]=2^{-d}$ and $\E[s_C^2]=\bigl(1/4+\E[(p-1/2)^2]\bigr)^d$.
Since the clauses with $\xi_C=0$ and $\xi_C=1$ form two independent $\operatorname{Pois}(D/2)$ families, the Poisson moment formulas $\E[W_b(\bar p)]=\exp(-\frac D2\E[s_C])=\ee^{-\lambda}$ and $\E[W_b(\bar p)^2]=\exp(\frac D2\E[(1-s_C)^2-1])=\ee^{-2\lambda}\exp(\frac D2\E[s_C^2])$ give
\[
 \operatorname{Var}\!\left[\ee^\lambda F(\bar p)/2\right]
 =\tfrac12\left(\exp\!\left(\tfrac D2\bigl(1/4+\E[(p-1/2)^2]\bigr)^d\right)-1\right).
\]
By $\E[G(p)]\le4$, we have $\E[(p-1/2)^2]\le1/16$ and hence $D\bigl(1/4+\E[(p-1/2)^2]\bigr)^d/2\le\lambda(5/8)^d$.
Let $\mathbb P_0$ be the law of $(\bar p,\bar Q,N,\xi)$ before reweighting; the ordinary and planted reweighted laws $\mathbb P_{\mathrm{ord}}$ and $\mathbb P_{\mathrm{pl}}$ have $\mathbb P_0$-densities $F(\bar Q)/F(\bar p)$ and $\ee^\lambda F(\bar Q)/2$, and under both laws the output marginal equals $W_1(\bar Q)/F(\bar Q)$ almost surely.
By the data-processing inequality and $\E_{\mathbb P_0}[F(\bar Q)\mid\bar p,N,\xi]=F(\bar p)$, $\|\operatorname{Law}(Q^+)-\operatorname{Law}(Q_{\mathrm{pl}}^+)\|_{\mathrm{TV}}\le\|\mathbb P_{\mathrm{ord}}-\mathbb P_{\mathrm{pl}}\|_{\mathrm{TV}}$ is at most
\[
 \tfrac12\E_{\mathbb P_0}\bigl[\bigl|1-\ee^\lambda F(\bar p)/2\bigr|\bigr]
 \le\tfrac12\sqrt{\operatorname{Var}[\ee^\lambda F(\bar p)/2]}
 \le\tfrac1{\sqrt 8}\sqrt{\ee^{\lambda(5/8)^d}-1}
 \le\eta_k
\]
for a suitable $\eta_k=\ee^{-\Omega(k)}$ uniformly over $0<\lambda\le k$.

For part~\textup{(ii)}, let $(Q,Q')$ be an optimal $\mathsf W_1$-coupling of $\operatorname{Law}(Q)$ and $\operatorname{Law}(Q')$, and draw $N\sim\operatorname{Pois}(D)$, i.i.d.\ copies $(Q_{C,i},Q'_{C,i})$ of $(Q,Q')$, and shared root signs $\xi_C$ and child literal signs to form $(\bar Q_{C,i},\bar Q'_{C,i})$.

For each clause $C$ with products $s_C=\prod_{i=1}^d\bar Q_{C,i}$ and $s'_C=\prod_{i=1}^d\bar Q'_{C,i}$, since $\E[\bar Q_{C,i}]=\E[\bar Q'_{C,i}]=1/2$ and $|\bar Q_{C,i}-\bar Q'_{C,i}|=|Q_{C,i}-Q'_{C,i}|$, independence and telescoping the product give $\E[|s_C-s'_C|]\le d2^{-(d-1)}\mathsf W_1(\operatorname{Law}(Q),\operatorname{Law}(Q'))$.
Conditioned on the number $j\sim\operatorname{Pois}(D/2)$ of clauses with $\xi_C=b$, each clause factor $1-s_C$ has expectation $1-2^{-d}$, so telescoping the product in $W_b(\bar Q)$ and $W_b(\bar Q')$ gives $\E[|W_b(\bar Q)-W_b(\bar Q')|\mid j]\le j(1-2^{-d})^{j-1}\E[|s_C-s'_C|]$, and averaging over $j\sim\operatorname{Pois}(D/2)$ yields
\[
 \E\bigl[|W_b(\bar Q)-W_b(\bar Q')|\bigr]\le\frac D2\ee^{-\lambda}\E[|s_C-s'_C|]\le 2d\lambda\ee^{-\lambda}\mathsf W_1(\operatorname{Law}(Q),\operatorname{Law}(Q')).
\]
For any $1$-Lipschitz function $\phi:[0,1]\to\mathbb R$ with $\phi(0)=0$, define $\Psi(w_0,w_1)=(w_0+w_1)\phi(w_1/(w_0+w_1))$ for $w_0+w_1>0$ and $\Psi(0,0)=0$.
For $f=w_0+w_1>0$ and $f'=w'_0+w'_1>0$ with $q=w_1/f$ and $q'=w'_1/f'$, since $|\phi(q)|\le q\le 1$ and $f'(q-q')=(1-q)(w_1-w'_1)-q(w_0-w'_0)$, writing $\Psi(w_0,w_1)-\Psi(w'_0,w'_1)=(f-f')\phi(q)+f'(\phi(q)-\phi(q'))$ gives $|\Psi(w_0,w_1)-\Psi(w'_0,w'_1)|\le 2(|w_0-w'_0|+|w_1-w'_1|)$ (which also holds when $f=0$ or $f'=0$ as $|\Psi(w_0,w_1)|\le w_1$).
Because $\E[\phi(Q_{\mathrm{pl}}^+)]=\frac12\ee^\lambda\E[\Psi(W_0(\bar Q),W_1(\bar Q))]$, we obtain
\[
 \bigl|\E[\phi(Q_{\mathrm{pl}}^+)]-\E[\phi((Q')_{\mathrm{pl}}^+)]\bigr|
 \le\ee^\lambda\sum_{b=0}^1\E\bigl[|W_b(\bar Q)-W_b(\bar Q')|\bigr]
 \le4d\lambda\,\mathsf W_1(\operatorname{Law}(Q),\operatorname{Law}(Q')).
\]
Taking the supremum over $1$-Lipschitz $\phi$ proves part~\textup{(ii)}.
\end{proof}

\paragraph{One-step Hellinger comparison.}
To contract boundary and pinning discrepancies across tree updates in both \pref{sec:boundary-transition} and \pref{sec:ordinary-local-law}, we next bound the one-step root-weight discrepancy in squared Hellinger distance.

Write $\mathsf H^2$ for squared Hellinger distance between finite probability laws; for the Bernoulli case, write
\[
 h^2(p,q)=\left(\sqrt p-\sqrt q\right)^2
          +\left(\sqrt{1-p}-\sqrt{1-q}\right)^2.
\]
For nonnegative finite measures $\xi,\zeta$ with positive total masses $|\xi|=\sum_z\xi(z)$ and $|\zeta|=\sum_z\zeta(z)$, put
\[
 x=\sqrt{|\xi|/|\zeta|},\qquad
 c=\sum_z\sqrt{\xi(z)\zeta(z)}/\sqrt{|\xi|\,|\zeta|}\in[0,1].
\]
Then $x^2+1-2xc\ge1-c^2\ge1-c$, so
\begin{equation}\label{eq:reference-hellinger-normalization}
 \mathsf H^2\left(\frac{\xi}{|\xi|},\frac{\zeta}{|\zeta|}\right)
 \le\frac2{|\zeta|}\sum_z\left(\sqrt{\xi(z)}-\sqrt{\zeta(z)}\right)^2.
\end{equation}
At fixed $N\ge0$, take independent copies $(p_{C,i},q_{C,i})$ of a pair $(p,q)\in[0,1]\times(0,1)$ for $C\in[N],i\in[d]$, apply a shared fair literal sign to each pair to form $(\bar p_{C,i},\bar q_{C,i})$ as in \pref{def:ordinary-tree-update}, and draw independent fair root signs $\xi_C\in\{0,1\}$.
Write $W_{b,p}=W_b(\bar p)$, $W_{b,q}=W_b(\bar q)$,
$F_q=W_{0,q}+W_{1,q}$, and $q^+=W_{1,q}/F_q\in(0,1)$.

\begin{lemma}[Hellinger comparison of one-step updates]\label{lem:local-star}
For $0\le v_0\le1/6$, there exist $A_N(v_0)\ge0$ ($N\ge0$) such that whenever $\E[G(q)]\le4$ and $\E[p(1-p)]\ge v_0$,
\[
 \E\left[\frac1{F_q}\sum_{b\in\{0,1\}}
       \left(\sqrt{W_{b,p}}-\sqrt{W_{b,q}}\right)^2\right]
 \le A_N(v_0)\,\E[h^2(p,q)],
\]
and for $N\sim\operatorname{Pois}(D)$ with $0\le\lambda\le k$,
\begin{equation}\label{eq:local-envelope}
 \E[G(q^+)]\le\frac72,\qquad
 \E[A_N(v_0)]=O\bigl(d^2\lambda\ee^{-2(d-1)v_0}\bigr).
\end{equation}
\end{lemma}
\begin{proof}
Set $\mathcal D=\E[h^2(p,q)]$, $a=1/2-v_0+3/(d-1)>1/3$, and
\[
 A_N(v_0)=Cd^2a^{d-1}(N+N^2 3^{-d})\exp(CdN6^{-d/2})
\]
for a sufficiently large absolute constant $C$.
For a single sign-flipped copy $(\bar p,\bar q)$ of $(p,q)$, we have $\E[\bar p^2]=1/2-\E[p(1-p)]\le 1/2-v_0$, while $\E[G(q)]\le4$ implies $\E[(q-1/2)^2]\le1/16$ and hence $\E[\bar q^2]\le5/16$.
For $x\in\{p,q\}$ and $0\le\theta\le1/(d-1)$, since $\bar x^2\le1$ and $1\le G(q)^\theta\le1+\theta(G(q)-1)$ by concavity of $z\mapsto z^\theta$, the hypothesis $\E[G(q)]\le4$ gives
\[
 \E[\bar x^2G(q)^\theta]
 \le\E[\bar x^2]+\E[G(q)^\theta-1]
 \le\E[\bar x^2]+\frac3{d-1}\le a,
\]
with $\E[\bar q^2G(q)^\theta]\le 5/16+3/(d-1)\le 1/3$ for large $k$.

For one clause with child variables $(\bar p_i,\bar q_i)_{i=1}^d$, put $s_p=\prod_{i=1}^d\bar p_i$,
$s_q=\prod_{i=1}^d\bar q_i$, and $H=(1-s_q)^{-1}$.
Since $1-s_q\ge1-\bar q_j\ge G(q_j)^{-1}$ for every $j\in[d]$,
we have $H\le G(q_j)$ for every $j$, and hence, for every
child $i\in[d]$,
\[
 H\le\prod_{j=1}^d G(q_j)^{1/d},\qquad
 H\le\prod_{j\ne i}G(q_j)^{1/(d-1)}.
\]
The first bound with $\theta=1/d$ and independence give $\E[Hs_q^2]\le\bigl(\E[\bar q^2G(q)^{1/d}]\bigr)^d\le 3^{-d}$.
To bound $\E[H(\sqrt{1-s_p}-\sqrt{1-s_q})^2]$, telescope $s_p-s_q$ across $i\in[d]$ by setting $t_i=\prod_{j<i}\bar p_j\prod_{j>i}\bar q_j$.
Using $|\sqrt{1-tx}-\sqrt{1-tx'}|\le t|\sqrt{1-x}-\sqrt{1-x'}|$ for $0\le t,x,x'\le 1$ (since $\sqrt{1-tx}+\sqrt{1-tx'}\ge\sqrt{1-x}+\sqrt{1-x'}$), Cauchy--Schwarz across the $d$ summands, and the second bound on $H$ (which omits child $i$) gives
\begin{align*}
 \E\left[H\left(\sqrt{1-s_p}-\sqrt{1-s_q}\right)^2\right]
 &\le d\sum_{i=1}^d\E\left[t_i^2\prod_{j\ne i}G(q_j)^{1/(d-1)}\right]\,\E\left[\left(\sqrt{1-\bar p_i}-\sqrt{1-\bar q_i}\right)^2\right]\\
 &\le\frac{d^2}{2}a^{d-1}\mathcal D,
\end{align*}
since each of the $d-1$ factors $j\ne i$ has expectation at most $a$ and averaging the fair literal sign of child $i$ gives
\[
 \E\left[\left(\sqrt{1-\bar p_i}-\sqrt{1-\bar q_i}\right)^2\right]=\mathcal D/2.
\]

Put $R=(1-s_p)/(1-s_q)$ and write
\[
 m=\E[R],\qquad e=\E\left[\left(\sqrt R-1\right)^2\right],\qquad \delta=m-1.
\]
The clause estimate above gives $e\le(d^2/2)a^{d-1}\mathcal D$.
Since $\E[s_p]=\E[s_q]=2^{-d}$ and $|s_p-s_q|\le2|\sqrt{1-s_p}-\sqrt{1-s_q}|$, Cauchy--Schwarz gives
\[
 \delta=\E[Hs_q(s_q-s_p)],\qquad
 \delta^2\le\E[Hs_q^2]\E[H(s_p-s_q)^2]
 \le4\cdot3^{-d}e.
\]
In particular, $|\delta|\le Cd6^{-d/2}$, because
$\mathcal D\le2$ and $a^{d-1}\le2^{-(d-1)}\ee^6$.

For independent copies $R_1,\ldots,R_j$ ($0\le j\le N$) of $R$, the identity
\[
 \E[(X-1)^2]=\left(\sqrt{\E[X^2]}-1\right)^2
             +2\left(\sqrt{\E[X^2]}-\E[X]\right)
\]
applied to $X=\prod_{c=1}^j\sqrt{R_c}$ gives
\[
 \E\left[\left(\sqrt{\prod_{c=1}^j R_c}-1\right)^2\right]
 =(m^{j/2}-1)^2
   +2\bigl(m^{j/2}-(\E[\sqrt R])^j\bigr)
 \le\exp(N|\delta|)(Ne+N^2\delta^2),
\]
where the $j=1$ case shows $0\le\sqrt m-\E[\sqrt R]\le e/2$, while $|\sqrt m-1|\le|\delta|$ and $0\le x^j-y^j\le jx^{j-1}(x-y)$ for $0\le y\le x\le\max\{1,\sqrt m\}\le\ee^{|\delta|/2}$ yield the inequality.

Finally, $F_q\ge W_{b,q}>0$ pointwise, so
\[
 \frac1{F_q}\sum_b\left(\sqrt{W_{b,p}}-\sqrt{W_{b,q}}\right)^2
 \le\sum_b\left(\sqrt{\frac{W_{b,p}}{W_{b,q}}}-1\right)^2.
\]
Conditioned on the root signs $\xi_1,\ldots,\xi_N$, each ratio $W_{b,p}/W_{b,q}$ is a product of $j=|\{C\in[N]:\xi_C=b\}|\le N$ independent copies of $R$. Applying the preceding inequality
and the bounds on $e$ and $\delta$ proves the fixed-$N$ bound.

For $N\sim\operatorname{Pois}(D)$, averaging over a single clause's fair root sign $\xi\in\{0,1\}$ gives $\E_\xi[(1-s_q)^{1-2\xi}]=\frac12((1-s_q)+(1-s_q)^{-1})=1+\frac12 Hs_q^2$.
Since $G(q^+)=1+\max\{W_{0,q}/W_{1,q},W_{1,q}/W_{0,q}\}$ and $\frac D2\E[Hs_q^2]\le\frac12 D3^{-d}=o_k(1)$, the Poisson product formula gives $\E[W_{0,q}/W_{1,q}]=\E[W_{1,q}/W_{0,q}]=\ee^{D\E[Hs_q^2]/2}$ and hence $\E[G(q^+)]\le 1+2\ee^{D3^{-d}/2}\le 7/2$.
Likewise, with $z=Cd6^{-d/2}$, uniformly for $D\le k2^k$ we have $Dz=o_k(1)$ and $D3^{-d}=o_k(1)$, so
\[
 \E[N\ee^{zN}]=D\ee^z\exp(D(\ee^z-1)),\qquad
 \E[N^2\ee^{zN}]=(D\ee^z+D^2\ee^{2z})\exp(D(\ee^z-1))
\]
give $\E[A_N(v_0)]=O(d^2D a^{d-1})$ after increasing $C$.
Since $D=2^{d+1}\lambda$ and $a=\frac12(1-2v_0+\frac6{d-1})\le\frac12\ee^{-2v_0+6/(d-1)}$, this implies \eqref{eq:local-envelope}.
\end{proof}

\subsection{Non-reconstruction below the critical threshold}\label{sec:boundary-transition}

Below $\gamma_*$, we establish non-reconstruction on the tree in two steps: first, we transfer the planted root-flow bound $\epsilon_k=(\ln k)^{10}/k$ of \pref{thm:local}(iii) via \pref{lem:tree-update-properties} to lower-bound the expected conditional root variance $\E[Q_{L_*,P}(1-Q_{L_*,P})]$ at a bounded depth $L_*$; second, we show that once this expected conditional variance is bounded away from zero, each subsequent tree update contracts the squared Hellinger distance between $Q_{r,P}$ and $p_{r,P}$ by \pref{lem:local-star}.

\begin{lemma}[Depth-$L_*$ expected conditional variance lower bound]\label{lem:depth-L-seed}
There is an integer $L_*=O((\ln\ln k)^{1/16})$ depending only on $k$ such that throughout \eqref{eq:main-connectivity-range} with $\alpha>0$,
\begin{equation}\label{eq:depth-L-seed}
 \E[Q_{L_*,P}(1-Q_{L_*,P})]\ge\frac{\epsilon_k}{8}.
\end{equation}
\end{lemma}
\begin{proof}
For $\alpha>0$ in \eqref{eq:main-connectivity-range}, \pref{thm:local}(iii) at parameter $\lambda_+=\lambda$ supplies an integer $L\ge1$ and $\vartheta\in[0,1]$, with $1\le L\le L_*$ for a common integer $L_*=O((\ln\ln k)^{1/16})$ depending only on $k$. Let $(\Phi,x)$ be sampled to depth $L$ from $\mathsf{Tree}_{\mathrm{pl}}(\lambda)$, and let $y$ be the output of \pref{alg:local-update} on $(\Phi,x)$.
Let $\widehat Q_r$ be the depth-$r$ boundary-conditioned root marginal on $\mathsf{Tree}_{\mathrm{pl}}(\lambda)$, whose law evolves by \pref{def:tree-updates} starting from $\widehat Q_0\sim\operatorname{Bernoulli}(1/2)$, and let $\delta_r=\mathsf W_1(\operatorname{Law}(\widehat Q_r),\operatorname{Law}(Q_{r,P}))$.
Since $\delta_0=0$ and $\mathsf W_1\le\|\cdot\|_{\mathrm{TV}}$ on $[0,1]$,
\pref{lem:tree-update-properties} gives
\[
 \delta_{r+1}\le\eta_k+4d\lambda\delta_r,
 \qquad
 \delta_L\le L\eta_k(1+4d\lambda)^L=\ee^{-\Omega(k)},
\]
using $L\ln k=o(k)$ and $\lambda\le k$.
In \pref{alg:local-update}, the revealed set $R$ contains the depth-$L$ boundary $\partial\Phi\defeq\{v\in V(\Phi):\operatorname{depth}(v)=L\}$ and is constructed by inspecting $x$ only on variables already in $R$, so for any fixed realization of the retention variables $(\eta_v)$ in \pref{alg:local-update}, every $x'\in\Omega_\Phi$ extending $x_R$ produces the same $R$; since $x\sim\mu_\Phi$ given $\Phi$ (\pref{def:planted-root-rate}) and $y\sim\mu_{\Phi,x_R}$, conditional on $(\Phi,R,x_R)$ the pair $(x,y)$ consists of independent draws from $\mu_{\Phi,x_R}$.
Writing $q_o=\mu_{\Phi,x_R}(\sigma_o=1)$, we have $\Pr(x_o\ne y_o\mid\Phi,R,x_R)=2q_o(1-q_o)$ and, by the tower property under $\mu_\Phi$ since $\partial\Phi\subseteq R$, $\E[q_o\mid\Phi,x_{\partial\Phi}]=\mu_\Phi(\sigma_o=1\mid\sigma_{\partial\Phi}=x_{\partial\Phi})=\widehat Q_L$.
Since $\max\{1,\Ham(x,y)\}\ge1$, \pref{thm:local}(iii) and Jensen's inequality for the concave function $q\mapsto q(1-q)$ give
\[
 \epsilon_k\le\Pr(x_o\ne y_o)
 =2\E[q_o(1-q_o)]
 \le2\E[\widehat Q_L(1-\widehat Q_L)].
\]
Since $q(1-q)$ is $1$-Lipschitz on $[0,1]$ and $L\le L_*$, posterior monotonicity \eqref{eq:tree-posterior-monotonicity} yields
\[
 \E[Q_{L_*,P}(1-Q_{L_*,P})]
 \ge\E[Q_{L,P}(1-Q_{L,P})]
 \ge\epsilon_k/2-\delta_L\ge\epsilon_k/8,
\]
which proves the stated bound.
\end{proof}

Combining the depth-$L_*$ expected conditional variance bound of \pref{lem:depth-L-seed} with the one-step Hellinger contraction of \pref{lem:local-star} now yields the non-reconstruction part of \pref{thm:main-reconstruction}.

\begin{proof}[Proof of \pref{thm:main-reconstruction} (Non-reconstruction)]
Use the deterministic limiting law $P$ from \pref{lem:ordinary-local-law}, and put $\mathcal D_r=\E[h^2(Q_{r,P},p_{r,P})]$.
For $\lambda=0$, $T_r$ has no clauses and $Q_{r,P}=p_{r,P}=1/2$ for $r\ge1$, so $\mathcal D_r=0$.
For $\lambda>0$ and $r\ge L_*$, \eqref{eq:tree-posterior-monotonicity} and
\pref{lem:depth-L-seed} give $\E[Q_{r,P}(1-Q_{r,P})]\ge\E[Q_{L_*,P}(1-Q_{L_*,P})]\ge v_0\defeq\epsilon_k/8$.

Let $(p_{r,P},Q_{r,P})$ evolve to $(p^+,Q^+)\stackrel{d}{=}(p_{r+1,P},Q_{r+1,P})$ by \pref{def:ordinary-tree-update}.
For $F(\bar Q)>0$, the normalization inequality
\eqref{eq:reference-hellinger-normalization}, applied with
$\xi(b)=W_b(\bar p)$ and $\zeta(b)=W_b(\bar Q)$, gives
\[
 F(\bar Q)h^2(Q^+,p^+)
 \le 2\sum_{b=0}^1
 \left(\sqrt{W_b(\bar Q)}-\sqrt{W_b(\bar p)}\right)^2.
\]
The inequality is trivial when $F(\bar Q)=0$.
Since the ordinary reweighting has density $F(\bar Q)/F(\bar p)$ (with zero contribution on $F(\bar Q)=0$), applying \pref{lem:local-star} to the input pair $(p,q)=(Q_{r,P},p_{r,P})$ at parameter $v_0=\epsilon_k/8$ (whose hypotheses hold because $\operatorname{Law}(p_{r,P})=P$ satisfies $\E_P[G(p)]\le4$ by \pref{lem:ordinary-local-law}(i)) and \eqref{eq:local-envelope} gives, for $r\ge L_*$ and sufficiently large $k$,
\[
 \mathcal D_{r+1}
 \le 2\E\left[\frac1{F(\bar p)}\sum_{b=0}^1\left(\sqrt{W_b(\bar Q)}-\sqrt{W_b(\bar p)}\right)^2\right]
 \le 2\E[A_N(v_0)]\mathcal D_r
 \le \frac18\mathcal D_r,
\]
since $v_0=(\ln k)^{10}/(8k)$ and $\lambda\le k$ imply $2\E[A_N(v_0)]=O(d^2\lambda\ee^{-(d-1)\epsilon_k/4})\le1/8$.
Thus $\mathcal D_r\le2\cdot8^{-(r-L_*)}$ for all $r\ge L_*$ and all $\alpha$ in \eqref{eq:main-connectivity-range}.
Because any sequence of clause counts $m$ with $m/n$ in the compact interval \eqref{eq:main-connectivity-range} has a subsequence with $m/n\to\alpha$ in \eqref{eq:main-connectivity-range}, along which \pref{lem:ordinary-local-law}(ii), $|q-p|\le h(q,p)$, and Cauchy--Schwarz give
\[
 \lim_{n\to\infty}\E[|q_{v,r}-p_v|]
 =\E[|Q_{r,P}-p_{r,P}|]
 \le\sqrt{\mathcal D_r}
 \le\sqrt{2}\cdot 8^{-(r-L_*)/2},
\]
the bound $\E[|q_{v,r}-p_v|]\le\sqrt{2}\cdot 8^{-(r-L_*)/2}+o_n(1)$ holds uniformly over \eqref{eq:main-connectivity-range} for each fixed $r\ge L_*$.
Letting $n\to\infty$ and then $r\to\infty$ establishes non-reconstruction throughout
\eqref{eq:main-connectivity-range}.
\end{proof}

\subsection{Reconstruction beyond the critical threshold}\label{sec:reconstruction-beyond}

Whereas below $\gamma_*$ the boundary-conditioned marginal $Q_{r,P}$ concentrates around $p_{r,P}$, above $\gamma_*$ it concentrates near $\{0,1\}$ at every depth. We establish this in \pref{lem:tree-prediction} by comparing the ordinary recursion directly with subcritical soft-field iterates via square-root test functions and absorbing the planted-approximation error into the recursion parameter.

Recall from \pref{sec:critical} that for a symmetric random marginal $Q\in[0,1]$ ($Q\stackrel{d}{=}1-Q$), its $f$-weighted measure $\nu$ at scale $S=d\lambda$ in \eqref{eq:posterior-functionals} satisfies $\langle\psi,\nu\rangle=d\lambda\,\E[\psi(Q)]$ for bounded Borel $\psi:[0,1]\to\mathbb R$ vanishing at $0,1$, with $f(q)=2\sqrt{q(1-q)}$, $f$-weighted mass $m(\nu)=\langle f,\nu\rangle$, and scaled unfrozen probability $u(\nu)=\langle\mathbb{1}_{(0,1)},\nu\rangle=d\lambda\Pr(0<Q<1)$.

\begin{lemma}[Tree prediction error]\label{lem:tree-prediction}
For every fixed $\varepsilon>0$ and sufficiently large $k$, if
$\lambda\ge\ln k+\ln\ln k+\gamma_*+\varepsilon$, then
\[
 \E[\min\{Q_{r,P},1-Q_{r,P}\}]\le\frac1{2d\lambda}
 \qquad(r\ge0).
\]
\end{lemma}
\begin{proof}
For $a=(a_0,a_1),b=(b_0,b_1)\in[0,\infty)^2$, define the nonnegative concave function $\psi_{a,b}:[0,1]\to[0,\infty)$ by
\[
 \psi_{a,b}(q)=\sqrt{[a_0(1-q)+a_1q][b_0(1-q)+b_1q]}
 -(1-q)\sqrt{a_0b_0}-q\sqrt{a_1b_1},
\]
which vanishes at $q=0$ and $q=1$. For symmetric measures $\nu,\nu'$ on $\mathbb R$ with $u(\nu),u(\nu')<\infty$ (so that $\langle\psi_{a,b},\nu\rangle\le\|\psi_{a,b}\|_\infty u(\nu)<\infty$), write
$\nu\preceq_{\mathrm{sq}}\nu'$ if $\langle\psi_{a,b},\nu\rangle\le\langle\psi_{a,b},\nu'\rangle$ for all $a,b\in[0,\infty)^2$.
Since $f=2\psi_{(1,0),(0,1)}$, the order $\nu\preceq_{\mathrm{sq}}\nu'$ implies $m(\nu)\le m(\nu')$.

For $\psi=\psi_{a,b}$ and $w=(w_0,w_1)\in[0,\infty)^2$, define $\Psi(w_0,w_1)=(w_0+w_1)\psi(w_1/(w_0+w_1))$ for $w_0+w_1>0$ and $\Psi(0,0)=0$, and for $s\in[0,1]$ define the one-clause difference operator
\[
 B_s\Psi(w)\defeq\frac12\bigl[\Psi((1-s)w_0,w_1)+\Psi(w_0,(1-s)w_1)\bigr]-(1-s)\Psi(w).
\]
Setting $A=a_0w_0+a_1w_1$ and $B=b_0w_0+b_1w_1$, direct substitution gives $\Psi((1-s)w_0,w_1)-(1-s)\Psi(w)=g_0(s)$ and $\Psi(w_0,(1-s)w_1)-(1-s)\Psi(w)=g_1(s)$, so $B_s\Psi(w)=\frac12\bigl(g_0(s)+g_1(s)\bigr)$, where for each $i\in\{0,1\}$,
\[
 g_i(s)\defeq\sqrt{(A-a_iw_is)(B-b_iw_is)}-(1-s)\sqrt{AB}-s w_{1-i}\sqrt{a_{1-i}b_{1-i}}
 =\psi_{a^{(i)},b^{(i)}}(s)
\]
with $a^{(i)}=(A,a_{1-i}w_{1-i})$ and $b^{(i)}=(B,b_{1-i}w_{1-i})$.
Furthermore, each $g_i(s)$ has a nonnegative series expansion $\sum_{\ell\ge2}c_\ell(s-s^\ell)$ with $c_\ell\ge0$ on $[0,1]$.
Indeed, if $AB=0$ then $g_i(s)=0$; if $AB>0$, dividing $g_i(s)$ by $\sqrt{AB}$ gives $\sqrt{(1-xs)(1-ys)}-(1-s)-s\sqrt{(1-x)(1-y)}$ with $x=a_iw_i/A\in[0,1]$ and $y=b_iw_i/B\in[0,1]$.
Since $(1-xs)^{-3/2}$ and $(1-ys)^{-3/2}$ have nonnegative binomial-series coefficients around $s=0$, the second derivative
\[
 \frac{\mathrm d^2}{\mathrm ds^2}\sqrt{(1-xs)(1-ys)}
 =-\frac{(x-y)^2}{4(1-xs)^{3/2}(1-ys)^{3/2}}
\]
has nonpositive power-series coefficients around $s=0$, so all Taylor coefficients of $\sqrt{(1-xs)(1-ys)}$ of degree $\ell\ge2$ are nonpositive; subtracting its linear interpolant between $s=0$ and $s=1$ gives $\sum_{\ell\ge2}c_\ell(s-s^\ell)$ with $c_\ell\ge0$, where continuity at $s=1$ yields $\sum_{\ell\ge2}c_\ell<\infty$ and uniform convergence on $[0,1]$.

For any finite intensity measure $\rho$ on $(0,1)$, let $s=(s_1,\ldots,s_N)$ be a Poisson process on $(0,1)$ with intensity $\rho$ and independent fair root signs $\xi_1,\ldots,\xi_N\in\{0,1\}$, and define
\[
 I(\rho;\Psi)\defeq\ee^{\int_{(0,1)}s\,\rho(\mathrm{d}s)}\,\E_\rho\bigl[\Psi(W_0(s),W_1(s))\bigr].
\]
Conditioning on $N\sim\operatorname{Pois}(\rho((0,1)))$ expands $\ee^{\rho((0,1))}\E_\rho[\Psi(W_0(s),W_1(s))]$ as the convergent series $\sum_{n\ge0}\frac1{n!}\int_{(0,1)^n}\E_\xi[\Psi(W_0,W_1)]\,\mathrm{d}\rho^{\otimes n}$.
Since adjoining a point $s_{n+1}=s\in(0,1)$ with an independent fair root sign $\xi_{n+1}\in\{0,1\}$ changes $\E_\xi[\Psi(W_0,W_1)]$ by $\E_\xi[B_s\Psi(W_0,W_1)-s\Psi(W_0,W_1)]$, differentiating the series term-by-term along $\rho_t=(1-t)\rho_0+t\rho_1$ ($t\in[0,1]$) for finite intensity measures $\rho_0,\rho_1$ on $(0,1)$ gives
\[
 \frac{\mathrm d}{\mathrm dt}I(\rho_t;\Psi)
 =\ee^{\int_{(0,1)}s\,\rho_t(\mathrm{d}s)}\,\E_{\rho_t}\!\left[\int_{(0,1)}B_s\Psi(W_0,W_1)\,(\rho_1-\rho_0)(\mathrm{d}s)\right].
\]
Set $\beta=d\lambda\ee^{-\lambda}$. For a symmetric marginal $Q\in[0,1]$ with $f$-weighted measure $\nu$ at scale $S=d\lambda$ and planted output $Q_{\mathrm{pl}}^+$ with $f$-weighted measure $\nu_{\mathrm{pl}}^+$, each signed child marginal $\bar Q_{C,i}$ in \pref{def:tree-updates} has the law of $Q$; let $\rho$ be the restriction to $(0,1)$ of $D$ times the law of $s=\prod_{i=1}^d Q_i$ for independent copies $Q_1,\ldots,Q_d$ of $Q$.
Under \pref{def:tree-updates}, the planted density $\ee^\lambda F/2$ gives $\langle\psi,\nu_{\mathrm{pl}}^+\rangle=\frac12 d\lambda\ee^\lambda\E[\Psi(W_0,W_1)]$: clauses with $s_C=0$ do not affect $(W_0,W_1)$, while any clause with $s_C=1$ forces $W_0=0$ or $W_1=0$ where $\Psi$ vanishes (including at $(0,0)$, where the planted density also vanishes); since $s_C=1$ occurs with Poisson mean $\rho_h=D\,\E[s]-\int_{(0,1)}s\,\rho(\mathrm{d}s)=2\lambda-\int_{(0,1)}s\,\rho(\mathrm{d}s)$ independently of clauses with $s_C\in(0,1)$, conditioning on no clause having $s_C=1$ (probability $\ee^{-\rho_h}$) gives $\langle\psi,\nu_{\mathrm{pl}}^+\rangle=\frac\beta2 I(\rho;\Psi)$.
Meanwhile, for any symmetric measure $\nu'$ on $\mathbb R$ with $u(\nu')<\infty$, the intensity measure $\rho_{\nu'}$ on $(0,1)$ in \eqref{eq:reference-clause-average} satisfies $\int_{(0,1)}\phi(s)\,\rho_{\nu'}(\mathrm{d}s)=4\langle\phi(1-\cdot),\nu'\rangle$ by \eqref{eq:posterior-functionals} (with $\int_{(0,1)}s\,\rho_{\nu'}(\mathrm{d}s)=2u(\nu')$), so \eqref{eq:reference-clause-average} gives $\langle\psi,\mathcal R_\beta\nu'\rangle=\frac\beta2 I(\rho_{\nu'};\Psi)$.
Since $Q^\ell\le Q$ gives $x_\ell\defeq 2\E[Q^\ell]\in[0,1]$ and $s-s^\ell$ vanishes at $\{0,1\}$, for every $\ell\ge2$ we have
\[
 \int_{(0,1)}(s-s^\ell)\,\rho(\mathrm{d}s)=2\lambda(1-x_\ell^d)
 \le2d\lambda(1-x_\ell)=\int_{(0,1)}(s-s^\ell)\,\rho_\nu(\mathrm{d}s).
\]
Whenever $\nu\preceq_{\mathrm{sq}}\nu'$, combining the nonnegative series expansion of $g_0,g_1$ in $s-s^\ell$ with closure of $\{\psi_{a,b}\}$ under $s\mapsto 1-s$ gives, for every $w\in[0,\infty)^2$,
\begin{align*}
 \int_{(0,1)}B_s\Psi(w)\,\rho(\mathrm{d}s)
 &\le\int_{(0,1)}B_s\Psi(w)\,\rho_\nu(\mathrm{d}s)
 =4\langle B_{1-\cdot}\Psi(w),\nu\rangle \\
 &\le4\langle B_{1-\cdot}\Psi(w),\nu'\rangle
 =\int_{(0,1)}B_s\Psi(w)\,\rho_{\nu'}(\mathrm{d}s).
\end{align*}
Integrating $\frac{\mathrm d}{\mathrm dt}I(\rho_t;\Psi)\ge0$ from $\rho_0=\rho$ to $\rho_1=\rho_{\nu'}$ therefore yields $\nu_{\mathrm{pl}}^+\preceq_{\mathrm{sq}}\mathcal R_\beta\nu'$ (and taking $\rho_0=0$, where $\frac12 I(0;\Psi)=\frac12\Psi(1,1)=\psi(1/2)=\langle\psi,\delta_0\rangle$ for $\delta_0\defeq\mathcal R_1(0)$, gives $\delta_0\preceq_{\mathrm{sq}}\mathcal R_1\nu'$).

Now let $\nu_r$ and $\nu_{r,\mathrm{pl}}^+$ be the $f$-weighted measures at scale $S=d\lambda$ of $Q_{r,P}$ and $(Q_{r,P})_{\mathrm{pl}}^+$.
Since both marginals are symmetric around $1/2$ and each $\psi=\psi_{a,b}$ is nonnegative and concave on $[0,1]$ with $0\le\frac12(\psi(q)+\psi(1-q))\le\psi(1/2)=\langle\psi,\delta_0\rangle$, \eqref{eq:graph-state-bias} and \pref{lem:tree-update-properties}(i) give $\|\operatorname{Law}(Q_{r+1,P})-\operatorname{Law}((Q_{r,P})_{\mathrm{pl}}^+)\|_{\mathrm{TV}}\le\eta_k$ and hence, with $\tilde\eta_k\defeq d\lambda\eta_k=\ee^{-\Omega(k)}$,
\[
 \langle\psi,\nu_{r+1}\rangle-\langle\psi,\nu_{r,\mathrm{pl}}^+\rangle
 =d\lambda\bigl(\E[\psi(Q_{r+1,P})]-\E[\psi((Q_{r,P})_{\mathrm{pl}}^+)]\bigr)
 \le\tilde\eta_k\langle\psi,\delta_0\rangle.
\]
Define $\nu_r'\defeq\mathcal R_{\beta+\tilde\eta_k}^{\,r}0$, which has $u(\nu_r')<\infty$ for every $r\ge0$ since $u(\mathcal R_{\beta+\tilde\eta_k}\nu')=(\beta+\tilde\eta_k)\ee^{u(\nu')}$ by \eqref{eq:reference-clause-average}.
Assuming $\nu_r\preceq_{\mathrm{sq}}\nu_r'$ (which holds at $r=0$ since $\nu_0=\nu_0'=0$), we obtain
\[
 \nu_{r+1}\preceq_{\mathrm{sq}}\nu_{r,\mathrm{pl}}^++\tilde\eta_k\delta_0
 \preceq_{\mathrm{sq}}\mathcal R_\beta\nu_r'+\tilde\eta_k\mathcal R_1\nu_r'
 =\mathcal R_{\beta+\tilde\eta_k}\nu_r'=\nu_{r+1}',
\]
so $\nu_r\preceq_{\mathrm{sq}}\mathcal R_{\beta+\tilde\eta_k}^{\,r}0$ for all $r\ge0$ by induction.
Throughout the stated range of $\lambda$, for sufficiently large $k$,
\[
 \beta+\tilde\eta_k\le\ee^{-\gamma_*-\varepsilon/2}+\tilde\eta_k
 <\ee^{-\gamma_*}.
\]
Thus \pref{thm:critical-threshold} gives $m(\nu_r)\le m(\nu_r')\le1$ at every
depth, and $2\min\{q,1-q\}\le f(q)$ proves the prediction-error bound.
\end{proof}

Combining the small tree prediction error of \pref{lem:tree-prediction} with the marginal regularity bound $\E_P[G(p)]\le4$ in \pref{lem:ordinary-local-law}(i) now completes the proof of \pref{thm:main-reconstruction}.

\begin{proof}[Proof of \pref{thm:main-reconstruction} (Reconstruction)]
Fix $\varepsilon>0$. For any $\alpha$ in \eqref{eq:main-reconstruction-bounds}, \pref{lem:ordinary-local-law} and \pref{lem:tree-prediction} give $\E[\min\{Q_{r,P},1-Q_{r,P}\}]\le1/(2d\lambda)\le1/(k\ln k)$ for all $r\ge1$, while \eqref{eq:graph-state-bias} gives $\E_P[(p-1/2)^2]\le1/16$ and hence $\E_P[|p-1/2|]\le1/4$ by Cauchy--Schwarz.
Since $|q-p|\ge|q-1/2|-|p-1/2|=1/2-\min\{q,1-q\}-|p-1/2|$, we also have $\E[|Q_{r,P}-p_{r,P}|]\ge 1/4-1/(k\ln k)>0$.
Because any sequence of clause counts $m$ with $m/n$ in the compact interval \eqref{eq:main-reconstruction-bounds} has a subsequence with $m/n\to\alpha$ in \eqref{eq:main-reconstruction-bounds}, along which $(p_v,q_{v,r})\Rightarrow(p_{r,P},Q_{r,P})$ by \pref{lem:ordinary-local-law}(ii), the bounds
\[
 \E[\min\{q_{v,r},1-q_{v,r}\}]\le\frac1{k\ln k}+o_n(1),\qquad
 \E[|q_{v,r}-p_v|]\ge\frac14-\frac1{k\ln k}-o_n(1)
\]
hold uniformly over \eqref{eq:main-reconstruction-bounds} for every fixed $r\ge1$.
On satisfiable formulas (which have probability $1-o_n(1)$ by \pref{lem:norm-lower}), $\min\{q_{v,r},1-q_{v,r}\}$ is the conditional error probability of the predictor $\mathbb{1}\{q_{v,r}\ge1/2\}$ given $\sigma_{\overline B_r(v)}$; since the lower bound on $\liminf_{n\to\infty}\E[|q_{v,r}-p_v|]$ is independent of $r$,
taking $r\to\infty$ establishes reconstruction throughout
\eqref{eq:main-reconstruction-bounds} together with the required prediction-error bound.
Combining with non-reconstruction proves
\eqref{eq:main-reconstruction-headline}.
\end{proof}

\subsection{Proof of local weak convergence of the Gibbs distribution}\label{sec:ordinary-local-law}

\paragraph{Conditional factorization under random pinning.}
\Needspace{6\baselineskip}
The following discrete form of Montanari's pinning argument~\cite[Lemma~3.1 and Theorem~2.2]{Mon08} gives conditional factorization after pinning a small random set of coordinates; the one-step tree contraction in \pref{lem:local-star} against free-boundary tree marginals then removes this pinning and identifies the limiting marginal law $P$.

For a probability measure $\mu$ on $\{0,1\}^n$ and integers $M\ge1$ and $\ell\ge0$, independently sample $\sigma\sim\mu$, $U\sim\operatorname{Unif}(\{0,\ldots,M-1\})$, and i.i.d.\ uniform coordinates $I_1,\ldots,I_\ell,J_1,J_2,\ldots\in[n]$.
Write $\mu^{(u)}=\mu(\cdot\mid\sigma_{J_1},\ldots,\sigma_{J_u})$ and $\mu_I^{(u)}$ for its joint marginal on $I=(I_1,\ldots,I_\ell)$.

\begin{lemma}[Conditional factorization under random pinning]\label{lem:pin-factorization}
For every probability measure $\mu$ on $\{0,1\}^n$ and integers $M\ge1$ and $\ell\ge0$,
\begin{equation}\label{pinCV:factorization}
 \E\left[\left\|\mu_I^{(U)}-\bigotimes_{a=1}^\ell\mu_{I_a}^{(U)}\right\|_{\mathrm{TV}}\right]
 \le\sqrt{\frac{\ell(\ell-1)\ln2}{4M}}.
\end{equation}
\end{lemma}
\begin{proof}
For $\ell\le1$ the left-hand side is zero. For $\ell\ge2$, write $\mathcal H(p)=-p\ln p-(1-p)\ln(1-p)$ for binary entropy (with $0\ln 0=0$), and let $\mathcal H_u=\E[\mathcal H(\mu_{I_1}^{(u)}(\sigma_{I_1}=1))]$ be the expected conditional entropy of a uniformly chosen coordinate given $u$ pinned coordinates.
By the chain rule for Shannon entropy, the expected relative entropy $\E\bigl[D_{\mathrm{KL}}\bigl(\mu_I^{(u)}\,\|\,\bigotimes_{a=1}^\ell\mu_{I_a}^{(u)}\bigr)\bigr]$ equals $\sum_{s=0}^{\ell-1}(\mathcal H_u-\mathcal H_{u+s})$, since conditioning sequentially on $\sigma_{I_1},\ldots,\sigma_{I_s}$ augments $J_1,\ldots,J_u$ by $s$ additional i.i.d.\ uniform pinned coordinates.
Since $0\le\mathcal H_u\le\ln2$ and $\mathcal H_u$ is nonincreasing in $u$,
\[
 \frac1M\sum_{u=0}^{M-1}\sum_{s=0}^{\ell-1}(\mathcal H_u-\mathcal H_{u+s})
 \le\frac{\ell(\ell-1)\ln2}{2M}.
\]
Pinsker's inequality and Cauchy--Schwarz give \eqref{pinCV:factorization}.
\end{proof}

\paragraph{The cavity method and contraction of pinned marginals.}
Let $\pi_t=\operatorname{Law}(p_{t,\delta_{1/2}})$ be the depth-$t$ free-boundary tree marginal law, so $\pi_0=\delta_{1/2}$ and $\pi_{t+1}$ is obtained from $\pi_t$ by the unconditioned update $q\mapsto q^+$ of \pref{def:ordinary-tree-update}.
Since $G(1/2)=2$, induction on $t$ using \eqref{eq:local-envelope} in \pref{lem:local-star} shows that $\pi_t$ is supported on $(0,1)$ and satisfies
\begin{equation}\label{eq:free-reference-regularity}
 \E_{\pi_t}[G(q)]\le\frac72<4
 \qquad(t\ge0).
\end{equation}
For a finite formula, let $\widehat p_v^{\,t}=\mu_{B_t(\Phi,v)}(\sigma_v=1)$ be the unconditioned root marginal of the subformula $B_t(\Phi,v)$ (equivalently, \pref{def:ordinary-tree} on $B_t(\Phi,v)$ with constant boundary weights $1/2$) when this neighborhood is a tree, and set $\widehat p_v^{\,t}=1/2$ otherwise (in particular, $\widehat p_v^{\,0}=1/2$).
For $u\ge0$ and a tuple $\boldsymbol J=(J_1,\ldots,J_u)$ of independent uniform coordinates in $[n]$, define the pinned weights $w_y$, pinned marginals $p_{v,y}$, and \emph{depth-$t$ pinned Hellinger discrepancy} $\mathcal D_t(\Phi,u)$ by
\[
 w_y=\mu_\Phi(\sigma_{\boldsymbol J}=y),\qquad
 p_{v,y}=\mu_\Phi(\sigma_v=1\mid\sigma_{\boldsymbol J}=y),\qquad
 \mathcal D_t(\Phi,u)=\E_{\boldsymbol J}\left[\frac1n\sum_v\sum_yw_y
       h^2(p_{v,y},\widehat p_v^{\,t})\right],
\]
omitting zero-weight pinned assignments $y\in\{0,1\}^u$ and setting $\mathcal D_t=0$ on unsatisfiable formulas.
For bounded random variables $X_{n,m}$ on $\Phi(k,n,m)$, define the \emph{asymptotic upper expectation} $\overline\E[X]$ at density $\alpha$ and the \emph{asymptotic pinned Hellinger discrepancy} $\mathcal U_t$ at depth $t$ by
\[
 \overline\E[X]=\limsup_{\substack{n\to\infty\\m/n\to\alpha}}
      \E[X_{n,m}],\qquad
 \mathcal U_t=\limsup_{M\to\infty}\overline\E\left[\frac1M\sum_{u<M}\mathcal D_t(\Phi,u)\right].
\]
The next lemma bounds $\mathcal U_{t+1}$ from $\mathcal U_t$; each step is based on the cavity method~\cite{MPZ02,MZ02,MMZ06,KMRTSZ07,COP18,CMR22}, which removes a variable and its incident clauses from $\Phi$ to relate its pinned marginal to those in the resulting cavity formula.

\begin{lemma}[Contraction of pinned marginals]\label{lem:pinned-marginal-contraction}
For $0<\alpha\le2^k(\ln2-1/\ln k)$, there is $\kappa_k=\exp(-\Omega(k/(\ln k)^2))$ such that
\[
 \mathcal U_{t+1}\le\kappa_k\mathcal U_t,\qquad
 \mathcal U_t\le2\kappa_k^t
 \qquad(t\ge0).
\]
\end{lemma}
\begin{proof}
Since $\mathcal U_0\le2$, it suffices to prove $\mathcal U_{t+1}\le\kappa_k\mathcal U_t$.
Fix integer parameters $\bar N,M\ge1$, $0\le u<M$, and $R>4$.
Choose a uniform root $v\in[n]$, let $D_v=\deg_\Phi(v)$, and set $V'_v\defeq[n]\setminus\{v\}$.

Fix an integer $0\le d_0\le\bar N$ and condition on $D_v=d_0$ (which has probability $\Pr(D_v=d_0)\to\Pr(N=d_0)>0$ for $N\sim\operatorname{Pois}(D)$): the $m-d_0$ clauses not containing $v$ form an independent uniform cavity formula $\Phi'_v\sim\Phi(k,n-1,m-d_0)$ on $V'_v$, while the $d_0$ clauses $C_1,\ldots,C_{d_0}$ containing $v$ have independent uniform $d$-subsets of $V'_v$ as their remaining variable sets and independent fair literal signs.

Couple these $d_0$ child $d$-subsets with an array $W=(w_{C,i})_{C\in[d_0],i\in[d]}$ of $d_0d$ i.i.d.\ uniform vertices in $V'_v$ so that the two coincide on the event $\mathcal E_{\mathrm{distinct}}$ that all $d_0d$ coordinates of $W$ are distinct, and couple $\boldsymbol J\sim\operatorname{Unif}([n]^u)$ with $\boldsymbol J'\sim\operatorname{Unif}((V'_v)^u)$ so that $\boldsymbol J=\boldsymbol J'$ on $\{v\notin\boldsymbol J\}$; by union bounds, $\Pr(\mathcal E_{\mathrm{distinct}}^c\mid D_v=d_0)\le(d_0d)^2/(n-1)$ and $\Pr(v\in\boldsymbol J)\le u/n\le M/n$.

For each $w\in V'_v$, let $\widehat p'_w\defeq\widehat p_w^{\,t}(\Phi'_v)$ be its depth-$t$ free-boundary tree marginal in $\Phi'_v$, and define its clipped version $q_w\defeq\min\{\max\{\widehat p'_w,R^{-1}\},1-R^{-1}\}\in[R^{-1},1-R^{-1}]$.
Using the coupled child array $W$ and the literal signs of $C_1,\ldots,C_{d_0}$, form the signed child array $\bar q=(\bar q_{C,i})$, normalizer $F_q\defeq W_0(\bar q)+W_1(\bar q)$, and one-step update $q^+\defeq W_1(\bar q)/F_q$.

Since every clause factor $1-\prod_{i=1}^d\bar q_{C,i}$ in $W_b(\bar q)$ is at least $c_R\defeq 1-(1-R^{-1})^d>0$, we have $F_q\ge2c_R^{d_0}$ deterministically.
Because $G(q_w)=\min\{G(\widehat p'_w),R\}\le R$ is a bounded function of $B_t(\Phi'_v,w)$ and $\E_{\pi_t}[G(q)]\le7/2<4$ by \eqref{eq:free-reference-regularity}, as $n\to\infty$ with $(m-d_0)/(n-1)\to\alpha$, \pref{lem:norm-lower} and \pref{lem:poisson-local-limit}(i) imply that the $\Phi'_v$-measurable event
\begin{equation}\label{eq:empirical-reference-regularity}
 \mathcal E_{\mathrm{base}}\defeq\left\{\ln Z_{\Phi'_v}\ge\frac{n-1}{2\ln k},\quad \frac1{n-1}\sum_{w\in V'_v}G(q_w)\le4\right\}
\end{equation}
holds with conditional probability $1-o_n(1)$ given $D_v=d_0$.
On $\mathcal E_{\mathrm{base}}$, at most a $4/R$ fraction of vertices $w\in V'_v$ have $q_w\ne\widehat p'_w$, because every such vertex has $G(q_w)=R$.

Let $\mathcal E_{\mathrm{tree}}$ be the event that $\mathcal E_{\mathrm{base}}\cap\mathcal E_{\mathrm{distinct}}\cap\{v\notin\boldsymbol J\}$ holds, $\Phi$ is satisfiable, $B_{t+1}(\Phi,v)$ is a tree, and $q_{w_{C,i}}=\widehat p'_{w_{C,i}}$ for all $(C,i)\in[d_0]\times[d]$.

On $\mathcal E_{\mathrm{tree}}$, the $d_0d$ child variables $w_{C,i}$ are the depth-$1$ variables of $B_{t+1}(\Phi,v)$ and their depth-$t$ balls $B_t(\Phi'_v,w_{C,i})$ in $\Phi'_v$ are the disjoint subtrees of $B_{t+1}(\Phi,v)$ rooted at $w_{C,i}$, so Bayes' rule on $B_{t+1}(\Phi,v)$ gives $\widehat p_v^{\,t+1}=q^+$.

Since $\Pr(D_v=d_0)$ is bounded away from zero and $\Omega_\Phi=\varnothing$ or $B_{t+1}(\Phi,v)$ not being a tree has unconditional probability $o_n(1)$ by \pref{lem:norm-lower} and \pref{lem:poisson-local-limit}(i), a union bound over the $d_0d$ uniform children $w_{C,i}\in V'_v$ on $\mathcal E_{\mathrm{base}}$ gives $\Pr(\mathcal E_{\mathrm{tree}}^c\mid D_v=d_0)\le 4d_0d/R+O_{d_0,M}(n^{-1})+o_n(1)$.
Since $h^2(p,q)\le2$ for all $p,q\in[0,1]$ and $\widehat p_v^{\,t+1}=q^+$ on $\mathcal E_{\mathrm{tree}}$, we obtain
\[
 \sum_yw_yh^2(p_{v,y},\widehat p_v^{\,t+1})
 \le\mathbb{1}\{\mathcal E_{\mathrm{tree}}\}\cdot\sum_yw_yh^2(p_{v,y},q^+)+2\mathbb{1}\{\mathcal E_{\mathrm{tree}}^c\}.
\]

On $\mathcal E_{\mathrm{base}}$, write $w'_y=\mu_{\Phi'_v}(\sigma_{\boldsymbol J'}=y)$ and $p'_{w,y}=\mu_{\Phi'_v}(\sigma_w=1\mid\sigma_{\boldsymbol J'}=y)$ for the pinned weights and marginals in $\Phi'_v$, and let $\bar p'_y=(\bar p'_{w_{C,i},y})$ be the corresponding signed child marginals.
Letting $\mathcal S_b\subseteq\{0,1\}^{V'_v}$ be the event that an assignment on $V'_v$ satisfies $C_1,\ldots,C_{d_0}$ when $\sigma_v=b$ (with $w_{C,1},\ldots,w_{C,d}$ in place of the child variables of $C$), define on $\{0,1\}^u\times\{0,1\}$ the nonnegative measures
\[
 \xi(y,b)=w'_y\mu_{\Phi'_v}(\mathcal S_b\mid\sigma_{\boldsymbol J'}=y),\qquad
 \widehat\xi(y,b)=w'_yW_b(\bar p'_y),\qquad
 \zeta(y,b)=w'_yW_b(\bar q).
\]
On $\mathcal E_{\mathrm{tree}}$, we have $\boldsymbol J=\boldsymbol J'$ and $\xi(y,b)=|\{\sigma\in\Omega_\Phi:\sigma_{\boldsymbol J}=y,\sigma_v=b\}|/Z_{\Phi'_v}$, so $|\xi|=\sum_{b=0}^1\mu_{\Phi'_v}(\mathcal S_b)=Z_\Phi/Z_{\Phi'_v}>0$ and $\tilde\xi(y,b)\defeq\xi(y,b)/|\xi|=w_y p_{v,y}^b(1-p_{v,y})^{1-b}$ is the joint law of $(\sigma_{\boldsymbol J},\sigma_v)$ under $\mu_\Phi$, while $\zeta$ has total mass $|\zeta|=F_q$ and normalized law $\tilde\zeta(y,b)\defeq\zeta(y,b)/F_q=w'_y(q^+)^b(1-q^+)^{1-b}$.
Moreover, because $\mathcal S_0$ and $\mathcal S_1$ depend on $\sigma_{V'_v}$ only through $\sigma_W$,
\[
 \|\xi-\widehat\xi\|_1\le2\delta_{\mathrm{fac}},
 \qquad
 \delta_{\mathrm{fac}}\defeq\sum_yw'_y\left\|\mu_{\Phi'_v,W}(\cdot\mid\sigma_{\boldsymbol J'}=y)-\bigotimes_{C,i}\mu_{\Phi'_v,w_{C,i}}(\cdot\mid\sigma_{\boldsymbol J'}=y)\right\|_{\mathrm{TV}}.
\]
For each $y$ with $w_y>0$, applying \eqref{eq:reference-hellinger-normalization} on $\{0,1\}$ with second argument $\tilde\xi(y,\cdot)$ (of total mass $w_y$) and first argument $\tilde\zeta(y,\cdot)$ gives
\[
 w_y h^2(p_{v,y},q^+)
 \le 2\sum_{b=0}^1
       \left(\sqrt{\tilde\xi(y,b)}-\sqrt{\tilde\zeta(y,b)}\right)^2.
\]
Summing over $y$ and applying \eqref{eq:reference-hellinger-normalization} on $\{0,1\}^u\times\{0,1\}$ to $\xi$ and $\zeta$, together with
\[
 \left(\sqrt{\xi}-\sqrt{\zeta}\right)^2
 \le 2|\xi-\widehat\xi|
      +2\left(\sqrt{\widehat\xi}-\sqrt{\zeta}\right)^2,
\]
gives on $\mathcal E_{\mathrm{tree}}$
\begin{equation}\label{eq:finite-reference-transfer}
 \sum_yw_y h^2(p_{v,y},q^+)
 \le\frac{8}{F_q}\left[2\delta_{\mathrm{fac}}
       +\sum_yw'_y\sum_{b=0}^1
          \left(\sqrt{W_b(\bar p'_y)}-\sqrt{W_b(\bar q)}\right)^2\right].
\end{equation}
Since $\mathcal E_{\mathrm{tree}}\subseteq\mathcal E_{\mathrm{base}}$ and the right-hand side of \eqref{eq:finite-reference-transfer} is nonnegative on $\mathcal E_{\mathrm{base}}$, we may bound $\mathbb{1}\{\mathcal E_{\mathrm{tree}}\}\cdot\sum_y w_y h^2(p_{v,y},q^+)$ by $\mathbb{1}\{\mathcal E_{\mathrm{base}}\}$ times the right-hand side of \eqref{eq:finite-reference-transfer}.

Because $\mathcal E_{\mathrm{base}}$ is $\Phi'_v$-measurable and $\boldsymbol J'\sim\operatorname{Unif}((V'_v)^u)$ and $W\sim\operatorname{Unif}((V'_v)^{d_0d})$ are independent of $(\Phi'_v,D_v=d_0)$, using $F_q\ge2c_R^{d_0}$ and applying \pref{lem:pin-factorization} to $\mu_{\Phi'_v}$ on $\{0,1\}^{V'_v}$ with $\ell=d_0d$ gives
\[
 \frac1M\sum_{u<M}\E\left[\mathbb{1}\{\mathcal E_{\mathrm{base}}\}\cdot\frac{\delta_{\mathrm{fac}}}{F_q}\,\middle|\,D_v=d_0\right]
 \le\frac{c_R^{-d_0}}{2}\sqrt{\frac{d_0d(d_0d-1)\ln2}{4M}}.
\]
For the root-weight sum in \eqref{eq:finite-reference-transfer}, on $\mathcal E_{\mathrm{base}}$ we have $\ln Z_{\Phi'_v}\ge(n-1)/(2\ln k)$.
For every pinned assignment $y\in\{0,1\}^u$ with $w'_y\ge\exp(-(n-1)/(4\ln k))$, subadditivity of entropy for $\mu_{\Phi'_v}(\cdot\mid\sigma_{\boldsymbol J'}=y)$ together with $\mathcal H(p)\le2\sqrt{p(1-p)}$ and Cauchy--Schwarz gives
\[
 \frac1{n-1}\sum_{w\in V'_v}\mathcal H(p'_{w,y})\ge\frac{\ln Z_{\Phi'_v}+\ln w'_y}{n-1}\ge\frac1{4\ln k},
 \qquad
 \E_w\bigl[p'_{w,y}(1-p'_{w,y})\bigr]\ge\frac1{64(\ln k)^2}=:v_0,
\]
where $\E_w$ denotes expectation over a uniform $w\in V'_v$.
Since $\E_w[G(q_w)]\le4$ on $\mathcal E_{\mathrm{base}}$, conditional on $(\Phi'_v,D_v=d_0,\boldsymbol J')$ the i.i.d.\ child pairs $(p'_{w_{C,i},y},q_{w_{C,i}})$ satisfy the hypotheses of \pref{lem:local-star} at parameter $v_0$, so taking the conditional expectation over $W$ and the clause signs yields
\begin{align*}
 \E_{W,\mathrm{signs}}\!\left[\frac1{F_q}\sum_{b=0}^1\left(\sqrt{W_b(\bar p'_y)}-\sqrt{W_b(\bar q)}\right)^2\,\middle|\,\Phi'_v,D_v=d_0,\boldsymbol J'\right]
 &\le A_{d_0}(v_0)\,\E_w\bigl[h^2(p'_{w,y},q_w)\bigr] \\
 &\le A_{d_0}(v_0)\left(\E_w\bigl[h^2(p'_{w,y},\widehat p'_w)\bigr]+\frac8R\right),
\end{align*}
using that $q_w=\widehat p'_w$ on all but at most a $4/R$ fraction of $w\in V'_v$ and $h^2\le2$.
Meanwhile, the remaining assignments $y\in\{0,1\}^u$ with $w'_y<\exp(-(n-1)/(4\ln k))$ have $\sum_y w'_y\le 2^M\exp(-(n-1)/(4\ln k))$.
Since
\[
 \sum_{b=0}^1\left(\sqrt{W_b(\bar p'_y)}-\sqrt{W_b(\bar q)}\right)^2\le2,
 \qquad F_q\ge2c_R^{d_0},
\]
their contribution to \eqref{eq:finite-reference-transfer} is at most $8c_R^{-d_0}2^M\exp(-(n-1)/(4\ln k))=o_n(1)$.

On $\mathcal E_{\mathrm{base}}$, averaging $\sum_y w'_y\E_w[h^2(p'_{w,y},\widehat p'_w)]$ over $\boldsymbol J'\sim\operatorname{Unif}((V'_v)^u)$ equals $\mathcal D_t(\Phi'_v,u)$ on the cavity formula $\Phi'_v\sim\Phi(k,n-1,m-d_0)$.
Summing over $0\le d_0\le\bar N$ with weights $\Pr(D_v=d_0)$, bounding the tail $\{D_v>\bar N\}$ by $2\Pr(D_v>\bar N)$, and taking limits first as $n\to\infty$ with $m/n\to\alpha$ (where $D_v\Rightarrow N\sim\operatorname{Pois}(D)$) and then as $M\to\infty$ yields
\[
 \mathcal U_{t+1}
 \le 8\sum_{d_0\le\bar N}\Pr(N=d_0)A_{d_0}(v_0)\left(\mathcal U_t+\frac8R\right)
       +\frac{8\bar Nd}{R}+2\Pr(N>\bar N).
\]
Letting $R\to\infty$ and then $\bar N\to\infty$, \eqref{eq:local-envelope} gives
\[
 \mathcal U_{t+1}\le 8\E[A_N(v_0)]\mathcal U_t
 \le\exp(-\Omega(k/(\ln k)^2))\mathcal U_t,
\]
proving the claimed contraction.
\end{proof}

\begin{proof}[Proof of \pref{lem:ordinary-local-law}]
When $\alpha=0$, $m=o(n)$ clauses touch at most $km=o(n)$ variables; on the $1-o_n(1)$ satisfiability event of \pref{lem:norm-lower}, all remaining variables are isolated and have independent $\operatorname{Bernoulli}(1/2)$ spins under $\mu_\Phi$, giving the claim with $P=\delta_{1/2}$.
For $\alpha>0$, \pref{lem:pinned-marginal-contraction} gives $\mathcal U_t\le2\kappa_k^t\to0$.

For independent uniform coordinates $I=(I_1,\ldots,I_\ell)\in[n]^\ell$ and $u$ pinned coordinates $\boldsymbol J=(J_1,\ldots,J_u)\in[n]^u$, on satisfiable formulas $\Phi$ we have $\mu_I=\sum_y w_y\mu_I^{(u)}(\cdot\mid\sigma_{\boldsymbol J}=y)$ with $\mu_{I_a}^{(u)}(\cdot\mid\sigma_{\boldsymbol J}=y)=\operatorname{Bernoulli}(p_{I_a,y})$.
We have
\[
 \left\|\bigotimes_{a=1}^\ell\operatorname{Bernoulli}(p_{I_a,y})
       -\bigotimes_{a=1}^\ell\operatorname{Bernoulli}(\widehat p_{I_a}^{\,t})\right\|_{\mathrm{TV}}
 \le\sum_{a=1}^\ell|p_{I_a,y}-\widehat p_{I_a}^{\,t}|.
\]
Together with $|p-q|\le h(p,q)$, the triangle inequality and Cauchy--Schwarz give
\[
 \E_{\boldsymbol J,I}\left[
 \left\|\mu_I-\bigotimes_{a=1}^\ell
       \operatorname{Bernoulli}(\widehat p_{I_a}^{\,t})
 \right\|_{\mathrm{TV}}\right]
 \le
 \E_{\boldsymbol J,I,\sigma}\left[
 \left\|\mu_I^{(u)}-\bigotimes_{a=1}^\ell
       \mu_{I_a}^{(u)}
 \right\|_{\mathrm{TV}}\right]
 +\ell\sqrt{\mathcal D_t(\Phi,u)}.
\]
Here $\sigma\sim\mu_\Phi$, independently of $\boldsymbol J,I$,
supplies the pinned values in the first term on the right.

Averaging over $u<M$, applying \pref{lem:pin-factorization} (with unsatisfiable formulas contributing $o_n(1)$ by \pref{lem:norm-lower}), and taking $n\to\infty$ followed by $M\to\infty$ yields
\[
 \overline\E\left[\E_I\left[\left\|\mu_I-
      \bigotimes_{a=1}^\ell\operatorname{Bernoulli}(\widehat p_{I_a}^{\,t})
       \right\|_{\mathrm{TV}}\right]\right]
 \le\ell\sqrt{\mathcal U_t}.
\]
For $\ell=1$, since $\|\mu_v-\operatorname{Bernoulli}(\widehat p_v^{\,t})\|_{\mathrm{TV}}=|p_v-\widehat p_v^{\,t}|$, this bounds $\frac1n\sum_{v\in[n]}|p_v-\widehat p_v^{\,t}|$ and hence $\mathsf W_1(P_n,\widehat P_n^{\,t})$, where $\widehat P_n^{\,t}\defeq\frac1n\sum_{v\in[n]}\delta_{\widehat p_v^{\,t}}\to\pi_t$ in probability by \pref{lem:poisson-local-limit}(i); thus
\[
 \overline\E[\mathsf W_1(P_n,\pi_t)]\le\sqrt{\mathcal U_t}.
\]
Consequently, $(\pi_t)_{t\ge0}$ is Cauchy in $\mathsf W_1$ with a deterministic limit $P$ depending only on $(k,\alpha)$, and $P_n\to P$ in probability as $m/n\to\alpha$.
By lower semicontinuity and \eqref{eq:free-reference-regularity}, $P$ satisfies $\E_P[G(p)]\le7/2\le4$, so it is supported on $(0,1)$ and satisfies \eqref{eq:graph-state-bias}, proving part~\textup{(i)}.

To establish part~\textup{(ii)}, fix $r\ge0$ and a finite rooted signed tree $T$ of depth at most $r$ with $\Pr(T_r\cong T)>0$, depth-$r$ variable boundary $\partial T$, $s=|V(T)\setminus\partial T|$ interior variables, $\ell=|\partial T|$ boundary variables, and $m_0$ clauses.

By \pref{def:factor-graph}, a clause of $\Phi$ belongs to the radius-$2r$ ball $B_r(\Phi,v)$ if and only if its scope intersects the interior variable set $V_{\mathrm{int}}\defeq V(B_r(\Phi,v))\setminus\partial B_r(\Phi,v)$.
Condition on a realization of the labeled subgraph $B_r(\Phi,v)$ that is isomorphic to $T$, and fix a rooted isomorphism identifying $V(B_r(\Phi,v))$ with $V(T)$ and $\partial B_r(\Phi,v)$ with $\partial T$.

Under this conditioning, $V_{\mathrm{int}}$ and the boundary tuple $\partial B_r(\Phi,v)$ of $\ell$ distinct vertices in $V_{\mathrm{ext}}\defeq[n]\setminus V_{\mathrm{int}}$ are fixed, while the remaining $m-m_0$ clauses of $\Phi$ are constrained only to have scopes disjoint from $V_{\mathrm{int}}$; hence they form a uniform formula $\Phi_{\mathrm{ext}}\sim\Phi(k,n-s,m-m_0)$ on $V_{\mathrm{ext}}$, with density $(m-m_0)/(n-s)\to\alpha$.

Because the law of $\Phi_{\mathrm{ext}}$ is invariant under permutations of $V_{\mathrm{ext}}$, the conditional law of its Gibbs marginal $\mu_{\Phi_{\mathrm{ext}}}|_{\partial B_r(\Phi,v)}$ on $\{0,1\}^{\partial T}$ coincides with the law of $\mu_{\Phi_{\mathrm{ext}},I'}$ at an independent uniformly sampled $\ell$-tuple $I'$ of distinct vertices in $V_{\mathrm{ext}}$, which in turn couples within $O(\ell^2/n)$ in total variation with $\ell$ i.i.d.\ uniform vertices $I=(I_1,\ldots,I_\ell)\in V_{\mathrm{ext}}^\ell$.

For each fixed $t\ge0$, conditional on $\Phi_{\mathrm{ext}}$ the values $\widehat p_{I_1}^{\,t}(\Phi_{\mathrm{ext}}),\ldots,\widehat p_{I_\ell}^{\,t}(\Phi_{\mathrm{ext}})$ are i.i.d.\ draws from the empirical law $\widehat P_n^{\,t}(\Phi_{\mathrm{ext}})\to\pi_t$, so $\bigl(\widehat p_{I_1}^{\,t}(\Phi_{\mathrm{ext}}),\ldots,\widehat p_{I_\ell}^{\,t}(\Phi_{\mathrm{ext}})\bigr)\Rightarrow\pi_t^{\otimes\ell}$ as $n\to\infty$, and $\pi_t^{\otimes\ell}\Rightarrow P^{\otimes\ell}$ as $t\to\infty$.

Applying the $\ell$-tuple total-variation bound above to $\Phi_{\mathrm{ext}}$ and letting $n\to\infty$ and then $t\to\infty$ shows that, conditional on $B_r(\Phi,v)\cong T$, the exterior marginal $\mu_{\Phi_{\mathrm{ext}}}|_{\partial B_r(\Phi,v)}$ on $\{0,1\}^{\partial T}$ converges in distribution to $\bigotimes_{u\in\partial T}\operatorname{Bernoulli}(p_u)$ with $\boldsymbol p=(p_u)_{u\in\partial T}\sim P^{\otimes\partial T}$.

For every satisfiable formula $\Phi$ with $B_r(\Phi,v)\cong T$, each clause of $\Phi$ either lies in $B_r(\Phi,v)$ or has all its variables in $V_{\mathrm{ext}}$, with $V(B_r(\Phi,v))\cap V_{\mathrm{ext}}=\partial B_r(\Phi,v)$; hence
\[
 \mu_\Phi|_{V(B_r(\Phi,v))}(\sigma)
 \propto \mathbb{1}\{\sigma\models T\}\cdot\mu_{\Phi_{\mathrm{ext}}}|_{\partial B_r(\Phi,v)}(\sigma_{\partial T}).
\]
Since $T$ is a finite tree and $P$ is supported on $(0,1)$, the limiting normalizing constant under $\bigotimes_{u\in\partial T}\operatorname{Bernoulli}(p_u)$ is positive almost surely.

By continuity of normalization, conditional on $B_r(\Phi,v)\cong T$, the local Gibbs law $\mu_\Phi|_{V(B_r(\Phi,v))}$ on $\{0,1\}^{V(T)}$ therefore converges in distribution to $\mu_{T,\boldsymbol p}$ (\pref{def:ordinary-tree}).
Because $\partial B_r(\Phi,v)$ separates $v$ from $\overline B_r(v)$ on $B_r(\Phi,v)\cong T$, we have $q_{v,r}=\mu_\Phi(\sigma_v=1\mid\sigma_{\partial B_r(\Phi,v)})$.
For any probability measure $\nu$ on $\{0,1\}^{V(T)}$ and any bounded continuous function $\phi:[0,1]^2\to\mathbb R$, writing $w_\nu(\tau,b)=\nu(\sigma_{\partial T}=\tau,\sigma_o=b)$ and $w_\nu(\tau)=w_\nu(\tau,0)+w_\nu(\tau,1)$, the expectation
\[
 \E_{\sigma\sim\nu}\bigl[\phi\bigl(\nu(\sigma_o=1),\nu(\sigma_o=1\mid\sigma_{\partial T})\bigr)\bigr]
 =\sum_{\tau:\,w_\nu(\tau)>0} w_\nu(\tau)\,\phi\left(\nu(\sigma_o=1),\,\frac{w_\nu(\tau,1)}{w_\nu(\tau)}\right)
\]
is a continuous function of $\nu\in\mathcal P(\{0,1\}^{V(T)})$ since $|\phi|$ is bounded and each summand vanishes as $w_\nu(\tau)\to0$.
By the continuous mapping theorem, conditional on $B_r(\Phi,v)\cong T$, the pair $(p_v,q_{v,r})$ converges in distribution to the law of $(p_{r,P},Q_{r,P})$ conditioned on $T_r\cong T$.
Since $\Pr(B_r(\Phi,v)\cong T)\to\Pr(T_r\cong T)$ for every finite rooted signed tree $T$ by \pref{lem:poisson-local-limit}(i) and $\sum_T\Pr(T_r\cong T)=1$, summing over $T$ yields $(p_v,q_{v,r})\Rightarrow(p_{r,P},Q_{r,P})$.
Finally, since $p_v\Rightarrow p_{r,P}$ and $\operatorname{Law}(p_v)=\E[P_n]\Rightarrow P$, we have $\operatorname{Law}(p_{r,P})=P$, proving part~\textup{(ii)}.
\end{proof}

\section{Proof of single-variable connectivity}
\label{sec:transfer-section}
We prove \pref{thm:main} in \pref{sec:transfer-global-section}.
The algorithm combines coupled Glauber dynamics, local repair, and greedy routing, whose guarantees are proved in \pref{sec:coupled-glauber}, \pref{sec:repair-appendix}, and \pref{sec:find-path}, respectively.
Finally, in \pref{sec:companion-local-connectivity}, we apply the repair construction to prove local connectivity up to the looseness threshold (\pref{thm:main-looseness}\ref{item:main-local-connectivity-part}).

\subsection{Path construction and correctness of \texorpdfstring{$\mathsf{Connect}$}{Connect}}
\label{sec:transfer-global-section}\label{sec:stability}
\begin{figure}[!ht]
\centering
\begin{tikzpicture}[
  x=0.84cm, y=1cm,
  >=Stealth,
  every node/.style={font=\footnotesize},
  nd/.style={circle, draw=black!65, fill=white, semithick, inner sep=1pt, minimum size=10mm, text=black!90},
  sol/.style={circle, draw=black, fill=white, thick, inner sep=1pt, minimum size=10mm, text=black, font=\footnotesize\bfseries},
  dots/.style={font=\normalsize, inner xsep=1pt, inner ysep=0pt},
  legendtext/.style={anchor=west, font=\scriptsize, text=black!85, inner sep=0pt, text height=1.5ex, text depth=.25ex},
  blk/.style={{Stealth[length=1.6mm]}-{Stealth[length=1.6mm]}, semithick, densely dashed, black!55},
  cwalk/.style={-{Stealth[length=2mm]}, thick, densely dashed, Navy!85},
  ewalk/.style={-{Stealth[length=2.2mm]}, very thick, Navy!85},
  rpath/.style={-{Stealth[length=2mm]}, thick, densely dashed, Green!75!black},
  gpath/.style={-{Stealth[length=2.2mm]}, very thick, red!75!black}
]
  \fill[black!6, rounded corners=6pt]
    (-0.85,2.25) -- (1.45,2.25) -- (1.45,-0.45) -- (8.95,-0.45) --
    (8.95,2.25) -- (11.25,2.25) -- (11.25,-2.48) -- (-0.85,-2.48) -- cycle;
  \draw[black!28, semithick, rounded corners=6pt]
    (-0.85,2.25) -- (1.45,2.25) -- (1.45,-0.45) -- (8.95,-0.45) --
    (8.95,2.25) -- (11.25,2.25) -- (11.25,-2.48) -- (-0.85,-2.48) -- cycle;
  \node[anchor=south, inner sep=0pt, text=black!80, font=\footnotesize\bfseries] at (5.2,-2.38) {$\Omega_\Phi$};

  \node[sol, label={[font=\scriptsize]above:$\sigma$}] (z0) at (0.3,1.25) {$z_0$};
  \node[nd] (z1) at (2.5,1.75) {$z_1$};
  \node[nd] (z2) at (5.2,1.98) {$z_2$};
  \node[dots, text=black!65] (zdots) at (6.55,1.98) {$\cdots$};
  \node[nd] (zd1) at (7.9,1.75) {$z_{b_\star-1}$};
  \node[sol, label={[font=\scriptsize]above:$\tau$}] (zd) at (10.1,1.25) {$z_{b_\star}$};

  \draw[blk, bend left=10] (z0) to (z1);
  \draw[blk, bend left=6] (z1) to (z2);
  \draw[-{Stealth[length=1.6mm]}, semithick, densely dashed, black!55, bend left=3] (z2) to (zdots);
  \draw[{Stealth[length=1.6mm]}-, semithick, densely dashed, black!55, bend left=3] (zdots) to (zd1);
  \draw[blk, bend left=10] (zd1) to (zd);

  \node[nd] (x1) at (2.5,0.22) {$x_1$};
  \node[nd] (x2) at (5.2,0.22) {$x_2$};
  \node[dots, text=black!65] at (6.55,0.22) {$\cdots$};
  \node[nd] (xd1) at (7.9,0.22) {$x_{b_\star-1}$};

  \node[sol, label={[font=\scriptsize, text=black!80]below:$=x_0$}] (y0) at (0.3,-1.5) {$y_0$};
  \node[sol] (y1) at (2.5,-1.5) {$y_1$};
  \node[sol] (y2) at (5.2,-1.5) {$y_2$};
  \node[dots, font=\normalsize\bfseries, text=red!75!black] (ydots) at (6.55,-1.5) {$\cdots$};
  \node[sol] (yd1) at (7.9,-1.5) {$y_{b_\star-1}$};
  \node[sol, label={[font=\scriptsize, text=black!80]below:$=x_{b_\star}$}] (yd) at (10.1,-1.5) {$y_{b_\star}$};

  \draw[ewalk] (z0) -- node[left, inner sep=2pt, font=\scriptsize\bfseries, text=Navy!85] {$\pi_0$} (y0);
  \draw[cwalk] (z1) -- (x1);
  \draw[cwalk] (z2) -- (x2);
  \draw[cwalk] (zd1) -- (xd1);
  \draw[ewalk] (yd) -- node[right, inner sep=2pt, font=\scriptsize\bfseries, text=Navy!85] {$\pi_{b_\star}$} (zd);

  \draw[rpath] (x1) -- (y1);
  \draw[rpath] (x2) -- (y2);
  \draw[rpath] (xd1) -- (yd1);

  \draw[gpath] (y0) -- node[above, inner sep=2pt, font=\scriptsize\bfseries, text=red!75!black] {$\gamma_0$} (y1);
  \draw[gpath] (y1) -- node[above, inner sep=2pt, font=\scriptsize\bfseries, text=red!75!black] {$\gamma_1$} (y2);
  \draw[gpath] (y2) -- (ydots);
  \draw[gpath] (ydots) -- (yd1);
  \draw[gpath] (yd1) -- node[above, inner sep=2pt, font=\scriptsize\bfseries, text=red!75!black] {$\gamma_{b_\star-1}$} (yd);

  \fill[white, rounded corners=5pt] (11.65,-1.5) rectangle (17.7,1.5);
  \draw[black!28, rounded corners=5pt] (11.65,-1.5) rectangle (17.7,1.5);

  \draw[blk] (11.95,1.08) -- (12.95,1.08);
  \node[legendtext] at (13.2,1.08) {Block interpolation};

  \draw[cwalk] (11.95,0.49) -- (12.95,0.49);
  \draw[ewalk] (11.95,0.23) -- (12.95,0.23);
  \node[legendtext] at (13.2,0.36) {Coupled Glauber};

  \draw[rpath] (11.95,-0.36) -- (12.95,-0.36);
  \node[legendtext] at (13.2,-0.36) {Local repair ($\mathsf{Repair}$)};

  \draw[gpath] (11.95,-1.08) -- (12.95,-1.08);
  \node[legendtext] at (13.2,-1.08) {Greedy routing ($\mathsf{FindPath}$)};
\end{tikzpicture}
\caption{Overview of $\mathsf{Connect}(\sigma,\tau)$ (\pref{alg:connect}). The shaded region represents $\Omega_\Phi$; solid arrows denote satisfying paths inside $\Omega_\Phi$, and dashed arrows mark intermediate construction steps that may leave $\Omega_\Phi$.}
\label{fig:connect-overview}
\end{figure}
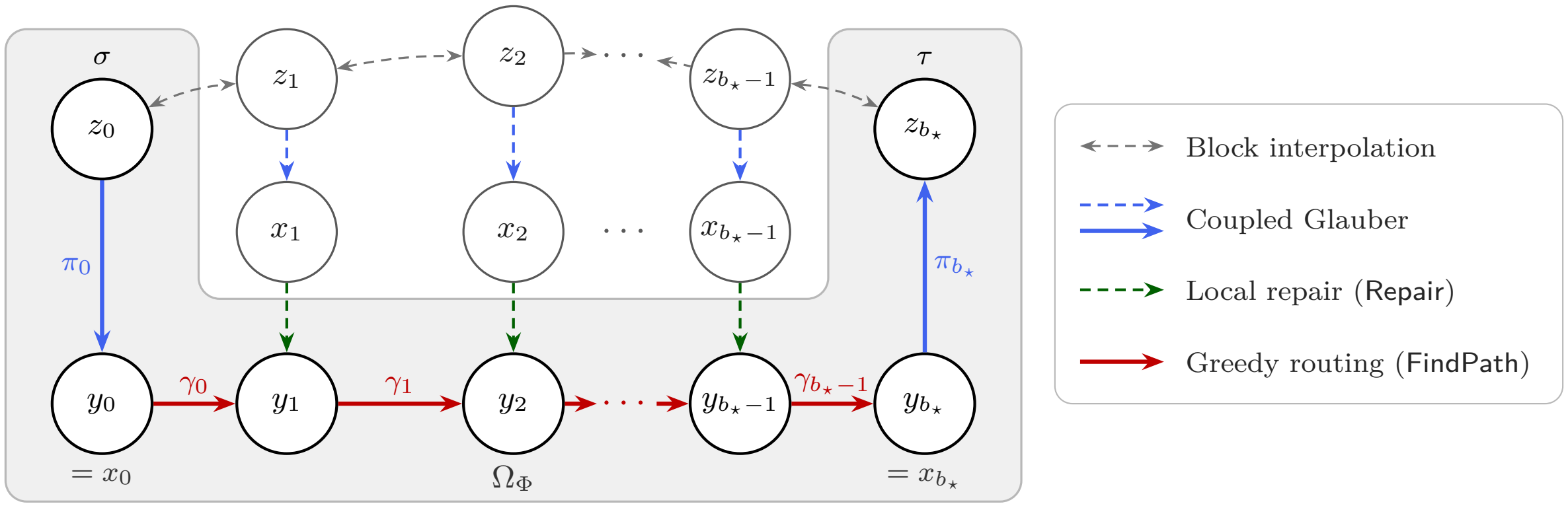

\paragraph{Overview of $\mathsf{Connect}$.}
Let $\Phi$ have $m$ clauses and let $\sigma,\tau\in\Omega_\Phi$.
We construct a sequence of nearby solutions and join consecutive ones through a common repair target, as illustrated in \pref{fig:connect-overview}.
The full procedure is given in \pref{alg:connect}, where $\operatorname{rev}(\gamma)$ denotes the reversal of a sequence $\gamma$.
Any failed subroutine call makes $\mathsf{Connect}$ return \textnormal{\textsc{Fail}}.

We first sample an independent $\Phi_{\mathrm{aux}}\sim\Phi(k,n,m)$ and set $\Phi_+=\Phi\cup\{C\in\Phi_{\mathrm{aux}}:C(\sigma)=0\text{ or }C(\tau)=0\}$.
This augmentation lets us transfer failure bounds proved for a Poisson formula to the original model conditional on $\sigma,\tau\in\Omega_\Phi$ (\pref{lem:transfer-extension}).
In the Poisson model, the subformulas formed by clauses satisfied and falsified by a fixed assignment are independent, simplifying the analysis of the walks and repairs.

\begin{enumerate}[label=\textup{(\arabic*)}]
\item \emph{Block interpolation.}
We choose a block count $b_\star=b_\star(k)$ and, for $n\ge b_\star$, partition $[n]$ into the consecutive blocks
\[
 B_i=\{\lfloor(i-1)n/b_\star\rfloor+1,\ldots,\lfloor in/b_\star\rfloor\},
 \qquad 1\le i\le b_\star.
\]
For $0\le i\le b_\star$, let the interpolating assignment $z_i$ agree with $\tau$ on $B_1\cup\dots\cup B_i$ and with $\sigma$ elsewhere.
Thus $z_0=\sigma$, $z_{b_\star}=\tau$, and consecutive assignments differ on at most $\lceil n/b_\star\rceil$ variables; the intermediate assignments may violate $\Phi$.

\item \emph{Coupled Glauber dynamics.}
For a formula $\Psi$ and an assignment $x$, let $\Psi[x]=\{C\in\Psi:C(x)=1\}$, and set $\Phi_i=\Phi_+[z_i]$.
We run $N=\Theta_k(n)$ steps of Glauber dynamics on $\Phi_i$ from its known satisfying assignment $z_i$, obtaining $x_i$.
At each step, we choose a variable uniformly at random and, keeping all other variables fixed, resample its value uniformly subject to satisfying $\Phi_i$.
All walks use the same random update sequence, sampled independently of the formulas.
For a repair scale $\rho=\rho(k)\in(0,1)$, we bound $\Ham(x_i,x_{i+1})$ by $\rho^2n/8$ and the number of clauses in $\Phi_+\setminus\Phi_i$ with at most one true literal under $x_i$ by $\rho^4n$ (\pref{lem:coupled-terminal}).
Intuitively, these two properties allow us to repair the assignments and connect consecutive ones by short satisfying paths.
The parameters $\rho$, $b_\star$, and $N$ will be chosen below.

\item \emph{Repairing the individual states.}
For $1\le i<b_\star$, we call $\mathsf{Repair}(\Phi_+,\Phi_i,x_i,\varnothing)$ to obtain $(U_i,y_i)$, where $y_i\in\Omega_{\Phi_+}\subseteq\Omega_\Phi$ and $y_i$ agrees with $x_i$ outside $U_i$.
We prove that $|U_i|\le\rho^2n/8$ and that every solution of $\Phi$ agreeing with $x_i$ outside $U_i$ can reach $y_i$ by satisfying single-variable flips, changing each disagreeing coordinate once (\pref{lem:transfer-terminal}).
The walks from $\sigma$ and $\tau$ stay in $\Omega_\Phi$, so we set $(U_0,y_0)=(\varnothing,x_0)$ and $(U_{b_\star},y_{b_\star})=(\varnothing,x_{b_\star})$.

\item \emph{Connecting consecutive solutions.}
To construct a satisfying path between $y_i$ and $y_{i+1}$, we make a second repair call with $W_i=U_i\cup U_{i+1}\cup\Delta(x_i,x_{i+1})$.
The close terminal states and small first repairs give $|W_i|\le3\rho^2n/8\le\rho^2n/2$.
The second repair enlarges $W_i$ to $V_i$ with $|V_i|\le\rho n/2$ and supplies a common target $t_i$.
Since both solutions agree with $x_i$ outside $W_i$, the repair's path guarantee lets both reach $t_i$ inside $\Omega_\Phi$ (\pref{lem:transfer-terminal}).
Then $\mathsf{FindPath}(y_i,y_{i+1},t_i)$ joins the path from $y_i$ to $t_i$ with the reversal of the path from $y_{i+1}$ to $t_i$, producing a satisfying path $\gamma_i$ of length at most $2|V_i|\le\rho n$ (\pref{lem:repair-regularity}).
\end{enumerate}
Since $\Phi\subseteq\Phi_0\cap\Phi_{b_\star}$, the Glauber dynamics on $\Phi_0$ starting at $\sigma$ and on $\Phi_{b_\star}$ starting at $\tau$ stay in $\Omega_\Phi$.
Omitting consecutive repeated assignments and reversing the trajectory starting at $\tau$ gives satisfying paths $\pi_0:\sigma\to y_0$ and $\pi_{b_\star}:y_{b_\star}\to\tau$.
Concatenating $\pi_0,\gamma_0,\ldots,\gamma_{b_\star-1},\pi_{b_\star}$ gives a satisfying path from $\sigma$ to $\tau$.

\begin{algorithm}[!ht]
\caption{$\mathsf{Connect}(\sigma,\tau)$: Single--Variable Path Construction in $\Phi$}\label{alg:connect}
\SetKwInOut{Input}{Input}
\SetKwInOut{Output}{Output}
\Input{A formula $\Phi$ and two solutions $\sigma,\tau\in\Omega_\Phi$.}
\Output{A satisfying path from $\sigma$ to $\tau$ in $\Omega_\Phi$, or \textnormal{\textsc{Fail}}.}
\If{$m=0$}{\Return the path flipping the coordinates in $\Delta(\sigma,\tau)$ in order\;}
Sample $\Phi_{\mathrm{aux}}\sim\Phi(k,n,m)$ and set $\Phi_+\gets\Phi\cup\{C\in\Phi_{\mathrm{aux}}:C(\sigma)=0\text{ or }C(\tau)=0\}$\;
Sample a random update sequence of length $N$\;
\For{$i\gets0$ \KwTo $b_\star$}{
    $\Phi_i\gets\Phi_+[z_i]$; run $N$ steps of Glauber dynamics $(X_0^{(i)},\dots,X_N^{(i)})$ on $\Phi_i$ from $X_0^{(i)}=z_i$ using this sequence, and set $x_i\gets X_N^{(i)}$\label{line:connect-walk}\;
}
Let $\pi_0$ and $\pi_{b_\star}$ be the satisfying paths of $(X_0^{(0)},\dots,X_N^{(0)})$ and $\operatorname{rev}(X_0^{(b_\star)},\dots,X_N^{(b_\star)})$\;
$(U_0,y_0)\gets(\varnothing,x_0)$,\quad $(U_{b_\star},y_{b_\star})\gets(\varnothing,x_{b_\star})$\;
\For{$i\gets1$ \KwTo $b_\star-1$}{
    $(U_i,y_i)\gets\mathsf{Repair}(\Phi_+,\Phi_i,x_i,\varnothing)$ (\pref{alg:common-repair})\tcp*[r]{repair $x_i\to y_i\in\Omega_\Phi$}\label{line:connect-empty-repair}
}
\For{$i\gets0$ \KwTo $b_\star-1$}{
    $W_i\gets U_i\cup U_{i+1}\cup\Delta(x_i,x_{i+1})$\;
    $(V_i,t_i)\gets\mathsf{Repair}(\Phi_+,\Phi_i,x_i,W_i)$ (\pref{alg:common-repair})\tcp*[r]{common target $t_i\in\Omega_\Phi$}\label{line:connect-pair-repair}
    $\gamma_i\gets\mathsf{FindPath}(y_i,y_{i+1},t_i)$ (\pref{alg:find-path})\;
}
\Return $\pi_0\circ\gamma_0\circ\dots\circ\gamma_{b_\star-1}\circ\pi_{b_\star}$\;
\end{algorithm}

We now collect the properties needed to analyze $\mathsf{Connect}$, then use them to prove \pref{thm:main}.
Recall that $\Phi_+$ augments $\Phi$ with the clauses of an independent $\Phi_{\mathrm{aux}}$ falsified by $\sigma$ or $\tau$.
For fixed $\sigma,\tau$, the next lemma compares the law of $\Phi_+$ conditional on $\sigma,\tau\in\Omega_\Phi$ with a Poisson formula law.
In the proofs below, we may therefore assume that $\Phi_+$ is sampled from the Poisson model.
We use the next lemma to return to the original model, then apply \pref{lem:norm-transfer} to transfer the bounds for fixed $\sigma,\tau$ to uniform solution pairs.

Under the Poisson law, we first sample $M\sim\operatorname{Pois}(m)$ and then $\Phi_+\sim\Phi(k,n,M)$.
We write $\Pr_{\operatorname{Pois}(m)}$ for probability under this law and recall that $\epsilon_0=2^{-k}$ is the probability that a fixed assignment falsifies a random clause.

\begin{lemma}[Poisson comparison]\label{lem:transfer-extension}
For an integer $m\ge1$, $\sigma,\tau\in\{0,1\}^n$, and any formula event $\mathcal B$ determined by the clause multiset,
\begin{equation*}
 \Pr(\Phi_+\in\mathcal B\mid\sigma,\tau\in\Omega_\Phi)
 \le15\sqrt m\,\ee^{4m\epsilon_0^2}
       \Pr_{\operatorname{Pois}(m)}(\Phi_+\in\mathcal B).
\end{equation*}
\end{lemma}
\begin{proof}
We compare the numbers of clauses satisfied by both $\sigma$ and $\tau$ and of the remaining clauses.
Let $C$ be a random clause with a uniform $k$-element scope and independent fair literal signs.
Let $A$ be the set of clauses satisfied by both $\sigma$ and $\tau$, $s=\Pr(C\in A)\in[1-2\epsilon_0,1-\epsilon_0]$, and $q=1-s\le2\epsilon_0\le1/2$.
Conditional on $\sigma,\tau\in\Omega_\Phi$, $\Phi$ consists of $m$ independent clauses conditioned to lie in $A$, independently of the $\operatorname{Bin}(m,q)$ clauses of $\Phi_+\setminus\Phi$ sampled from $A^c$.
By Stirling's approximation, we obtain, for $0\le b\le m$,
\[
\begin{gathered}
 \frac{\Pr(\operatorname{Bin}(m,q)=b)}{\Pr(\operatorname{Pois}(mq)=b)}
 =\frac{\Pr(\operatorname{Pois}(ms)=m-b)}{\Pr(\operatorname{Pois}(m)=m)}
 \le\frac{3}{\sqrt s}<5,\\
 \Pr(\operatorname{Pois}(ms)=m)\ge\frac{\exp(-m(s-1-\ln s))}{3\sqrt m}\ge\frac{\ee^{-4m\epsilon_0^2}}{3\sqrt m},
\end{gathered}
\]
using $-q-\ln(1-q)\le q^2$.
Replacing $\operatorname{Bin}(m,q)$ by $\operatorname{Pois}(mq)$ makes the clause multiset equal in law to that of a $\operatorname{Pois}(m)$ formula conditioned to have $m$ clauses in $A$.
Under the Poisson law, let $N_A\sim\operatorname{Pois}(ms)$ count the clauses in $A$.
The preceding bounds give
\[
\begin{aligned}
 \Pr(\Phi_+\in\mathcal B\mid\sigma,\tau\in\Omega_\Phi)
 &\le5\Pr_{\operatorname{Pois}(m)}(\Phi_+\in\mathcal B\mid N_A=m)\\
 &\le\frac{5\Pr_{\operatorname{Pois}(m)}(\Phi_+\in\mathcal B)}
              {\Pr_{\operatorname{Pois}(m)}(N_A=m)}
 \le15\sqrt m\,\ee^{4m\epsilon_0^2}
       \Pr_{\operatorname{Pois}(m)}(\Phi_+\in\mathcal B).
\end{aligned}
\]
This proves the lemma.
\end{proof}

We next state the structural conditions used in the repair guarantees.
For the rest of the section, set $\rho=k^{-100\lceil(\ln k)^6\rceil}$ and $\eta=(\ln k)^{10}/(64k)$.
The degree cutoff $D_k\in\mathbb N$ will be chosen in the proof of \pref{lem:unsigned-envelope}.

\begin{definition}[Formula regularity]\label{def:unsigned-envelope}
Define the following two events for a formula $\Phi$ on $[n]$:
\begin{enumerate}[label=\textup{(\roman*)}]
\item\label{item:regularity-neighborhoods}
\emph{Local tree structure.} $\mathsf{LT}(\Phi)$ holds if at least $(1-\eta)n$ variables $v\in[n]$ are $(2k+4,D_k)$-locally tree-structured in $\Phi$ (as defined in \pref{lem:ball-sensitivity}).
\item\label{item:regularity-sparsity}
\emph{Local sparsity.} $\mathsf{LS}(\Phi)$ holds if $|\{C\in\Phi:\operatorname{vbl}(C)\subseteq A\}|<2|A|/k$ for every $A\subseteq[n]$ with $0<|A|\le\rho n$.
\end{enumerate}
We call $\Phi$ \emph{regular} when $\mathsf{Reg}(\Phi):=\mathsf{LT}(\Phi)\cap\mathsf{LS}(\Phi)$ holds.
\end{definition}

\begin{definition}[Good Repair]\label{def:repair}
For $\Phi_0\subseteq\Phi_+$, $x\in\Omega_{\Phi_0}$ and $W\subseteq[n]$, an \emph{$(x,W)$-repair} is a pair $(U,y)$ with $U\supseteq W$, $y\in\Omega_{\Phi_+}$, and $y_{[n]\setminus U}=x_{[n]\setminus U}$ satisfying the following path property.
For every subformula $\Psi\subseteq\Phi_+$ and every $\sigma\in\Omega_\Psi$ agreeing with $x$ outside $U$, there is a satisfying path in $\Omega_\Psi$ from $\sigma$ to $y$ of length $\Ham(\sigma,y)$.
The repair is \emph{good} if $|U|\le\frac{1}{2\rho}|W|+\frac{\rho^2n}{8}$.
\end{definition}

The parameter $W$ specifies variables that the repair set must contain and the path property allows several solutions to share the same repair target.
We prove the following guarantees for $\mathsf{Repair}$ (\pref{alg:common-repair}) in \pref{sec:repair-appendix} using the complete forcing trees supplied by \pref{lem:set-avoiding-trees}.
Recall that $\operatorname{sat}_x(C)$ consists of the variables whose literals in $C$ are true under $x$ (\pref{def:support}).

\begin{lemma}[Correctness of $\mathsf{Repair}$]\label{lem:transfer-terminal}
For any $\Phi_0\subseteq\Phi_+$, $x\in\Omega_{\Phi_0}$ and $W\subseteq[n]$, $\mathsf{Repair}(\Phi_+,\Phi_0,x,W)$ returns an $(x,W)$-repair or \textnormal{\textsc{Fail}} in $O_k(n(n+|\Phi_+|))$ time.

For sufficiently large $k$, fix $m\ge0$ satisfying
\[
 m\le\frac{2^kn}{k}\left(\ln k+\ln\ln k+1+
 \frac{\ln\ln k+1/2}{\ln k}-\frac{6(\ln\ln k)^2}{(\ln k)^2}\right).
\]
Fix $z\in\{0,1\}^n$ and an integer $t\ge0$, sample $M\sim\operatorname{Pois}(m)$ and $\Phi_+\sim\Phi(k,n,M)$, and let $X$ be the state after $t$ steps of Glauber dynamics on $\Phi_0=\Phi_+[z]$ from $z$.
With probability at least $1-\ee^{-\rho^8n}$, if $\Phi_+$ is regular and at most $\rho^4n$ clauses in $\Phi_+\setminus\Phi_0$ have at most one true literal under $X$, then $\mathsf{Repair}(\Phi_+,\Phi_0,X,W)$ returns a good $(X,W)$-repair for every $W\subseteq[n]$ with $|W|\le\rho^2n/2$.
\end{lemma}

We then use $\mathsf{FindPath}$ to connect two solutions through the repair target, as proved in \pref{sec:find-path}.

\begin{lemma}[Correctness of $\mathsf{FindPath}$]\label{lem:repair-regularity}
Let $(U,t)$ be an $(x,W)$-repair and $\Phi\subseteq\Phi_+$.
For $y,y'\in\Omega_\Phi$ agreeing with $x$ outside $U$, $\mathsf{FindPath}(y,y',t)$ (\pref{alg:find-path}) returns a satisfying path from $y$ to $y'$ in $\Omega_\Phi$ of length $\Ham(y,t)+\Ham(y',t)\le2|U|$ in $O_k(n(n+|\Phi|))$ time.
\end{lemma}

We next verify the regularity required by \pref{lem:transfer-terminal}.

\begin{lemma}[Formula regularity]\label{lem:unsigned-envelope}
For sufficiently large $k$ and $n\ge n_0(k)$, if $\Phi$ consists of at most $2^kn$ clauses on $[n]$ with independent uniform scopes, then
$\Pr(\mathsf{Reg}(\Phi)^c)\le n^{-4}$.
\end{lemma}
\begin{proof}
At $2^kn$ clauses, the limiting tree in \pref{lem:poisson-local-limit}\textup{(i)} has finite expected size through variable depth $2k+5$.
Choose $D_k$, depending only on $k$, large enough that this expectation is at most $\eta D_k/8$.
Markov's inequality and local convergence therefore bound the expected fraction of variables failing the tree or degree condition by $\eta/8+o_n(1)$.
By \pref{lem:ball-sensitivity}, inserting or deleting one clause changes the number of such variables by $O_k(1)$.
The fixed-count bound in \pref{lem:prob-poisson-lipschitz} gives $\Pr(\mathsf{LT}(\Phi)^c)\le\ee^{-\Omega_k(n)}$; clause-deletion monotonicity extends this to smaller formulas.

For $\mathsf{LS}(\Phi)$, sets of size $a<k$ contain no clause, and for $k\le a\le\rho n$ with $q=a/n$, each clause lies in a fixed $a$-set with probability $(a)_k/(n)_k\le q^k$. A union bound gives
\[
 \Pr(\exists A:|A|=a,\ |\{C:\operatorname{vbl}(C)\subseteq A\}|\ge2a/k)
 \le\left(\frac{\ee}{q}\right)^a\left(\frac{\ee2^knq^k}{2a/k}\right)^{2a/k}
 \le\left(\ee^4q^{1-2/k}\right)^a.
\]
For $k\le a\le\sqrt n$, the summand is at most $(\ee^4n^{-1/2+1/k})^a$, so its sum is at most $2\ee^{4k}n^{1-k/2}$ for sufficiently large $n$.
For $\sqrt n<a\le\rho n$, we have $\ee^4\rho^{1-2/k}\le1/2$, giving a total of at most $n2^{-\sqrt n}$.
For $k\ge12$, the first bound is at most $2\ee^{4k}n^{-5}$.
Together with the exponentially small failure probability of $\mathsf{LT}(\Phi)$, these bounds sum to at most $n^{-4}$ for sufficiently large $n\ge n_0(k)$.
\end{proof}

The following lemma bounds the distance between consecutive walks and the number of omitted clauses with at most one true literal; its proof is in \pref{sec:coupled-glauber}.
Coupled walks use the same random update sequence.

\begin{lemma}[Bounds for Glauber dynamics]\label{lem:coupled-terminal}
For sufficiently large $k$, there are constants $C_k>0$ and $T_k\in\mathbb N$, depending only on $k$, such that the following hold for sufficiently large $n$.
\begin{enumerate}[label=\textup{(\roman*)}]
\item\label{item:coupled-proximity}
\textup{\textbf{Uniform scope guarantee.}}\par
Let $\Psi$ have at most $2^kn$ clauses with independent uniform scopes, and fix $B\subseteq[n]$, an integer $t\ge0$, and $0<r\le n/4$.
With probability at least $1-\exp(|B|-r/(C_k\ee^{C_kt/n}))$, for all $F\subseteq\Psi$ and $z,z'$ agreeing outside $B$, the coupled $t$-step walks from $z,z'$ on $F[z],F[z']$ end within Hamming distance $r$.
\item\label{item:coupled-exceptions}
\textup{\textbf{Poisson guarantee.}}\par
For fixed $z\in\{0,1\}^n$, sample $M\sim\operatorname{Pois}(m)$ and then $\Phi_+\sim\Phi(k,n,M)$.
Let $X$ be the $t$-step Glauber state on $\Phi_+[z]$ from $z$, and let $\mathcal D$ be the clauses in $\Phi_+\setminus\Phi_+[z]$ with at most one true literal under $X$.
Assume $m/n$ lies in \eqref{eq:main-connectivity-range} and $t\ge T_kn$.
For every $s>0$, with probability at least $1-\ee^{-4\rho^{10}n}-(\ee\rho^5n/s)^s$, if $\Phi_+$ is regular, then $|\mathcal D|\le s$.
\end{enumerate}
\end{lemma}

\begin{proof}[Proof of \pref{thm:main}]
The case $m=0$ uses the direct coordinate-flip path, so assume $m\ge1$.
We set $N=T_kn$ and $b_\star=\lceil48C_k\rho^{-2}\ee^{C_kT_k}\rceil$, using the constants in \pref{lem:coupled-terminal}.
We claim that, with probability at least $1-1/n$ over $\Phi$, the formula is satisfiable and, conditional on $\Phi$, the following three events hold simultaneously with probability at least $1-1/n$ over independent uniform choices of $\sigma,\tau\in\Omega_\Phi$ and the algorithm's randomness:
\begin{enumerate}[label=\textup{(\roman*)}]
\item\label{item:connect-good-regularity} $\mathsf{Reg}(\Phi_+)$ holds.
\item\label{item:connect-good-closeness} $\max_{0\le i<b_\star}\Ham(x_i,x_{i+1})\le\rho^2n/8$.
\item\label{item:connect-good-repairs} For every $0\le i\le b_\star$, at most $\rho^4n$ clauses in $\Phi_+\setminus\Phi_i$ have at most one true literal under $x_i$, and $\mathsf{Repair}(\Phi_+,\Phi_i,x_i,W)$ returns a good $(x_i,W)$-repair for every $0\le i<b_\star$ and every set $W\subseteq[n]$ with $|W|\le\rho^2n/2$.
\end{enumerate}
We first derive the conclusions of the theorem and bound the running time, then prove this claim.

We now construct a satisfying path from $\sigma$ to $\tau$, assuming the three good events hold.
The repair calls with $W=\varnothing$ give $|U_i|\le\rho^2n/8$; the bound also holds for $U_0=U_{b_\star}=\varnothing$.
Thus $|W_i|\le3\rho^2n/8\le\rho^2n/2$, so every repair call with $W=W_i$ succeeds and satisfies $|V_i|\le|W_i|/(2\rho)+\rho^2n/8\le\rho n/2$.
Each $y_i$ lies in $\Omega_\Phi$ and agrees with $x_i$ outside $U_i$; hence both $y_i,y_{i+1}$ agree with $x_i$ outside $W_i\subseteq V_i$.
By \pref{lem:repair-regularity}, $\gamma_i$ connects $y_i$ to $y_{i+1}$ in $\Omega_\Phi$ and has length at most $2|V_i|\le\rho n$.
Since $\Phi\subseteq\Phi_0\cap\Phi_{b_\star}$, the walks from $\sigma$ and $\tau$ stay in $\Omega_\Phi$.
Concatenating $\pi_0,\gamma_0,\ldots,\gamma_{b_\star-1},\pi_{b_\star}$ yields a satisfying path from $\sigma$ to $\tau$ of length at most $2N+b_\star\rho n=O_k(n)$.
By \pref{lem:transfer-terminal}, the same reasoning verifies every returned path, with length at most $2N+2b_\star n$.

Then, we show that a connected component contains at least a $1-1/n$ fraction of all solutions.
Set $L=2N+2b_\star n=O_k(n)$.
For each solution $x$, let $S_x$ consist of the solutions reachable from $x$ by a satisfying path of length at most $L$.
For every formula satisfying the claim, averaging over $\sigma$ gives an $x$ with $\mu_\Phi(S_x)\ge1-1/n$; set $S=S_x$.
Shortest satisfying paths from $x$ to members of $S$ stay in $S$.
Thus any two members can be joined through $x$ within $S$ in at most $2L=O_k(n)$ flips.

We next bound the running time.
Since $|\Phi_+|\le2m$, \pref{lem:transfer-terminal} and \pref{lem:repair-regularity} give $O_k(n(n+m))$ time in total for the $O_k(1)$ calls to $\mathsf{Repair}$ and $\mathsf{FindPath}$.
The $O_k(n)$ Glauber steps and path verification take the same time bound by clause scans; any failed check returns \textnormal{\textsc{Fail}}.
Hence, the total running time is $O_k(n(n+m))$.

We now bound the failure probabilities of the three good events.
For~\ref{item:connect-good-regularity}, before conditioning on $\sigma,\tau\in\Omega_\Phi$, $\Phi\cup\Phi_{\mathrm{aux}}\supseteq\Phi_+$ consists of $2m\le2^kn$ clauses with independent uniform scopes.
By \pref{lem:unsigned-envelope}, $\Pr(\mathsf{Reg}(\Phi\cup\Phi_{\mathrm{aux}})^c)\le n^{-4}$.
Since clause deletion preserves regularity, $\Phi_+$ is then regular for every choice of $\sigma,\tau$.
For~\ref{item:connect-good-closeness}, we apply \pref{lem:coupled-terminal}\ref{item:coupled-proximity} with $t=N$, $r=\rho^2n/8$, $\Psi=\Phi\cup\Phi_{\mathrm{aux}}$, $F=\Phi_+$, and $(B,z,z')=(B_{i+1},z_i,z_{i+1})$.
Its simultaneous guarantee over subformulas and starting assignments permits these choices, which depend on $\sigma,\tau$, directly in the original model.
Since $|B_{i+1}|\le2n/b_\star$, the failure probability for each block is at most $\exp(2n/b_\star-\rho^2n/(8C_k\ee^{C_kN/n}))\le\ee^{-4n/b_\star}$.
A union bound over the $b_\star$ blocks shows that closeness holds for all $\sigma,\tau$ outside a single event of probability at most $b_\star\ee^{-4n/b_\star}=\ee^{-\Omega_k(n)}$.

We next bound the probability that~\ref{item:connect-good-repairs} fails while~\ref{item:connect-good-regularity} holds, starting with the Poisson model.
We then use \pref{lem:transfer-extension} to return to the original model conditional on $\sigma,\tau\in\Omega_\Phi$, and \pref{lem:norm-transfer} to obtain bounds for uniform solution pairs.
We fix $\sigma,\tau\in\{0,1\}^n$, so the interpolating assignments $z_i$ are fixed, and sample $\Phi_+$ under the Poisson law.
Let $\mathcal D_i=\{C\in\Phi_+\setminus\Phi_i:|\operatorname{sat}_{x_i}(C)|\le1\}$ for $0\le i\le b_\star$.
Let $\mathcal B_{\rm rep}$ be the event that $\mathsf{Reg}(\Phi_+)$ holds but~\ref{item:connect-good-repairs} fails.
For each $i$, \pref{lem:coupled-terminal}\ref{item:coupled-exceptions} with $t=N$ and $s=\rho^4n$ shows that, except with probability $2\ee^{-4\rho^{10}n}$, regularity of $\Phi_+$ implies $|\mathcal D_i|\le\rho^4n$; here $(\ee\rho)^{\rho^4n}\le\ee^{-4\rho^{10}n}$.

Since $\gamma_*<1$ by \pref{thm:critical-threshold}, \eqref{eq:main-connectivity-range} satisfies the density bound in \pref{lem:transfer-terminal}.
For each $i<b_\star$, we apply \pref{lem:transfer-terminal} with starting assignment $z_i$ and $t=N$.
Except with probability $\ee^{-\rho^8n}$, regularity and $|\mathcal D_i|\le\rho^4n$ guarantee good repairs for every $W\subseteq[n]$ with $|W|\le\rho^2n/2$.
This simultaneous guarantee covers the $W_i$ that depend on earlier walks and repairs.
A union bound gives Poisson probability at most $2(b_\star+1)\ee^{-4\rho^{10}n}+b_\star\ee^{-\rho^8n}\le b_\star\ee^{-2\rho^{10}n}$ for $\mathcal B_{\rm rep}$ when $n$ is large.

For fixed $\sigma,\tau$ and update sequence, $\mathcal B_{\rm rep}$ depends only on the clause multiset, by the choice rules in \pref{sec:repair-appendix}.
We apply \pref{lem:transfer-extension} for each fixed update sequence, then average over that sequence to return to the original model conditional on $\sigma,\tau\in\Omega_\Phi$.
Recalling $\alpha=m/n$ and $\Gamma=\alpha k^3\epsilon_0^2$, we obtain
\begin{equation}\label{eq:Bfix-bound}
 \Pr(\mathcal B_{\rm rep}\mid\sigma,\tau\in\Omega_\Phi)
 \le15\sqrt m\,b_\star\exp(-2\rho^{10}n+4m\epsilon_0^2)
 \le\ee^{-\rho^{10}n},
\end{equation}
since $\Gamma+\alpha\epsilon_0(2/3)^k=o_k(\rho^{10})$.

We now average over uniform solution pairs.
Let $\mathcal Z$ and $L_Z=40n\ee^{\Gamma n}$ be as in \pref{sec:normalization}; on $\mathcal Z$, $\Phi$ is satisfiable.
Given $\Phi$, let $p_\Phi$ be the probability that at least one of the three good events fails, over independent uniform $\sigma,\tau\in\Omega_\Phi$ and the algorithm's randomness; set $p_\Phi=1$ when $\Omega_\Phi=\varnothing$.
The regularity and distance estimates hold simultaneously for all $\sigma,\tau$, so their failure bounds also apply to independent $\sigma,\tau\sim\mu_\Phi$.
For $\mathcal B_{\rm rep}$, we apply \pref{lem:norm-transfer} with $r=2$ and $b=\ee^{-\rho^{10}n}$ from \eqref{eq:Bfix-bound}; its hypothesis $\alpha\le2^k/512$ holds throughout \eqref{eq:main-connectivity-range} for sufficiently large $k$.
Combining these bounds, we obtain, uniformly in $m$,
\[
 \E[\mathbb{1}\{\mathcal Z\}\cdot p_\Phi]
 \le n^{-4}+\ee^{-\Omega_k(n)}
       +2L_Z^2\exp((2\alpha\epsilon_0(2/3)^k-\rho^{10})n)
 \le\frac{1}{4n^2}.
\]
The current density range also satisfies the hypothesis of \pref{lem:norm-lower}, so $\Pr(\mathcal Z^c)\le1/(2n)$.
By Markov's inequality, we obtain $\Pr_\Phi(\Phi\notin\mathcal Z\text{ or }p_\Phi>1/n)\le\frac{1}{2n}+n\E[\mathbb{1}\{\mathcal Z\}\cdot p_\Phi]\le\frac{1}n$.
\end{proof}

\subsection{Coupled Glauber dynamics}\label{sec:coupled-glauber}
We prove \pref{lem:coupled-terminal}, which gives the distance and exceptional-clause bounds for the coupled walks.
For part~\ref{item:coupled-proximity}, we bound how disagreements spread when the walks use the same random updates.
For part~\ref{item:coupled-exceptions}, we first show that a walk moves far from its starting assignment, then use the Poisson formula law to bound $|\mathcal D|$.

\begin{proof}[Proof of \pref{lem:coupled-terminal}\ref{item:coupled-proximity}]
We first construct a set $I_j$ containing all disagreements until its size reaches $r$.
The probability bound is trivial when $|B|\ge r$, so assume $|B|<r$.
Let $v_j$ be the coordinate chosen at step $j$ of the shared random update sequence.
We set $I_0=B$ and let $\mathcal C_0=\{C\in\Psi:\operatorname{vbl}(C)\cap I_0\ne\varnothing\}$ be the clauses meeting $I_0$.
For $0\le j<t$, let $U_j=I_j\cup\bigcup_{C\in\mathcal C_j}\operatorname{vbl}(C)$ be the variables in $I_j$ or a clause meeting $I_j$; if $|I_j|<r$ and $v_{j+1}\in U_j$, set
\[
 I_{j+1}=I_j\cup\{v_{j+1}\},\qquad
 \mathcal C_{j+1}=\{C\in\Psi:\operatorname{vbl}(C)\cap I_{j+1}\ne\varnothing\},
\]
and otherwise set $(I_{j+1},\mathcal C_{j+1})=(I_j,\mathcal C_j)$.
For any $F\subseteq\Psi$ and $z,z'$ agreeing outside $B$, let $Y_j,Y'_j$ be the two states after $j$ steps on $F[z],F[z']$ using this sequence.
Every clause in $F[z]\triangle F[z']$ meets $I_0$.
If $Y_j,Y'_j$ agree outside $I_j$ and $v_{j+1}\notin U_j$, every clause of $F[z]\cup F[z']$ containing $v_{j+1}$ lies in $(F[z]\cap F[z'])\setminus\mathcal C_j$ and has the same variable values in both chains.
Their conditional marginals at $v_{j+1}$ therefore agree, and the shared update bit gives the same new value.
By induction, $\Delta(Y_t,Y'_t)\subseteq I_t$ whenever $|I_t|<r$, for every choice of update bits.
It therefore suffices to bound $\Pr(|I_t|\ge r)$.

We next control the growth of this set at each update of the dynamics.
Let $D_j=|I_j|+\sum_{C\in\mathcal C_j}|\operatorname{vbl}(C)\setminus I_j|$ count its variables and the incidences from clauses in $\mathcal C_j$ to variables outside $I_j$; thus $D_j\ge|U_j|$.
We reveal the scopes of clauses meeting $I_j$ as elements enter $I_j$, and let $\mathcal F_j$ be the information consisting of the revealed scopes, their clause indices, and $v_1,\dots,v_j$.
Conditional on $\mathcal F_j$, the $|\Psi\setminus\mathcal C_j|$ unrevealed scopes are independent uniform $k$-subsets of $[n]\setminus I_j$.
For any $v\in U_j\setminus I_j$ with $|I_j|<r\le n/4$, the number of clauses in $\Psi\setminus\mathcal C_j$ containing $v$ is $B_j\sim\operatorname{Bin}(|\Psi\setminus\mathcal C_j|,k/(n-|I_j|))$ of mean at most $k2^{k+1}$.
Thus $\E[\ee^{kaB_j}\mid\mathcal F_j,v_{j+1}=v]\le\exp(k2^{k+1}(\ee^{ka}-1))$ for $a\ge0$.
Conditional on $\mathcal F_j$, the event $I_{j+1}\ne I_j$ has probability at most $D_j/n$ and increases $D_j$ by at most $kB_j$.
With $C_k=2^{k+4}k^2$, we have
\[
 \exp(k2^{k+1}(\ee^{ka}-1))-1\le C_ka
 \qquad(0\le a\le2/C_k).
\]
Taking $a_j=C_k^{-1}\ee^{-C_kj/n}$, we obtain
\[
 \E[\ee^{a_{j+1}D_{j+1}}\mid\mathcal F_j]
 \le\ee^{a_{j+1}D_j}(1+C_ka_{j+1}D_j/n)
 \le\ee^{a_jD_j}.
\]

We finally use Markov's inequality to turn this growth estimate into a tail bound.
At $j=0$, each clause belongs to $\mathcal C_0$ independently with probability at most $k|B|/n$.
Since $D_0\le|B|+k|\mathcal C_0|$, the initial binomial bound gives $\E[\ee^{a_0D_0}]\le\exp(|B|[a_0+k2^{k+1}(\ee^{ka_0}-1)])\le\ee^{|B|}$.
Iterating the preceding conditional inequality, we obtain
\[
 \Pr(|I_t|\ge r)
 \le\ee^{-a_tr}\E[\ee^{a_tD_t}]
 \le\exp\left(|B|-\frac{r}{C_k\ee^{C_kt/n}}\right).
\]
This event depends only on the clause scopes and update coordinates, so the bound holds for all $F\subseteq\Psi$ and assignments $z,z'$ agreeing outside $B$.
\end{proof}

For \pref{lem:coupled-terminal}\ref{item:coupled-exceptions}, we first show that Glauber dynamics preserves the conditional uniform satisfying distribution, then bound the distance from the starting assignment.
\begin{lemma}\label{lem:glauber-stationarity}
For $m\ge0$, sample $M\sim\operatorname{Pois}(m)$ and $\Phi_+\sim\Phi(k,n,M)$, and independently choose $z$ uniformly from $\{0,1\}^n$.
Let $X_t$ be the state after $t$ steps of Glauber dynamics on $\Phi_0=\Phi_+[z]$ from $z$.
For every integer $t\ge0$, we have $(\Phi_0,X_t)\overset{\mathrm d}=(\Phi_0,z)$.
\end{lemma}
\begin{proof}
We claim that, conditional on $\Phi_0$, $z$ is uniform on $\Omega_{\Phi_0}$.
To see this, note that conditional on $z$, the numbers of copies of possible clauses satisfied by $z$ are independent Poisson variables with the same mean, and the expected total number $m(1-2^{-k})$ does not depend on $z$.
Thus, for any fixed formula $F$, $\Pr(\Phi_0=F\mid z)$ is constant over $z\in\Omega_F$ and zero otherwise.
Since $z$ is initially uniform on $\{0,1\}^n$, the claim follows from Bayes' formula.

For a fixed formula $F$, every legal single-variable flip has Glauber transition probability $1/(2n)$ in both directions, so the transition matrix is symmetric and preserves $\mu_F$.
Thus $X_t$ and $z$ have the same conditional distribution given $\Phi_0$, proving the identity.
\end{proof}

Recall that $\rho=k^{-100\lceil(\ln k)^6\rceil}$.
Choose $T_k\in\mathbb N$ sufficiently large depending only on $k$, and set $N=T_kn$.
\begin{lemma}[Hamming distance after the Glauber walk]\label{lem:escape-signed}
Assume $m/n$ lies in \eqref{eq:main-connectivity-range}.
For fixed $z\in\{0,1\}^n$, sample $\Phi_+$ as in \pref{lem:glauber-stationarity} and let $(X_t)_{t\ge0}$ be Glauber dynamics on $\Phi_0=\Phi_+[z]$ from $z$.
Then, for every integer $t\ge N$, with probability at least $1-\ee^{-4\rho^{10}n}$, if $\mathsf{LT}(\Phi_+)$ holds, then
\[
 \Ham(z,X_t)>8\ln(1/\rho)n/k.
\]
\end{lemma}
\begin{proof}
Fix an integer $t\ge N$.
We first obtain small-set expansion on a suitable set of states by comparing local updates with Glauber moves.
We then show that, with high probability, $\mathsf{LT}(\Phi_+)$ implies that the walk stays in this set through time $N$ and ends outside any prescribed small set.
Finally, we apply these bounds to a Hamming ball.

\paragraph{Small-set expansion by path comparison.}
For each variable, the number of clauses in $\Phi_0=\Phi_+[z]$ forcing it under $z$ has mean $\lambda_+=mk/(n2^k)$.
By \pref{thm:local}\textup{(iii)}, we choose $L=O((\ln\ln k)^{1/16})$ and $\vartheta\in[0,1]$ for \pref{alg:local-update} at this $\lambda_+$ with planted root flow at least $\epsilon_k=64\eta$; for large $k$, $L\le2k+4$.
Let $K_v$ be the transition matrix of this update on $B_L(\Phi_0,v)$, keeping all variables at depth $L$ or outside the ball unchanged, when $v$ is $(2k+4,D_k)$-locally tree-structured; otherwise, it leaves the current assignment unchanged.
Each $K_v$ is symmetric on $\Omega_{\Phi_0}$ by \pref{thm:local}\textup{(i)}.
Let $G(x)$ be the total local-update flow of these kernels, and let $\Omega_{\mathrm{flow}}$ contain the states where it is at least $2\eta n$:
\[
 G(x)=\sum_{v,y\ne x}K_v(x,y)
       \frac{\mathbb{1}\{x(v)\ne y(v)\}}{\Ham(x,y)},
 \qquad \Omega_{\mathrm{flow}}=\{x\in\Omega_{\Phi_0}:G(x)\ge2\eta n\}.
\]
For any nonempty $A\subseteq\Omega_{\Phi_0}$, we claim that
\[
 \sum_{x\in A}\sum_{\substack{y\in A\\y\ne x}}\sum_v
 K_v(x,y)\frac{\mathbb{1}\{x(v)\ne y(v)\}}{\Ham(x,y)}
 \le2|A|\log_2|A|.
\]
Indeed, under a uniform coordinate permutation, $v\in\Delta(x,y)$ is the first disagreement with probability $1/\Ham(x,y)$.
For a fixed permutation, we group assignments of $A$ that agree on the preceding coordinates, then split each group into $A_0,A_1$ according to the value at the next coordinate $v$.
Symmetry and unit row sums give
$\sum_{x\in A_0,y\in A_1}(K_v(x,y)+K_v(y,x))\le2\min\{|A_0|,|A_1|\}$.
We charge at most two units to each assignment in the smaller part, resolving ties consistently.
Whenever an assignment is charged, its group shrinks by at least half, so it is charged at most $\log_2|A|$ times.
Averaging proves the bound.
Consequently, if $A\subseteq\Omega_{\mathrm{flow}}$ and $|A|\le2^{\eta n/2}$, the local-update flow leaving $A$ is at least $\eta n|A|$.

To transfer this bound to Glauber moves, for each local transition $(v,x,y)$ from $A$ to $A^c$ with $x(v)\ne y(v)$, fix a satisfying single-flip path inside $B_L(\Phi_0,v)$ using \pref{thm:local}\textup{(i)}.
Charge $K_v(x,y)/\Ham(x,y)$ to its first directed edge leaving $A$.
Set $B_k=M_{2k+4,D_k}$ and fix a directed edge $e=(a,b)$ flipping $u$.
Encode a contributing transition by $(v,x|_{V(B_L(\Phi_0,v))})$.
Since the ball contains $u$, \eqref{eq:ball-sensitivity} gives at most $B_k$ choices for $v$, each with at most $2^{B_k}$ restrictions of $x$.
The code and $a$ determine $x$, which agrees with $a$ outside the ball.
For each code, summing over $y$ charges at most $\sum_yK_v(x,y)=1$.
Thus the total weight charged to any directed edge is at most $B_k2^{B_k}$.
Let $F$ be the transition matrix of Glauber dynamics, which assigns probability $1/(2n)$ to each legal directed single-flip edge.
Dividing the outgoing local weight by this bound on the charge per edge, we obtain
\[
 \sum_{x\in A}F(x,A^c)\ge\frac{\eta|A|}{4B_k2^{B_k}}
 \qquad(A\subseteq\Omega_{\mathrm{flow}},\ |A|\le2^{\eta n/2}).
\]

\paragraph{Staying in $\Omega_{\mathrm{flow}}$ and avoiding small sets.}
For every fixed set $B\subseteq\{0,1\}^n$, we show that, except with probability $(N+1)\ee^{-10\rho^8n}+\sqrt{3|B|}\,2^{-\eta n/4}$, $\mathsf{LT}(\Phi_+)$ implies that the walk stays in $\Omega_{\mathrm{flow}}$ through time $N$ and $X_t\notin B$.
We prove these two guarantees in turn.

First, we claim that, with probability at least $1-(N+1)\ee^{-10\rho^8n}$,
\begin{equation}\label{eq:glauber-stay}
 \mathsf{LT}(\Phi_+)\ \Longrightarrow\
 X_j\in\Omega_{\mathrm{flow}}\quad\text{for all }0\le j\le N.
\end{equation}
For $y\in\Omega_{\Phi_0}$, let $g_v(\Phi_0,y)$ be the local-update flow from \pref{sec:local-formulas}, with the same $L,\vartheta$.
It uses only the subformula $\mathcal L_y$ of clauses with one or two $y$-true literals and is zero unless $v$ is $(L,k^4)$-locally tree-structured in $\mathcal L_y$.
The comparison in \pref{lem:root-flow}\textup{(i)} bounds $g_v$ by the corresponding contribution to $G$ at each $(2k+4,D_k)$-locally tree-structured root of $\Phi_0$.
Since $0\le g_v\le1$, on $\mathsf{LT}(\Phi_0)$ we have $G(y)\ge\sum_vg_v(\Phi_0,y)-\eta n$.
Apply \pref{lem:root-flow}\textup{(ii)} with planted density $m(1-2^{-k})/n$, whose forcing mean is the same $\lambda_+$ used to choose $L,\vartheta$.
With $b_k=\exp(-C(L+1)\ln k)$, we have $3\eta<\epsilon_k/2$ and $b_k\gg\rho^8$, so $\Pr(\sum_vg_v(\Phi_0,z)<3\eta n)\le3\ee^{-b_kn}\le\ee^{-10\rho^8n}$.
Simultaneous flips of variable values and literal signs preserve the model and the events we estimate, so we may take $z$ uniform.
By \pref{lem:glauber-stationarity}, the same bound holds with $z$ replaced by $X_j$ at each $j$.
Since $\mathsf{LT}(\Phi_+)$ implies $\mathsf{LT}(\Phi_0)$, a union bound over $0\le j\le N$ proves \eqref{eq:glauber-stay}.
Sign symmetry gives this guarantee for every fixed $z$.

Let $\mathcal E_N=\{X_j\in\Omega_{\mathrm{flow}}\text{ for all }0\le j\le N\}$.
We next show that, for every fixed $\Phi_0$,
\begin{equation}\label{eq:glauber-small-set}
 \Pr(X_t\in B,\ \mathcal E_N\mid\Phi_0)
 \le\sqrt{3|B|}\,2^{-\eta n/4}.
\end{equation}
For functions on $\Omega_{\mathrm{flow}}$, we use $\langle f,g\rangle=\sum_x f(x)g(x)$ and the corresponding $\ell^p$ norms.
Let $Q=F|_{\Omega_{\mathrm{flow}}\times\Omega_{\mathrm{flow}}}$ and $\mathcal E_Q(f)=\langle f,(I-Q)f\rangle$.
Thus $Q^j(x,y)$ is the probability of reaching $y$ at step $j$ without leaving $\Omega_{\mathrm{flow}}$.
The update at a fixed coordinate averages over its allowed values, so its operator is an orthogonal projection.
Thus $F$, and its principal submatrix $Q$, have all eigenvalues in $[0,1]$.
For $x\in\Omega_{\mathrm{flow}}$, let $f_j(y)=Q^j(x,y)$ and $g_j(y)=\max\{f_j(y)-2^{-\eta n/2},0\}$.
The total probability $\|f_j\|_1$ is at most one, so $g_j$ is nonzero on at most $2^{\eta n/2}$ states and $\|g_j\|_2^2\ge\|f_j\|_2^2-2^{1-\eta n/2}$.
Extending these functions by zero outside $\Omega_{\mathrm{flow}}$, we have $\mathcal E_Q(f)=\frac12\sum_{y,z\in\Omega_{\Phi_0}}F(y,z)(f(y)-f(z))^2$.
For $s\ge0$, let $A_s=\{y\in\Omega_{\mathrm{flow}}:g_j(y)^2>s\}$.
Applying the preceding bound to each $A_s$ and integrating over $s$, then using Cauchy--Schwarz, we obtain
\[
 \frac{\eta\|g_j\|_2^2}{4B_k2^{B_k}}
 \le\frac{1}{2}\sum_{y,z\in\Omega_{\Phi_0}}F(y,z)|g_j(y)^2-g_j(z)^2|
 \le\sqrt{2\mathcal E_Q(g_j)}\,\|g_j\|_2.
\]
Since $|g_j(y)-g_j(z)|\le|f_j(y)-f_j(z)|$, we obtain
\[
 \mathcal E_Q(f_j)\ge\mathcal E_Q(g_j)
 \ge\frac{\eta^2}{32B_k^2 4^{B_k}}
          (\|f_j\|_2^2-2^{1-\eta n/2}).
\]
Since the eigenvalues of $Q$ lie in $[0,1]$, we have
$\|f_{j+1}\|_2^2\le\|f_j\|_2^2-\mathcal E_Q(f_j)$.
Iterating from $\|f_0\|_2^2=1$ gives $\|f_N\|_2^2\le2^{1-\eta n/2}+\exp(-\eta^2N/(32B_k^2 4^{B_k}))\le3\cdot2^{-\eta n/2}$ for sufficiently large $T_k$.
For $t\ge N$, paths staying in $\Omega_{\mathrm{flow}}$ through time $N$ contribute $Q^N(x,\cdot)F^{t-N}$.
The $\ell^2$ contraction of $F$ and Cauchy--Schwarz bound this contribution on any set $B$ by $\sqrt{3|B|}\,2^{-\eta n/4}$.
For $x\notin\Omega_{\mathrm{flow}}$ this contribution is zero, proving \eqref{eq:glauber-small-set}.
Averaging over $\Phi_0$ and combining with \eqref{eq:glauber-stay} proves the stated guarantee.

\paragraph{Application to a Hamming ball.}
Put $q=8\ln(1/\rho)/k<1/2$ and $\mathcal H(q)=-q\ln q-(1-q)\ln(1-q)$.
The radius-$qn$ Hamming ball about $z$ has at most $(n+1)\ee^{\mathcal H(q)n}$ elements, and $\mathcal H(q)=O((\ln k)^8/k)=o(\eta)$.
Applying \eqref{eq:glauber-stay} and \eqref{eq:glauber-small-set} to this ball, the claimed implication fails with probability at most
\[
 (N+1)\ee^{-10\rho^8n}
 +\sqrt{3(n+1)}\exp\left(\left(\frac{\mathcal H(q)}{2}-\frac{\eta\ln2}{4}\right)n\right)
 \le\ee^{-4\rho^{10}n}.\qedhere
\]
\end{proof}

\begin{proof}[Proof of \pref{lem:coupled-terminal}\ref{item:coupled-exceptions}]
The walk depends only on $\Phi_0$ and its independent updates.
By Poisson thinning, conditional on $(z,\Phi_0,X)$, the clauses of $\Phi_+\setminus\Phi_0$ are therefore independent uniform clauses falsified by $z$, with a Poisson number of mean $m2^{-k}$.
In each such clause, the literals true under $X$ are precisely those on variables in $\Delta(z,X)$.
Having at most one true literal requires at least $k-1$ of its variables to avoid $\Delta(z,X)$; a union bound over the remaining position bounds this probability by $k(1-\Ham(z,X)/n)^{k-1}\le k\ee^{-(k-1)\Ham(z,X)/n}$.
For $\Ham(z,X)\ge8\ln(1/\rho)n/k$, the conditional law of $|\mathcal D|$ is therefore Poisson with mean at most $mk2^{-k}\rho^{8(1-1/k)}\le\rho^5n$.
For every $s>0$, the conditional probability of $|\mathcal D|>s$ is then at most $(\ee\rho^5n/s)^s$ by a Poisson tail bound.
By \pref{lem:escape-signed}, with probability at least $1-\ee^{-4\rho^{10}n}$, $\mathsf{LT}(\Phi_+)$ implies the required distance bound.
Since regularity implies $\mathsf{LT}(\Phi_+)$, a union bound proves the claim.
\end{proof}

\subsection{Correctness of \texorpdfstring{$\mathsf{Repair}$}{Repair}}\label{sec:repair-appendix}\label{sec:entropy}
We prove \pref{lem:transfer-terminal}, starting with the description of $\mathsf{Repair}$ (\pref{alg:common-repair}).
Given $\Phi_0\subseteq\Phi_+$, $x\in\Omega_{\Phi_0}$, and $W\subseteq[n]$, the algorithm enlarges $W$ to a \emph{repair set} $U$ and constructs a solution of $\Phi_+$ agreeing with $x$ outside $U$.
It uses two operations, \emph{tree repair} and \emph{pair-matching repair}, to handle clauses whose true literals under $x$ involve only variables in $U$.

Tree repair uses \emph{small complete forcing trees} (\pref{def:forcing-tree}) in $\Phi_0$ under $x$, where \emph{small} means having at most $(\ln k)^6$ variables.
Recall that each clause in a forcing tree forces its parent variable and has one selected child variable, whose literal is false under $x$.
A complete forcing tree contains every clause of $\Phi_0$ forcing any of its variables.
For a clause $C$ with $\operatorname{sat}_x(C)\subseteq U$, we look for such a tree $T_C$ rooted in $\operatorname{vbl}(C)$ whose variable set satisfies $V(T_C)\cap U=\varnothing$.
If one exists, we retain it together with its associated clause $C$ and add $V(T_C)$ to $U$.
Its root lies outside the previous $U$ and therefore has an $x$-false literal in $C$.
Let $\mathcal F$ be the family of \emph{retained trees}, with variable set $V(\mathcal F)=\bigcup_{T\in\mathcal F}V(T)$.
The set $\mathcal A$ of \emph{clauses assigned to tree repair} consists of all clauses in these trees and all their associated clauses.
At termination, we flip all variables in the retained trees; each associated clause is then satisfied by its tree's root and each tree clause by its child variable.

If no such tree exists, pair-matching repair assigns two variables to $C$ using a pair matching.
Let $\mathcal J$ be the set of \emph{clauses assigned to pair-matching repair}, and let $M$ record their assigned variable pairs.

\begin{definition}[Pair matching]\label{def:fix-primitives}
A \emph{pair matching} is a pair $M=(\mathcal J,(M_C)_{C\in\mathcal J})$, where $\mathcal J\subseteq\Phi_+$ and the sets $M_C\subseteq\operatorname{vbl}(C)$ have size two and are pairwise disjoint.
Its matched variable set is $V(M)=\bigcup_{C\in\mathcal J}M_C$.
\end{definition}

We reindex clauses lexicographically by their increasing variable lists, literal truth values under $x$, and membership in $\Phi_0$, assigning consecutive indices to identical copies within each class.
All remaining choices use fixed lexicographic orders on variable labels and these clause indices.
We maintain $M$ as a matching in the factor graph (\pref{def:factor-graph}) with copies $1,2$ of each clause in $\mathcal J$, both connected to its original variables.
An augmenting path starts at an unmatched clause copy, alternates unmatched and matched edges, and ends at an unmatched variable; exchanging matched and unmatched edges along the path matches that copy.
For $C\in\Phi_+\setminus\mathcal J$, $\mathsf{Augment}(M,C)$ adds its two copies and matches them in order, each along the first shortest augmenting path, returning an updated pair matching with clause set $\mathcal J\cup\{C\}$ or \textnormal{\textsc{Fail}} if either augmentation fails.
Each augmentation keeps all previously matched variables matched and adds one new variable to $V(M)$.
At termination, we make both literals on $M_C$ true for each $C\in\mathcal J$.

In \pref{alg:common-repair}, a clause is \emph{assigned} if it belongs to $\mathcal A\cup\mathcal J$, and \emph{unassigned} otherwise.
We process an unassigned clause when $\operatorname{sat}_x(C)\subseteq U$, since none of its literals outside $U$ is true under $x$.

\begin{algorithm}[!htb]
\caption{$\mathsf{Repair}(\Phi_+,\Phi_0,x,W)$: Local repair}\label{alg:common-repair}
\SetKwInOut{Input}{Input}
\SetKwInOut{Output}{Output}
\Input{A formula $\Phi_+$, a subformula $\Phi_0\subseteq\Phi_+$, an assignment $x\in\Omega_{\Phi_0}$, and an initial variable set $W\subseteq[n]$.}
\Output{$U\supseteq W$ and $y\in\Omega_{\Phi_+}$ with $y_{[n]\setminus U}=x_{[n]\setminus U}$, or \textnormal{\textsc{Fail}}.}
$U\gets W$,\quad $\mathcal F\gets\varnothing$,\quad $\mathcal A\gets\varnothing$,\quad $\mathcal J\gets\varnothing$\;
Initialize the empty pair matching $M=(\mathcal J,(M_C)_{C\in\mathcal J})$\;
\While{there is an unassigned $C\in\Phi_+$ with $\operatorname{sat}_x(C)\subseteq U$}{
    Choose the first such clause $C$\;
    \eIf{some small complete forcing tree rooted in $\operatorname{vbl}(C)$ avoids $U$}{
        \tcp{tree repair}
        Retain the first such tree $T_C$ with its associated clause $C$, and add $T_C$ to $\mathcal F$\;
        Add $C$ and all clauses of $T_C$ to $\mathcal A$; $U\gets U\cup V(T_C)$\;
    }{
        \tcp{pair-matching repair}
        $M\gets\mathsf{Augment}(M,C)$\quad(\Return \textnormal{\textsc{Fail}} if the augmentation fails)\;
        $\mathcal J\gets\mathcal J\cup\{C\}$,\quad $U\gets U\cup V(M)$\;
        Remove from $\mathcal F$ every tree meeting $V(M)$, and remove its clauses and associated clause from $\mathcal A$\;
    }
}
Set $y\gets x$, flip all variables in $V(\mathcal F)$, and make both literals of each $C\in\mathcal J$ on $M_C$ true\;
\Return $(U,y)$\;
\end{algorithm}

When a tree is removed, its variables remain in $U$ and its clauses, including the associated clause, become unassigned.

\begin{proof}[Proof of the correctness guarantee in \pref{lem:transfer-terminal}]
We show by induction that the retained trees and the pair matching use disjoint variables and handle disjoint clause sets.
At every completed iteration, $U$ contains $W$, $V(M)$, and all retained-tree variables, and only grows.
The retained trees are pairwise variable-disjoint and avoid $W\cup V(M)$; their clause sets, each including its associated clause, are pairwise disjoint and avoid $\mathcal J$.
Every assigned clause $C$ satisfies $\operatorname{sat}_x(C)\subseteq U$.
These properties hold initially.
A new tree avoids the old $U$, and each of its clauses forces a parent outside $U$.
Such a clause is neither already assigned nor the selected clause $C$, since both have all their $x$-true literals on variables in $U$.
A matching update also processes an unassigned clause and removes every tree meeting the updated $V(M)$.
Thus both operations preserve the invariants.

Suppose the procedure returns $(U,y)$.
The invariants make $y$ well-defined, with $U\supseteq W$ and $y=x$ outside $U$.
Every clause with an $x$-true literal outside $U$ keeps that literal true; every other clause is assigned at termination.
For an assigned clause define
\[
 U_C=\begin{cases}
 M_C,&C\in\mathcal J,\\
 \{\text{root of }T_C\},&C\text{ is associated with a retained tree }T_C,\\
 \{\text{child of }C\text{ in the tree}\},&C\text{ belongs to a retained tree}.
 \end{cases}
\]
Each variable in $U_C$ satisfies $C$ when set to its $y$-value.
Each retained-tree variable serves just one clause: its associated clause if it is the root, and its parent clause otherwise.
Consequently, the $U_C$ are pairwise disjoint, with size one inside $V(\mathcal F)$ or size two outside it, and $y\in\Omega_{\Phi_+}$.

We now construct the required path for any $\Psi\subseteq\Phi_+$ and $\sigma\in\Omega_\Psi$ agreeing with $x$ outside $U$.
Starting from $z=\sigma$, let $D=\Delta(z,y)\subseteq U$ be the current disagreement set; we flip only coordinates in $D$.
If $D\setminus V(\mathcal F)\ne\varnothing$, any clause $C\in\Psi$ forcing a variable $v\in D$ under $z$ has no $x$-true literal outside $U$, hence is assigned.
Also $U_C\subseteq D$: a variable in $U_C\setminus D$ would already satisfy $C$ at its $y$-value and cannot be $v$, whose literal is false under $y$.
Disjointness bounds the number of such forcing clauses by
\[
 |D\cap V(\mathcal F)|+\tfrac12|D\setminus V(\mathcal F)|<|D|.
\]
Each clause forces at most one variable, so some variable of $D$ is unforced and can be flipped legally.
It need not lie outside the trees.
Repeat until $D\subseteq V(\mathcal F)$; each step decreases $|D|$.

Now process retained trees in reverse addition order, and within each tree from leaves to root, skipping variables already equal to $y$.
Every remaining flip changes an $x$-value to its opposite.
A clause of a retained tree can lose a true literal only when its parent variable flips, by which time its child in the tree has reached its satisfying $y$-value.
For an associated clause $C$, every variable in $\operatorname{sat}_x(C)$ was already in $U$ when $T_C$ was added.
Any such variable in a retained tree belongs to an earlier tree, so the root of $T_C$ reaches its $y$-value before any still-changing $x$-true literal of $C$ is lost.
Clauses $C\in\mathcal J$ already have both literals on $M_C$ true, and unassigned clauses retain a true literal outside $U$.
Both phases therefore preserve $\Psi$ and flip each disagreement once, giving length $\Ham(\sigma,y)$.

Finally, we bound the running time.
We first order clauses by radix sorting and list the clauses of $\Phi_0$ forcing each variable under $x$, in $O_k(n+|\Phi_+|)$ total time.
A complete forcing tree must include every clause in the list of each of its variables, so a tree of size at most $s=\lfloor(\ln k)^6\rfloor$ requires at most $s-1$ child choices, each with at most $k-1$ options.
Rejecting branches that repeat a variable or exceed this size, we enumerate all small complete forcing trees in $O_k(1)$ time per root.
Each tree addition grows $U$, and each successful pair augmentation grows $V(M)$, so there are $O(n)$ iterations.
Clause and tree scans and at most two breadth-first searches take $O_k(n+|\Phi_+|)$ time per iteration, yielding the claimed bound.
\end{proof}

To bound the size of a repair, we estimate how many clauses lack an available tree and how many have both true and false literals on variables in the repair set.
For $x\in\Omega_{\Phi_0}$, let $\mathcal C_{\mathrm{force}}=\{C\in\Phi_0:|\operatorname{sat}_x(C)|=1\}$ and $\mathcal C_{\mathrm{multi}}=\{C\in\Phi_+:|\operatorname{sat}_x(C)|\ge2\}$.
Here $\mathcal C_{\mathrm{force}}$ is the clause collection of $(\Phi_0)_x^{(1)}$ from \pref{def:support}.
We call the clauses in $\mathcal C_{\mathrm{multi}}$ \emph{multi-true} under $x$.
For $B,U\subseteq[n]$, define the following numbers, keeping $x,\Phi_0,\Phi_+$ fixed.
Let $\mathsf{NoTree}(B)$ count clauses $C\in\mathcal C_{\mathrm{force}}\cup\mathcal C_{\mathrm{multi}}$ with $\operatorname{sat}_x(C)\subseteq B$ such that no small complete forcing tree in $\Phi_0$ under $x$, rooted at a variable whose literal in $C$ is false, avoids $B$.
Let $\mathsf{ForcingLink}(U)$ count clauses in $\mathcal C_{\mathrm{force}}$ whose forced variable lies in $U$ and which also have an $x$-false literal on a variable in $U$.
Let $\mathsf{MultiLink}(U)$ count clauses $C\in\mathcal C_{\mathrm{multi}}$ with $\operatorname{sat}_x(C)\subseteq U$ and an $x$-false literal on a variable in $U$.

\begin{lemma}[Small-set bounds for repair]\label{lem:repair-small-sets}
Let $k$ be sufficiently large and assume
\[
 0\le m\le\frac{2^kn}{k}\left(\ln k+\ln\ln k+1+
 \frac{\ln\ln k+1/2}{\ln k}-\frac{6(\ln\ln k)^2}{(\ln k)^2}\right).
\]
Fix $x\in\{0,1\}^n$.
Let $\Phi_0$ consist of $\operatorname{Pois}(m(1-2^{-k}))$ independent uniform $k$-clauses satisfied by $x$, and independently let $G$ consist of $\operatorname{Pois}(m)$ independent uniform $k$-clauses.
With probability at least $1-\ee^{-\rho^8n}$, the following hold for every $\Phi_+$ with $\Phi_0\subseteq\Phi_+\subseteq\Phi_0\cup G$:
\begin{enumerate}[label=\textup{(\roman*)}]
\item\label{item:repair-blocked-trees}
\emph{Clauses without an available forcing tree:} $\mathsf{NoTree}(B)<\max\{\rho^4n,|B|/(1024(\ln k)^{20})\}$ for every $B\subseteq[n]$ with $|B|\le\rho n$;
\item\label{item:repair-forcing-clauses}
\emph{Clauses forcing variables in $B$:} $|\{C\in\mathcal C_{\mathrm{force}}:\operatorname{sat}_x(C)\subseteq B\}|\le(\ln k)^8|B|$ for every $B\subseteq[n]$ with $\rho^4n\le|B|\le\lceil2\rho n\rceil$;
\item\label{item:repair-internal-clauses}
$\mathsf{ForcingLink}(U)+2\mathsf{MultiLink}(U)<(1+1/(4(\ln k)^6))|U|$ for every $U\subseteq[n]$ with $\rho^4n\le|U|\le\rho n$.
\end{enumerate}
\end{lemma}
\begin{proof}
It suffices to prove the bounds for $\Phi_+=\Phi_0\cup G$.
Deleting clauses outside $\Phi_0$ preserves its forcing trees, $\mathcal C_{\mathrm{force}}$, and $\mathsf{ForcingLink}$, and can only decrease $\mathsf{NoTree}$ and $\mathsf{MultiLink}$.
Poisson thinning gives $\mathcal C_{\mathrm{force}}\sim\mathsf{Pois}_1(x,\lambda)$ with $\lambda=mk/(n2^k)\le2\ln k$.
The set $\mathcal C_{\mathrm{multi}}$ is independent of $\mathcal C_{\mathrm{force}}$ and has the law obtained by retaining clauses with at least two $x$-true literals from a Poisson formula of mean $2m$.

\smallskip\noindent\emph{(i) Clauses without an available forcing tree.}
For~\ref{item:repair-blocked-trees}, the density assumption places $\lambda$ in the range of \pref{lem:set-avoiding-trees}.
Also, $2\rho\le(16k^{2\lceil(\ln k)^6\rceil})^{-1}$, so that lemma allows every avoided set of size at most $2\rho n$.
Let $\mathcal P$ be the event that the lemma's conclusion holds for every subformula; its failure probability is at most $\ee^{-10\rho^8n}$.
If~\ref{item:repair-blocked-trees} fails, we set $b=\lceil\max\{\rho^4n,|B|/(1024(\ln k)^{20})\}\rceil$ and enlarge $B$ to a set $B'$ of size $\lceil1024b(\ln k)^{20}\rceil\le2\rho n$.
For each fixed $B'$, we expose only the clauses of $\mathcal C_{\mathrm{force}}$ forcing variables outside $B'$.
If $\mathcal P$ holds, this reduced formula has at least $n/2$ roots of small complete forcing trees avoiding $B'$.
These trees remain valid after restoring clauses forcing variables in $B'$.
Every clause counted by $\mathsf{NoTree}(B)$ is also counted by $\mathsf{NoTree}(B')$ whenever $B\subseteq B'$.
We do not condition on $\mathcal P$: for each reduced formula admitting at least $n/2$ such roots, choose them deterministically and bound the remaining independent clauses.
Each such clause has all its true literals on variables in $B'$ and no false literal on any of these roots.
Among the $2n$ signed literals, only the $2|B'|$ literals on $B'$ and the single $x$-false literal on each variable outside $B'$ and the root set are permitted, at most $n/2+|B'|$ in total.
By \pref{lem:sampling-products}, a uniform $k$-clause with fair signs has all its literals in this set with probability at most $2^{-k}(1/2+|B'|/n)^k$.
The remaining forcing clauses and $\mathcal C_{\mathrm{multi}}$ are independent Poisson processes conditional on the reduced formula, so the number satisfying these restrictions has mean at most $2m2^{-k}(1/2+|B'|/n)^k\le n(3/5)^k$.
A Poisson tail bound and a union over $B'$ therefore give
\[
 \sum_{b=\lceil\rho^4n\rceil}^{\lceil\rho n/(1024(\ln k)^{20})\rceil}
 \binom n{\lceil1024b(\ln k)^{20}\rceil}
 \left(\frac{\ee n(3/5)^k}{b}\right)^b
 \le\sum_{b\ge\rho^4n}\ee^{-kb/4}
 \le\ee^{-10\rho^8n},
\]
since $1024(\ln k)^{20}\ln(1/\rho)=O((\ln k)^{27})=o(k)$.

\smallskip\noindent\emph{(ii) Clauses forcing variables in $B$.}
For~\ref{item:repair-forcing-clauses}, the count for a fixed set $B$ is Poisson of mean $\lambda|B|$.
Summing over the possible sizes of $B$, we bound the failure probability by
\[
 \sum_{a=\lceil\rho^4n\rceil}^{\lceil2\rho n\rceil}
 \binom na\left(\frac{2\ee}{(\ln k)^7}\right)^{a(\ln k)^8}\le\ee^{-10\rho^8n},
\]
using $(\ln k)^8\ln((\ln k)^7/(2\ee))\gg\ln(1/\rho)$.

\smallskip\noindent\emph{(iii) Bounding $\mathsf{ForcingLink}(U)+2\mathsf{MultiLink}(U)$.}
For~\ref{item:repair-internal-clauses}, we fix $U\subseteq[n]$ with $\rho^4n\le|U|\le\rho n$.
The first count requires the forced variable and a variable with a false literal in $U$; the second requires at least two variables with true literals and one with a false literal in $U$.
Hence the independent Poisson counts $\mathsf{ForcingLink}(U)$ and $\mathsf{MultiLink}(U)$ satisfy
\[
 \E[\mathsf{ForcingLink}(U)]\le\frac{\lambda k|U|^2}{n},\qquad
 \E[\mathsf{MultiLink}(U)]\le\frac{2\lambda k^2|U|^3}{n^2}.
\]
Indeed, the first bound follows by exposing the forced variable.
For the second, choosing one variable in $U$ with a false literal and summing over the $j\ge2$ variables with true literals in $U$, we bound the Poisson mean by
\[
 2m2^{-k}\sum_{j=2}^{k-1}\binom kj(k-j)\left(\frac{|U|}{n}\right)^{j+1}
 \le\frac{\lambda k^2|U|^3}{n^2}\ee^{k|U|/n}.
\]
With $u=|U|$ and $\tau=\ln(n/(k^4u))>0$, independence and the Poisson generating functions give
\[
 \ln\E[\ee^{\tau(\mathsf{ForcingLink}(U)+2\mathsf{MultiLink}(U))}]
 \le\frac{\lambda ku^2}{n}\ee^\tau
       +\frac{2\lambda k^2u^3}{n^2}\ee^{2\tau}
 =\left(\frac{\lambda}{k^3}+\frac{2\lambda}{k^6}\right)u\le u.
\]
Consequently, a union bound over $U$ gives failure probability at most
\[
 \sum_{u=\lceil\rho^4n\rceil}^{\lfloor\rho n\rfloor}
 \left[\ee^2k^{4+1/(\ln k)^6}\left(\frac{u}{n}\right)^{1/(4(\ln k)^6)}\right]^u
 \le\sum_{u\ge \lceil\rho^4n\rceil}k^{-10u}
 \le\ee^{-10\rho^8n},
\]
since $\rho^{1/(4(\ln k)^6)}\le k^{-25}$.
The union of the four exceptional events has probability at most $4\ee^{-10\rho^8n}\le\ee^{-\rho^8n}$ for large $n$.
\end{proof}

We next show that these three bounds, together with regularity and $|\mathcal D|\le\rho^4n$, guarantee that $\mathsf{Repair}$ succeeds.

\begin{lemma}\label{lem:repair-deterministic}
Let $k$ and $n\ge n_0(k)$ be sufficiently large, fix $\Phi_0\subseteq\Phi_+$ and $x\in\Omega_{\Phi_0}$, and put $\mathcal D=\{C\in\Phi_+\setminus\Phi_0:|\operatorname{sat}_x(C)|\le1\}$.
Suppose $\Phi_+$ is regular, $|\mathcal D|\le\rho^4n$, and the three bounds in \pref{lem:repair-small-sets} hold for $(\Phi_+,\Phi_0,x)$.
Then $\mathsf{Repair}(\Phi_+,\Phi_0,x,W)$ returns a good $(x,W)$-repair for every $W\subseteq[n]$ with $|W|\le\rho^2n/2$.
\end{lemma}
\begin{proof}
Fix any $W\subseteq[n]$ with $|W|\le\rho^2n/2$.
Here $\mathcal D=\Phi_+\setminus(\mathcal C_{\mathrm{force}}\cup\mathcal C_{\mathrm{multi}})$ by the definitions of $\mathcal C_{\mathrm{force}},\mathcal C_{\mathrm{multi}}$.
At a completed stage with $|U|\le\rho n$, let $R=U\setminus V(\mathcal F)$ be the repair variables outside the retained trees.
We first bound $|U|$ in terms of $|R|$, then control $|R|$ through the pair matching.
Both $U$ and $R$ only grow: new trees avoid $U$, and removed trees leave their variables in $U$.
Each retained tree has one fewer clause than variables, and also contributes its associated clause to $\mathcal A$, so $|\mathcal A|=|V(\mathcal F)|=|U|-|R|$.
Every clause in $\mathcal A\cap\mathcal C_{\mathrm{force}}$ is counted by $\mathsf{ForcingLink}(U)$, and every clause in $\mathcal A\cap\mathcal C_{\mathrm{multi}}$ is counted by $\mathsf{MultiLink}(U)$.
The weight two on the latter count leaves a positive contribution from multi-true assigned clauses.
Since $\mathcal C_{\mathrm{force}},\mathcal C_{\mathrm{multi}},\mathcal D$ partition $\Phi_+$, we obtain
\[
 \mathsf{ForcingLink}(U)+2\mathsf{MultiLink}(U)
 \ge |U|-|R|+|\mathcal A\cap\mathcal C_{\mathrm{multi}}|-\lceil\rho^4n\rceil.
\]
\pref{lem:repair-small-sets}\ref{item:repair-internal-clauses} therefore gives $|\mathcal A\cap\mathcal C_{\mathrm{multi}}|\le |R|+\lceil\rho^4n\rceil+|U|/(4(\ln k)^6)$; this also holds if $|U|<\rho^4n$, since then $|\mathcal A\cap\mathcal C_{\mathrm{multi}}|\le |U|<\rho^4n$.
When a clause of $\mathcal C_{\mathrm{force}}$ is selected, its forced variable lies in $U$ and outside the retained trees: a retained tree contains every clause of $\Phi_0$ forcing one of its variables, and those clauses are already assigned.
This variable remains in $R$, so \pref{lem:repair-small-sets}\ref{item:repair-forcing-clauses}, enlarging $R$ if needed, bounds the number of retained trees associated with such clauses by $(\ln k)^8(|R|+\lceil\rho^4n\rceil)$.
Each remaining retained tree is associated with a clause in $\mathcal C_{\mathrm{multi}}$ or $\mathcal D$.
Since each tree has at most $(\ln k)^6$ variables and $|U|=|R|+|V(\mathcal F)|$, substituting these bounds gives
\[
 |U|\le\tfrac32(\ln k)^{14}(|R|+\lceil\rho^4n\rceil)+|U|/4,
\]
and hence $|U|\le2(\ln k)^{14}(|R|+\lceil\rho^4n\rceil)$.

We next bound $|R|$ through the size of the pair matching.
Each matching update adds two new matched variables.
Previously matched variables remain matched and avoided by retained trees, so only these two additions can remove trees, at most two in total.
Thus $|R|\le|W|+4(\ln k)^6|\mathcal J|$.
Every clause of $\mathcal J\setminus\mathcal D$ is counted by $\mathsf{NoTree}(U)$, since $U$ only grows.
By \pref{lem:repair-small-sets}\ref{item:repair-blocked-trees}, we obtain
\[
 |R|\le |W|+4(\ln k)^6\bigl(\rho^4n+\mathsf{NoTree}(U)\bigr)
 \le |W|+8(\ln k)^6\rho^4n+\frac{|U|}{256(\ln k)^{14}}.
\]
Combining the last two bounds, with $\lceil\rho^4n\rceil\le2\rho^4n$ for large $n$, we obtain
\begin{equation}\label{eq:repair-size-linear}
 |U|\le4(\ln k)^{14}|W|+32(\ln k)^{20}\rho^4n.
\end{equation}

We finally show that every augmentation succeeds and the algorithm terminates.
For $|U|\le\rho n$, just before augmenting with a selected clause $C$, every clause of $(\mathcal J\cup\{C\})\setminus\mathcal D$ is counted by $\mathsf{NoTree}(U)$.
Hence $2|\mathcal J\cup\{C\}|\le4\rho^4n+|U|/(512(\ln k)^{20})<\rho n$.
If the clause copies could not all be matched to distinct variables, Hall's theorem would give some $\mathcal K\subseteq\mathcal J\cup\{C\}$ with variable union $A$ satisfying $|A|<2|\mathcal K|\le2|\mathcal J\cup\{C\}|<\rho n$.
By $\mathsf{LS}(\Phi_+)$ in \pref{def:unsigned-envelope}\ref{item:regularity-sparsity}, we have $|\mathcal K|<2|A|/k<4|\mathcal K|/k$, a contradiction for $k\ge4$.
Thus both augmentations succeed whenever the current $|U|\le\rho n$.
For $|W|\le\rho^2n/2$, the right-hand side of \eqref{eq:repair-size-linear} is less than $\rho n/4$ for sufficiently large $k$.
A first stage with $|U|\ge\rho n/2$ would still have $|U|<\rho n$, since each step adds at most $\max\{(\ln k)^6,2\}$ vertices, contradicting \eqref{eq:repair-size-linear}.
Starting from $|W|\le\rho^2n/2$, we therefore remain below $\rho n/2$, and every augmentation succeeds.
For sufficiently large $k$, the right-hand side of \eqref{eq:repair-size-linear} is at most $|W|/(2\rho)+\rho^2n/8$.
Hence the output is a good $(x,W)$-repair for every $W\subseteq[n]$ with $|W|\le\rho^2n/2$.
\end{proof}

\begin{proof}[Proof of the probabilistic guarantee in \pref{lem:transfer-terminal}]
We first take $z$ independently uniform on $\{0,1\}^n$ and fix $t$.
Conditional on $z$, the omitted clauses $\Phi_+\setminus\Phi_0$ are independent of $\Phi_0$ and the walk.
We may therefore sample a Poisson formula $G$ of mean $m$, independently of $(z,\Phi_0,X)$, and realize $\Phi_+=\Phi_0\cup(G\setminus G[z])$.
By \pref{lem:glauber-stationarity}, conditional on $X=x$, $\Phi_0$ has the planted law in \pref{lem:repair-small-sets}, independently of $G$.
Since $\Phi_0\subseteq\Phi_+\subseteq\Phi_0\cup G$, that lemma's three bounds hold for $(\Phi_+,\Phi_0,x)$ with conditional probability at least $1-\ee^{-\rho^8n}$.

We apply \pref{lem:repair-deterministic} whenever $\Phi_+$ is regular and at most $\rho^4n$ omitted clauses have at most one true literal under $x$, and average over $X$.
Simultaneously flipping variable values and literal signs preserves the formula law, Glauber dynamics, and the prescribed choices in repair.
The failure probability is therefore the same for every starting assignment $z$, proving the claim for fixed $z$.
\end{proof}

\subsection{Greedy routing and correctness of \texorpdfstring{$\mathsf{FindPath}$}{FindPath}}\label{sec:find-path}
For $x\in\{0,1\}^n$ and $v\in[n]$, let $x^{\oplus v}$ denote $x$ with coordinate $v$ flipped. Given solutions $y,y'\in\Omega_\Phi$ and a common target $t\in\Omega_\Phi$, $\mathsf{FindPath}(y,y',t)$ (\pref{alg:find-path}) greedily flips disagreeing coordinates from $y$ to $t$ and from $y'$ to $t$ while preserving satisfaction of $\Phi$, and concatenates the first path with the reversal of the second.

\begin{algorithm}[ht]
\caption{$\mathsf{FindPath}(y,y',t)$: Greedy routing in $\Phi$}\label{alg:find-path}
\SetKwInOut{Input}{Input}
\SetKwInOut{Output}{Output}
\Input{A formula $\Phi$ and solutions $y,y',t\in\Omega_\Phi$ satisfying the hypotheses of \pref{lem:repair-regularity}.}
\Output{A satisfying path from $y$ to $y'$ in $\Omega_\Phi$.}
\ForEach{$a\in\{y,y'\}$}{
    $z\gets a$,\quad $\gamma_a\gets(a)$\;
    \While{$z\ne t$}{
        $v\gets\min\{u\in[n]:z(u)\ne t(u),\ z^{\oplus u}\in\Omega_\Phi\}$,\quad $z\gets z^{\oplus v}$,\quad $\gamma_a\gets\gamma_a\circ(z)$\;
    }
}
\Return $\gamma_y\circ\operatorname{rev}(\gamma_{y'})$\;
\end{algorithm}

\begin{proof}[Proof of \pref{lem:repair-regularity}]
Every current assignment $z$ in either greedy path satisfies $\Phi$ and agrees with $x$ outside $U$.
By the definition of a repair, there is a satisfying path from $z$ to $t$ of length $\Ham(z,t)$.
If $z\ne t$, its first step flips a disagreeing coordinate, so the set in \pref{alg:find-path} is nonempty.
Any chosen legal flip preserves these conditions and decreases $\Ham(z,t)$ by one.
Thus the two greedy paths reach $t$ in $\Ham(y,t)$ and $\Ham(y',t)$ flips, and concatenating the first with the reversal of the second proves the lemma.
At each step, scanning the clauses identifies all forced variables, after which we choose the first unforced variable disagreeing with $t$, in $O_k(n+|\Phi|)$ time.
There are at most $2n$ flips, and writing the path takes $O(n^2)$ time, giving the stated running time.
\end{proof}

\subsection{Local connectivity up to the looseness threshold}
\label{sec:companion-local-connectivity}
We finish this section by proving local connectivity below the $*$-core threshold~\pref{thm:main-looseness}\ref{item:main-local-connectivity-part}.
To connect two nearby solutions, the repair construction is applied to their disagreement set.
Both solutions can then reach the repair target along short satisfying paths.
Recall that $\rho=k^{-100\lceil(\ln k)^6\rceil}$.

\begin{lemma}[Local single-variable connectivity]
\label{lem:companion-local-connectivity}
Up to the lower bound in \pref{eq:main-looseness-headline}, with probability at least $1-1/n$ over $\Phi$, the formula is satisfiable and all but an $\ee^{-\Omega_k(n)}$ fraction of its solutions $x$ can be joined to every $y\in\Omega_\Phi$ with $\Ham(x,y)\le\rho^2n/2$ by a satisfying path of at most $\rho n$ single-variable flips, constructible in deterministic $O_k(n(n+m))$ time.
More precisely, with $c_k=\rho^8/4$, the probability that $\Phi\in\mathcal Z$ but fewer than a $1-\ee^{-c_kn}$ fraction of its solutions satisfy the stated algorithmic guarantee is at most $n^{-4}+\ee^{-c_kn}$.
\end{lemma}

\begin{proof}
The case $m =0$ is trivial, so assume $m>0$.
For $x\in\Omega_F$, let $\mathcal B(F,x)$ be the event that $F$ is regular but $\mathsf{Repair}(F,F,x,W)$ fails to return a good $(x,W)$-repair for some $W\subseteq[n]$ with $|W|\le\rho^2n/2$.
If $F$ is regular and $\mathcal B(F,x)$ does not occur, every such call returns a good repair $(U,t)$.
For $y\in\Omega_F$ with $\Ham(x,y)\le\rho^2n/2$ and $W=\Delta(x,y)$, $\mathsf{FindPath}(x,y,t)$ connects $x$ to $y$ using at most $2|U|\le |W|/\rho+\rho^2n/4\le\rho n$ flips by \pref{lem:repair-regularity}.
Both calls are deterministic and each takes $O_k(n(n+|F|))$ time.

For the original random formula $\Phi\sim\Phi(k,n,m)$, regularity fails with probability at most $n^{-4}$ by \pref{lem:unsigned-envelope}.

Fix $x$ and let $F$ consist of $\operatorname{Pois}(m)$ independent clauses conditioned to be satisfied by $x$ (the Poissonized planted law).
For sufficiently large $k$, $m/(1-2^{-k})$ satisfies the density bound in \pref{lem:repair-small-sets}.
Applying that lemma with this parameter and $\Phi_+=\Phi_0=F$, we obtain its three bounds with probability at least $1-\ee^{-\rho^8n}$.
Since no clauses are omitted from $F$, \pref{lem:repair-deterministic} gives $\Pr(\mathcal B(F,x))\le\ee^{-\rho^8n}$.

We now transfer this bound to the original formula $\Phi\sim\Phi(k,n,m)$ using \pref{lem:norm-transfer} with $r=1$, obtaining $\E_\Phi[\mathbb{1}\{\mathcal Z\}\cdot\mu_\Phi(\mathcal B)]\le3\sqrt m\,L_Z\ee^{-\rho^8n}\le\ee^{-\rho^8n/2}$, where the last inequality uses $\Gamma=O(k^2\ln k\,2^{-k})=o_k(\rho^8)$.
With $c_k=\rho^8/4$, Markov's inequality bounds the probability of $\mathcal Z\cap\{\mu_\Phi(\mathcal B)>\ee^{-c_kn}\}$ by $\ee^{-c_kn}$.
The event $\mathcal Z$ ensures satisfiability, and the probability that it holds but the stated local connectivity guarantee fails is at most $n^{-4}+\ee^{-c_kn}$.
Adding $\Pr_\Phi(\mathcal Z^c)\le1/(2n)$ from \pref{lem:norm-lower} gives total failure probability at most $1/(2n)+n^{-4}+\ee^{-c_kn}\le1/n$.
\end{proof}

\begin{proof}[Proof of local connectivity in \pref{thm:main-looseness}\ref{item:main-local-connectivity-part}]
Put $c_k=\rho^8/4$.
By \pref{eq:looseness-failure-probability}, \pref{lem:companion-local-connectivity}, and \pref{lem:norm-lower}, there is a formula event $\mathcal G\subseteq\mathcal Z$ with $\Pr_\Phi(\mathcal G^c)\le1/(2n)+2n^{-4}+4/n^2+\ee^{-c_kn}\le1/n$.
For every $\Phi\in\mathcal G$, the looseness and local connectivity constructions have their stated guarantees for all but fractions $1/(2n)$ and $\ee^{-c_kn}$ of the solutions, respectively.

Fix $\Phi\in\mathcal G$ and let $S\subseteq\Omega_\Phi$ consist of the solutions $x$ for which the looseness construction succeeds for every requested variable and the local connectivity construction succeeds for every $y\in\Omega_\Phi$ with $\Ham(x,y)\le\rho^2n/2$.
Then $\mu_\Phi(S)\ge1-1/(2n)-\ee^{-c_kn}\ge1-1/n$.
Thus every $x\in S$ satisfies both parts~\ref{item:main-looseness-part} and~\ref{item:main-local-connectivity-part}, with local connectivity radius $\rho^2n/2=\Theta_k(n)$ and the stated path lengths.
Since $m=O_k(n)$, the running times are $O_k(n+m)=O_k(n)$ and $O_k(n(n+m))=O_k(n^2)$, respectively.
\end{proof}

\section{Proof of shattering}\label{sec:shattering-below-one}
In this section, we prove \pref{thm:main-shattering}, extending the
shattering regime below the $*$-core threshold. Following the
statistical-physics and metastability perspective on random constraint
satisfaction problems~\cite{KMRTSZ07,AC08,ACRT11,BDH16,Haj88},
we view the number of violated clauses as an energy landscape on the
hypercube $\{0,1\}^n$ and track both Hamming separation and single-flip
energy barriers through a common sublevel graph.

\begin{definition}[Energy-sublevel graph]
\label{def:sbo-sublevel-graph}
Given a $k$-CNF formula $\Phi$ on $n$ variables, define the following
quantities:
\begin{itemize}
\item For $y\in\{0,1\}^n$, the \emph{energy} $E_\Phi(y)$ is the number
of clauses of $\Phi$ violated by $y$, counting multiplicities (so
$\Omega_\Phi=\{y\in\{0,1\}^n:E_\Phi(y)=0\}$).
\item For an integer step size $D\ge0$ and an energy threshold $h\ge0$,
the \emph{energy-sublevel graph} $\mathcal G_{D,h}(\Phi)$ is the graph
with vertex set $\{y\in\{0,1\}^n:E_\Phi(y)\le h\}$ in which distinct
vertices $y,y'$ are adjacent whenever $|\Delta(y,y')|\le D$.
\end{itemize}
\end{definition}

\begin{theorem}[Low-energy component bound]\label{thm:shattering-below-one}
For every sufficiently large fixed $k$, there exist $\delta_k,\zeta_k>0$
such that, uniformly over \eqref{eq:main-shattering-range}, for all
sufficiently large $n$, with probability at least $1-1/n$,
$\Phi\sim\Phi(k,n,\lfloor\alpha n\rfloor)$ is satisfiable and
\[
  \mu_\Phi(\mathcal A\cap\Omega_\Phi)\le\ee^{-\zeta_kn}
  \quad\text{for every component $\mathcal A$ of }
  \mathcal G_{D,h}(\Phi),\qquad
  D=\lfloor\delta_kn\rfloor,\quad h=\lfloor n/k^{60}\rfloor.
\]
\end{theorem}

\begin{proof}[Proof of \pref{thm:main-shattering}]
On the high-probability event of \pref{thm:shattering-below-one}, the
nonempty intersections of the connected components of
$\mathcal G_{D,h}(\Phi)$ with $\Omega_\Phi$ form the claimed partition of
$\Omega_\Phi$. Solutions in distinct regions lie in different components
of $\mathcal G_{D,h}(\Phi)$ and therefore have Hamming distance strictly
greater than $D$. Moreover, for sufficiently large $n$ we have $D\ge1$,
so every single-variable-flip path between distinct regions must exit
$\mathcal G_{D,h}(\Phi)$ and thus visit an assignment of energy strictly
greater than $h$.
\end{proof}

By \eqref{eq:critical-gap},
$\gamma_*\le1-(3-2\sqrt2)/\ee<\ln(34/13)$,
so for sufficiently large $k$ this range lies strictly above
the reconstruction threshold.

We prove \pref{thm:shattering-below-one} using the prefix $\Phi_0$ at the
lowest density in \eqref{eq:main-shattering-range}: every path in
$\mathcal G_{D,h}(\Phi)$ is also a path in $\mathcal G_{D,h}(\Phi_0)$.
In \pref{sec:sbo-aux-confinement}, auxiliary cores and DAG trimming turn
a large core jump along a low-energy path into a witness. In
\pref{sec:sbo-planted-witness}, a first-moment bound rules out such
witnesses with high probability under the planted law. Finally,
\pref{sec:sbo-energy-barriers} transfers this bound to the full formula,
confines the component to a small Hamming ball, and bounds its mass using
the full formula's solution count.

\subsection{Auxiliary cores and component localization}
\label{sec:sbo-aux-confinement}
The disagreement set of two solutions, restricted to the $*$-core, is
$*$-flippable~\cite{MR12} (\pref{obs:core-difference-flippable}). Below
the $*$-core threshold, unforced variables can drive acyclic forcing
cascades. To control changes among variables forced under a satisfying
assignment $x\in\Omega_\Phi$, we define an auxiliary core relative to
each disagreement set on those variables.

\begin{definition}[Auxiliary core]\label{def:sbo-core}
Fix a formula $\Phi$ and a satisfying assignment $x\in\Omega_\Phi$:
\begin{itemize}
\item Let $U_x\subseteq[n]$ be the set of variables unforced under $x$
(\pref{def:support}), and let $H$ be the directed graph on the forced
variables $[n]\setminus U_x$, with an edge $v\to w$ whenever some clause
forcing $v$ under $x$ contains $w\ne v$.
\item For $W\subseteq[n]\setminus U_x$, call a clause $C$
\emph{exceptional} under $W$ if
$\operatorname{sat}_x(C)\subseteq U_x\cup W$ and
$|\operatorname{vbl}(C)\cap(U_x\cup W)|\ge3$. Define the
\emph{exceptional set}
\[
  W_{\ge3}
  =\bigcup_{\substack{C\in\Phi:\,C\text{ exceptional under }W}}
       (\operatorname{vbl}(C)\cap W)\subseteq W.
\]
The \emph{auxiliary core} $W_{\mathrm{aux}}$ is the largest subset
$K\subseteq W$ such that every $v\in K\setminus W_{\ge3}$ has an
incoming edge from $K$ in $H[W]$.
\end{itemize}
\end{definition}

Equivalently, $W_{\mathrm{aux}}$ is obtained from $W$ by repeatedly
deleting vertices in $W\setminus W_{\ge3}$ of indegree zero in the
remaining subgraph while keeping all vertices of $W_{\ge3}$ fixed, so
that $H_0:=H[W\setminus W_{\mathrm{aux}}]$ is a directed acyclic graph.
An edge $u\to v$ records that changing $v$ can repair a clause forcing
$u$; the exceptional set accounts for clauses involving several changed
or unforced variables.

For the next lemma, fix $\Phi$, $x\in\Omega_\Phi$, $h\ge0$,
$\epsilon\in(0,6/7)$, $\kappa>0$, $N\ge2/\epsilon$, and an integer
$D\ge0$, using the notation of \pref{def:sbo-core}. For an assignment
$y$, write $W_y=\Delta(x,y)\setminus U_x$. Let $\operatorname{src}(G)$
denote the indegree-zero vertices of a directed graph $G$, including
isolated vertices. Call $S\subseteq[n]\setminus U_x$ an \emph{auxiliary-core
witness}, or simply a \emph{witness}, relative to $\Phi,x,N,\epsilon,\kappa$ if
\begin{equation}\label{eq:sbo-forbidden}
  (6/7-\epsilon)N\le|S|\le(6/7+\epsilon)N,\qquad
  |\operatorname{src}(H[S])\setminus S_{\ge3}|\le\kappa N.
\end{equation}
Unlike Hamming distance or overlap with a fixed $*$-core, the size of
the auxiliary core can change by more than $D$ when the disagreement
set changes by at most $D$: a small step may attach a large acyclic
cascade.

\begin{lemma}[Component localization via DAG trimming]
\label{lem:sbo-confinement}
Suppose $D\le\epsilon N/2$ and every set of at most $D$ variables is
incident to at most $\epsilon\kappa N/[4(k-1)]$ clauses of $\Phi$. If
no assignment $y$ with $E_\Phi(y)\le h$ has $W_y$ as a witness, then
every assignment $y$ in the
$\mathcal G_{D,h}(\Phi)$-component of $x$ satisfies
$|(W_y)_{\mathrm{aux}}|<(6/7-\epsilon)N$.
\end{lemma}

To prove this lemma, we track the auxiliary core along a path from $x$.
The incidence hypothesis bounds the sources created when a step causes
a large core jump; the following trimming lemmas turn part of the
resulting acyclic cascade into a witness without increasing the energy.

For the following auxiliary lemmas, fix a formula $\Phi$ and
$x\in\Omega_\Phi$, using the notation of \pref{def:sbo-core}.
Call a vertex subset of a directed graph \emph{outgoing-closed} if it
contains all outgoing neighbors of its vertices.
Maximality of $W_{\mathrm{aux}}$ implies
that $W_{\ge3}\subseteq W_{\mathrm{aux}}$ and that $W_{\mathrm{aux}}$ is
outgoing-closed in $H[W]$. For $A\subseteq[n]$, write $x^{\oplus A}$ for
the assignment obtained by flipping the coordinates in $A$ (extending
the notation $x^{\oplus v}$ of \pref{sec:find-path}).

Fix $W\subseteq[n]\setminus U_x$, $B\subseteq U_x$, and an outgoing-closed
$A\subseteq W\setminus W_{\mathrm{aux}}$ in $H_0$, and put
$S=W_{\mathrm{aux}}\cup A$.
\begin{lemma}[Energy-nonincreasing trimming]\label{lem:sbo-closed-trimming}
\begin{equation}\label{eq:sbo-energy-trimming}
  E_\Phi(x^{\oplus(S\cup B)})\le E_\Phi(x^{\oplus(W\cup B)})
\end{equation}
and
\begin{equation}\label{eq:sbo-trimming}
  \operatorname{src}(H[S])\setminus S_{\ge3}
  \subseteq\operatorname{src}(H_0[A]).
\end{equation}
\end{lemma}
\begin{proof}
$S$ is outgoing-closed in $H[W]$. Suppose a clause $C$ is satisfied by
$x^{\oplus(W\cup B)}$ but violated by $x^{\oplus(S\cup B)}$. Then all
its $x$-true variables lie in $S\cup B$, and some $x$-false variable
$v\in W\setminus S$ was flipped before trimming. If $C$ has at least two
$x$-true variables, these and $v$ make $C$ exceptional under $W$, so
$v\in W_{\ge3}\subseteq S$, a contradiction. Otherwise $C$ forces its
unique $x$-true variable $u$, which lies in $S$ since $B\subseteq U_x$.
The edge $u\to v$ contradicts the outgoing-closure of $S$. Thus every
previously satisfied clause remains satisfied, proving
\eqref{eq:sbo-energy-trimming}. Moreover, because
$\operatorname{vbl}(C)\cap W\subseteq W_{\ge3}\subseteq S$ for every
clause exceptional under $W$, we have $S_{\ge3}=W_{\ge3}$. Since every
vertex of $W_{\mathrm{aux}}\setminus W_{\ge3}$ has an incoming edge from
$W_{\mathrm{aux}}$ in $H[S]$, \eqref{eq:sbo-trimming} follows.
\end{proof}

For $W,W'\subseteq[n]\setminus U_x$, put $Q=W'\setminus W$ and let
$A_0$ be the smallest outgoing-closed subset of $H_0$ containing
$(W'_{\mathrm{aux}}\cap W)\setminus W_{\mathrm{aux}}$.
\begin{lemma}[Sources created by a core change]\label{lem:sbo-core-jump}
We have $|W_{\mathrm{aux}}\cup A_0|\ge|W'_{\mathrm{aux}}|-|Q|$.
Every source of $H_0[A_0]$ lies outside $Q$ in a clause incident to $Q$.
\end{lemma}
\begin{proof}
The size bound follows from
$W'_{\mathrm{aux}}\cap W\subseteq W_{\mathrm{aux}}\cup A_0$.
Every source $z$ of $H_0[A_0]$ belongs to
$(W'_{\mathrm{aux}}\cap W)\setminus W_{\mathrm{aux}}$ and hence lies
outside $W_{\ge3}$. Since $z\in W'_{\mathrm{aux}}$, either
$z\in W'_{\ge3}$ (so $z$ lies in a clause incident to $Q$) or $z$ has an
incoming edge in $H[W']$ from some $u\in W'_{\mathrm{aux}}$. Such a
vertex $u$ cannot lie in $W_{\mathrm{aux}}$ (as $W_{\mathrm{aux}}$ is
outgoing-closed in $H[W]$) or in $A_0$ (as $z$ is a source of
$H_0[A_0]$), so $u\in Q$. Thus every source of $H_0[A_0]$ lies outside
$Q$ in a clause incident to $Q$.
\end{proof}

Let $G$ be a finite DAG with $s$ sources and maximum outdegree at most
$d$, fix $\epsilon\in(0,6/7)$ and an integer $1\le M\le|V(G)|$, and put
$L=\lceil\log_2(4/\epsilon)\rceil\le4/\epsilon$.
\begin{lemma}[DAG trimming]\label{lem:sbo-dag-trimming}
$G$ has an outgoing-closed subset $A$ with
\[
 M-\lfloor M/2^L\rfloor\le|A|\le M,\qquad
 |\operatorname{src}(G[A])|\le L\max\{s,d\}.
\]
\end{lemma}
\begin{proof}
For a vertex $u$ of $G$, let $\operatorname{desc}_G(u)$ be the vertices
reachable from $u$ in $G$, including $u$. For every integer
$1\le t\le|V(G)|$, $G$ contains an
outgoing-closed vertex subset of size in $[t,2t-1]$ that induces at most
$\max\{s,d\}$ sources: if $|\operatorname{desc}_G(u)|<t$ for all
$u\in\operatorname{src}(G)$, take a minimal union of sets
$\operatorname{desc}_G(u)$ ($u\in\operatorname{src}(G)$) of size at
least $t$; otherwise, choose $u\in V(G)$ with
$|\operatorname{desc}_G(u)|\ge t$ such that every outgoing neighbor $w$
of $u$ has $|\operatorname{desc}_G(w)|<t$, and take
$\operatorname{desc}_G(u)$ if $|\operatorname{desc}_G(u)|=t$, or a
minimal union of sets $\operatorname{desc}_G(w)$ over outgoing neighbors
$w$ of $u$ of size at least $t$ if $|\operatorname{desc}_G(u)|>t$.

Define outgoing-closed
sets $\varnothing=A_0'\subseteq A_1'\subseteq\dots\subseteq A_L'\subseteq V(G)$
by setting $A_i'=A_{i-1}'$ if $|A_{i-1}'|=M$, and otherwise applying the
single-step bound to $G[V(G)\setminus A_{i-1}']$ with
$t=\lceil(M-|A_{i-1}'|)/2\rceil$ and adjoining the resulting subset to
$A_{i-1}'$. Because $A_{i-1}'$ is outgoing-closed in $G$, no edge of $G$
leaves $A_{i-1}'$, so
$\operatorname{src}(G[V(G)\setminus A_{i-1}'])\subseteq\operatorname{src}(G)$
and $M-|A_i'|\le\lfloor(M-|A_{i-1}'|)/2\rfloor$. Consequently,
$A=A_L'$ has the stated properties.
\end{proof}

\begin{proof}[Proof of \pref{lem:sbo-confinement}]
With the parameters of \pref{lem:sbo-confinement}, consider a path from
$x$ in $\mathcal G_{D,h}(\Phi)$, and write
each assignment along it as $x^{\oplus(W'\cup B')}$ with
$W'\subseteq[n]\setminus U_x$ and $B'\subseteq U_x$. Suppose toward a
contradiction that the auxiliary core reaches size $(6/7-\epsilon)N$,
and let $W\cup B\to W'\cup B'$ be the first step crossing this threshold.
Since the initial core is empty, $|W_{\mathrm{aux}}|<(6/7-\epsilon)N$.
If $|W'_{\mathrm{aux}}|\le(6/7+\epsilon)N$, then
\pref{lem:sbo-closed-trimming} with $A=\varnothing$ shows that
$S'=W'_{\mathrm{aux}}$ is a witness with
$E_\Phi(x^{\oplus(S'\cup B')})\le h$, a contradiction. Thus
$|W'_{\mathrm{aux}}|>(6/7+\epsilon)N$. Put $Q=W'\setminus W$; then
$|Q|\le D$. Moreover, $Q\ne\varnothing$: otherwise
$W'\subseteq W$, which by monotonicity of the exceptional sets and the
auxiliary core would give $W'_{\mathrm{aux}}\subseteq W_{\mathrm{aux}}$.
In particular, $D\ge1$.

Let $H_0,A_0$ be as in \pref{lem:sbo-core-jump} for these $W,W'$.
That lemma, the clause-incidence hypothesis, and $D\le\epsilon N/2$
show that the outgoing-closed subgraph $G=H_0[A_0]$ has
$|V(G)|\ge|W'_{\mathrm{aux}}|-|Q|-|W_{\mathrm{aux}}|
>(6/7+\epsilon/2)N-|W_{\mathrm{aux}}|$ and at most
$(k-1)\cdot\epsilon\kappa N/[4(k-1)]=\epsilon\kappa N/4$ sources.
Every $v\in A_0\subseteq W\setminus W_{\mathrm{aux}}$ lies outside
$W_{\ge3}$, so each clause forcing $v$ has at most one other variable in
$U_x\cup W$ and contributes at most one outgoing edge from $v$ in $G$.
Applying the incidence hypothesis to $\{v\}$ (since $D\ge1$) bounds the
maximum outdegree of $G$ by $\epsilon\kappa N/[4(k-1)]$.

Set $M=\lfloor6N/7\rfloor-|W_{\mathrm{aux}}|$ and
$L=\lceil\log_2(4/\epsilon)\rceil$. The inequalities above and
$N\ge2/\epsilon$ ensure $1\le M\le|V(G)|$. Applying
\pref{lem:sbo-dag-trimming} to $G=H_0[A_0]$ yields an outgoing-closed
subset $A\subseteq A_0$ of $H_0$.
By \eqref{eq:sbo-energy-trimming} and \eqref{eq:sbo-trimming}, the set
$S=W_{\mathrm{aux}}\cup A$ satisfies
$E_\Phi(x^{\oplus(S\cup B)})\le h$ and
\[
  (6/7-\epsilon)N\le\lfloor 6N/7\rfloor-N/2^L\le|S|\le\lfloor 6N/7\rfloor,\qquad
  |\operatorname{src}(H[S])\setminus S_{\ge3}|
  \le|\operatorname{src}(H_0[A])|\le L\,\frac{\epsilon\kappa N}{4}\le\kappa N,
\]
where we used $2^{-L}\le\epsilon/4$ and $N\ge2/\epsilon$. This
contradicts the no-witness hypothesis and completes the proof.
\end{proof}

For the remainder of this section, set
\[
 \begin{gathered}
  \alpha_0=\frac{2^k}{k}\bigl(\ln k+\ln\ln k+\ln(34/13)\bigr),
  \quad m_0=\lfloor\alpha_0n\rfloor,\\
  \lambda=\frac{km_0}{n(2^k-1)},\quad p=\ee^{-\lambda},\quad N=np,\\
  \delta_k=2^{-k}k^{-3},\quad D=\lfloor\delta_kn\rfloor,\quad
  h=\lfloor n/k^{60}\rfloor.
 \end{gathered}
\]
For fixed $x\in\{0,1\}^n$, under the Poissonized planted law with mean
$m_0$ clauses, each variable is forced at rate $\lambda$ and remains
unforced with probability $p$, so $N=\E[|U_x|]$. For sufficiently large
$k$ and $n$,
$1/(8k\ln k)\le p\le1/40$.
The following incidence bound verifies the corresponding hypothesis of
\pref{lem:sbo-confinement} for an ordinary random formula with $m_0$
clauses.

\begin{lemma}[Clause incidence]\label{lem:sbo-formula-properties}
Fix $\epsilon\in(0,6/7)$ and $\kappa>0$. For sufficiently large fixed
$k$, we have $D\le\epsilon N/2$ and,
for sufficiently large $n$, with probability at least
$1-\ee^{-c n/k^2}$, every set of at most $D$ variables is incident to at
most $\epsilon\kappa N/[4(k-1)]$ clauses of
$\Phi\sim\Phi(k,n,m_0)$, where $c>0$ depends only on $\epsilon,\kappa$.
\end{lemma}
\begin{proof}
Since $N\ge n/(8k\ln k)$, we have
$D\le\epsilon N/2$ for large $k$. For a fixed set $Q$ of at most $D$
variables, the number of incident clauses is binomial with mean at most
$km_0D/n\le 2n\ln k/k^3$. The required threshold is at least
$\epsilon\kappa n/(32k^2\ln k)$, so a Chernoff bound gives failure
probability at most $\ee^{-c' n/k^2}$ for some $c'>0$ depending only on
$\epsilon,\kappa$. The number of choices of $Q$ is at most
$\exp(O(2^{-k}n/k^2))$; a union bound completes the proof.
\end{proof}

\subsection{Planted witness bounds}
\label{sec:sbo-planted-witness}
For fixed $x\in\{0,1\}^n$, let $\Phi$ consist of $\operatorname{Pois}(m_0)$
independent uniform clauses satisfied by $x$; interpret $U_x$ and $W_y$
relative to this $\Phi$ and $x$.

\begin{lemma}[Low-energy auxiliary-core witnesses]
\label{lem:sbo-census}\label{lem:sbo-energy-census}
There exist absolute constants $\epsilon\in(0,3/28)$ and
$\kappa,\eta>0$ such that, uniformly over $x\in\{0,1\}^n$, for
sufficiently large fixed $k$ and sufficiently large $n$,
\[
 \Pr\bigl(\exists y:E_\Phi(y)\le h\text{ and }W_y
 \text{ is a witness}\bigr)\le\ee^{-\eta N}.
\]
\end{lemma}

To count witnesses, we encode their forcing clauses in a matrix whose
rows are the forced variables and whose columns also include their active
unforced partners. The witness constraints relate the numbers of zero
rows and columns to exceptional clauses and energy; the following matrix
and exponential-rate bounds control their contribution at $|S|\approx6N/7$.

For integers $q,\ell\ge1$, $\rho>0$, and $0<\xi<1$, let
\[
 M_{q,\ell}(\rho,\xi)=
 \sum_{\mathbf d\in\mathbb Z_{\ge0}^{q\times\ell}}
 \xi^{z_{\rm row}(\mathbf d)+z_{\rm col}(\mathbf d)}\prod_{i,j}\frac{\rho^{d_{ij}}}{d_{ij}!},
\]
where $\mathbf d=(d_{ij})$ and $z_{\rm row}(\mathbf d),z_{\rm col}(\mathbf d)$ count its
zero rows and columns.
For $t>0$ with $\ell t<1$, set
\[
 A_\ell(t)=\frac{\ell!t^\ell}{\prod_{i=1}^{\ell}(1-it)}.
\]

\begin{lemma}[Weighted matrix bound]\label{lem:sbo-weighted-matrix}
For $u>\ell\rho$,
\[
 M_{q,\ell}(\rho,\xi)\le
 (\ee^u-1+\xi)^q(1-\xi)^{\ell-u/\rho}A_\ell(\rho/u).
\]
\end{lemma}
\begin{proof}
The beta integral, with the substitution $s=\ee^{-tv}$, gives
\[
 A_\ell(t)=\int_0^\infty\ee^{-v}(\ee^{tv}-1)^\ell\,\dd v.
\]
Labeling the $J=\sum_{i=1}^q\sum_{j=1}^{\ell}d_{ij}$ occurrences and factoring the row
and column choices bounds $M_{q,\ell}(\rho,\xi)$ by
\[
 \sum_{J\ge0}J!\rho^J[z^J](\ee^z-1+\xi)^q
                         [w^J](\ee^w-1+\xi)^\ell.
\]
Bounding $[z^J](\ee^z-1+\xi)^q\le u^{-J}(\ee^u-1+\xi)^q$ and using
$J!=\int_0^\infty\ee^{-v}v^J\,\dd v$ yields
\begin{equation}\label{eq:sbo-matrix}
 (\ee^u-1+\xi)^q
 \int_0^\infty\ee^{-v}(\ee^{\rho v/u}-1+\xi)^\ell\,\dd v.
\end{equation}
The substitution $s=(1-\xi)\ee^{-\rho v/u}$ bounds the integral by
$(1-\xi)^{\ell-u/\rho}A_\ell(\rho/u)$.
\end{proof}

Write $\mathcal H(b)=-b\ln b-(1-b)\ln(1-b)$ for $b\in[0,1]$,
with $\mathcal H(0)=\mathcal H(1)=0$. For $a>0$, $b\in[0,1]$, and
$0<\beta<1$, define
\[
  \Psi_\beta(a,b)=a(1-\ln a)+\mathcal H(b)-\frac\beta2(a+b)^2
                 +a\ln(\ee^{a+b}-1)+(a+b)\bigl[\ln\beta+(\beta^{-1}-1)\ln(1-\beta)\bigr].
\]
\begin{lemma}[Exponential-rate bound]\label{lem:sbo-pressure}
\[
 \max_{0\le b\le1}\Psi_{13/34}(6/7,b)<-\frac{99}{2000000}.
\]
\end{lemma}
\begin{proof}
For $a>0$, $0<\beta<1$, and $0<b<1$,
$\partial_b^2\Psi_\beta(a,b)=-1/[b(1-b)]-\beta-a\ee^{a+b}/(\ee^{a+b}-1)^2\le-4$,
and this strong concavity extends to $[0,1]$ by continuity. At
$\beta=13/34$ and $(a,b)=(6/7,2/7)$, expanding $\ee^u$ through degree
$8$ at $u=8/7$ and
\[
  \ln s=2\sum_{j\ge0}\frac{\vartheta^{2j+1}}{2j+1},
  \qquad \vartheta=\frac{s-1}{s+1},
\]
through $j=8$, with respective absolute remainders bounded by
\[
  \frac{u^9}{9!(1-u/10)}
  \qquad\text{and}\qquad
  \frac{2|\vartheta|^{19}}{19(1-\vartheta^2)},
\]
yields
\[
  -0.000066<\Psi_{13/34}(6/7,2/7)<-0.000059,\qquad
  -0.00198<\partial_b\Psi_{13/34}(6/7,2/7)<-0.00196.
\]
Strong concavity in $b$ therefore gives
\[
  \max_{0\le b\le1}\Psi_{13/34}(6/7,b)
  <-\frac1{20000}+\frac{(1/500)^2}{8}
  =-\frac{99}{2000000}<0.
\]
\end{proof}

We first record the constraints imposed by a witness. Fix
$\epsilon\in(0,6/7)$ and $\kappa>0$ for the next lemma. For
$S\subseteq[n]\setminus U_x$ and $B\subseteq U_x$, write
$U=U_x\cup S$ and $y=x^{\oplus(S\cup B)}$, so $W_y=S$. For $v\in S$ and
$w\in U\setminus\{v\}$, let $d_{v,w}$ count the clauses forcing $v$
that meet $U\setminus\{v\}$ only at $w$, and set $d_{v,v}=0$.
Put
\[
 R=\left\{w\in B:\sum_{v\in S}d_{v,w}>0\right\}.
\]
Let $z_{\rm row}$ and $z_{\rm col}$ be the numbers of zero rows and
columns of the $S\times(S\cup R)$ matrix $(d_{v,w})$, and put
$I_{\ge3}=\sum_C|\operatorname{vbl}(C)\cap U|$, where the sum runs over
clauses exceptional under $S$. Let $V_f$ count clauses forcing a variable
in $S$ that have no other variable in $U$ or exactly one other variable
in $U$, lying in $U_x\setminus R$. Let $V_2$ count clauses with exactly two
true variables under $x$, both in $S\cup R$, and all other variables
outside $U$.

\begin{lemma}[Constraints imposed by a witness]
\label{lem:sbo-witness-constraints}
If $E_\Phi(y)\le h$ and $S$ is a witness, then
\[
 V_f+V_2\le h,\qquad
 z_{\rm row}+z_{\rm col}\le2I_{\ge3}+\kappa N+h.
\]
\end{lemma}
\begin{proof}
Consider a clause counted by $V_f$, forcing $v\in S$. If it has no other
variable in $U$, none of its other variables is flipped. Otherwise its
unique other variable $w$ in $U$ lies in $U_x\setminus R$; if $w\in B$
then $d_{v,w}>0$ and $w\in R$, a contradiction. Thus the clause is
violated by $y$. Every clause counted by $V_2$ is also violated, since $S\cup R$ is flipped and its remaining
variables lie outside $U$. These pools are disjoint, giving the first
inequality.

A zero row indexed by $v\in S\setminus S_{\ge3}$ requires a clause
counted by $V_f$: the variable $v$ is forced under $x$, but no
exceptional forcing clause is available at $v$, and all its row entries
indexed by $S\cup R$ vanish. Hence each clause forcing $v$ meets
$U\setminus\{v\}$ in at most one variable; if present, that variable
lies in $U_x\setminus R$. Every column in $R$ is occupied by definition.
A column indexed by
$w\in S\setminus(S_{\ge3}\cup\operatorname{src}(H[S]))$ is also
occupied, since an incoming edge of $w$ in $H[S]$ comes from a forcing
clause meeting $U$ only at its forced variable and $w$ when
$w\notin S_{\ge3}$. Distinct zero rows give distinct forcing clauses.
Thus
\[
 z_{\rm row}\le |S_{\ge3}|+V_f,\qquad
 z_{\rm col}\le |S_{\ge3}|+|\operatorname{src}(H[S])\setminus S_{\ge3}|.
\]
The witness condition and $|S_{\ge3}|\le I_{\ge3}$ give the second
inequality.
\end{proof}

For the next lemma, fix $\epsilon\in(0,3/28)$, $\kappa>0$,
$\delta\in(0,1/10)$, and $T\ge\ln2$, and condition on the set $U_x$ of unforced variables
with $||U_x|-N|\le\delta N$.
Under the Poissonized planted law, the clause families forcing each variable
are independent Poisson processes of rate $\lambda$, independent also
of the clauses with at least two $x$-true literals. Conditioning on $U_x$
therefore makes the forcing pools empty on $U_x$ and nonempty outside it,
without changing the multi-true clause process.
For integers $q,r$ satisfying $|q/N-6/7|\le\epsilon$ and
$0\le r\le|U_x|$, let $X_{q,r}$ count pairs $S\subseteq[n]\setminus U_x$, $R\subseteq U_x$
of sizes $q,r$ for which some $B\subseteq U_x$ gives $E_\Phi(x^{\oplus(S\cup B)})\le h$,
$S$ is a witness, and $R$ is the set determined by $S,B$ above. Set
\[
 \rho=\lambda\frac{\binom{n-|U_x|-q}{k-2}}{\binom{n-1}{k-1}},\quad
 u=\frac{q+r}{N},\quad \mu_2=\rho\binom{q+r}{2},\quad
 \tau=\lambda^{-2}.
\]

\begin{lemma}[Conditional witness-pair bound]
\label{lem:sbo-witness-first-moment}
Uniformly over all choices of $x,U_x,q,r$ above,
for sufficiently large fixed $k$
and sufficiently large $n$,
\begin{align}
 \E[X_{q,r}\mid U_x]
 &\le\binom{n-|U_x|}{q}\binom{|U_x|}{r}
 \left(\frac p{1-p}\right)^q
 (\ee^u-1)^q A_{q+r}(\rho/u)\notag\\
 &\quad\cdot\exp\!\left\{-\mu_2+T\kappa N+O_T(N/\lambda)
       +O(\ee^{-T})N+(T+2\ln\lambda)h
       +\tau(q\lambda+\mu_2)\right\}.
 \label{eq:sbo-energy-first-moment}
\end{align}
\end{lemma}
\begin{proof}
Let $\E_0$ denote expectation under the Poissonized planted law before
conditioning on the unforced set, with $U_x$ and $S$ treated as fixed
sets. Under this law, the $d_{v,w}$ with
$w\ne v$ are independent $\operatorname{Pois}(\rho)$ variables, and
$V_2$ has mean $\mu_2$ independently of the forcing clauses. For a
candidate pair $(S,R)$, put
$\xi=\ee^{-T}$. By \pref{lem:sbo-witness-constraints}, we may insert the
weight $\xi^{z_{\rm row}+z_{\rm col}}$ at multiplicative cost
$\exp(2TI_{\ge3}+T\kappa N+Th)$.

Allowing diagonal entries and empty columns in $R$ enlarges the weighted
matrix sum. Applying \pref{lem:sbo-weighted-matrix} with
$\ell=q+r$ and $u=(q+r)/N$ bounds it by
\[
 (\ee^u-1)^q A_{q+r}(\rho/u)\exp(O(\ee^{-T})N).
\]
Indeed, $\xi\le1/2$, $u$ is bounded away from zero, and uniformly over
the stated sizes,
\[
 N\rho=\ell\rho/u=k\lambda\ee^{-\lambda}(1+o_k(1)+o_n(1))
       =13/34+o_k(1)+o_n(1),
\]
which is bounded away from both $0$ and $1$ for sufficiently large $k,n$.

Conditional on $U_x$, each family forcing a variable in $S$ contributes
$p/(1-p)$ times its factorial weights, subject to being nonempty.
Dropping the non-emptiness constraint enlarges the sum, and $p^q$
contains the Poisson normalization for all forcing pools in $S$.
For $U=U_x\cup S$, we have $|U|\le3N$. If a clause has $j\ge3$
variables in $U$, there are $2^j-1$ planted sign patterns with all
$x$-true literals in $U$: $j$ have one true literal and the others have
at least two. Thus the mean number of exceptional clauses with $j$
variables in $U$ under the Poissonized planted law before conditioning on
$U_x$ is
\[
 \frac{n\lambda}{k}(2^j-1)
 \frac{\binom{|U|}{j}\binom{n-|U|}{k-j}}{\binom nk}.
\]
Let $\mu_{\ge3}=O(N/\lambda)$ be the mean number of exceptional clauses
forcing variables in $S$ before conditioning on $U_x$. Their Poisson
normalization $\ee^{-\mu_{\ge3}}$ is already included in $p^q$;
compensating for this normalization and bounding the moment over all
exceptional clauses gives
\[
 \ee^{\mu_{\ge3}}\E_0[\ee^{2TI_{\ge3}}]
 \le\exp\!\left(O(N/\lambda)+\frac{n\lambda}{k}\sum_{j\ge3}
 \frac{(2k|U|\ee^{2T}/n)^j}{j!}\right)
 =\exp(O_T(N/\lambda)),
\]
uniformly for $|U|\le3N$, since $kp=O(1/\lambda)$ and hence
$\lambda(kp)^2=O(1/\lambda)$.

For sufficiently large $k,n$, we have $\lambda>1$ and hence $0<\tau<1$.
By \pref{lem:sbo-witness-constraints},
\[
 \mathbb{1}\{V_f+V_2\le h\}\le\tau^{-h}\tau^{V_f+V_2}.
\]
The total unconditioned intensity $\mu_f$ of $V_f$ is at most
$q\lambda$. Since its normalization $\ee^{-\mu_f}$ is already in
$p^q$, summing its tilted factorial weights gives $\exp(\tau\mu_f)$.
The independent count $V_2$ contributes
$\E[\tau^{V_2}]=\exp(-(1-\tau)\mu_2)$. The clause types counted by
$V_f$ and $V_2$ are disjoint from each other, the matrix $(d_{v,w})$,
and the exceptional clauses. Combining these contributions with the matrix and exceptional
bounds, and summing over the possible $(S,R)$, proves the claim.
\end{proof}

\begin{proof}[Proof of \pref{lem:sbo-energy-census}]
Return to the Poissonized planted law without conditioning on $U_x$. Fix $\epsilon\in(0,3/28)$,
$\delta\in(0,1/10)$, and $\kappa>0$ to be chosen below. Since $|U_x|\sim\operatorname{Bin}(n,p)$, the event
$||U_x|-N|>\delta N$ has probability $\ee^{-\Omega(N)}$.

On its complement, take $q,r$ in the ranges of
\pref{lem:sbo-witness-first-moment}, set $a=q/N$ and $b=r/|U_x|$, and
write $\beta=13/34$. Then $N\rho=\beta+o_k(1)+o_n(1)$ and
$h/N\le8\ln k/k^{59}$. For each fixed $T\ge\ln2$,
\[
 (T+2\ln\lambda)h+\tau(q\lambda+\mu_2)=o_k(N).
\]
Stirling's formula and a Riemann-sum approximation give
\[
 \frac1N\ln A_{q+r}(\rho/u)
 =u\ln\beta+u(\beta^{-1}-1)\ln(1-\beta)+o_k(1)+o_n(1).
\]
Since $u=a+b+O(\delta)$, \pref{lem:sbo-witness-first-moment} and
Stirling's formula imply that $N^{-1}\ln\E[X_{q,r}\mid U_x]$ is at most
\[
 \Psi_\beta(6/7,b)
 +O(\epsilon+\delta+\ee^{-T}+T\kappa)
 +O_T(1/\lambda)+o_k(1)+o_n(1).
\]
By \pref{lem:sbo-pressure}, the leading term is uniformly negative in $b$.
Choose $\epsilon,\delta$ sufficiently small, then $T$ large and
$\kappa$ small, and finally $k,n$ large. Every assignment $y$ with
$E_\Phi(y)\le h$ for which $W_y$ is a witness determines a pair counted
by some $X_{q,r}$. Markov's inequality, a
sum over the at most $(n+1)^2$ choices of $(q,r)$, and the bound on
$\Pr(||U_x|-N|>\delta N)$ prove the claim after reducing $\eta>0$.
\end{proof}

\subsection{Completion of the proof}\label{sec:sbo-energy-barriers}
Confinement bounds only the auxiliary core of a forced disagreement set.
To bound the Hamming distance, we show that every sufficiently large
disagreement set puts at least $N$ variables in its exceptional set,
which lies in the auxiliary core. We prove this expansion property for a
single planted solution at the endpoint density; for related expansion
arguments, see~\cite{AC08}.

\begin{lemma}[Planted forcing expansion]\label{lem:sbo-planted-expansion}
For fixed $x\in\{0,1\}^n$, let $\Phi$ consist of $m_0$ independent
uniform clauses satisfied by $x$. Put $r_k=(\ln k)^{-2}$. For sufficiently
large fixed $k$ and sufficiently large $n$, with probability at least
$1-\ee^{-n/(5\ln k)}$, every $A\subseteq[n]$ with $|A|\ge r_kn$ contains
at least $N$ variables forced under $x$ by clauses containing at least
two other variables of $A$.
\end{lemma}
\begin{proof}
Fix $A$ and $A'\subseteq A$. If $v\in A'$, the probability that a
uniform $(k-1)$-subset of $[n]\setminus\{v\}$ contains at least two
members of $A$ is at least $1/2$ for sufficiently large $k,n$: its mean
is at least $kr_k/2\to\infty$ and its variance is at most its mean. The
events that a clause forces different
variables are disjoint, so a planted clause forces a variable in $A'$
and contains at least two other variables of $A$ with probability at
least $k|A'|/[2n(2^k-1)]$. Thus the probability that none of the $m_0$
clauses does so is at most $\ee^{-\lambda|A'|/2}$.

For large $k$, we have $N\le|A|/4$. If fewer than $N$ variables in $A$
are forced by such clauses, some $A'\subseteq A$ of size at least
$3|A|/4$ has no such variable. Taking a union bound over $A'$ and $A$
gives a failure probability at most
\[
 \sum_{s\ge r_kn}\binom ns 2^s\ee^{-3\lambda s/8}
 \le \sum_{s\ge r_kn}\ee^{-\lambda s/4}
 \le\ee^{-n/(5\ln k)},
\]
for sufficiently large $k,n$: here
$\ln(2en/s)\le\ln(2e)+2\ln\ln k\le\lambda/8$ and $\lambda\ge\ln k$.
\end{proof}

\begin{proof}[Proof of \pref{thm:shattering-below-one}]
Let $\Phi\sim\Phi(k,n,\lfloor\alpha n\rfloor)$, with $\alpha$ in
\eqref{eq:main-shattering-range}, and let $\Phi_0$ be its first $m_0$
clauses. Put $r_k=(\ln k)^{-2}$, and let $\epsilon,\kappa,\eta$ be the
constants from \pref{lem:sbo-energy-census}, decreasing $\eta$ if
necessary. By \pref{lem:sbo-formula-properties}, except with probability
$\ee^{-\Omega_k(n)}$, the incidence and numerical hypotheses of
\pref{lem:sbo-confinement} hold for $\Phi_0$ and $D$; in particular,
$N\ge2/\epsilon$ for sufficiently large $n$.

Let $\mathcal B_\Phi\subseteq\Omega_\Phi$ contain the solutions $x$ for
which either $\Phi_0$ has an assignment of energy at most $h$ whose
forced disagreement set is a witness relative to $\Phi_0,x$, or the
expansion property of
\pref{lem:sbo-planted-expansion} fails for $\Phi_0$ and $x$. Conditional
on any fixed $x$ satisfying $\Phi$, the clauses of $\Phi_0$ are
independent uniform clauses satisfied by $x$. De-Poissonizing
\pref{lem:sbo-energy-census} by conditioning its Poisson clause count to
equal $m_0$ costs at most $3\sqrt{m_0}$. Together with
\pref{lem:sbo-planted-expansion}, this bounds the conditional probability
that $x\in\mathcal B_\Phi$ by the two terms below. Applying the
one-assignment form of \pref{lem:norm-transfer} to the full formula gives
\[
 \E_\Phi[\mathbb{1}\{\mathcal Z\}\mu_\Phi(\mathcal B_\Phi)]
 \le L_Z\left(3\sqrt{m_0}\,\ee^{-\eta N}
                 +\ee^{-n/(5\ln k)}\right)
 \le\ee^{-\eta N/2}.
\]
Indeed, throughout \eqref{eq:main-shattering-range},
$\Gamma=O(k^3 2^{-k})=o(p)$, while
$N=\Theta(n/(k\ln k))$. By Markov's inequality, except with probability
$\ee^{-\eta N/4}$, on $\mathcal Z$ we have
$\mu_\Phi(\mathcal B_\Phi)\le\ee^{-\eta N/4}$.

Now fix $x\in\Omega_\Phi\setminus\mathcal B_\Phi$. Every path in
$\mathcal G_{D,h}(\Phi)$ is also a path in
$\mathcal G_{D,h}(\Phi_0)$, since $\Phi_0$ is a subformula of $\Phi$.
For any assignment $y$ in the component of $x$, compute $W_y$ and its
auxiliary core relative to $\Phi_0$ and $x$. Then
\pref{lem:sbo-confinement} gives $|(W_y)_{\mathrm{aux}}|<(6/7-\epsilon)N<N$. If
$A=\Delta(x,y)$ had size at least $r_kn$, the expansion property would
supply at least $N$ variables of $A$ forced under $x$ by clauses
containing two other variables of $A$. These variables lie in
$(W_y)_{\ge3}\subseteq(W_y)_{\mathrm{aux}}$, a contradiction. Thus the component
is contained in the Hamming ball of radius $r_kn$ about $x$.

On $\mathcal Z$, \pref{lem:norm-lower} gives
$Z_\Phi\ge\ee^{n/(2\ln k)}$, while the ball contains at most
$\exp(n\mathcal H(r_k))\le\ee^{n/(6\ln k)}$ assignments for large $k$.
Consequently, each component containing a solution outside
$\mathcal B_\Phi$ has solution mass at most $\ee^{-n/(3\ln k)}$; a
component whose solutions all lie in $\mathcal B_\Phi$ has mass at most
$\ee^{-\eta N/4}$. Since $N\ge n/(8k\ln k)$, the stated bound follows
with the chosen $\delta_k$ and
$\zeta_k=\min\{1,\eta/4\}/(8k\ln k)$. The failure probability of
$\mathcal Z$ is at most $1/(2n)$ by \pref{lem:norm-lower}, and all other
failure probabilities are exponentially small in $n$ for each fixed
$k$. Their sum is at most $1/n$ for sufficiently large $n$.
\end{proof}

\section*{AI methodology}\label{sec:ai-methodology}

During the preparation of this manuscript, the authors used ChatGPT
(5.6~Sol, 6~Sol, 6~Astra, 6.1~Sol, and 6.1~Terra),
Gemini~4~Argon, and Dots for substantive mathematical research
and manuscript preparation.

AI tools were used extensively to propose and develop algorithms
and proof arguments, explore sharper bounds, and identify gaps
in candidate proofs.
They materially contributed to the mathematical development
of the main results and also assisted with numerical checks,
literature comparisons, proof simplification, and manuscript
drafting and revision.

Through sustained interaction with the AI assistants, the authors
interpreted intermediate findings, developed conceptual insights,
and progressively formulated and refined the research questions.
They first pursued sharper single-variable connectivity bounds
(\pref{thm:main-theorem}), then investigated looseness and its
transition to rigidity
(\pref{thm:main-looseness}\ref{item:main-looseness-part}
and~\ref{item:main-rigidity-part}).
Subsequent investigations addressed the reconstruction threshold
governed by $\gamma_*$
(\pref{thm:main-reconstruction}),
local connectivity throughout the proved looseness range
(\pref{thm:main-looseness}\ref{item:main-local-connectivity-part}),
and shattering coexisting with local accessibility
(\pref{thm:main-shattering}).
These evolving questions shaped the research directions
and the interpretation of the results.

The authors assessed alternative approaches and reviewed
and revised the arguments incorporated into the manuscript.
They take full responsibility for the correctness, originality,
and integrity of the work, including its proofs and references.

\bibliographystyle{alpha}
\bibliography{references}

@inproceedings{AC08,
  author    = {Dimitris Achlioptas and Amin {Coja-Oghlan}},
  title     = {Algorithmic barriers from phase transitions},
  year      = {2008},
  booktitle = {Proceedings of the 49th Annual IEEE Symposium on Foundations of Computer Science},
  pages     = {793--802},
  doi       = {10.1109/FOCS.2008.11},
  url       = {https://arxiv.org/abs/0803.2122v2}
}

@incollection{Ach09,
  author    = {Dimitris Achlioptas},
  title     = {Random Satisfiability},
  booktitle = {Handbook of Satisfiability},
  editor    = {Armin Biere and Marijn Heule and Hans van Maaren and Toby Walsh},
  series    = {Frontiers in Artificial Intelligence and Applications},
  volume    = {185},
  chapter   = {8},
  publisher = {IOS Press},
  year      = {2009},
  pages     = {245--270},
  doi       = {10.3233/978-1-58603-929-5-245},
  url       = {https://ebooks.iospress.nl/volumearticle/4992}
}

@article{ACRT11,
  author  = {Dimitris Achlioptas and Amin {Coja-Oghlan} and Federico Ricci-Tersenghi},
  title   = {On the solution-space geometry of random constraint satisfaction problems},
  journal = {Random Structures \& Algorithms},
  volume  = {38},
  number  = {3},
  pages   = {251--268},
  year    = {2011},
  doi     = {10.1002/rsa.20323},
  url     = {https://chimera.roma1.infn.it/FEDERICO/Publications_files/2011_RSA_38_251.pdf}
}

@article{AP04,
  author  = {Dimitris Achlioptas and Yuval Peres},
  title   = {The threshold for random {$k$-SAT} is {$2^k\log 2-O(k)$}},
  year    = {2004},
  journal = {Journal of the American Mathematical Society},
  volume  = {17},
  number  = {4},
  pages   = {947--973},
  doi     = {10.1090/S0894-0347-04-00464-3}
}

@inproceedings{AR06,
  author    = {Dimitris Achlioptas and Federico Ricci-Tersenghi},
  title     = {On the solution-space geometry of random constraint satisfaction problems},
  booktitle = {Proceedings of the 38th Annual ACM Symposium on Theory of Computing},
  pages     = {130--139},
  year      = {2006},
  doi       = {10.1145/1132516.1132537},
  url       = {https://arxiv.org/abs/cs/0611052}
}

@inproceedings{BH22,
  author    = {Guy Bresler and Brice Huang},
  title     = {The algorithmic phase transition of random {$k$-SAT} for low degree polynomials},
  booktitle = {2021 IEEE 62nd Annual Symposium on Foundations of Computer Science (FOCS)},
  pages     = {298--309},
  publisher = {IEEE},
  year      = {2022},
  doi       = {10.1109/FOCS52979.2021.00038},
  url       = {https://arxiv.org/abs/2106.02129}
}

@article{BMZ05,
  author  = {Alfredo Braunstein and Marc M\'ezard and Riccardo Zecchina},
  title   = {Survey propagation: an algorithm for satisfiability},
  journal = {Random Structures \& Algorithms},
  volume  = {27},
  number  = {2},
  pages   = {201--226},
  year    = {2005},
  doi     = {10.1002/rsa.20057},
  url     = {https://arxiv.org/abs/cs/0212002}
}

@article{BS20,
  author  = {Louise Budzynski and Guilhem Semerjian},
  title   = {The Asymptotics of the Clustering Transition for Random Constraint Satisfaction Problems},
  journal = {Journal of Statistical Physics},
  volume  = {181},
  number  = {5},
  pages   = {1490--1522},
  year    = {2020},
  doi     = {10.1007/s10955-020-02635-8},
  url     = {https://doi.org/10.1007/s10955-020-02635-8}
}

@article{CCGPRZZ26,
  author  = {Arnab Chatterjee and Amin {Coja-Oghlan} and Catherine Greenhill and Vincent Pfenninger and Maurice Rolvien and Pavel Zakharov and Kostas Zampetakis},
  title   = {The random {$k$-SAT} {Gibbs} uniqueness threshold revisited},
  journal = {Combinatorics, Probability and Computing},
  volume  = {35},
  number  = {4},
  pages   = {433--485},
  year    = {2026},
  doi     = {10.1017/S0963548326100406},
  url     = {https://doi.org/10.1017/S0963548326100406}
}

@article{CGGGHMM24,
  author  = {Zongchen Chen and Andreas Galanis and Leslie Ann Goldberg and Heng Guo and Andr\'es Herrera-Poyatos and Nitya Mani and Ankur Moitra},
  title   = {Fast sampling of satisfying assignments from random {$k$-SAT} with applications to connectivity},
  year    = {2024},
  journal = {SIAM Journal on Discrete Mathematics},
  volume  = {38},
  number  = {4},
  pages   = {2750--2811},
  doi     = {10.1137/23M1595722},
  url     = {https://doi.org/10.1137/23M1595722}
}

@inproceedings{CLWYY25,
  author    = {Zongchen Chen and Aditya Lonkar and Chunyang Wang and Kuan Yang and Yitong Yin},
  title     = {Counting random {$k$-SAT} near the satisfiability threshold},
  year      = {2025},
  booktitle = {Proceedings of the 57th Annual ACM Symposium on Theory of Computing},
  pages     = {867--878},
  doi       = {10.1145/3717823.3718163},
  url       = {https://arxiv.org/abs/2411.02980v1},
  note      = {Theorem references follow the numbering in arXiv:2411.02980v1}
}

@inproceedings{CMM23,
  author    = {Zongchen Chen and Nitya Mani and Ankur Moitra},
  title     = {From algorithms to connectivity and back: finding a giant component in random {$k$-SAT}},
  year      = {2023},
  booktitle = {Proceedings of the 2023 Annual ACM-SIAM Symposium on Discrete Algorithms (SODA)},
  pages     = {3437--3470},
  doi       = {10.1137/1.9781611977554.ch132},
  url       = {https://arxiv.org/abs/2207.02841v2},
  note      = {Theorem and conjecture references follow the numbering in arXiv:2207.02841v2}
}

@article{CMR22,
  author  = {Amin {Coja-Oghlan} and No{\"e}la M{\"u}ller and Jean B. Ravelomanana},
  title   = {Belief propagation on the random {$k$-SAT} model},
  journal = {The Annals of Applied Probability},
  volume  = {32},
  number  = {5},
  pages   = {3718--3796},
  year    = {2022},
  doi     = {10.1214/21-AAP1772},
  url     = {https://doi.org/10.1214/21-AAP1772}
}

@article{CO10,
  author  = {Amin {Coja-Oghlan}},
  title   = {A better algorithm for random {$k$-SAT}},
  year    = {2010},
  journal = {SIAM Journal on Computing},
  volume  = {39},
  number  = {7},
  pages   = {2823--2864},
  doi     = {10.1137/09076516X}
}

@article{COP16,
  author  = {Amin {Coja-Oghlan} and Konstantinos Panagiotou},
  title   = {The asymptotic {$k$-SAT} threshold},
  journal = {Advances in Mathematics},
  volume  = {288},
  pages   = {985--1068},
  year    = {2016},
  doi     = {10.1016/j.aim.2015.11.007},
  url     = {https://arxiv.org/abs/1310.2728}
}

@article{COP18,
  author  = {Amin {Coja-Oghlan} and Will Perkins},
  title   = {Belief propagation on replica symmetric random factor graph models},
  journal = {Annales de l'Institut Henri Poincar{\'e} D, Combinatorics, Physics and their Interactions},
  volume  = {5},
  number  = {2},
  pages   = {211--249},
  year    = {2018},
  doi     = {10.4171/AIHPD/53},
  url     = {https://ems.press/journals/aihpd/articles/15519}
}

@article{CPP12,
  author  = {Amin {Coja-Oghlan} and Angelica Y. Pachon-Pinzon},
  title   = {The decimation process in random {$k$-SAT}},
  year    = {2012},
  journal = {SIAM Journal on Discrete Mathematics},
  volume  = {26},
  number  = {4},
  pages   = {1471--1509},
  doi     = {10.1137/110842867},
  url     = {https://doi.org/10.1137/110842867}
}

@article{DSS22,
  author  = {Jian Ding and Allan Sly and Nike Sun},
  title   = {Proof of the satisfiability conjecture for large {$k$}},
  journal = {Annals of Mathematics},
  volume  = {196},
  number  = {1},
  pages   = {1--388},
  year    = {2022},
  doi     = {10.4007/annals.2022.196.1.1},
  url     = {https://annals.math.princeton.edu/2022/196-1/p01}
}

@article{Fri99,
  author  = {Ehud Friedgut},
  title   = {Sharp thresholds of graph properties, and the {$k$-SAT} problem},
  journal = {Journal of the American Mathematical Society},
  volume  = {12},
  number  = {4},
  pages   = {1017--1054},
  year    = {1999},
  doi     = {10.1090/S0894-0347-99-00305-7}
}

@article{GKMP09,
  author  = {Parikshit Gopalan and Phokion G. Kolaitis and Elitza Maneva and Christos H. Papadimitriou},
  title   = {The connectivity of {Boolean} satisfiability: Computational and structural dichotomies},
  journal = {SIAM Journal on Computing},
  volume  = {38},
  number  = {6},
  pages   = {2330--2355},
  year    = {2009},
  doi     = {10.1137/07070440X},
  url     = {https://arxiv.org/abs/cs/0609072}
}

@inproceedings{GM07,
  author    = {Antoine Gerschenfeld and Andrea Montanari},
  title     = {Reconstruction for models on random graphs},
  booktitle = {Proceedings of the 48th Annual IEEE Symposium on Foundations of Computer Science},
  pages     = {194--204},
  year      = {2007},
  publisher = {IEEE},
  doi       = {10.1109/FOCS.2007.58},
  url       = {https://doi.org/10.1109/FOCS.2007.58}
}

@article{KMRTSZ07,
  author  = {Florent Krzakala and Andrea Montanari and Federico Ricci-Tersenghi and Guilhem Semerjian and Lenka Zdeborov\'a},
  title   = {Gibbs States and the Set of Solutions of Random Constraint Satisfaction Problems},
  journal = {Proceedings of the National Academy of Sciences},
  volume  = {104},
  number  = {25},
  pages   = {10318--10323},
  year    = {2007},
  doi     = {10.1073/pnas.0703685104},
  url     = {https://arxiv.org/abs/cond-mat/0612365v2}
}

@article{KS94,
  author  = {Scott Kirkpatrick and Bart Selman},
  title   = {Critical behavior in the satisfiability of random {Boolean} expressions},
  journal = {Science},
  volume  = {264},
  number  = {5163},
  pages   = {1297--1301},
  year    = {1994},
  doi     = {10.1126/science.264.5163.1297},
  url     = {https://www.cs.cornell.edu/selman/papers/pdf/science-94-kirkpatrick-selman.pdf}
}

@book{LP18,
  author    = {G{\"u}nter Last and Mathew Penrose},
  title     = {Lectures on the {Poisson} Process},
  publisher = {Cambridge University Press},
  year      = {2018},
  doi       = {10.1017/9781316104477},
  url       = {https://doi.org/10.1017/9781316104477}
}

@incollection{McD89,
  author    = {Colin McDiarmid},
  title     = {On the method of bounded differences},
  booktitle = {Surveys in Combinatorics, 1989},
  editor    = {Johannes Siemons},
  series    = {London Mathematical Society Lecture Note Series},
  volume    = {141},
  pages     = {148--188},
  publisher = {Cambridge University Press},
  year      = {1989},
  doi       = {10.1017/CBO9781107359949.008}
}

@article{MM06,
  author  = {Marc M\'ezard and Andrea Montanari},
  title   = {Reconstruction on trees and spin glass transition},
  journal = {Journal of Statistical Physics},
  volume  = {124},
  number  = {6},
  pages   = {1317--1350},
  year    = {2006},
  doi     = {10.1007/s10955-006-9162-3},
  url     = {https://arxiv.org/abs/cond-mat/0512295}
}

@article{MMW07,
  author    = {Elitza Maneva and Elchanan Mossel and Martin J. Wainwright},
  title     = {A new look at survey propagation and its generalizations},
  journal   = {Journal of the ACM},
  volume    = {54},
  number    = {4},
  pages     = {17},
  year      = {2007},
  articleno = {17},
  numpages  = {41},
  doi       = {10.1145/1255443.1255445},
  url       = {https://arxiv.org/abs/cs/0409012}
}

@article{MMZ05,
  author  = {Marc M\'ezard and Thierry Mora and Riccardo Zecchina},
  title   = {Clustering of Solutions in the Random Satisfiability Problem},
  journal = {Physical Review Letters},
  volume  = {94},
  number  = {19},
  pages   = {197205},
  year    = {2005},
  doi     = {10.1103/PhysRevLett.94.197205},
  url     = {https://arxiv.org/abs/cond-mat/0504070}
}

@article{MMZ06,
  author  = {Stephan Mertens and Marc M\'ezard and Riccardo Zecchina},
  title   = {Threshold values of random {$K$-SAT} from the cavity method},
  journal = {Random Structures \& Algorithms},
  volume  = {28},
  number  = {3},
  pages   = {340--373},
  year    = {2006},
  doi     = {10.1002/rsa.20090},
  url     = {https://arxiv.org/abs/cs/0309020}
}

@article{Mol05,
  author  = {Michael Molloy},
  title   = {Cores in random hypergraphs and {Boolean} formulas},
  journal = {Random Structures \& Algorithms},
  volume  = {27},
  number  = {1},
  pages   = {124--135},
  year    = {2005},
  doi     = {10.1002/rsa.20061}
}

@article{Mon08,
  author  = {Andrea Montanari},
  title   = {Estimating random variables from random sparse observations},
  journal = {European Transactions on Telecommunications},
  volume  = {19},
  number  = {4},
  pages   = {385--403},
  year    = {2008},
  doi     = {10.1002/ett.1289},
  url     = {https://doi.org/10.1002/ett.1289}
}

@article{MPZ02,
  author  = {Marc M\'ezard and Giorgio Parisi and Riccardo Zecchina},
  title   = {Analytic and Algorithmic Solution of Random Satisfiability Problems},
  journal = {Science},
  volume  = {297},
  number  = {5582},
  pages   = {812--815},
  year    = {2002},
  doi     = {10.1126/science.1073287}
}

@inproceedings{MR12,
  author    = {Michael Molloy and Ricardo Restrepo},
  title     = {Frozen variables in random {Boolean} constraint satisfaction problems},
  booktitle = {Proceedings of the Twenty-Fourth Annual ACM-SIAM Symposium on Discrete Algorithms},
  year      = {2013},
  pages     = {1306--1318},
  doi       = {10.1137/1.9781611973105.95},
  url       = {https://doi.org/10.1137/1.9781611973105.95}
}

@article{MRT11,
  author  = {Andrea Montanari and Ricardo Restrepo and Prasad Tetali},
  title   = {Reconstruction and clustering in random constraint satisfaction problems},
  journal = {SIAM Journal on Discrete Mathematics},
  volume  = {25},
  number  = {2},
  pages   = {771--808},
  year    = {2011},
  doi     = {10.1137/090755862},
  url     = {https://doi.org/10.1137/090755862}
}

@article{MRTS08,
  author  = {Andrea Montanari and Federico Ricci-Tersenghi and Guilhem Semerjian},
  title   = {Clusters of solutions and replica symmetry breaking in random {$k$}-satisfiability},
  journal = {Journal of Statistical Mechanics: Theory and Experiment},
  volume  = {2008},
  number  = {04},
  pages   = {P04004},
  year    = {2008},
  doi     = {10.1088/1742-5468/2008/04/P04004},
  url     = {https://arxiv.org/abs/0802.3627v2}
}

@inproceedings{MSL92,
  author    = {David Mitchell and Bart Selman and Hector J. Levesque},
  title     = {Hard and Easy Distributions of {SAT} Problems},
  booktitle = {Proceedings of the Tenth National Conference on Artificial Intelligence},
  year      = {1992},
  pages     = {459--465},
  url       = {https://cdn.aaai.org/AAAI/1992/AAAI92-071.pdf}
}

@article{MZ02,
  author  = {Marc M\'ezard and Riccardo Zecchina},
  title   = {Random {$K$}-satisfiability problem:
             From an analytic solution to an efficient algorithm},
  journal = {Physical Review E},
  volume  = {66},
  number  = {5},
  pages   = {056126},
  year    = {2002},
  doi     = {10.1103/PhysRevE.66.056126},
  url     = {https://arxiv.org/abs/cond-mat/0207194}
}

@article{MZ96,
  author  = {R\'emi Monasson and Riccardo Zecchina},
  title   = {Entropy of the {$K$}-Satisfiability Problem},
  journal = {Physical Review Letters},
  volume  = {76},
  number  = {21},
  pages   = {3881--3885},
  year    = {1996},
  doi     = {10.1103/PhysRevLett.76.3881}
}

@article{Sem08,
  author  = {Guilhem Semerjian},
  title   = {On the freezing of variables in random constraint satisfaction problems},
  year    = {2008},
  journal = {Journal of Statistical Physics},
  volume  = {130},
  number  = {2},
  pages   = {251--293},
  doi     = {10.1007/s10955-007-9417-7},
  url     = {https://arxiv.org/abs/0705.2147}
}

@article{sly2009reconstruction,
  author  = {Allan Sly},
  title   = {Reconstruction of Random Colourings},
  journal = {Communications in Mathematical Physics},
  volume  = {288},
  number  = {3},
  pages   = {943--961},
  year    = {2009},
  doi     = {10.1007/s00220-009-0783-7},
  url     = {https://arxiv.org/abs/0802.3487}
}

@misc{sly2016reconstruction,
  title  = {Reconstruction of colourings without freezing},
  author = {Allan Sly and Yumeng Zhang},
  year   = {2016},
  doi    = {10.48550/arXiv.1610.02770},
  howpublished = {arXiv preprint arXiv:1610.02770},
  url    = {https://arxiv.org/abs/1610.02770}
}

@article{Zhou10,
  author  = {Haijun Zhou},
  title   = {Solution space heterogeneity of the random
             {$K$}-satisfiability problem: Theory and simulations},
  journal = {Journal of Physics: Conference Series},
  volume  = {233},
  number  = {1},
  pages   = {012011},
  year    = {2010},
  doi     = {10.1088/1742-6596/233/1/012011},
  url     = {https://arxiv.org/abs/1001.2951}
}

@article{Haj88,
  author  = {Bruce Hajek},
  title   = {Cooling Schedules for Optimal Annealing},
  journal = {Mathematics of Operations Research},
  volume  = {13},
  number  = {2},
  pages   = {311--329},
  year    = {1988},
  doi     = {10.1287/moor.13.2.311}
}

@book{BDH16,
  author    = {Anton Bovier and Frank {den Hollander}},
  title     = {Metastability: A Potential-Theoretic Approach},
  series    = {Grundlehren der mathematischen Wissenschaften},
  volume    = {351},
  publisher = {Springer},
  year      = {2016},
  doi       = {10.1007/978-3-319-24777-9}
}

\appendix
\section{Solution-count estimates}\label{app:normalization}
The lower bound uses the weighted second-moment approach of
Achlioptas--Peres~\cite{AP04}; related solution-count estimates appear
in~\cite{ACRT11} and~\cite[Corollary~14]{CPP12}.
We include direct proofs for completeness and to obtain the explicit
failure bound $1/(2n)$ in \pref{lem:norm-lower}.
The pair-transfer bound follows an unweighted overlap calculation.
We use $\epsilon_0,M,\Gamma,L_Z$ from \pref{sec:normalization}.

\begin{proof}[Proof of \pref{lem:norm-lower}]
We choose clause weights whose first three nonconstant Fourier levels
vanish. For truth signs $s\in\{-1,1\}^k$, write
$s_A=\prod_{i\in A}s_i$, $S=\sum_{j=1}^3\binom kj$, and
$\bar w=1-\epsilon_0(1+S)$. Define
\begin{equation}\label{eq:norm-weight}
 w(s)=1-\epsilon_0\prod_{i=1}^k(1-s_i)
      -\epsilon_0\sum_{1\le|A|\le3}
                    \left(1+(-1)^{|A|+1}s_A\right).
\end{equation}
Each summand in the last sum is zero or two and vanishes on the
all-false vector. Thus $w=0$ on unsatisfied clauses and
$0<1-2\epsilon_0S\le w\le1$ on satisfied clauses for large $k$.
The Fourier expansion
\[
 w(s)=\bar w+\epsilon_0\sum_{|A|\ge4}(-1)^{|A|+1}s_A
\]
gives $\E[w]=\bar w$ and $\E[w^2]=\bar w^2+\epsilon_0\bar w$
under independent fair truth signs.

Let $W,V$ sum the products of clause weights and their squares,
respectively, over all assignments. Only solutions contribute, so
$W^2\le Z_\Phi V$ by Cauchy--Schwarz. Clause independence gives
\begin{equation}\label{eq:norm-weighted-ratio}
 \frac{(\E[W])^2}{\E[V]}
 =2^n\left(\frac{\bar w}{\bar w+\epsilon_0}\right)^m
 =M\left(1+\frac{\epsilon_0^2S}{\bar w}\right)^{-m}
 \ge M\ee^{-\Gamma n},
\end{equation}
using $\bar w\ge1/2$ and $2S=(k^3+5k)/3\le k^3$.
It suffices to show $\operatorname{Var}[W]/(\E[W])^2<1/(10n)$:
Chebyshev and Markov then give
$W\ge\E[W]/2$ and $V\le10n\E[V]$ simultaneously with probability
at least $1-1/(2n)$, implying $Z_\Phi\ge M/L_Z$.

\smallskip\noindent\emph{Overlap representation.}
For two independent uniform assignments, let $R$ be the fraction of
agreements minus the fraction of disagreements, and let $H_j(R)$ be
the mean product of $j$ agreement signs sampled without replacement.
Averaging literal signs cancels distinct Fourier monomials;
matching degree-$j$ monomials give $H_j(R)$. Hence
\begin{equation}\label{eq:norm-overlap-identity}
 \frac{\E[W^2]}{(\E[W])^2}=\E[(1+f(R))^m],\qquad
 f(R)=\frac{\epsilon_0^2}{\bar w^2}
                 \sum_{j=4}^k\binom kj H_j(R).
\end{equation}
Set $y=m|f(R)|$. Since $1+f(R)\ge0$, the relative variance is
at most $\E[\ee^y-1]$.

\smallskip\noindent\emph{Central overlaps.}
For $|R|\le1/2$ and $n\ge32k$, put $X=\sqrt nR$ and
$t=|R|+\sqrt{2k/n}\le3/4$. Sampling without replacement gives
$H_0=1$, $H_1=R$, and
$(n-j)H_{j+1}=nRH_j-jH_{j-1}$. Since
\[
 (n-j)t^2-n|R|t-j
 =\sqrt{2kn}\,t-j(1+t^2)\ge2(k-j)\ge0,
\]
induction yields
\begin{equation}\label{eq:norm-hypergeom-bounds}
 |H_j(R)|\le t^j\qquad(0\le j\le k).
\end{equation}
Using $\alpha\epsilon_0\le1$ and $\bar w\ge1/2$, we obtain
\[
 y\le4n\epsilon_0t^4\sum_{j=4}^k\binom kj(3/4)^{j-4}
 \le16n(7/8)^kt^4\le\frac{a_k(X^4+1)}n,
 \qquad a_k=512k^2(7/8)^k.
\]
Pairing indices in the normalized sum $X$ of fair signs gives
$\E[X^{2r}]\le(2r-1)!!\le2^rr!$, hence
$\E[(X^4+1)\ee^{X^2/4}]\le66$ by expansion of the exponential.
For large $k$, $a_k\le1$ and $y\le X^2/4+a_k/n$ on this range.
Thus $\ee^y-1\le y\ee^y$ gives
\begin{equation}\label{eq:norm-central}
 \E[(\ee^y-1)\mathbb{1}\{|R|\le1/2\}]
 \le66a_k\ee^{a_k/n}/n<1/(20n).
\end{equation}

\smallskip\noindent\emph{Noncentral overlaps.}
For $r=|R|>1/2$, put $t=(1+r)/2$. Counting samples whose agreement
signs are all positive gives the exact identity
\[
 0\le\sum_{j=0}^k\binom kjH_j(R)
 =2^k\frac{\binom{n(1+R)/2}{k}}{\binom nk}
 \le2^kt^k.
\]
Since $|H_j(R)|\le1$ and $t>3/4$, it follows that
\[
 y\le\frac{\alpha\epsilon_0}{\bar w^2}\,nt^k
       \bigl[1+(1+S)(2/3)^k\bigr]
 \le cnt^k,\qquad c=\ln2-\frac3{4\ln k},
\]
where the last inequality uses the density bound and
$\bar w=1-O(k^3\epsilon_0)$.
Each value of $R$ has probability at most $\ee^{-nI(r)}$, where
$I(r)=\ln2-\mathcal H((1-r)/2)$ and
$\mathcal H(q)=-q\ln q-(1-q)\ln(1-q)$.
For large $k$,
\[
 I(r)-ct^k\ge\frac1{2\ln k}.
\]
Indeed, if $r\le1-2\ln k/k$, then $t^k\le1/k$ and
$I(r)\ge r^2/2>1/8$. Otherwise,
$\mathcal H((1-r)/2)\le\mathcal H(\ln k/k)\le1/(4\ln k)$.
Summing over the $n+1$ possible overlaps bounds this contribution
by $(n+1)\ee^{-n/(2\ln k)}<1/(20n)$ for large $n$.
Together with \eqref{eq:norm-central}, this proves the variance bound.

Finally, the density hypothesis gives
\[
 \frac1n\ln\frac M{L_Z}
 \ge\frac1{\ln k}-O(k^3\epsilon_0)-\frac{\ln(40n)}n
 \ge\frac1{2\ln k}
\]
for large $k$ and $n$, proving the exponential lower bound.
\end{proof}

\begin{proof}[Pair bound in \pref{lem:norm-transfer}]
Let $Q\sim\operatorname{Bin}(n,1/2)/n$ be the relative distance
between two independent uniform assignments. A clause is false at
both exactly when its scope lies in their agreement set and every
literal is false there. Thus its joint satisfaction probability is
\[
 s_q=1-2\epsilon_0+
       \epsilon_0\frac{\binom{n-qn}{k}}{\binom nk}
 \le1-2\epsilon_0+\epsilon_0(1-q)^k,
\]
where $\binom ak=0$ for integers $a<k$. Hence
$s_q/(1-\epsilon_0)^2\le\exp(2\epsilon_0(1-q)^k)$.
Clause independence and Hoeffding's bound
$\Pr(Q<1/3)\le\ee^{-n/18}$ give
\[
 \frac{\E[Z_\Phi^2]}{M^2}
 =\E\left[\left(\frac{s_Q}{(1-\epsilon_0)^2}\right)^m\right]
 \le\ee^{2m\epsilon_0(2/3)^k}
      +\ee^{(2\alpha\epsilon_0-1/18)n}.
\]
Since $2\alpha\epsilon_0\le1/256<1/18$, the right-hand side is
at most $2\exp(2\alpha\epsilon_0(2/3)^kn)$, as required.
\end{proof}

\end{document}